\documentclass[aps,nofootinbib,floatfix]{revtex4}
\usepackage{bm}
\usepackage{graphicx,tabularx}
\usepackage{latexsym,color}
\usepackage{physics}
\usepackage{centernot}
\usepackage{relsize}
\usepackage{mathtools}
\usepackage{stmaryrd}
\usepackage{rotating}
\usepackage{amssymb}
\usepackage[T1]{fontenc}
\usepackage[latin1]{inputenc}
\usepackage{natbib}
\usepackage{amsmath,amssymb,amsthm,mathrsfs,amsfonts,dsfont}
\usepackage{color}
\usepackage[hyperfootnotes=true]{hyperref}

\usepackage{enumitem}

\begin{document}
\title{Kerr metric bundles.\\\emph{Killing horizons confinement,  {light-surfaces} and horizons replicas}}
\author{Daniela  Pugliese and Hernando Quevedo}
\email{d.pugliese.physics@gmail.com}
\affiliation{\vspace{3mm}
Institute of Physics, Faculty of Philosophy \& Science,
  Silesian University in Opava,
 Bezru\v{c}ovo n\'{a}m\v{e}st\'{i} 13, CZ-74601 Opava, Czech Republic \\
          Dipartimento di Fisica, Universit\`a di Roma ``La Sapienza", I-00185 Roma, Italy \\
					Instituto de Ciencias Nucleares, Universidad Nacional Aut\'onoma de M\'exico,  AP 70543, M\'exico, DF 04510, Mexico \\
					Department of Theoretical and Nuclear Physics,
Kazakh National University,
Almaty 050040, Kazakhstan
}
\date{\today}
\begin{abstract}
We provide a complete  characterization of the  metric Killing  bundles (or  metric bundles) of {the} Kerr geometry.  Metric bundles, first introduced in \cite{remnants}  can be  generally defined for axially symmetric {spacetimes} with Killing horizons and, for the case of Kerr geometries,   are sets of black holes (\textbf{BHs}) or black holes   and naked singularities (\textbf{NSs}) geometries. Each metric of a bundle has an  equal limiting photon  (orbital) frequency, which defines the bundle {and coincides} with  the frequency of a Killing
{horizon}   in the extended plane. In this plane each bundle  is represented as a curve  tangent to the curve that represents the horizons, which  thus emerge as the envelope surfaces of the metric bundles.
   We show that the horizons frequency can be used to establish a connection between \textbf{BHs} and \textbf{NSs}, providing an alternative representation of such spacetimes in the extended plane and an alternative definition of the \textbf{BH} horizons.
  We introduce the concept of inner horizon confinement and horizons replicas and study the possibility of detecting their frequencies.
We study the bundle characteristic frequencies  constraining the  inner horizon confinement in  the outer region of the plane i.e. the possibility of detect  frequency related to  the inner horizon,  and the horizons replicas, structures which may be  detectable,  for example, from the  emission spectra of  \textbf{BHs} spacetimes.
It is shown that such observations can be performed close to the rotation axis of the Kerr geometry, depending on   the  \textbf{BH} spin.
We argue that these results could be used to further investigate black holes and their thermodynamic properties.
\end{abstract}

\maketitle

\date{\today}
\def\be{\begin{equation}}
\def\ee{\end{equation}}
\def\bea{\begin{eqnarray}}
\def\eea{\end{eqnarray}}
\newcommand{\bt}[1]{\mathbf{\mathtt{#1}}}
\newcommand{\tb}[1]{\textbf{{#1}}}
\newcommand{\rtb}[1]{\textcolor[rgb]{1.00,0.00,0.00}{\tb{#1}}}
\newcommand{\btb}[1]{\textcolor[rgb]{0.00,0.00,1.00}{\tb{#1}}}
\newcommand{\otb}[1]{\textcolor[rgb]{1.00,0.50,0.00}{\tb{#1}}}
\newcommand{\gtb}[1]{\textcolor[rgb]{0.00,.50,0.00}{\tb{#1}}}
\newcommand{\ptb}[1]{\textcolor[rgb]{0.70,0.00,0.70}{\tb{#1}}}
\newcommand{\il}{~}
\newcommand{\Qa}{\mathcal{Q}}
\newcommand{\la}{\mathcal{A}}
\newcommand{\laa}{\mathcal{L}}

\section{Introduction}\label{Sec:impe-null-i}
In this work, we present the   general  analysis of the metric Killing bundles (or metric bundles \textbf{MBs}) of  the Kerr geometry.
The definition of metric bundles was first introduced in  \cite{remnants}  for the Kerr geometries, framed   in the analysis of the
 Kerr black holes (\textbf{BHs}) and naked singularities (\textbf{NSs}) properties.  \textbf{MBs} can be  generally defined in spacetimes with Killing horizons.

{The idea is to bundle geometries according to some particular characteristic common to all the geometries of the bundles, which allows us to explore the properties of the bundled metrics from a special perspective. Bundles enlighten properties attributable to different points of spacetime and connect different geometries including,  for example, \textbf{BHs} and \textbf{NSs}.
These properties can be  measured through the observation of light-like  radii  and the analysis of the light surfaces implicated in many astrophysical phenomena such as \textbf{BH} shadows, accretion disks and  magnetospheres.
To this end, we introduce the concept of
extended plane which, in brief,  can be defined as a  graphic representing a collection of metrics related by a  common property.
  We specify below the details  of these definitions.}

A metric Killing  bundle of the Kerr geometry  is a collection of Kerr  spacetimes characterized by  a  particular frequency defined as  the photon (circular) orbital frequency, $\omega$, at which the four-velocity norm of a particular stationary observer vanishes.
It  is straightforward to prove that $\omega$ is also the frequency (angular velocity) of a particular \textbf{BH} horizon.
{A metric bundle  is
 represented by a curve in the  \emph{extended plane}, i.e.,  a plane $\mathcal{P}-r$, where $\mathcal{P}$ is a parameter of the Kerr spacetime and $r$ is the radial Boyer-Lindquist (BL) coordinate. Thus, an extended plane represents all the metrics of the Kerr family so that varying the parameter $\mathcal{P}$, we can extend a particular analysis to include all Kerr metrics.

Thus, the metrics of one metric Killing bundle with characteristic frequency $\bar{\omega}$ are all and the only Kerr  (\textbf{BHs} or \textbf{NSs}) spacetimes, where at some point $r$ the  limiting light-like orbital frequency is $\omega=\bar{\omega}$. In  the extended plane, all the  curves associated to the \textbf{MBs} (bundle curves) are tangent to the horizons curve (the  curve  representing   the  Killing horizons of all Kerr \textbf{BHs}). Then,  this tangency condition implies that  each bundle characteristic frequency $\bar{\omega}$ coincides with the frequency  $\omega_H$ of a Killing horizon. Consequently, the horizons in the extended plane emerge as the envelope surface of the collection of all the metric bundles.

The metric bundles of the Kerr spacetimes  contain either  \textbf{BHs} or \textbf{BHs} \emph{and}  \textbf{NSs}. Therefore, it is possible to find a  \textbf{BH-NS}  correspondence by using the fact that all bundles are tangent to the horizon. Moreover, the metric bundles analysis  provides also  an alternative  interpretation of \textbf{NSs} and  \textbf{BHs} horizons in the extended plane.

In fact, the exploration of \textbf{MBs} as   metric  structures   singles out some fundamental properties of  the \textbf{BHs} and  \textbf{NSs} solutions, which are related, in particular, to the thermodynamic properties of \textbf{BH} spacetimes and to the possibility to extract information about the \textbf{BH} horizons, i.e., to  detect properties which are directly attributable  to  the presence  of the  \textbf{BH} horizons. Particularly, through the study of the  Kerr metric bundles, we  define     the   "horizons replicas"  that   could  be detected,  for example, from the spectra of electromagnetic emissions coming from \textbf{BHs} and, in particular, from locations close to   the \textbf{BH}  rotational axis.
The horizon replicas are special orbits  of a Kerr spacetime  with limiting photon frequency equal to the \textbf{BH} (inner or outer) horizon frequency, which coincides with  the bundle characteristic frequency in the corresponding   point on the extended plane.
The representation of the \textbf{BH} solutions in the extended plane, as in Figs\il(\ref{Fig:Ly-b-rty}), can be used to highlight some properties of the \textbf{BH} horizons that could have a significant  impact  on the study of \textbf{BH} physics, on the interpretation of \textbf{NS} solutions, and on the investigation of \textbf{BH}  thermodynamics.


 Specifically, in the case of the
Kerr geometry on the equatorial plane considered in \cite{remnants}, it  turned out that weak  naked singularities (\textbf{WNSs}), for which the spin-mass ratio $a=J/M$ is close to the value of the extreme \textbf{BH}, are related to a portion of the inner horizon,  whereas strong naked singularities\footnote{Our definition of weak and strong naked singularities is
 related exclusively to the value of the spin parameter and has been
explored in several works \cite{Pugliese:2010he,Pu:Kerr,Pu:Neutral,Pu:Charged,
Pu:KN,ergon,Pu:class,observers}.
The \textbf{MBs} characteristics could provide also a different definition not exclusively dependent on the source spin.
We point out also that \textbf{NSs} can be  defined  differently as strong curvature singularities \cite{strong}.}
\textbf(\textbf{SNSs}) with $a>2M$ are related  to the outer  horizon.
	In addition, \textbf{WNSs} are characterized by the presence of Killing bottlenecks, which are defined as ``restrictions'' of the Killing throats  appearing in \textbf{WNSs}.
	Killing throats (tunnels) emerge   through the analysis of the radii
	of light surfaces (related to the \textbf{MBs} definition),
	which are functions of  the   spin parameter $a$ and the stationary observers frequency $\omega$ \cite{observers,remnants}.
Moreover,   Killing bottlenecks, interpreted
 as  ``horizons remnants'' in \cite{remnants} and related to metric bundles in
\cite{remnants,termo-BH}, are also connected with the concept of  pre-horizon regime introduced in
\cite{de-Felice1-frirdtforstati,de-Felice-first-Kerr}.
 The pre-horizon regime was analyzed in \cite{de-Felice-first-Kerr}.
It was concluded
that a gyroscope would conserve a memory of the static or
stationary initial state, leading to the gravitational
collapse of a mass distribution \cite{de-Felice3,de-Felice-mass,de-Felice4-overspinning,Chakraborty:2016mhx}.
Killing throats and bottlenecks    were also grouped in \cite{Tanatarov:2016mcs}  in   structures named ``whale  diagrams'' of the Kerr and Kerr-Newman spacetimes--see also
\cite{Mukherjee:2018cbu,Zaslavskii:2018kix,Zaslavskii:2019pdc}. For an analysis of the self-force corrections to gyroscope precession in the Kerr spacetime see \cite{Bini:2019lcd,Bini:2018ylh,Bini:2019lkm}.
In \textbf{NS} geometries,  a  Killing  throat
is a connected and bounded region in the $r-\omega$ plane,
containing  all the stationary observers allowed within
	two limiting   frequencies  $]\omega_-, \omega_+[$. On the other hand, in the case of
	\textbf{BHs}, a Killing  throat is either a disconnected  region
	or a region  bounded by  singular  surfaces in the  extreme Kerr \textbf{BH} spacetime.
	The \textbf{BH} extreme Kerr spacetime, therefore,  represents the  limiting case of the  Killing bottleneck   (as defined  in the BL frame), where the tunnel narrowing  closes on the \textbf{BH} horizon.
Metric bundles  are connected with the Killing bottleneck definition and therefore with horizons remnants.
In \cite{remnants}, we performed the  \textbf{MBs} analysis corresponding to the equatorial plane of the  Kerr, Reissner-Nordstr\"om and Kerr-Newman geometries.
In this work we address
also the off-equatorial case of  the Kerr spacetime.

Metric bundles are a  relatively new concept  and in this article we present   the general  analysis for the Kerr geometries.  We focus  particularly on  the \textbf{MBs} characteristics that can have an impact on  the  observational properties  associated to  \textbf{BH} formation and evolution. Thus,  below  we precise the  \textbf{MBs} definitions  for Kerr spacetimes and relate them explicitly to quantities of importance in \textbf{BH}  thermodynamics.
We then  discuss the relations between \textbf{MBs} and stationary observes and light surfaces, which are used to constrain  many processes associated to the physics of jet emission, accretion disks and  energy extraction from \textbf{BHs}.
We conclude this introduction with the article plan.

\textbf{The Kerr geometry and metric bundles}

The Kerr geometry in BL coordinates is described by the line element
\bea
\nonumber
&& ds^2=-\frac{\Delta-a^2 \sin ^2\theta}{\rho^2}dt^2+\frac{\rho^2}{\Delta}dr^2+\rho^2
d\theta^2+\frac{\sin^2\theta\left(\left(a^2+r^2\right)^2-a^2 \Delta \sin^2\theta\right)}{\rho^2}d\phi^2\\\label{alai}
&&-2\frac{a M \sin^2\theta \left(a^2-\Delta+r^2\right)}{\rho^2}d\phi dt\ ,
\\
&&
\Delta\equiv r^2-2Mr+a^2,\quad\mbox{and}\quad\rho^2\equiv r^2+a^2\cos^2\theta \ .
\eea
Alternately, it is convenient to write the line element (\ref{alai}) as follows
\bea\label{Eq:metric-1covector}
\dd s^2=-\alpha^2 \dd t^2+\frac{A \sigma}{\rho^2} (\dd \phi- \omega_z \dd t)^2+\frac{\rho^2}{\Delta}\dd r^2+\rho^2 \dd\theta^2,\quad \sigma\equiv\sin^2\theta,\quad A\equiv (r^2+a^2)^2-a^2 \Delta \sigma
\eea
where
$\alpha=\sqrt{(\Delta \rho^2/A)}$  and $\omega_z=2 a M r/A$ are the lapse function and the frequency of the zero angular momentum fiducial observer (\textbf{ZAMOS}) \cite{observers}, whose four velocity is $u^a=(1/\alpha,0,0,\omega_z/\alpha)$ orthogonal  to the surface   of constant $t$.
 This vacuum exact solution of the Einstein equations  describes an axisymmetric,   stationary, asymptotically flat spacetime, where
the parameter $M\geq0$  is      interpreted as  the mass  of the gravitational source.  The rotational parameter associated to  the central singularity is the spin (the  {specific} angular momentum) $a\equiv J/M $, while   $J$ is the
{total} angular momentum {(here related to the total ADM mass, while the product $aM$ is the total ADM
angular momentum;
for a  review on stationary  black holes see, for example, \cite{Chrusciel})}. For  $a=0$, the metric  (\ref{alai}) describes   the limiting static and spherically symmetric  Schwarzschild  geometry. In this work, we also consider this special case, which corresponds to  the "zeros"  of the Kerr \textbf{MBs} in the extended plane.

\textbf{Killing horizons, metric bundles and characteristic frequencies.}

The horizons and the inner and  outer  static limits for the {Kerr} geometry  are,
$
r_{\mp}=M\mp\sqrt{M^2-a^2}$ and $ r_{\epsilon}^{\mp}=M\mp\sqrt{M^2-a^2 \cos ^2\theta}\ ,
$ respectively.
The event horizons  of a spinning \textbf{BH}  are   Killing horizons   with respect to  the Killing field
$\mathcal{L}_H=\partial_t +\omega_H^{\pm} \partial_{\phi}$, where  $\omega_H^{\pm}$ is the angular velocity {(frequency)} of the horizons  representing   the \textbf{BH} rigid rotation\footnote{The event horizon of a stationary asymptotically flat solution with matter satisfying suitable hyperbolic equations  is a Killing horizon. The strong
rigidity theorem connects the event horizon with a Killing
horizon.}.
The vectors  $\xi_{t}=\partial_{t} $  and
$\xi_{\phi}=\partial_{\phi} $  are the stationary
  and axisymmetric  {Killing} fields,  respectively.
In the limiting case of spherically symmetric, static  spacetimes,
 the event horizons are  Killing horizons with {respect} to the  Killing vector
$\partial_t$ and  the
event, apparent, and Killing horizons  with respect to the  Killing field   $\xi_t$ coincide.
The results we discuss in this work follow from the   investigation of the properties of the Killing vector $\mathcal{L}=\partial_t +\omega \partial_{\phi}$.
The quantity  $\mathbf{\mathcal{L_N}}\equiv\mathcal{L}\cdot\mathbf{\mathcal{L}}$ becomes null  for photon-like
particles with rotational frequencies $\omega_{\pm}$.
Metric bundles correspond to the solutions of the condition
$\mathbf{\mathcal{L_N}}=0$.
The quantity $\omega$ will be called the characteristic \textbf{MB} frequency.  The vector $\mathcal{L}$, the frequency $\omega$, and the  limits
($\mathcal{L}_H,\omega_H^{\pm}$),  enter the definition of \textbf{BHs} horizons,  establishing relations between black holes, extreme black holes and their thermodynamic properties.
Therefore, the  Killing vector
$
\mathcal{L}_{\pm}\equiv \xi_{t}+\omega_{\pm}\xi_{\phi}
$
can be interpreted as generator of  null curves ($g_{\alpha\beta}\mathcal{L}^\alpha_{\pm}\mathcal{L}^\beta_{\pm}=0$)
as the Killing vectors $\mathcal{L}_{\pm} $ are also generators of Killing event  horizons.
The Kerr  horizons are,  therefore,  {null} (lightlike) hypersurfaces generated by the flow of a Killing vector,
whose {null} generators coincide with the orbits of an
one-parameter group of isometries, i.e., in general,    there exists a Killing field $\mathcal{L}$, which is normal to the null surface.

Notably, many quantities considered in this analysis are conformal invariants of the metric
and inherit some of the properties of the Killing vector $\mathcal{L}$, which  identifies a  Killing throat up to a conformal transformation.
The simplest case is when one considers a conformal expanded (or contracted) spacetime  where $\tilde{\mathbf{\xi}}^2\equiv
\tilde{\mathbf{g}}(L,L)=\Xi^2\mathbf{g}(L,L)$.
This holds also   for a "conformal expanded"  Killing tensor\footnote{{Concerning the  {conformal properties} of the \textbf{MBs}, we notice that the  \textbf{MBs} definition is invariant under  conformal transformations of the metric and  of the Killing vectors: $\bar{\laa}=\Xi_{\bullet}(\xi_t+\omega(r_\bullet)\xi_\phi)=\Xi_{\bullet}\laa$, where $r_{\bullet}$ is a set of variables that  do not contain  $t$ or $\phi$,  $\Xi_{\bullet}$  is non-null or null in a finite number of points. Concerning the relation with the spherically symmetric (and static) case we stress here that for each \textbf{BH} spacetime the hypersuperfice $S_+$  defined by $r=r_+$ is a  smooth null hypersurface. The null
generator $\mathcal{L}_+ $ of $S_+$  is the limit  of $\laa$ on the horizon ($r_+$),  the frequency is that of the horizon, with an abuse of notation here we consider$\laa=\laa_+$.
In the Kerr spacetime, on the other hand, the Killing vector $\xi_t$ is timelike only
outside the hypersurface $r_{\epsilon}^+$ on which it becomes null.
In the region $]r_+,r_{\epsilon}^+[$ , outer  ergoregion, $\xi_t$ is spacelike-- these hypersurfaces are represented in Figs\il(\ref{FIG:toa11},\ref{Fig:Ly-b-rty},\ref{FIG:disciotto1}). The vector $\xi_t$ is also  spacelike on and tangent to $S_+$ (except
on the  rotation axis where $\xi_t$ is again null).
Thus, the \textbf{BH} horizon $r_+$ is  a non-degenerate (bifurcate) Killing
horizon generated by the vector field $\mathcal{L}$.
In the case $a = 0$ (where, in this case, $\omega=0$)
$\xi_t$   (now generator of $r_+$) is hypersurface-orthogonal.
}} $\tilde{\mathcal{L}}\equiv \Xi \mathcal{L}$

The \textbf{MBs} definition is tightly  connected in the Kerr geometry to the (light-like and time-like) stationary observers definition. It also relates  \textbf{MBs} with several processes in which a \textbf{BH} interacts with its  environment such as the accretion disks and magnetospheres.
The vector $\laa$ appears in the description of certain \textbf{BH} evolution processes because it enters the definitions of thermodynamic variables and stationary observers.
We will see below the relation between the definition of stationary  observers, \textbf{MBs} and  \textbf{BH} thermodynamics.

\textbf{Stationary observers}

The vector  $\mathcal{L}$, the condition $\mathcal{L_N}=0$ and \textbf{MBs} are closely related to the definition of stationary observes, i. e., observers with
 a tangent vector which is  a  Killing vector.
 Their    four-velocity  $u^\alpha$ is thus   a
linear combination of the two Killing vectors $\xi_{\phi}$ and $\xi_{t}$; therefore,
$
u^\alpha=\gamma\mathcal{L}^{\alpha}= \gamma (\xi_t^\alpha+\omega \xi_\phi^\alpha$),  where
 $\gamma$ is a normalization factor and $d\phi/{dt}={u^{\phi}}/{u^t}\equiv\omega$.
The dimensionless quantity $\omega$  is the orbital frequency of the stationary observer.
Because of the spacetime  symmetries, the coordinates  $r$ and $\theta$  of a stationary observer are constants along  its worldline, i. e.,  a stationary observer does not see the spacetime changing along its trajectory.
Specifically, the causal structure defined by timelike stationary  observers is characterized by a frequency   bounded in the range $\omega\in]\omega_-,\omega_+[$--\cite{malament}.
On the other hand, static observers are defined by the limiting condition  $\omega=0$  and  cannot exist in the ergoregion\footnote{{Light surfaces,  in fact, play a relevant role for constraining the
energy extraction mechanism from \textbf{BHs}, which regulates the Blandford-Znajek process,  for example. They also constrain the properties of  accretion disks or the  Grad-Shafranov  equations for the force free magnetosphere around a \textbf{BH}. With reference to the metric in the form (\ref{Eq:metric-1covector}) we can define the quantities:
 $\upsilon = d\phi - \omega\dd t,$ and $
\varpi = \partial t  + \omega \partial \phi  $, which are
 clearly related to the Killing field $\laa$ and
 represent  the co-rotation 1-form and the co-rotation vector, respectively,  measuring the co-rotation  in the Grad-Shafranov  approach.
 Consequently,  $\upsilon$ and $\varpi$  define light surfaces  (
 where $\upsilon$ or $\varpi$ vanishes).
 The  energy extraction process can take place regulated by the light surfaces which  are, in general,  two  located outside the  horizon  $r_+$ and one  within the ergoregion $]r_+,r_{\epsilon}^+[$.
}}.
The limiting frequencies  $\omega_{\pm}$, which are photon orbital frequencies, solutions of the condition $\mathcal{L_{N}}=0$, determine
the frequencies $\omega_H^{\pm}$ of the Killing horizons.

\textbf{Black hole thermodynamics, metric bundles and the quantity
$\mathcal{L_{\mathcal{N}}}$}

The thermodynamic properties of black holes are related to the definition of metric bundles in a rather immediate way.
The \textbf{BH} surface gravity  $\kappa$, which is also a  conformal invariant of the metric,
may be defined as the  {rate} at which the norm   of the Killing vector $\mathcal{L}$
vanishes from
outside (i.e. $r>r_+$). In fact, $\kappa$ is in general defined through the relation
$\nabla^\alpha\mathcal{L_{N}}=-2\kappa \mathcal{L}^\alpha$ and for
the Kerr spacetime it becomes $\mathcal{\kappa}_{Kerr}= (r_+-r_-)/2(r_+^2+a^2)$.
The surface gravity   re-scales with the conformal Killing vector, i.e. it  is not the same on all generators but,  because of the symmetries,  it is constant along one specific generator.
The  \textbf{BH} event horizon of
stationary  solutions
has  constant surface gravity--or the surface gravity is constant on the horizon of stationary black holes, which is postulated as the zeroth \textbf{BH}  law-area theorem (see for example \cite{Chrusciel:2012jk,Wald:1999xu}).
More generally, the  \textbf{BH}  horizon area
is non-decreasing, a property which is considered as the second law of
\textbf{BH}  thermodynamics, establishing  the impossibility  to achieve
with any physical process a \textbf{BH} state with zero surface gravity.

Clearly, in the extreme Kerr  spacetime ($a=M$), where  $r_{\pm}=M$, the surface gravity  is zero. This implies that the temperature  is also null ($T_H = 0$),  with a non-vanishing  entropy\footnote{This fact has consequences
also regarding the stability
 against Hawking radiation: A  non-extremal
\textbf{BH} cannot reach the   extremal limit in a finite number of steps--third law.}\cite{Chrusciel:2012jk,Wald:1999xu,WW}. {
On the other hand, the condition (constance of )  $\nabla^a \laa=0$  when $\kappa=0$ substantially constitutes the definition of the  degenerate Killing horizon--degenerate \textbf{BH}--, in the case of Kerr geometries only the extreme \textbf{BH} case is degenerate; therefore, in the extended plane it corresponds to the point $a=M$ $r=M$. A fundamental theorem of Boyer shows that
degenerate horizons are closed.}
This fact also establishes a topological difference between black holes and extreme black holes. More generally, the
 first law $\delta M = (1/8\pi)\kappa \delta A + \omega_H \delta J$ relates the
variation of \textbf{BH} mass $\delta M$, \textbf{BH} horizon area $\delta A$, and angular momentum $ \delta J$
with the \textbf{BH} surface gravity  $\kappa$ and angular velocity $\omega_H$ on the outer horizon.
The term $(\omega_H \delta J)$ can be interpreted as the ``work''.
The (Hawking) temperature term is naturally related to the surface gravity, $T_{H}= {\hbar c\kappa }/{2\pi k_{B}}$  ($k_{B}$ is Boltzmann constant) and the horizon area $A$
to the entropy, $S= k_{B} A/{\cal L}_P^2$ (${\mathcal L}_P$ is the Planck length,  $\hbar$ reduced Planck constant, $c$ is the speed of light)\footnote{{If the \textbf{BH}
temperature  is $T= \kappa/(2\pi)$,  its entropy is
 $S= A/(4\hbar G)$, the  pressure-term is  $p= - \omega_H $,  where the
internal energy is $U$= GM ($M = c^2m/G $= mass, where $m$ is a mass term). The \textbf{BH} (horizon) area $A$ is clearly related to the outer horizon definition,  $A = 8\pi mr_+$, while the  volume term is $V= G  J/c^2$ (where $J = amc^3/G$).}
The \textbf{BH} horizon area will be considered here alternatively
in the extended plane in Figs\il\ref{Fig:Ly-b-rty}.}.
It is convenient
to re-express some of the concepts of \textbf{BH} thermodynamics in terms  of the norm
$\mathcal{L}_{\mathcal{N}}$, which defines  the
metric bundles. \textbf{(1)} The norm $\mathcal{L_{N}}\equiv \mathcal{L}\cdot\mathcal{L}$ is  constant on the horizon.
\textbf{(2)} The  surface gravity is the  constant $\kappa: \nabla^\alpha\mathcal{L_{N}}=-2\kappa \mathcal{L}^\alpha$,
 evaluated on the {outer} horizon $r_+$
 (equivalently,
  $\mathcal{L}^\beta\nabla_\alpha \mathcal{L}_\beta=-\kappa \mathcal{L}_\alpha$ and  $L_{\mathcal{L}}\kappa=0$, where $L_{\mathcal{L}}$ is the Lie derivative,-a non affine geodesic equation, i.e.,
$\kappa=$constant on the orbits of $\mathcal{L}$.).

\medskip

\textbf{Article overview}


This article is organized as follows. In Sec.\il(\ref{Sec:bundle-description}), we present  the main definitions and notations used in this work and introduce the concept of Kerr metric bundles.
Then, in Sec.\il(\ref{Sec:pri-photon-fre}),  we start the analysis of the \textbf{MBs} characteristic frequencies and their relation to the   photon orbital frequencies and to the horizon frequencies. This will allow us to introduce the concept of  horizon replicas, special {orbits}  with the   frequency equal  to the horizon frequencies.
{Among these special orbits,  there are  retrograde solutions with the   frequency equal in magnitude to the horizon frequency and defined in a supplement of the  extended plane.
 A systematic analysis of this case  is considered in Appendix \il(\ref{Sec:contro-orbits-omega}) and  deepened in
Appendix \il(\ref{Sec:allea-5Ste-cont}), focusing on the characterization of
the horizons frequencies as \textbf{MBs} frequencies.
 Concluding remarks follow in Sec.\il(\ref{Sec:conclu}).
}
 In Appendix \il(\ref{Sec:explci-exstremAomega}),  we present the  explicit form of several quantities which are significant for  the metric bundles:
{Explicit \textbf{MBs} and light surfaces are presented in}
Appendix \il(\ref{Sec:explci-exstremAomega}) and in  Appendix \il(\ref{Sec:wokers-super-cynd}),
respectively. The condition $\mathcal{L_{\mathcal{N}}}=0$ is studied in detail
  in Appendix \il(\ref{Sec:boad-eny-isol}).
General notes on the  \textbf{MBs} of the extended plane are given in
Appendix \il(\ref{Sec:gener}). In Appendix \il(\ref{Sec:contro-orbits-omega}), we discuss
the meaning of negative characteristic frequencies. In the extended plane,
there are certain regions which are bounded by special curves. The areas of these regions
are calculated in Appendix \il(\ref{Sec:areas}).
Finally, in Appendix \il(\ref{Sec:replicas}) we present some special characteristics of  the
horizons replicas.

{Throughout this work, we introduce a  number of symbols and notations  necessary to
explain all the results obtained  for these recently proposed concepts;  however, there is in fact a relatively small set of concepts that are listed for reference in Table\il\ref{Table:pol-cy-multi} and  constitute  the core of the \textbf{MBs} we analyze in this work.}
\begin{table*}
\caption{{Lookup table with the main symbols and relevant notations  used throughout the article with a brief description and reference to the first place where they appear.}}
\label{Table:pol-cy-multi}
\centering
\begin{tabular}{lll}
 \hline \hline
$\mathcal{L}$& null  Killing vector $\mathcal{L}=\partial_t +\omega \partial_{\phi}$
 (generators of Killing event  horizons)& Sec.\il(\ref{Sec:impe-null-i}) \\
$\mathbf{\mathcal{L_N}}$ &norm   $\mathbf{\mathcal{L_N}}\equiv\mathbf{ g } (\mathcal{L},\mathcal{L})=\mathcal{L}\cdot\mathbf{\mathcal{L}}$ of the   Killing vector $\mathcal{L}$& Eq.\il(\ref{Eq:bab-lov-what})-- Sec.\il(\ref{Sec:impe-null-i})-Sec.\il(\ref{Sec:explci-exstremAomega})\\
$ \omega_{\pm}$&   light-like   (solutions of $\mathbf{\mathcal{L_N}}=0$)  limiting frequencies  for stationary observers & Eq.\il(\ref{Eq:bab-lov-what})
\\
$a_{\pm} $&
{horizon curve} in the extended plane &Eq.\il(\ref{Eq:horiz-curve})---Figs\il(\ref{Fig:Ly-b-rty})
--Figs\il(\ref{FIG:disciotto1})
\\
$
 a_{\epsilon}^{\pm}$
&
ergosurfaces  curve in the extended plane & Eq.\il(\ref{Eq:ergos})--Figs\il(\ref{FIG:toa11})--Figs\il(\ref{Fig:Ly-b-rty})
--Figs\il(\ref{FIG:disciotto1})\\
$\omega_b$ &  bundle frequencies& Eq.\il(\ref{Eq:bab-lov-what1})--Figs\il(\ref{FIG:raisemK})
\\
$\omega_{H}^{\pm}$&horizons frequencies& Eq.\il(\ref{Eq:tae-fraieuy} )
\\
$r_g$&
{bundle curve tangent radius to  the horizon curve in the extended plane}& Eq.\il(\ref{Eq:rom-a-witho-felix})
\\
 $a_g(a_0)$
 &bundle curve tangent spin to  the horizon curve in the extended plane& Eq.\il(\ref{Eq:rom-a-witho-felix})
\\
$a_0$& bundle  origin spin in the extended plane& Eq.\il(\ref{Eq:bab-lov-what1})
\\
$\Gamma_{a_0}$  &{metric bundles with equal spin origin $a_0$}&
 Eq.\il(\ref{Eq:inoutrefere})
\\
 $\Gamma_{\sigma}$&{metric bundles  on the same plane $\theta$} ($\sigma=\sin^2\theta$)&
 Eq.\il(\ref{Eq:rgrbomegab})
\\
$\Gamma_{\omega_b}$ &{metric bundles  with equal bundle frequencies $\omega_b$}
& Eq.\il(\ref{Eq:trav7see})
\\
 $\Gamma_{a_g}$ &{metric bundles with equal bundle tangent spin $a_g$}&Eq.\il(\ref{Eq:enagliny})
 \\
 \hline\hline
\end{tabular}
\end{table*}
%




\section{Metric bundles of  Kerr spacetimes}
\label{Sec:bundle-description}
{We start this Section  by considering in Sec.\il(\ref{Sec:definitiosn}) explicitly the definitions of extended plane, metric bundles, horizons replicas,  horizons confinement, and causal balls. In Sec.\il(\ref{Sec:deep}), we deepen the discussion on   the metric Killing bundles concept for  the Kerr geometry.
The main characteristics of the metric bundles are the subject of Sec.\il(\ref{Sec:charCCT}).
This Section closes in Sec.\il(\ref{Sec:pri-photon-fre}) with the analysis of photon  orbital frequency and the horizons frequencies. }

\subsection{Extended plane, metric bundles, horizons replicas, horizons confinement and causal balls}\label{Sec:definitiosn}

{\bf Extended plane definition:}
An extended plane is a flat, two--dimensional surface, as in  Figs\il\ref{Fig:Ly-b-rty}.
To define an extended plane $\mathcal{P}-r$, it  is necessary  to select a  parameter  $\mathcal{P}$ of a spacetime in terms of a particular coordinate $r$ so that varying the value of the parameter $\mathcal{P}$, we can extend a specific analysis to include all possible metrics of that spacetime.
On an extended plane  $\mathcal{P}-r$, each curve can be of particular importance. For instance, horizontal lines and vertical lines represent particular members of the spacetime family.

To analyze the details of the information contained in an extended plane, we will consider in this work the particular case of the Kerr spacetime, but this definition can be applied to any spacetime, in principle.
In the Kerr spacetime,  we will consider two examples of extended planes, for which we consider the {BL} coordinates $\{t,r,\theta,\phi\}$ and assume that $M=1$.

$
\mbox{\textbf{Extended plane I:}}\quad \pi_a\equiv  a-r
$, in this case the parameter $\mathcal{P}=a$ is the dimensionless spin ($r$ also is dimensionless). This example has been analyzed in detail previously  in  \cite{remnants}.

$
\mbox{\textbf{Extended plane II:}}\quad \pi_\la\equiv \la-r
$, where $\la\equiv a\sqrt{\sigma}$- Figs\il(\ref{Fig:Ly-b-rty}). In this case, the dimensionless parameter is  $\mathcal{P}=\la=a\sqrt{\sigma}$ and $r$  is dimensionless too. The previous case is obtained for $\sigma=1$.
A particularly important curve in this plane is the horizons curve, which represents the horizons of the entire Kerr family.
The horizon curve  is obviously independent from the polar angle $\theta$; therefore, the horizons are represented by the same curve  in both planes.

Note that for a spacetimes family with a  number  $q$ of parameters $\mathcal{P}$,   the   extended plane can be defined as a $(1+q)$ dimensional surface in which the entire collection of metrics is contained.  For example, in  the case of Kerr-Newman \textbf{(KN)} spacetimes, as discussed in \cite{remnants},  one could identify  as parameter $\mathcal{P}$ the   spin-dimensionless parameter $a/M$,  the dimensionless electric change $Q/M$ or  the "total charge" $\Qa_{T}\equiv \sqrt{(Q/M)^2+(a/M)^2}$). Alternatively, one could also consider  the couple $(a/M,Q/M)$, leading in each case to different extended planes.

{\bf Metric bundle definition:}
A  metric bundle $\Gamma_{\omega}$ was  defined in \cite{remnants} as  the set   of all and only geometries with a given value  of the characteristic frequency $\omega$, which is defined from condition  $\mathcal{L}_{\mathcal{N}}(\omega)=0$. Each  bundle is represented as a curve in the extended plane $a/M-r/M$ or $\la/M-r/M$. However,
 it is also convenient to  consider a more general definition as follows. A metric bundle $\Gamma_{\mathbf{x}}$  is the set  of all and only geometries with a given value of the characteristic  quantity $\mathbf{x}:\quad \laa_{\mathcal{N}}(\mathbf{x})=0$.  Metric bundles correspond to curves  in the extended plane $a/M-r/M$ or $\la/M-r/M$.

In Table\il(\ref{Table:pol-cy-multi}),  we  list the main \textbf{MBs}  analyzed  in this work.
Clearly, we can consider bundles $\Gamma_{\odot}$  as the set  of all and only geometries  having a given value of the characteristic  quantities  $\odot \equiv \{\mathbf{x}_i\}_i:\quad\forall \textbf{x}_i\in  \odot: \laa_{\mathcal{N}}(\mathbf{x_i})=0$. The set $\Gamma_{\odot}$ corresponds to curves in the extended plane.
In Sec.\il(\ref{Sec:gamma-x-y}), we consider two  quantities $\{\mathbf{x_i}\}_i\equiv \{x,y\}$ so that
$\Gamma_{\mathbf{x};\mathbf{y}}$  is the set  of all and only geometries  having a give value of the characteristic  quantities  $\mathbf{x}:\quad \laa_{\mathcal{N}}(\mathbf{x})=0$ \emph{and}   $\mathbf{y}:\quad \laa_{\mathcal{N}}(\mathbf{y})=0$.
In general,   $\Gamma_{\mathbf{x;y}}$ cannot be represented as  $\Gamma_{\mathbf{x;y}} = \Gamma_{\mathbf{x}}\cap\Gamma_{\mathbf{y}}$, i.e., the  condition $\{\textbf{C}_\mathbf{x;y}\in\Gamma_{\mathbf{x;y}}:\mathbf{C}_\mathbf{x;y}\in \Gamma_{\mathbf{x}}\}$ is satisfied only in special cases.

\textbf{Horizons replicas}

Let $ \wp (\bar{r})$ be   the curve representing the horizon  in the extended plane.
We say that there is a replica of the horizon (on an  horizontal line $\mathcal{P}=$constant), if there exists a radius (orbit)
$r_\bullet<  \bar{r}$  such that
	$\wp(r_\bullet)\equiv \wp_\bullet= \wp (\bar{r})$, where $r_{\bullet}$ is a point of the horizon curve in the extended plane, for example,  the horizon  frequency $\omega$,  evaluated on $\bar{r}$.  We are, in fact, interested mostly in the case $\wp=\omega$, which  is the bundle characteristic frequency and horizon frequency.
	According to the definitions of metric bundles, there are clearly   replicas in different  geometries, i.e. there is  a pair of metric parameters values, $p\neq \bar{p}$, and a couple of points, $\bar{r}>r_\bullet$, such that  there is
	$\omega(r_\bullet(p),p)\equiv \omega_\bullet^p=\omega (\bar{r}(\bar{p}),\bar{p})$,
	where $p$ and $\bar{p}$ are  values of   the extended plane parameter, corresponding  therefore  to two different geometries, i.e. two horizontal lines in the extended plane.  To the two   points $(p,r_\bullet)$ and $(\bar{p},\bar{r})$ of the extended plane, there corresponds an equal  light--like  particle orbital  frequency. (It is clear that in the Kerr extended plane, the particular case $\omega_+=\omega_-$ holds only  on  the horizon point $a=M$ and $r_{\pm}=M$.)  We prove in this  analysis that  these structures   reveal their significance in the region proximal  to $(\theta\approx0, \sigma \approx0)$.  Examples of these orbits are in Figs\il(\ref{Fig:JirkPlottGerm}).}

{\bf Horizons confinement and causal balls}

{Opposite with respect to the horizon replicas, the (\textbf{MBs}) \textbf{horizon confinement}  is a concept   interpreted as due to  the presence of a  "local  causal ball" in the extended plane, which is   a region of the extended plane $\mathcal{P}-r$ where the \textbf{MBs}  $\Gamma_\mathbf{x}$ are entirely confined,  i.e.,  there are no horizons replicas  in any other   region of the extended plane outside the causal ball. Particularly,  we can restrict  this definition  to the  case $\mathcal{P}=$constant.
	For example,  in the Kerr extended plane a causal ball  is  the region upper bounded by a portion of the   inner horizon, which means that the horizons frequencies defined for these points of the inner horizons cannot be measured (locally)  outside this region.
	}

\subsection{Developing  the concept of metric Killing bundles of the Kerr geometry}\label{Sec:deep}
Metric bundles were first   introduced in \cite{remnants} as sets of Kerr geometries that can include \textbf{BHs} \emph{or} \textbf{BHs} \emph{and} \textbf{NSs}, where each spacetime of the bundle  has, at  a certain radius $r$
usually different for different geometries of the bundle,
an equal  limiting    photon  frequency $\omega_b\in \{\omega_+,\omega_-\}$,  {or  simply $\omega$}, called  the \emph{characteristic bundle frequency}.

\begin{figure}
 \includegraphics[width=7cm]{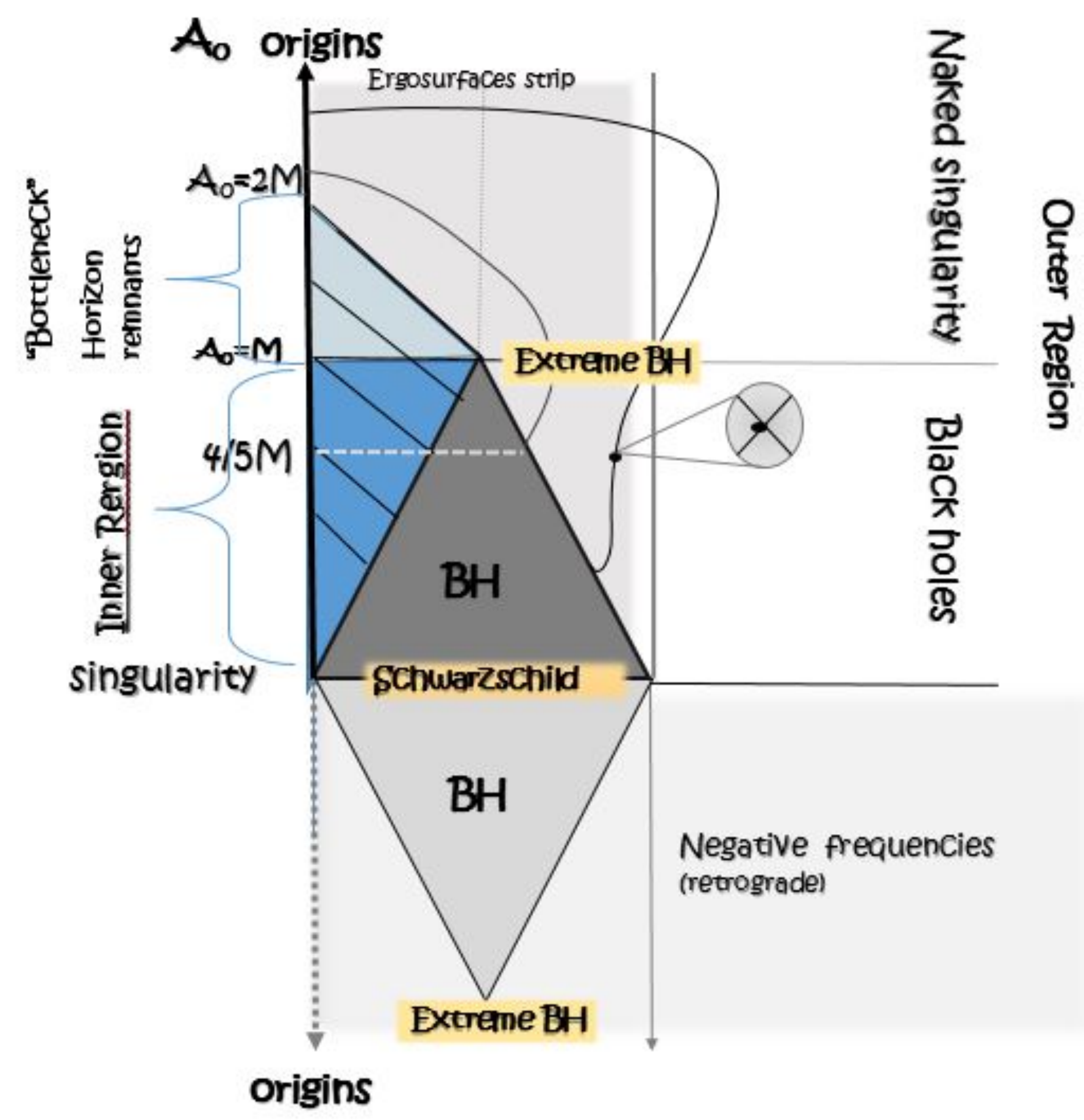}
 \includegraphics[width=4cm]{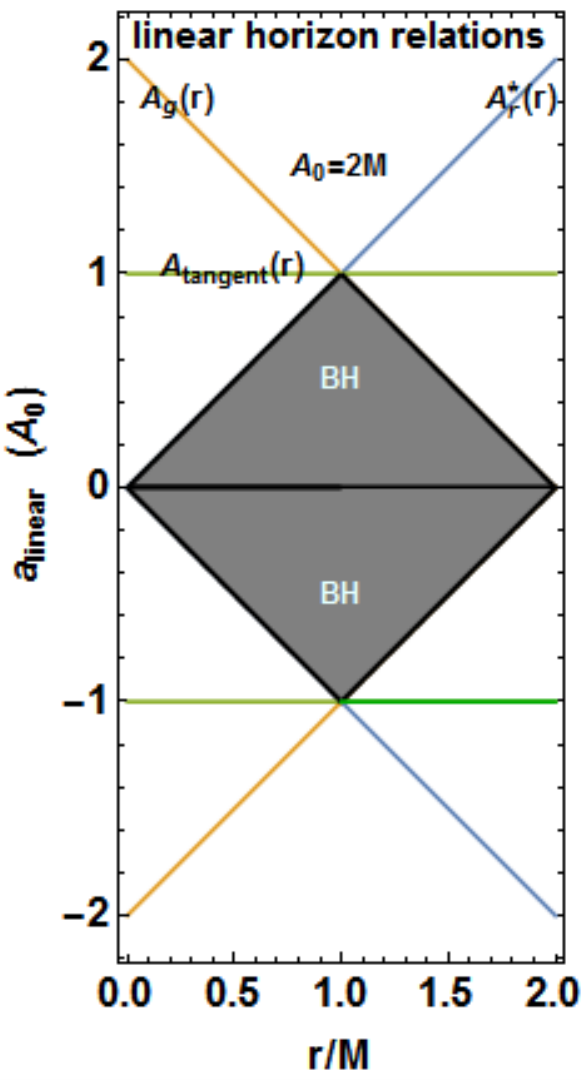}
 \includegraphics[width=4cm]{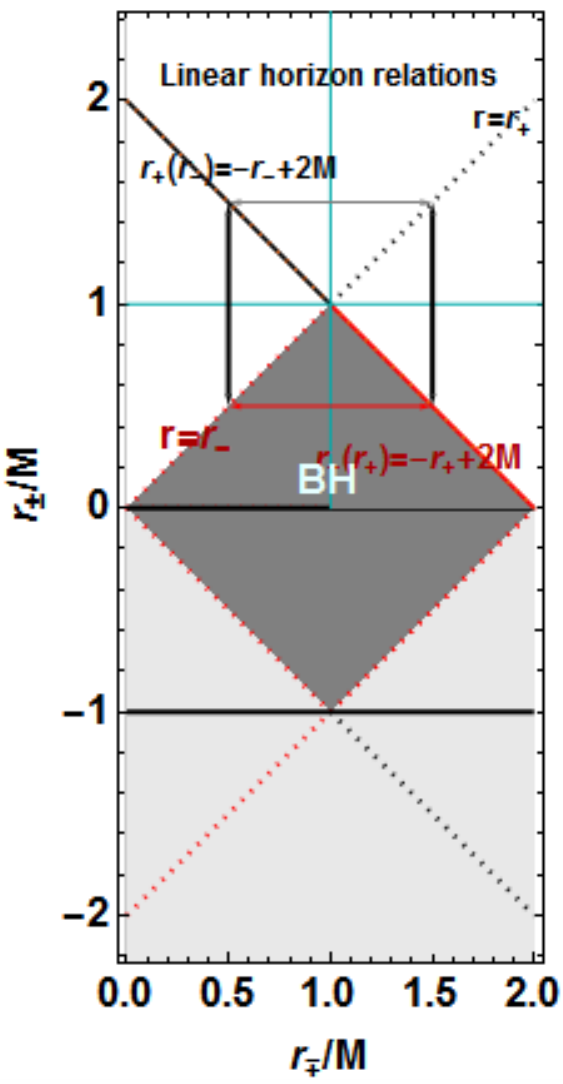}
  \caption{\textbf{Kerr extended plane.}  Horizons properties are independent of the plane $\sigma\equiv \sin^2\theta$. Here
	$\la\equiv a\sqrt{\sigma}$. The plane is split in the negative frequencies region $(a<0)$  (below part) and the upper part for positive bundle characteristic frequencies. Central gray triangle is the \textbf{BH} in the positive $a_0>0$ region of the extended plane.  This contains all the Kerr  \textbf{BH} geometries.   Each point of a bundle  is represented on the plane.
	In the left panel, each point of a bundle is a crossing point of two bundles. The  \emph{interior region}, \emph{exterior region} and the \emph{bottleneck} region are defined, for example, in Fig.\il(\ref{Fig:meniangu}).
	The transformations are given in Eq.\il(\ref{Eq:A0-A0-Atanget}). The linear relations for the horizons in the extended plane are obtained explicitly in \cite{remnants}, for example,  in the $(r-r)$ plane.
The center   panel represents the extended plane  $(a,r)$ with the transformations given in Eq.\il(\ref{Eq:A0-A0-Atanget})--Fig.\il(\ref{Fig:meniangu})). The right panel  shows the linear relations within the transformations $r_{\mp}(r_{\pm})$ of
\cite{remnants}.}\label{Fig:Ly-b-rty}
\end{figure}

Metric bundles are defined in  the Kerr extended plane. In general,  in an extended plane,  we can consider the entire collection of metrics of a parameterized family of solutions. Examples of extended planes are given in  Figs\il(\ref{FIG:funzplo},\ref{FIG:disciotto1},\ref{FIG:rccolonog},\ref{FIG:Aslongas}).
The characterization of the extended plane of Kerr geometry gives rise to the representation of \textbf{BH} solutions as in
 Figs\il(\ref{Fig:Ly-b-rty}).
A metric bundle, as  a set of spacetimes defined by one
characteristic photon orbital frequency $\omega$, is therefore  a curve on the
extended plane,  characterized
by a particular relation between the metrics parameters. These objects
turn out to establish a relation between \textbf{BHs} and \textbf{NSs}
the extended plane and  allow us  to reinterpret the Killing horizons and
to find connections between black holes and
naked singularities throughout the horizon curve. All the metric bundles are tangent to the
horizon curve in the extended plane. Then, the horizon curve
emerges as the envelope surface of the set of metric bundles.
As a consequence, in \cite{remnants} we defined weak  naked singularity  (\textbf{WNS}), related  to a part of
the inner horizon, whereas strong naked singularities (\textbf{SNSs})
were related to the outer horizon.

The definition of metric bundles is based upon the analysis of the condition
$\mathcal{L_N}=0$, which  depends on the even powers of   $\sin \theta$ or $\cos\theta$.
Therefore, it is convenient to introduce the quantity
$\sigma\equiv\sin \theta^2\in[0,1]$. Moreover, the condition   $\mathcal{L_N}=0$
is invariant under the transformation  $(a,\omega)$ in $(-a,-\omega)$; therefore, we focus on the case  $a\in \Re$ and $\omega\geq0$. However, in this work we also consider,  particularly in Sec.\il\ref{Sec:contro-orbits-omega},  the case of negative characteristic frequencies    $\omega_b$, which corresponds to an extended plane with $a_0<0$, see
Figs\il(\ref{Fig:Ly-b-rty}), where $a_0$ is the bundle origin--
Figs\il(\ref{FIG:disciotto1}).

In the section $a_0>0$ of the extended plane, each metric bundle is tangent to the horizon  curve; viceversa,  each point of the horizon is tangent  to a metric bundle. The  points
 $(r=0,a=0)$, $(r=2M,a=0)$  and $(r=2M, a=+\infty)$ of the extended plane are special limiting cases corresponding  to the static Schwarzschild geometry.
This property implies  that the bundle  frequency coincides with  a frequency $\omega_H^{\pm}$ of the horizon at the tangent point, i.e. $\omega_b\in \{\omega_+,\omega_-\}=\omega_{H}^{\pm}$.
As at a point $r$, in general, there are two  different   limiting  photon frequencies $\omega_{\pm}$, then it follows that  at each point of the extended plane (with the exception of the horizon curve) there have to be a maximum   of two different crossing  metric bundles (with the same value of  $\theta$).
The bundle frequency is in particular the frequency of the horizon at the  point of tangency with the bundle --see for example Figs\il(\ref{FIG:rccolonog}) for the representation of metric bundles tangent to the horizon curve in the extended plane.  The metric bundle is defined by the tangent point   on the horizon $r_g$, and the horizon defines all metric bundles (including \textbf{BHs} or \textbf{BHs} \emph{and} \textbf{NSs}) depending only on the plane $\theta$. On the other hand, as we shall see below, there are classes  $\Gamma_{\omega}$ of metric bundles  with equal frequencies $\omega_b$ (equal tangent radius $r_g$), but different bundle origins i.e. the point $(r=0,a=a_0)$-- Figs\il(\ref{FIG:rccolonog}).
The bundle tangency point ($a_g,r_g$),  and therefore the bundle  frequency $\omega_b$, is independent of the polar angle $\theta$.
 It follows  that the horizons frequencies $\omega_H^{\pm}$ in the extended plane are sufficient to define  the limiting frequencies $\omega_{\pm}$ at any point $(r,\theta,\varphi)$ in any geometry $a$. The tangency property is independent of
 $\theta$ and $\varphi$; this reflects the fact that  the horizon curve in the extended plane, which is a sphere of radius $r=M$ centered in the point  ($r=M, a=0$) with area $A_{\pm}=\pi M^2$, is  independent  of the plane $\theta$ and azimuthal angle $\varphi$ (a property related to the rigidity of the \textbf{BH} Killing horizons).

\begin{figure}
 \includegraphics[width=6cm]{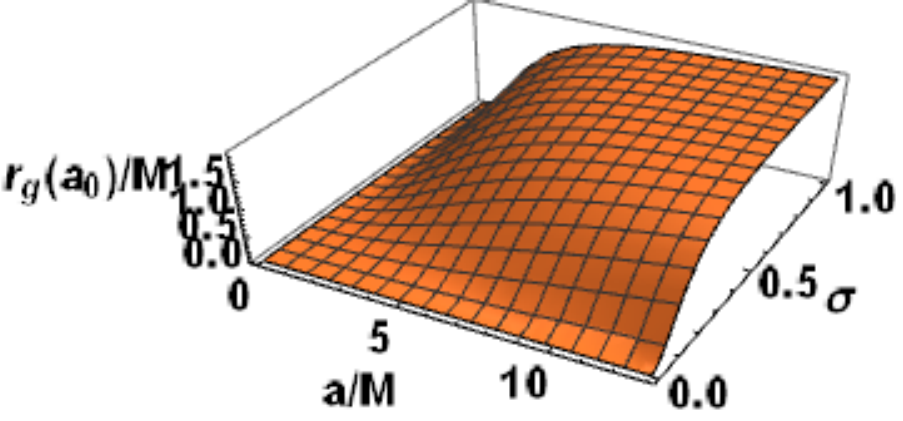}
            \includegraphics[width=6cm]{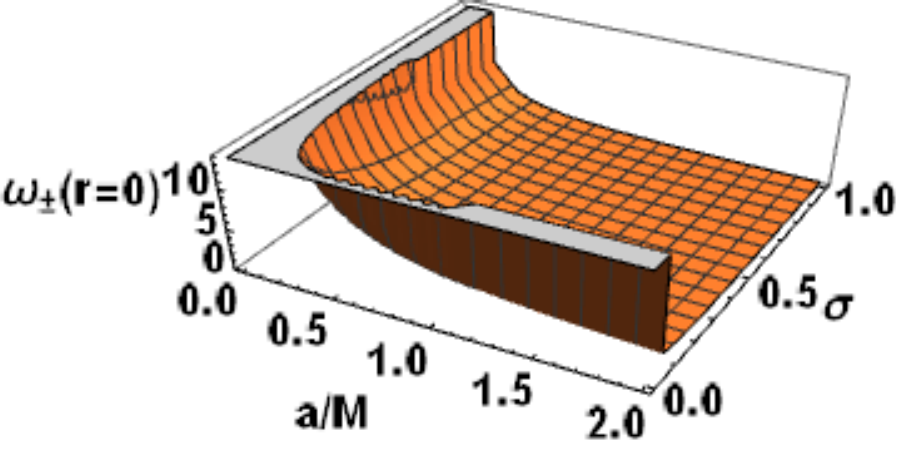}
            \includegraphics[width=5.6cm]{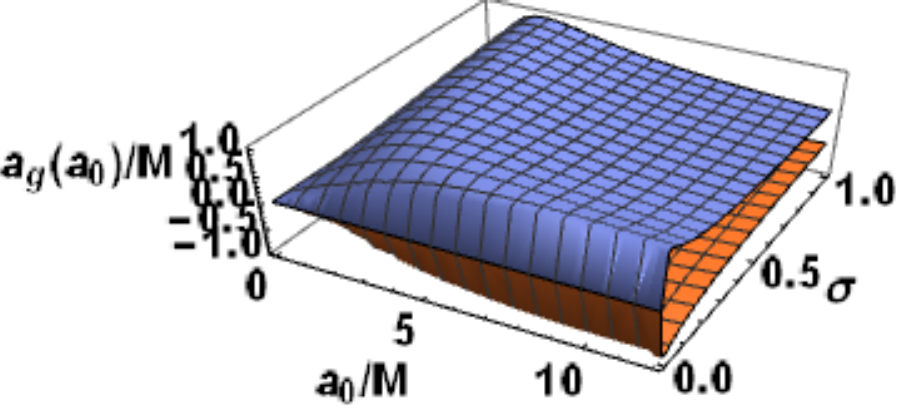} \\
              \includegraphics[width=5cm]{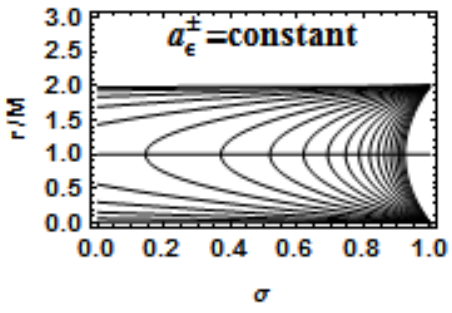}
               \includegraphics[width=5cm]{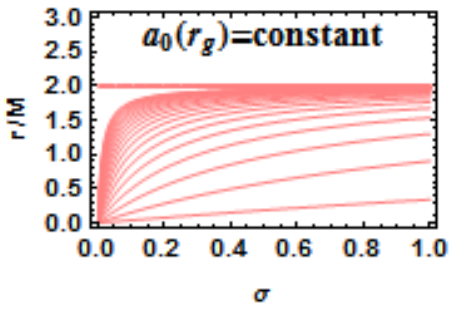}
                \includegraphics[width=5cm]{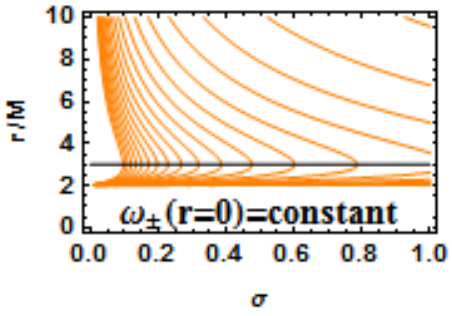}
 \caption{Upper panels: 3D plots of the bundle tangent radius $r_g(a_0)/M$  as function of $a/M$ and $\sigma\in[0,1]$ (left),
the origin frequencies (singularity frequencies) $\omega_{\pm}$=constant for $r=0$ as function of $(a/M, \sigma)$ (center),
the tangent spin $a_g({a_0})/{M}\in[0,1]$  as function of $(a/M, \sigma)$ (right). Below panels. Left panel: ergosurface $a_{\epsilon}^{\pm}=$constant of Eq.\il(\ref{Eq:wonder-wil}) in the extended plane as function of $r/M$ and $\sigma=\sin \theta^2$. Center panel: the bundle origin spin $a_0(r_g)$ of Eq.\il(\ref{Eq:in-our-hea}) as function of the tangent spin $r_g\in[0,2M]$ in the plane $(\sigma,r/M)$. Right panel: the origin frequencies (singularity frequencies) $\omega_{\pm}$=constant for $r=0$ in the plane $(\sigma,r/M)$ see Eq.\il(\ref{Eq:wonder-wil}).}\label{FIG:toa11}
\end{figure}

More precisely, in the extended plane, the horizon $a_{\pm}$    (solution of $ \Delta\equiv a^2+(r-2M) r=0$)   and the ergosurface curve $a_{\epsilon}^{\pm}$,  are
\be\label{Eq:horiz-curve}
a_{\pm}\equiv\sharp\sqrt{r(2-r)}, \quad \sharp = \pm
\ee
and
\be
\label{Eq:ergos}
a_{\epsilon}^{\pm}\equiv\pm\sqrt{\frac{(r-2) r}{\sigma -1}}=\pm\frac{a_{\pm}}{\sqrt{1-\sigma}},\quad \mbox{for}\quad
\sigma\neq1,\quad \mbox{thus}\quad\lim_{\sigma\rightarrow 1}a_{\epsilon}^{\pm}=\pm\infty,
\ee
respectively (see Figs\il\ref{FIG:toa11}).  {For the sake of simplicity here and  in the following sections  we  mainly adopt  dimensionless quantities where $r\rightarrow r/M$ and $a\rightarrow a/M$}.
Note that curve $a_{\epsilon}^{\pm}$ is defined for  $\sigma\in[0,1[$ in $r\in[0,M]$ (corresponding to the inner ergosurface $r_{\epsilon}^-$, with $r_{\epsilon}^-=0$ for $\sigma=1$), where  for $\sigma=1$  there is  $r=0$ with $a\geq0$ and    for  $\sigma\in[0,1[$ with  $r\in[M,2M]$ (corresponding to the outer ergosurface $r_{\epsilon}^+$), with $r_{\epsilon}^+=2M$ for $\sigma=1$. {In most cases, we will consider the sector of the extended plane $a>0$ with   $a_{\pm}>0$, and  therefore
$\sharp=+1$.}

Explicitly, we can define the
photon orbital limiting frequencies  $\omega_{\pm}$ as
\bea&&\nonumber
\mathcal{L_N}=\mathbf{\mathcal{L}\cdot \mathcal{L}}=0 \quad\mbox{for}\quad
\omega=\omega_{\pm}\equiv\frac{\pm\sqrt{\sigma  \Delta \Sigma^2}-2 a r \sigma }{\sigma  \left[a^2 \sigma  \Delta-\left(a^2+r^2\right)^2\right]}, \\\label{Eq:bab-lov-what}
&& \mbox{where}\quad  \Delta\equiv a^2+(r-2) r,\quad\mbox{and} \quad
 \Sigma\equiv r^2-a^2 (\sigma -1),\quad\sigma\equiv\sin^2\theta \nonumber\\
&& \mbox{with}\quad
 \lim_{r\rightarrow\infty}\omega_{\pm}=0,\quad \lim_{a\rightarrow\infty}\omega_{\pm}=0,\quad\lim_{a\rightarrow0}\omega_{\pm}=
\mp\frac{(r-2) r}{\sqrt{(r-2) r^5 \sigma }}.
\eea
On the origin $r=0$ of the extended plane, there is
\bea
&&\label{Eq:bab-lov-what1}
\omega_b=\omega^{\pm}_0\equiv\pm\frac{1}{a \sqrt{\sigma }}.\quad \mbox{The \textbf{bundle origin spin }is }\quad a_0(\omega,\sigma)\equiv\pm \frac{\csc (\theta )}{\omega}=\pm \frac{1}{\omega\sqrt{\sigma }},
\eea
Similarly to the case of equatorial bundles  (for  $\sigma=1$)
investigated in  \cite{remnants}, the  bundle frequency $\omega_b$
is constant along the curve that represents the bundle {(for any constant $\sigma$)}.  Thus, in particular,
the frequency of the origin $\omega_0$ coincides with the bundle frequency $\omega_b$; on the other hand,  we note that the origin spin $a_0$ depends on the plane $\sigma$ at fixed $\omega$.

\begin{figure}
 \includegraphics[width=5cm]{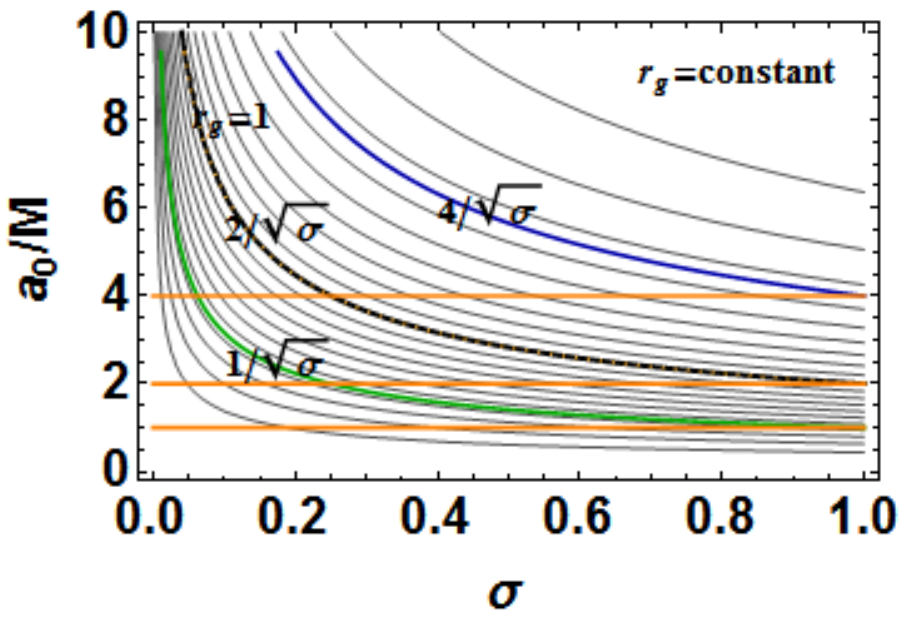}
 \includegraphics[width=5cm]{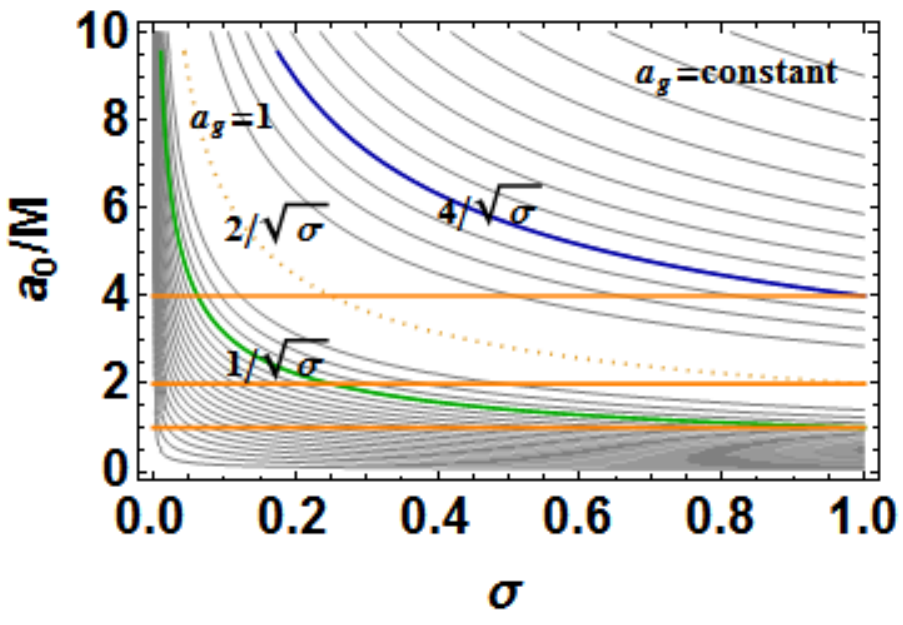}
  \caption{Plot of the constant bundle tangent radius $r_g/M$ and tangent spin $a_g/M$ on the plane $\sigma\in[0,M]$ and for $a_0>0$. Special values and curves are also show. These special values play an essential role for the characteristic frequencies of the bundles (see also Sec.\il\ref{Sec:allea-5Ste-cont}).  It refers to the analysis of Eq.\il(\ref{Eq:own-poi-tn-spea}).}\label{Fig:apappreterra}
\end{figure}

For a fixed radius $r$, there are two   limiting  photon frequencies
$\omega_{\pm}$. Then, it follows that  at each point of the extended plane
(with the exception of the horizon curve) there have to be a maximum
(for a fixed value of $\theta$) of two different crossing  metric bundles.
The bundle frequency coincides with the frequency of the horizon
at the  point where the bundle is tangent to the horizon curve,  as illustrated in
Figs\il(\ref{FIG:rccolonog}).
The fact that metric bundles are tangent to the horizon sphere in the extended plane has some significant properties. Explicitly, we can write
\bea&&\label{Eq:tae-fraieuy}
\forall \theta\neq0,\quad\omega_b\equiv\omega_{\pm}(a_{\pm})=
 \frac{\sqrt{(2-r) r}}{2 r}=\frac{a_{\pm}}{2r}\equiv \omega_H^{\pm}(r),
\\\nonumber
&&
\mbox{with \textbf{bundle tangent radius:} }\quad r_g(\omega)=\frac{2}{4 \omega ^2+1}\ ,\  \omega_H^\mp(a)\equiv \frac{r_\pm}{2a},
 \\\label{Eq:rom-a-witho-felix}
 &&\quad\mbox{and \textbf{bundle tangent spin:}}\quad a_g(a_0)\equiv\frac{4 a_0\sqrt{\sigma }}{a_0^2 \sigma +4}\ ,\quad\mbox{\textbf{bundle origin spin:}}\quad
a_0=\frac{1}{\sqrt{\sigma}\omega_b}.
\eea
Then, for $\sigma=1$ and $r=M$, we have that $\omega_b=1/2$, which corresponds to
the origin $ a_0=2M$. Moreover, for $ a_0=+\infty$, it holds that $r=2M$ and $\omega_b=0$, where $\omega_H^\mp$ are the  frequencies of the horizons $r_{\mp}$ respectively in the extended plane, $r_g$ is the radius at the tangent point of the metric bundle with the horizon in the extended plane--see Fig\il(\ref{FIG:toa11})--(\ref{FIG:toa8}).
The tangent spin $a_g$  and the tangent radius $r_g$ are respectively  the spin and the radius  defined from the tangency of the  metric bundle   with the  horizon curve in the extended plane. We note that  the bundle frequency in terms of the origins spin $a_0$, for  fixed plane $\sigma$,   is maximum for the limiting case $a=0$ and null for $a\rightarrow\infty$,  moreover there is $\omega_0^{\pm}=1/\sqrt{\sigma}$ for  the  extreme \textbf{BH}, and it is minimum on the equatorial plane where it is  $\omega_0^{\pm}=M/a$, in agreement with \cite{remnants}. (As  per  definition the metric bundle has constant frequency at any  point of the bundle in the extended plane,  then particularly the bundle characteristic frequency $\omega_b$ is the frequency of the bundle  origin $a_0$.)

Metric bundles constitute the basis for the   representation  of the Kerr extended plane in Figs\il(\ref{Fig:Ly-b-rty})
where the \textbf{BH} is represented by the  isosceles triangle with height
$h_{\pm}=M$, base $b_{\pm}=2M$, sides length $l_{\pm}=\sqrt{2}M$ and  area $A_{\pm}=M^2$ or, in the case of negative frequencies (corresponding to the extension $a_0<0$), by a rhombus.

In this work, we  discuss the  transformations that are needed to draw these  diagrams based on the  metric bundles explicitly discussed in Eq.\il(\ref{Eq:A0-A0-Atanget})  and  represented in  Figs\il(\ref{Fig:Ly-b-rty}), where we also define the   inner and outer  regions  of the extended plane and  the Killing bottleneck  relevant for the problem of the horizon confinement  discussed in \cite{remnants}.
The  bottleneck region  identifies the spin  interval connected to the emergence  of the Killing bottleneck   in the  Killing  throat  of weak naked singularities, structures emerging from the light surfaces as functions of the  orbital frequency  $\omega$   introduced   in \cite{observers,remnants}. The  bottleneck region is related to the concept of the horizon remnants, which  were also  highlighted as pre-horizons  of \cite{de-Felice-first-Kerr,de-Felice1-frirdtforstati}  and whale diagrams in \cite{Tanatarov:2016mcs,Mukherjee:2018cbu,Zaslavskii:2018kix,Zaslavskii:2019pdc}.
It is important to note that the quantities of Figs\il(\ref{Fig:Ly-b-rty}) are essentially defined according to the variable  $\la\equiv a\sqrt{\sigma}$, we will discuss the importance of this element in details below \footnote{{It has also  been  noted in \cite{PD} that quantity  $\la$,  and specifically $\bar{\ell}\equiv \ell/\la$ poses constraints on the relativistic (geometrically) thick accretion toroidal configurations in the Kerr spacetimes, where $\ell$ is the specific fluid angular momentum $\ell\equiv L/E$, (L,E) being the constat of motions associated to the Killing fields $(\xi_{\phi},\xi_t)$ respectively. This quantity is also related to the rotational version of the Killing vectors $\xi_t$ and $\xi_{\phi}$ i.e. the
canonical vector fields
$\tilde{V}\equiv(r^2+a^2)\partial_t +a\partial_{\phi}$
 and $\tilde{W}\equiv\partial_{\phi}+\la \partial_t$ then the contraction
 the geodesic four-velocity with $\tilde{W}$ leads to the (non-conserved) quantity
$L-E \la$,
 function of the conserved quantities $(E,L)$, the spacetime parameter  $a$ and the polar coordinate $\theta$.
When we consider the principal null congruence
$
\gamma_{\pm}\equiv\pm\partial_r+\Delta^{-1} \tilde{V}$
the angular momentum $L=\la$ that is $\bar{\ell}=1$ (and $E=+1$, in proper unit), every principal null geodesic is then characterized by $\bar{\ell}=1$, on the horizon it is
$L=E=0$.}}.

Although we consider  the  inside region  bounded by the   inner horizon, we are also interested in the  information contained in outer region in the whole extended plane.

Horizontal lines in the \textbf{BH} sector, $a_0\in[0,M]$, correspond to a particular
\textbf{BH} source so that a translation along the triangle (horizon curve) describes the evolution in the vicinity of the source (rigid in the sense  of the metric bundles \cite{remnants}). The  metric  bundles of the set $\Gamma_{a_g}$ considered in Eq.\il(\ref{Eq:enagliny}), have equal tangent spin (but, in general, different tangent radius $r_g$) and relate points on the horizontal lines on the triangle. In Figs\il(\ref{Fig:Ly-b-rty}), we illustrate the concept of   ``replicas" of the \textbf{BH} triangle.

 In Figs\il(\ref{Fig:JirkPlottGerm}), a horizontal line in the \textbf{BH} region shows particular intersections with the metric  bundles. The black hole is ``inaccessible'' to the metric bundles since they are tangent to the horizon (do not penetrate the sphere of \textbf{BH} region in the extended plane).
     Replicas of the  \textbf{BH} triangle and the set  $\Gamma_{a_g}$ allow us to establish a connection between the inner regions, bounded by  the  inner  horizon, and the region  outside the outer horizon.

The explicit expression  $a_{\omega}$, which determines metric bundles,
 can be found in  Sec.\il(\ref{Sec:explci-exstremAomega}). The condition $\mathcal{L}\cdot\mathcal{L}=0$ can also be solved for the radius $r$ (light surfaces  $r_{s}^i$ as functions of the frequencies  $\omega$  and the polar plane $\sigma$ for fixed values of
$a$), as explained  in Sec.\il(\ref{Sec:wokers-super-cynd}). Otherwise, the condition $\mathcal{L}\cdot\mathcal{L}=0$ can be solved for the polar plane $\sigma\in]0,1]$  in terms of  $a$, $\omega$ and $r$, as presented  in  Sec.\il(\ref{Sec:boad-eny-isol}).

%

%

\subsection{{Characteristics of the metric bundles}}\label{Sec:charCCT}
{We  explore  the  \textbf{MBs}   considering  the more general definition $\Gamma_\mathbf{x}$  and $\Gamma_{\mathbf{x;y}}$  introduced in Sec.\il(\ref{Sec:definitiosn}), for various  quantities $\mathbf{x}$  and $\mathbf{y}$ listed in Table\il(\ref{Table:pol-cy-multi}). This analysis is focused on the   characteristics and properties of bundles as curves in the extended plane $\mathcal{P}-r$, where $\mathcal{P}=\{\la,a\}$.  We start  in Sec.\il(\ref{Sec:static})  with the analysis of the limiting case of static geometry (the Schwarzschild background)  corresponding to  the zeros of the \textbf{MBs} curves, i.e.  the metric bundles curves  $\mathbf{C}$ intersections  with  the axis  $a=0$. The
main features of metric bundles are discussed in Sec.\il(\ref{Sec:main-quanti-isr}).
The crossing of \textbf{MBs} with notable  curves of  the extended plane is discussed in Sec.\il(\ref{Sec:main-quanti-isr}); we also study the intersections of the curves  $\mathbf{C}$ with the horizontal lines   $\mathcal{P}=$constant,  which represent a single geometry, exploring,  therefore,  the metric bundles characteristics in one specific spacetime.  The analysis of the intersections of the \textbf{MBs} curves with  the vertical lines of the plane  singles out a fixed orbit  $r=$constant. Finally, we explore the tangency conditions  of the bundles with the horizon curve.
 In Sec.\il(\ref{Sec:gamma-x-y}),
 we study in details the \textbf{MBs} $\Gamma_{\mathbf{x}}$ for different quantities $\textbf{x}$
as listed in Table\il(\ref{Table:pol-cy-multi})
This subsection  closes in Sec.\il(\ref{Sec:lyb-cross-polit}) with the investigation  of crossing
metric bundles, i.e.,  the intersections of \textbf{MBs} curves in the extended plane,  determining   the couple of  light-like orbital limiting frequencies for time-like stationary observers.}

\subsubsection{\textbf{The zeros of the metric bundles: the static geometry}}\label{Sec:static}

\begin{figure}
   \includegraphics[width=5cm]{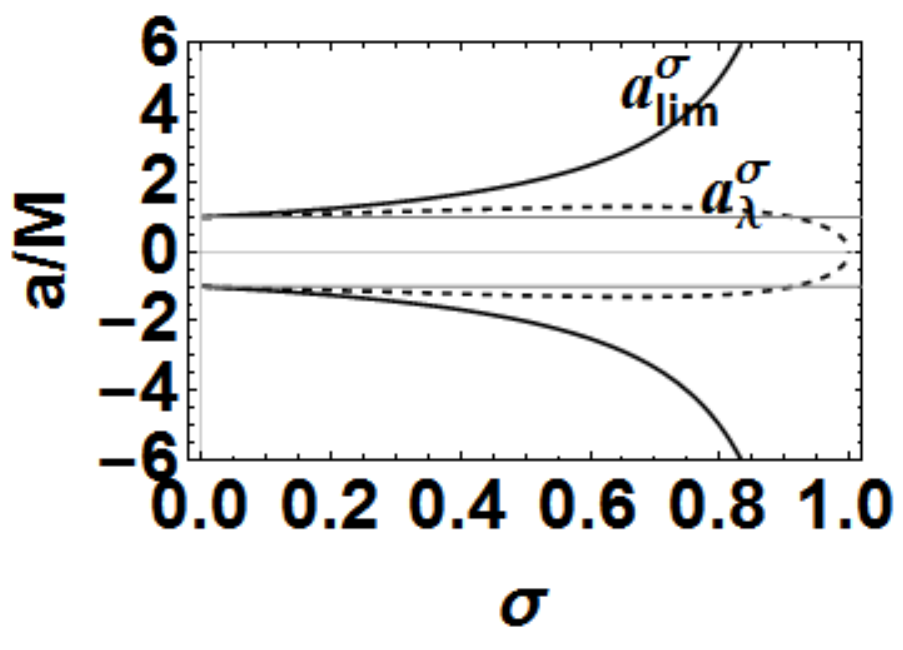}
  \includegraphics[width=5cm]{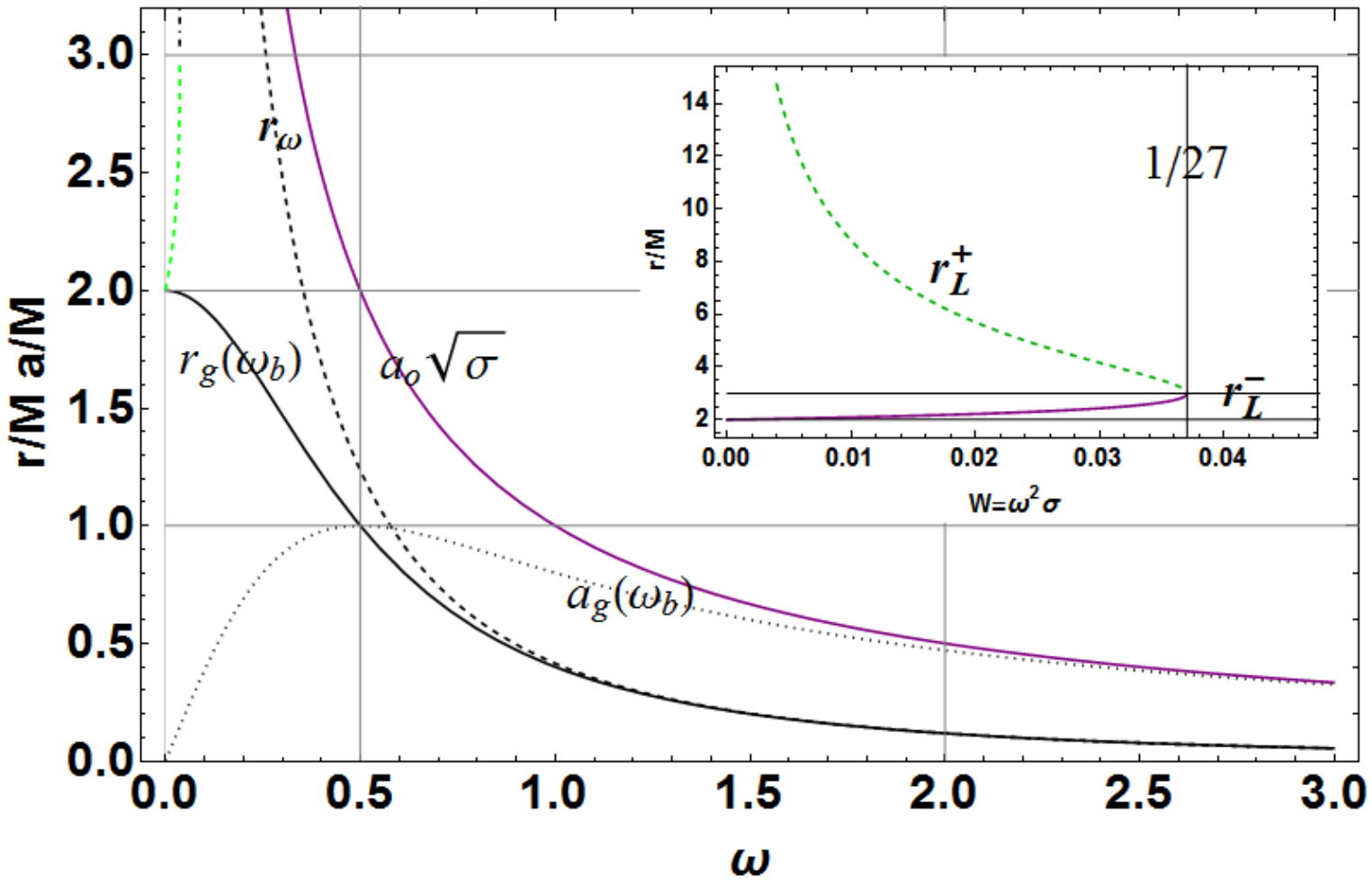}
    \includegraphics[width=7cm]{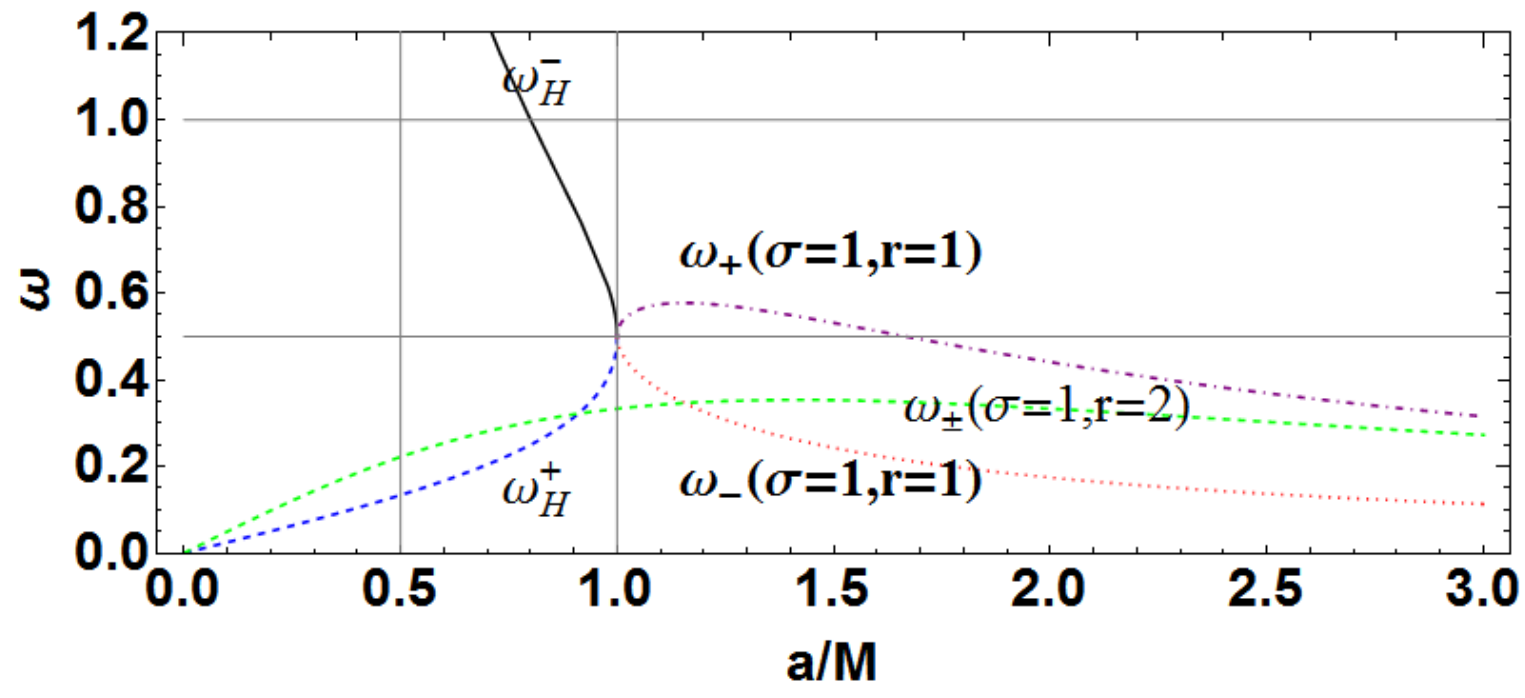}
      \includegraphics[width=6cm]{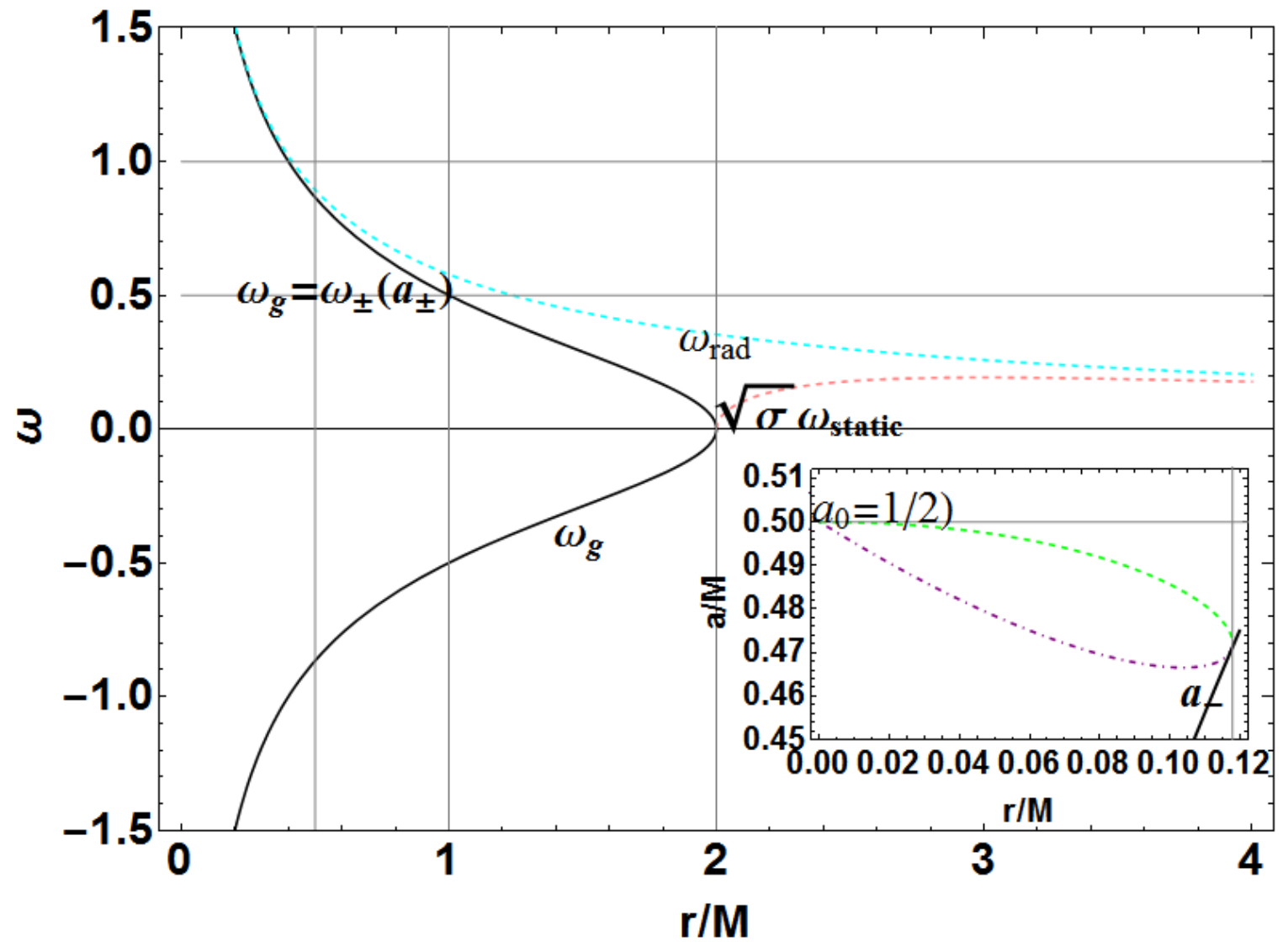}
        \includegraphics[width=6cm]{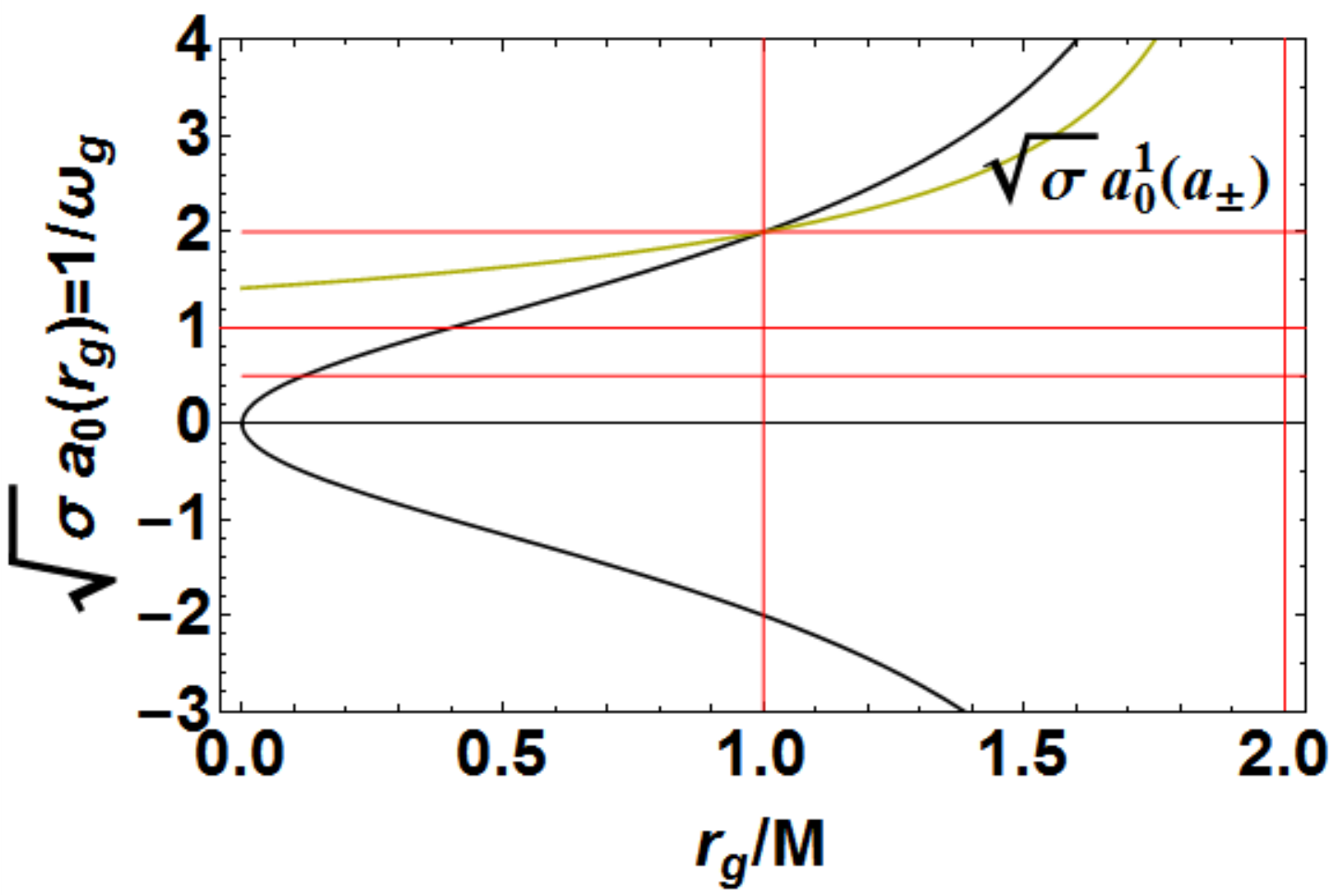}
 \caption{{Panels refer to the analysis of the  metric bundles zeros, determined by
 Eq.\il(\ref{Eq:rLmrLp}),  and to the results presented in Sec.\il(\ref{Sec:boad-eny-isol}) regarding
spins, frequencies  and  planes limiting  values, which are  relevant for the existence of  bundles.}
Upper panels. Left: spins $
a_{\lim}^{\sigma }$ and $
a_{\lambda }^{\sigma }$ of Eq.\il(\ref{Eq:sunt-more}) as functions of $\sigma=\sin\theta^2$. Center: Inner panel: $r_L^{\pm}$ as function of $W\equiv \omega^2\sigma$ of Eq.\il(\ref{Eq:rLmrLp}), the limiting value  $W=1/27$  is also shown.
Plot of  $r_{\omega}$ of Eq.\il(\ref{Eq:hid}).
The tangent radius $r_g(\omega_b)$ of Eq.\il(\ref{Eq:rgrbomegab}) as function of the  bundle frequency.
The bundle origin is $a_0\sqrt{\sigma}$ and the
tangent spin is $a_g (\omega_b)$ as given in  Eq.\il(\ref{Eq:rgrbomegab}).
Right panel: frequencies of Eq.\il(\ref{Eq:wonder-wil}) and the horizon frequencies $\omega_H^{\pm}$.
Bottom panels. Left: Horizon frequencies, as bundle tangent frequencies $\omega_b=\omega_{\pm}(a_\pm)$ of Eqs.\il(\ref{Eq:tae-fraieuy}), as functions of the radius $r/M$, $\omega_{static}$  of Eq.\il(\ref{Eq:rLmrLp}) and $\omega_{rad}$ of Eq.\il(\ref{Eq:sunt-more}). The inner panel is the bundle with origin $a_0=1/2$. Right panel: origin spins $a_0/M$.}
\label{FIG:toa8}
\end{figure}

The zeros of the metric bundles, i.e. solutions  $a_{\omega}(\sigma)=0$ explicitly given in Sec.\il(\ref{Sec:explci-exstremAomega}), correspond
to the static case described by the Schwarzschild metric. The frequencies are
(see Figs\il\ref{FIG:toa8})
\bea&&\label{Eq:rLmrLp}
\omega_{static}=\pm\frac{\sqrt{r-2}}{r^{3/2} \sqrt{\sigma }}.\quad\mbox{Equivalently, in terms of radii (light surfaces) }
 \\\nonumber
 &&r_L^+(W)\equiv\frac{2 \sqrt{\frac{1}{W}} \cos \left[\frac{1}{3} \cos ^{-1}\left(-\frac{3 \sqrt{3}}{\sqrt{\frac{1}{W}}}\right)\right]}{\sqrt{3}},\quad r_L^-(W)\equiv-\frac{2 \sqrt{\frac{1}{W}} \cos \left[\frac{1}{3} \left[\cos ^{-1}\left(-\frac{3 \sqrt{3}}{\sqrt{\frac{1}{W}}}\right)+\pi \right]\right]}{\sqrt{3}},
\\\nonumber
&& \mbox{where}\quad W\equiv\sigma  \omega ^2\geq0,\quad W_{\sigma=1}\equiv\frac{r-2}{r^3},\quad r_L^+(W)=r_L^-(W)=3M\quad \mbox{for}\quad W=W_{\max}\equiv \sigma  \omega ^2=1/27
\eea
which also depend on the polar angle and on the  metric bundle origin. The radii $r_L^{\pm}(W)$ are solutions of the condition
$\mathcal{L}\cdot\mathcal{L}=0$. The general form of these solutions
(light surfaces) for the stationary case
can be found in Sec.\il(\ref{Sec:wokers-super-cynd}).
It is clear, in fact, that the problem for the static case can be written in terms of the  variable $W\equiv\sigma  \omega ^2\geq0$. This quantity, in fact, defines the bundle origin
$a_0$ in terms of its frequency according to Eqs.\il(\ref{Eq:bab-lov-what1}).
The limiting values $W_{\max}=1/27$, which occurs for  $r=3M$,
corresponds to a  photon (last) circular orbit  and  is also an extremum for
	$\omega_{static}$,  where  $\omega_{static}={1}/{3 \sqrt{3} \sqrt{\sigma }}$.
Finally the condition $\mathcal{L}\cdot\mathcal{L}=0$ can  be solved for  the bundle
frequency $\omega(a)=\omega_0^{\pm}(a_s)$, leading to {relations between the spins $(a,a_s)$ as follows}
\bea&&
\mbox{from}\quad \omega(a)=\omega_0^{\flat}(a_s)\quad a_s^{\flat,\mp}\equiv\flat\frac{2 a r \sqrt{\sigma }\mp\sqrt{\Delta \Sigma^2}}{a^2 (\sigma -1)-(r-2) r},\quad\mbox{where}\quad \flat\equiv \pm
\eea

\subsubsection{\textbf{Main features of metric bundles}}\label{Sec:main-quanti-isr}

\begin{figure}
  \includegraphics[width=5cm]{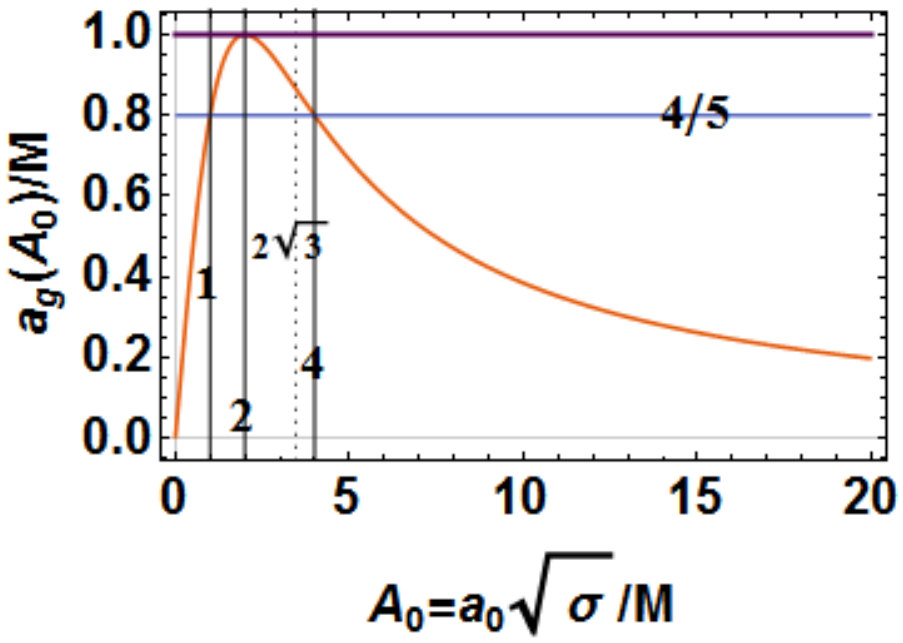}
    \includegraphics[width=5cm]{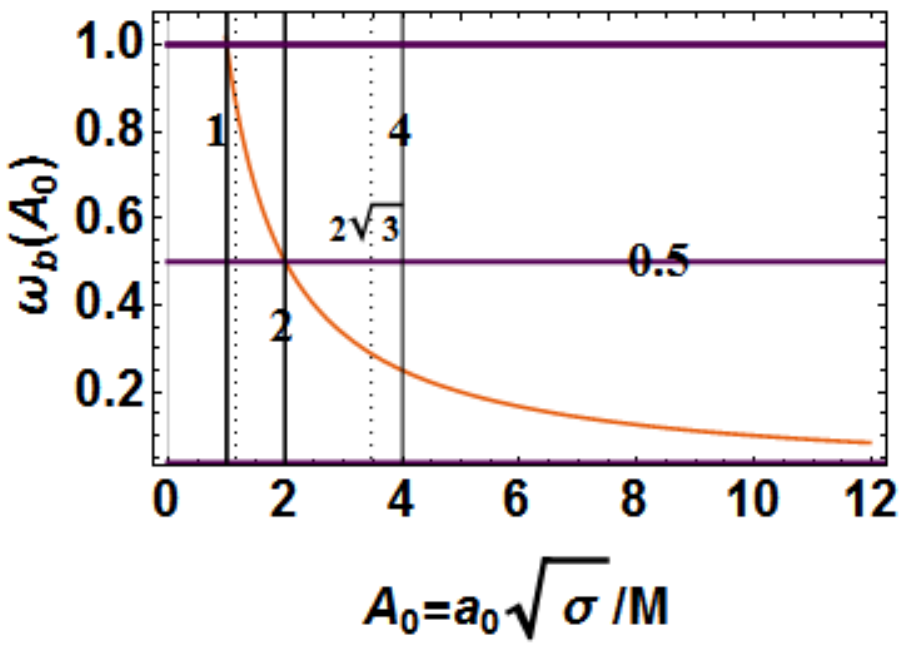}
  \includegraphics[width=5cm]{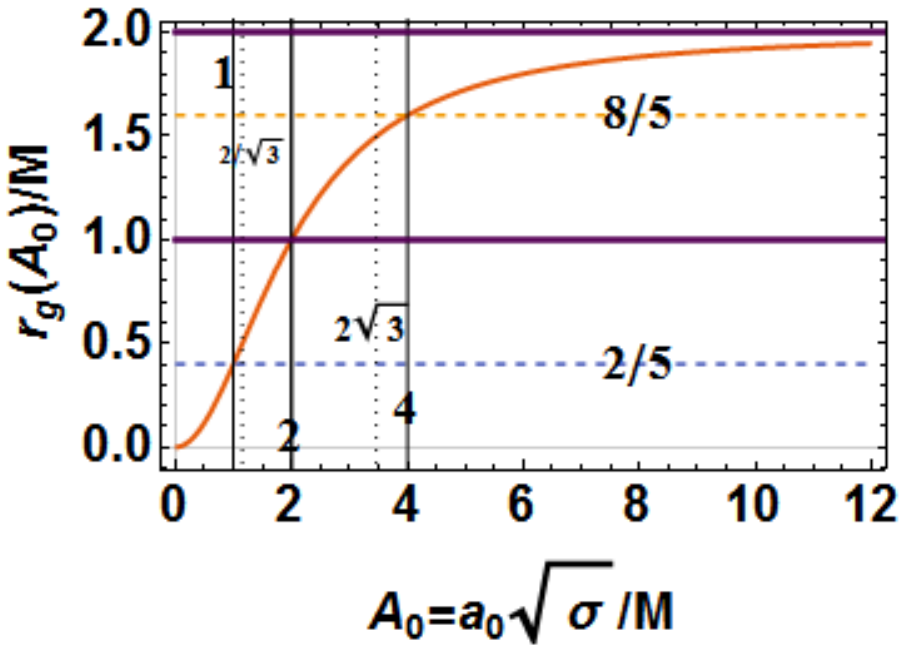}
 \caption{Tangent spin $a_g(\la_0)$ (left panel), bundle frequency $\omega_b(\la_0)$ (center panel)  and tangent radius $r_g(\la_0)$ (right panel)  of the bundle as functions of $\la_0\equiv a_0 \sqrt{\sigma}$,  where $a_0$ is the  bundle origin $\sigma=\sin \theta^2$--see  Eqs\il(\ref{Eq:rgrbomegab}). }\label{FIG:raisemK}
\end{figure}

The main  quantities describing metric bundles, as used in Figs\il(\ref{Fig:Ly-b-rty}),
are related in a  simple way as follows
\bea\label{Eq:summ-colo-quiri}
\mathbf{(1)}\quad \frac{r_g}{a_g}=\frac{a_0 \sqrt{\sigma}}{2},\quad \mathbf{(2)}\quad \frac{r_g}{a_g}=\frac{1}{2 \omega_b}\quad \mbox{from}\quad \mathbf{(3)}\quad a_0= \frac{1}{\sqrt{\sigma}\omega_b},\quad \mathbf{(4)}\quad a_0(a_g)=\frac{2 r_{\mp}(a_g)}{a_g \sqrt{\sigma }}=\frac{4\omega_H^{\pm}(a_g)}{\sqrt{\sigma }}.
\eea
Relation \textbf{(1)} relates the bundle tangent spin  $a_g$  and bundle origin $a_0$ with the tangent radius $r_g$ and depends on  $\sigma$. On the contrary, relation \textbf{(2)}  does not depend on the plane $\sigma$ as it relates $r_g$ and  $a_g$ which are defined at the horizon, with the bundle frequency  $\omega_b$.
We recall that the bundle frequency   is uniquely determined by the radius   $r_g$ (but not by the tangent spin $a_g$). Furthermore,  relation \textbf{(2)} is more general than relation  \textbf{(1)} since it  defines the class $\Gamma_{\omega_b}$ of bundles
with equal frequencies $\omega_b$ as studied in Eqs.\il(\ref{Eq:trav7see})  and
Figs\il(\ref{FIG:rccolonog}). However, according to Eq.\il(\ref{Eq:trav7see})
  for fixed  $\omega_b,\ a_g,$ and $ r_g$,   there is a class of bundles
	related by means of $\sigma$ planes (for a fixed $r_g$,  there is one an only one characteristic frequency $\omega_b$ independently of the plane $\sigma$).
Finally, regarding relation $\mathbf{(4)}$, which defines the bundle origin $a_0(a_g)$ as function of the bundle tangent point $a_g$, it is worth noting that this is the inverse relation of  $a_g(a_0,\sigma)$ given in Eq.\il(\ref{Eq:rom-a-witho-felix}) and shown
 in  Figs\il(\ref{FIG:raisemK}).

We now study some important properties of  metric bundles and introduce the concept of pairs of corresponding  bundles.

Note that each metric bundle corresponds to one and only one bundle frequency
$\omega_b$ and to one and only one tangent point $r_g$. To each frequency $\omega_b$ corresponds one and only one tangent spin and tangent radius and, viceversa, to a tangent  radius $r_g$ corresponds only one $\omega_b$. There is, therefore,
the class  $\Gamma_{\omega_b}$ composed by metric bundles with  equal characteristic  frequency (and equal tangent spin $a_g$ and radius $r_g$) having,  in general, different planes  $\sigma$ and, consequently, different origins $a_0$.
We study the class  $\Gamma_{\omega_b}$  in  Eqs.\il(\ref{Eq:trav7see}).
The class $\Gamma_{a_g}$  is  composed of metric bundles with equal tangent spin  $a_g$, but
different planes $\sigma$ and, therefore, different origins $a_0$,
with  the tangent radius $(r_g,r_g^1)\in r_{\pm}(a_g)$ and frequencies
$(\omega_b, \omega_b^1)\in\omega_H^{\pm}$. This case is studied in
Eqs.\il(\ref{Eq:trav7see}). Bundles of this class form the \textbf{BH} spacetime  horizon with spin $a_g$.
The class  $\Gamma_{\sigma}$  is composed by metric bundles on the same
plane $\sigma$; we study this case in Eq.\il(\ref{Eq:rgrbomegab}).
We note that the {condition  $\mathcal{L}_{\mathcal{N}}=0$}  depends  on the  angle
$\theta$; therefore, as we will also see in detail, many essential bundle properties can be described in terms of the variable $\la\equiv a \sqrt{\sigma}$,
except for the fact that bundles are tangent to the horizon which is independent of
$\theta$.
For instance, $\Gamma_{a_0}$  is the class of metric bundles with  the same origin $a_0$; this case is studied in Eq.\il(\ref{Eq:inoutrefere}).

\textbf{Notable  curves in the extended plane}

In the extended plane, {there are certain curves  representing}  geometries with similar properties. We consider the following notable curves:

{\textbf{(1)}
Vertical lines,  $\bar{r}=$constant can be used to analyze   properties of Kerr geometries,
for  all $a\in[0,M]$,   at the point $\bar{r}$ on different planes $\sigma$. At $\bar{r}$, for fixed $a=\bar{a}$,
there is an even number of bundles that intersect at $\bar{r}$, apart from  the horizon curve $a_{\pm}$}.

{
\textbf{(2)}  Horizontal lines $\bar{a}=$constant characterize properties of a fixed Kerr geometry.
The spin $\bar{a}$ can be considered as origin $a_0$, if $a_0>0$, or also as
tangent  spin $a_g$, if  $a\in[0,M]$.}

\subsubsection{\textbf{Crossing  of metric bundles: determination of  the orbital limiting frequencies}}\label{Sec:lyb-cross-polit}

Metric bundles cross on the extended plane at a point $(a,r)$.
The   two frequencies   $(\omega_1, \omega_2)$ of a \textbf{MB}  with fixed $(a,\sigma,r)$  are related as
\bea
\omega_2=-\frac{4 a r}{a^2 \sigma \Delta-\left(a^2+r^2\right)^2}-\omega_1,
\eea
where $\omega_1$ and $\omega_2$ are given as  $\omega_{\pm}$ in Eq.\il(\ref{Eq:bab-lov-what}).
The two origins of the \textbf{MBs}, as functions of any point in the bundle  $(r,a,\sigma)$, are
\bea&&\nonumber
a_0^1\equiv\frac{\sqrt{\sigma  \Delta \Sigma^2}+2 a r \sigma }{\sqrt{\sigma } \left[a^2 (\sigma -1)-(r-2) r\right]}\quad a_0^2\equiv\frac{\sqrt{\sigma } \left[a^4 (1-\sigma)+a^2 r [2 \sigma +r (2-\sigma)]+r^4\right]}{\sqrt{\sigma  \Delta   \Sigma^2}+2 a r \sigma },\quad\mbox{where}
\\\nonumber
&& a_0^1 a_0^2=\frac{\left(a^2+r^2\right)^2-a^2 \sigma  \Delta }{a^2 (\sigma -1)-(r-2) r},\quad\mbox{for}\quad (a,r,\sigma) \quad \mbox{at the bundle crossing}.
\\\label{Eq:desapata}
&&
\mbox{Then}\quad a_0^1=a_0^2\quad \mbox{for}\quad a=a_{\pm}(r)\quad (r=r_{\pm}(a)),\quad
\\\nonumber
&&{or}\quad  a=\tilde{a}_{\pm}=\pm \frac{r}{\sqrt{\sigma -1}},\quad(r=\tilde{r}_{\pm}=\pm a \sqrt{\sigma -1})\quad\mbox{or}\quad \sigma =\tilde{\sigma}\equiv\frac{a^2+r^2}{a^2}.
\eea
Conditions (\ref{Eq:desapata}) show \textbf{MBs} solutions  with the same origin $a_0$ and tangent point in  $(a,r,\sigma)$, that is,
the same spin $a$, equal  plane $\sigma$ and crossing  radius  $r$. This analysis is  related to  the horizon  curve, where
\bea
a_0^{1}(r_{\pm}(a))=\frac{2 r_{\pm}}{a \sqrt{\sigma }},\quad a_{0}^1(a_{\pm})=\pm \frac{2 r}{a_{\pm} \sqrt{\sigma }},\quad  a_0^1(\tilde{a}_{\pm})=
\pm\frac{r\sqrt{\sigma}}{\sqrt{\sigma -1}},\quad  a_0^1(\tilde{r}_{\pm})=a \sqrt{\sigma },\quad a_0^1(\tilde{\sigma})=a \sqrt{\frac{r^2}{a^2}+1}.
\eea
Obviously, the second frequency at a point of the bundle geometry is also  a horizon frequency, that is, if $\omega_b$ is a bundle frequency,  then at the point $r$ a photon will have  orbital frequency $\omega_b$, while  the second frequency of the couple
$\omega_{\pm}$ is associated with the  horizon frequency of the bundle crossing at the  point $(a,r)$.

If $(a_{\times},r_{\times})$ represent the crossing points of two \textbf{MBs}  with frequencies $\omega_b(a_g)<\omega^p_b(a^p_g)$,
it is clear that $\omega_b(a_g)=\omega_H^{+}(a_g)$ and $\omega^p_b(a^p_g)=\omega^-_H(a^p_g)$. It follows that the two crossing \textbf{MBs} are necessarily one tangent to an   outer  horizon and another one tangent to an  inner horizon. On the other hand,  the relation between
 \textbf{MBs} frequencies is independent of the angle  $\theta$ (equivalently $\sigma$) as the bundle characteristic  frequencies are horizon frequencies (they will, of course,  depend on the plane $\sigma$ when related to their  bundle origin spins).
 Now, it is possible to find   \textbf{MBs} crossing at a point $(r_{\times},a_{\times})$ of the extended plane,
which have  different planes $\sigma$. However, such \textbf{MBs}  relate, in the geometry  with crossing spin $a_{\times}$,
frequencies associated with the crossing  orbit $r_{\times}$,  but on different planes  $(\sigma,\sigma_p)$ of the same Kerr geometry.
It is possible to constraint this case considering  $\omega_{\pm}$ in Eq.\il(\ref{Eq:bab-lov-what}) as functions of $\sigma$ only or, viceversa, using the solutions
$\sigma_{\omega}^{\pm}$ of Eqs\il(\ref{Eq:sigma-omega-sol-pm}) as functions of $\omega$.

\begin{figure}
  \includegraphics[width=5.5cm]{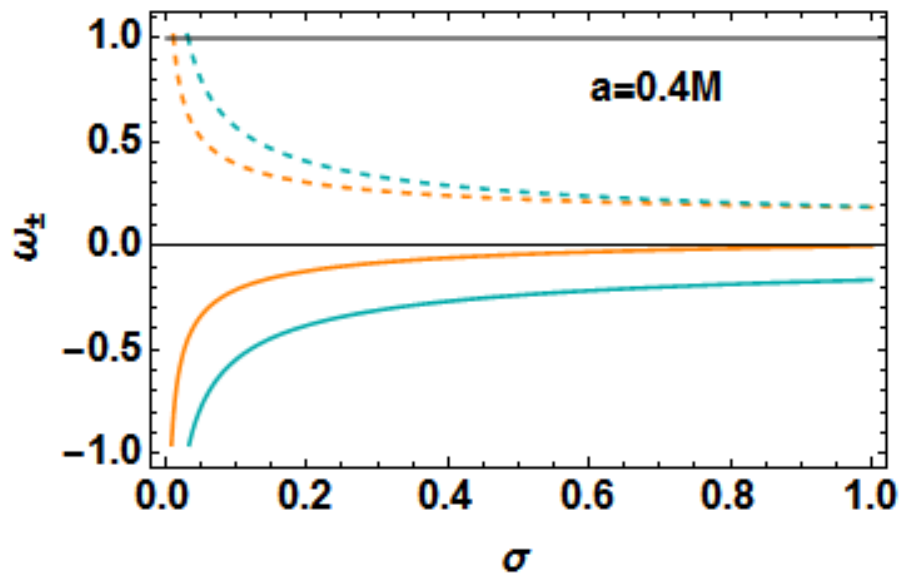}
\includegraphics[width=5.5cm]{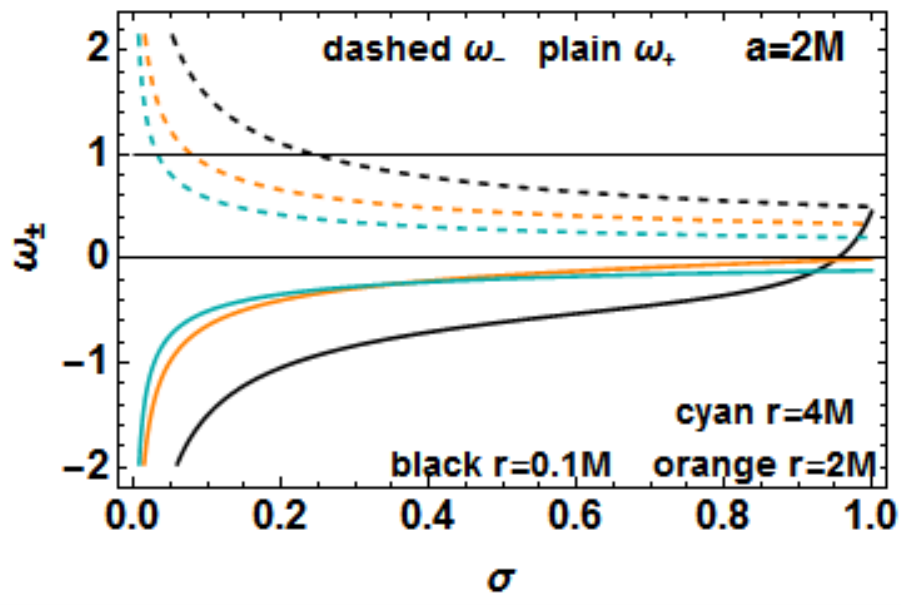}
\includegraphics[width=5.5cm]{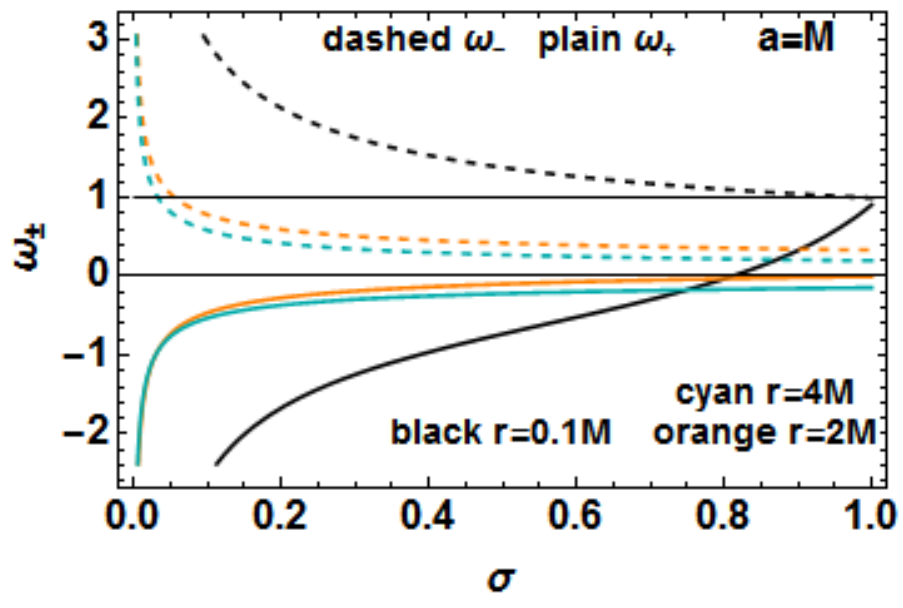}\\
\includegraphics[width=4.2cm]{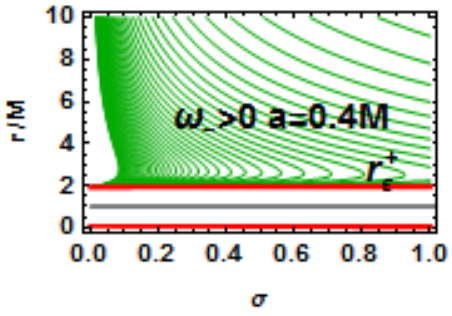}
\includegraphics[width=4.2cm]{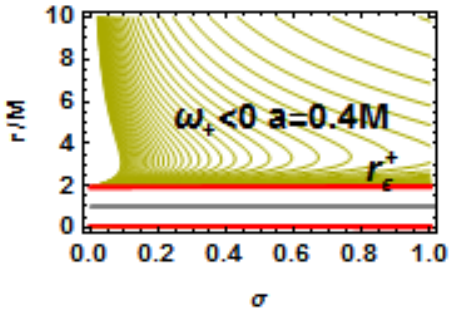}
\includegraphics[width=4.2cm]{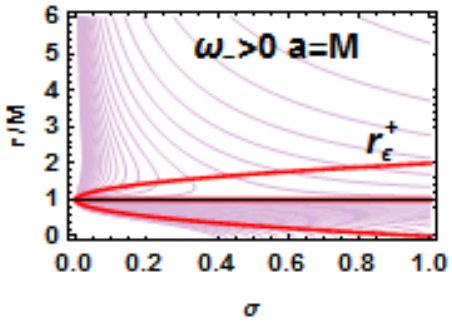}
\includegraphics[width=4.2cm]{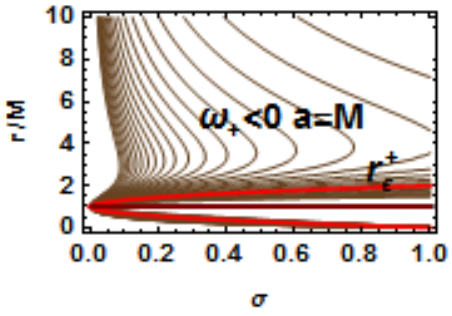}\\
\includegraphics[width=4.2cm]{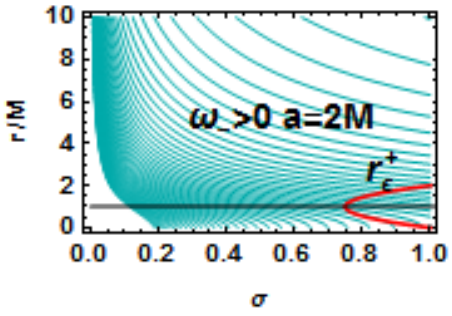}
\includegraphics[width=4.2cm]{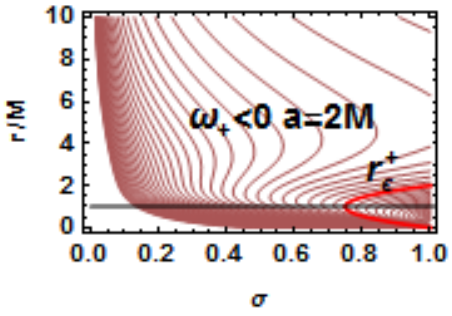}
\includegraphics[width=4.2cm]{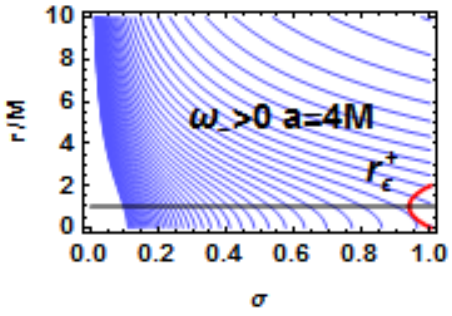}
\includegraphics[width=4.2cm]{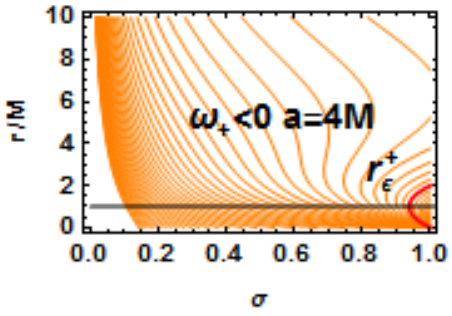}
  \caption{Upper lines: bundle frequencies $\omega_{\pm}$ as functions of the plane $\sigma\in[0,1]$ for fixed spin $a/M$ and radius $r/M$ of the extended plane. Note the frequencies signs (negative for retrograde photon motion)) and the decreasing  magnitude with the plane $\sigma$ at fixed $r$. The limiting roles of the horizons and ergosurfaces can also be noted.
	Below lines: solutions $\omega_{\pm}$=constant on the  $(r/M,\sigma)$ plane. Note the frequencies sign, always positive in the ergoregion (red curves are the ergosurfaces $r_{\epsilon}^{\pm}$), horizons are also show. The curves are studied for different spins $a/M>0$. See the analysis of Sec.\il(\ref{Sec:bundle-description})  on the crossing  of metric bundles and  determination of orbital limiting frequencies}\label{Fig:Altrosolag}
\end{figure}

However, we note that  for fixed $\sigma$ at  the point  $(a_{\times},r_{\times})$ there are  frequencies
$(\omega_H^{+}(a_g),\omega_H^{-}(a_g^p))$ for two tangent bundle spins $(a_g,a_g^p)$, respectively. Because of the existence of the two frequencies $\omega_{\pm}$ \emph{per} point \emph{per} spin and  \emph{per} plane, we infer that at each point of the extended plane there is an even number of crossing \textbf{MBs} {(a part of the limiting case of the horizon curve)}.
In particular, for the same plane  $\sigma$, we have that $(\omega_H^{+}(a_{g,1},\sigma^{1}),\omega_H^{-}(a^p_{g,1},\sigma^{1})$ and
the relations (\ref{Eq:rgrbomegab}) hold as special cases of $\Gamma_{\sigma}$.
In general, the following relations hold:
\bea
a_g= \frac{4 \omega_+(a_\times)}{4\omega_+(a_\times)^2+1}\quad a^p_g= \frac{4 \omega_-(a_\times)}{4 \omega_-(a_\times)^2+1}\quad a_0=\frac{1}{\sqrt{\sigma}\omega_+(a_\times)}=\frac{2  r_+(a_g)}{\sqrt{\sigma}a_g}\quad a^p_0=\frac{1}{\sqrt{\sigma_p} \omega_-(a_{\times})}=\frac{2  r_-(a^p_g)}{\sqrt{\sigma_p}a^p_g}.
\eea
In Fig.\il(\ref{Fig:Altrosolag}), we show the frequencies at a fixed point  $(a,r)$   of the extended plane for different  $\sigma$.
Notice the frequency signs (negative for retrograde photon motion) and the decreasing  behavior of the frequencies in terms of the plane
$\sigma$ at fixed $r$. Notice also the role of the  ergoregion.
In the next Section, we will focus particularly on \textbf{MBs} with fixed $a$ and  characteristic frequency equal to
$\omega_H^{\pm}(a)$. This is clearly the problem of the \textbf{MBs} confinement  with a frequency equal to the  horizon frequency.

\section{Extracting information from Kerr metric bundles: Photon  frequency and horizon frequencies}\label{Sec:pri-photon-fre}
We now  consider the condition   $\omega_x=\omega_y$, where $\omega_x\in\{\omega_+,\omega_-\}$ and
 $\omega_y\in\{\omega_H^-,\omega_H^-\}$. We look for all those solutions of $\mathcal{L}_{\mathcal{N}}\equiv\mathcal{L}\cdot\mathcal{L}=0$ associated to orbits $r$ different from the horizon,  but characterized by a photon  with orbital frequency  equal to that of the horizon  per equal $a$.
We are looking for solutions of the normalization condition $\mathcal{L}_{\mathcal{N}}=0$ when  $\omega=\omega_H(a)$ for a  geometry with spin  $a$  with a   $\bar{r}\neq r_{\pm}$. We are particularly interested in cases where $\bar{r}>r_+$, from which
the most relevant case is for $\bar{r}>r_+$ with  $\omega_{*}(\bar{r})=\omega_H^-$ for $\omega_{*}\in\{\omega_+,\omega_-\}$.

\begin{figure}
  \includegraphics[width=4cm]{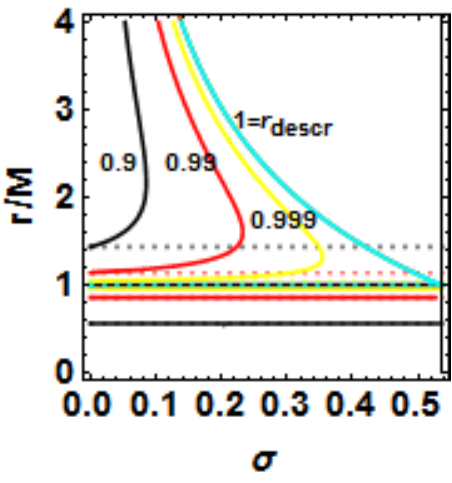}
  \includegraphics[width=4.5cm]{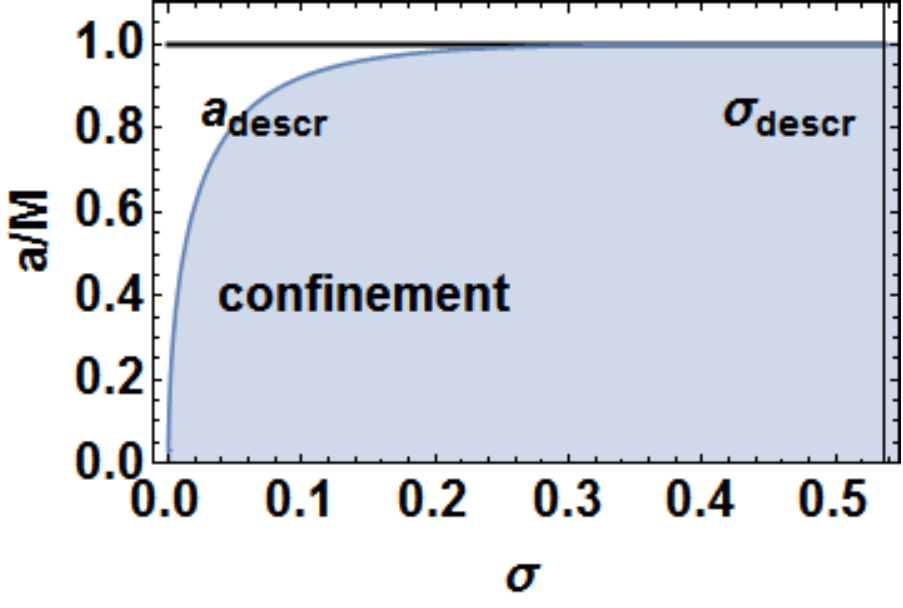}
  \includegraphics[width=4.5cm]{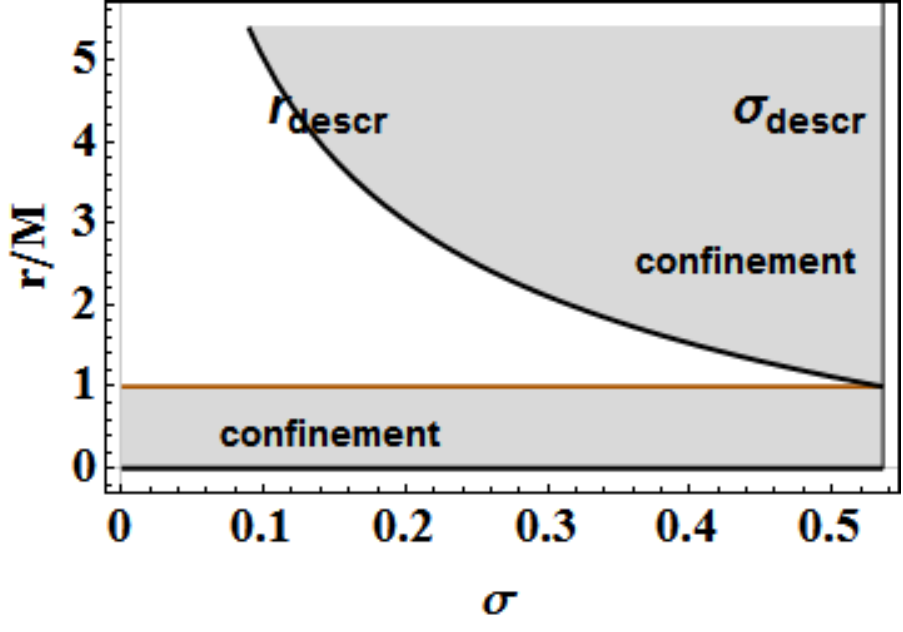}
  \includegraphics[width=4.5cm]{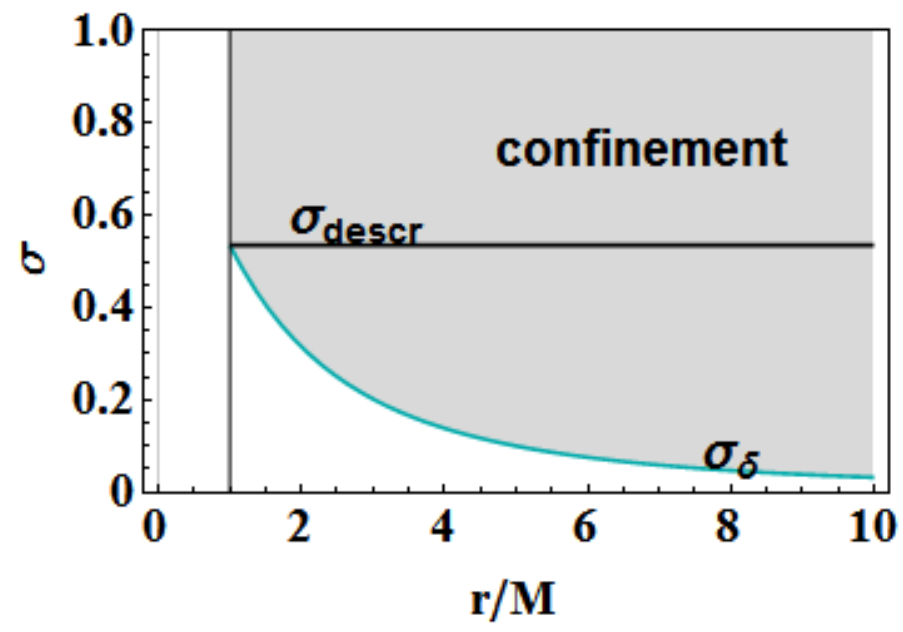}
  \includegraphics[width=4.5cm]{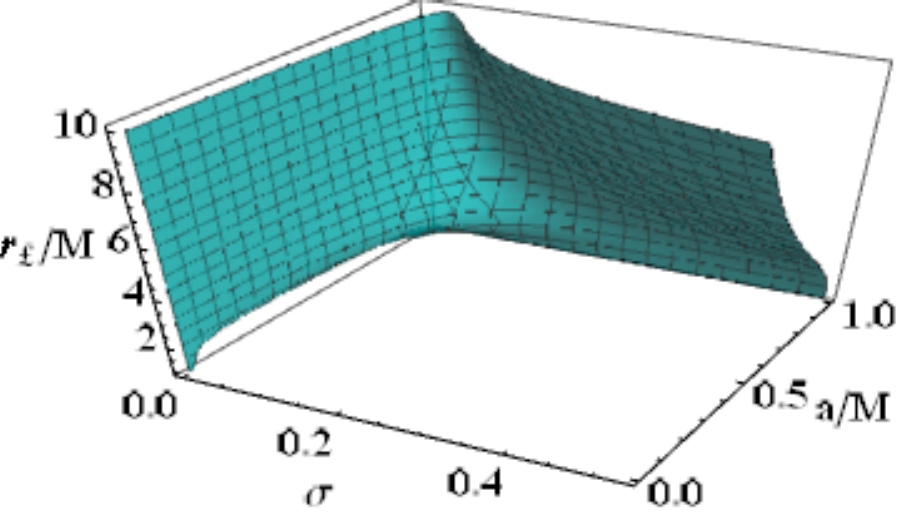}
  \caption{Inner horizon confinement   on the extended plane. Constraints on the existence of orbits of the exterior region
	on the extended plane  $r>r_+$ with frequencies $\omega_-=\omega_H^-$. See the analysis of Sec.\il(\ref{Sec:pri-photon-fre}) and Eq.\il(\ref{Eq:famu-set-cin}). Left panel: orbits $r_{\pounds}$ solutions of $\omega_-=\omega_H^-$ (photons defined by the condition $\mathcal{L}_{\mathcal{N}}=0$) as function of the plane $\sigma\in[0,1]$ for different spins $a/M$.
	The limiting value $a=M$, black curve coincident also with $r_{descr}(\sigma)$. The solutions of $\omega_-=\omega_H^-$ clearly include also the inner horizon $r_-(a)$. The horizons $r_{\pm}(a)$ are also showed (horizontal lines).
	The limiting value $\sigma_{descr}\approx0.53M$.  A solution is clearly the inner horizon $r_-$.
The larger is the \textbf{BH} spin, the larger can be the plane value function $\sigma\leq \sigma_{decr}$.
 Shaded regions are confinement regions.  Below:  3D plot is $r_{\pounds}$, containing a solution of  $\omega_-=\omega_H^-$ as function of $a/M$ and $\sigma$. Quantities are defined in Eq.\il(\ref{Eq:delta-a-delta},\ref{Eq:delta-a-delta1},\ref{Eq:delta-a-delta2},\ref{Eq:delta-a-delta3}). }\label{Fig:Pnoveplot}
\end{figure}

Combining  the  considerations  of the bundle crossing and inner horizon confinement problem, we note that the existence of an  ``external" orbit (at $r>r_+$), with one frequency equal to the inner horizon frequency  on the extended plane,  implies that in the region  $r>r_+$ bundles tangent to the inner and outer horizons intersect. We are then confronted with a two-sided problem:
\textbf{1.} The problem of finding a point $r>r_+(\bar{a})$ in $\bar{a}\in[0,M]$ with frequency  $\omega_H^-(a_*)$ for a general spin $a_*\neq \bar{a}$  such that  $\omega_\pm(\bar{a})=\omega_H^-(a_*)$.
\textbf{2.} The problem of finding a solution $\omega_\pm(\bar{a})=\omega_H^-(\bar{a})$ (i.e. $a_*= \bar{a}$).
We have met this kind of problems in several parts of our investigation. We will show that the second problem exists   on planes very close to the rotation axis. Consider
the spin $\bar{a}=$constant of a bundle with frequency $\omega_H^-(\bar{a})$ on a plane $\sigma$ and analyze the problem for $\sigma$
with $r>r_+(\bar{a})$ and $\omega_{\pm}(\bar{a})=\omega_H^-(\bar{a})$.
It can be proved that solutions of this problem are $r_{\pounds}$ (or, equivalently, in terms of the spin $a_{\pounds}$  and in terms  of the plane $\sigma_{\pounds}$), if the following conditions are satisfied
\bea&&\label{Eq:famu-set-cin}
\sigma\in[0,\sigma_{descr}],\quad \mbox{or}\quad\sigma\in[0,\sigma_{\delta}],\quad r\in[M,r_{descr}],\quad a\in[a_{\delta},M]
\\
&&\nonumber\mbox{where}
\quad
\sigma_{\delta}\equiv\frac{1}{2} \left[r (r+2)-\sqrt{(r+1)^2 [r (r+2)+9]}+5\right]; r_{descr}\equiv\sqrt{\sigma +\frac{4}{\sigma }-4}-1;
\eea
and $\sigma_{\pounds}$, $a_{\pounds}$  and $r_{\pounds}$  are solutions of the equations
\bea\nonumber
&&
\mathbf{\sigma_{\pounds}:}\quad \sigma ^4 \left[a^8+a^6 (r-2) r\right]-2 a^4 \sigma ^3 \left[a^4+2 a^2 \left(r^2-2\right)+r^4-8 r+8\right]+16 \left[a^4+a^2 (r-2) r\right]+8 \sigma  \left(a^6+2 a^4 \left(r^2-3\right)+\right.
\\\nonumber
&&\left.a^2 r \left(r^3-4 r+8\right)-2 r^4\right)+\sigma ^2 \left[a^8+a^6 (r-2) (3 r+8)+a^4 [(r-2) r (r+2) (3 r+4)+48]+\right.
\\\label{Eq:delta-a-delta1}
&&\left.a^2 r [r (r (r (r (r+2)+4)-8)+16)-32]\right)=0,
\\&&\nonumber
\\\nonumber
&&
\mathbf{a_{\pounds}:}\quad a^8 \left(\sigma ^4-2 \sigma ^3+\sigma ^2\right)+a^6 (\sigma -1) \sigma  \left[\sigma  \left[r^2 (\sigma -3)-2 r (\sigma +1)+8\right]-8\right]+a^4 \left(-2 \left(r^4-8 r+8\right) \sigma ^3+\right.
\\\nonumber
&&\left.16 \left(r^2-3\right) \sigma +[(r-2) r (r+2) (3 r+4)+48]\sigma ^2+16\right)+a^2 r \left(8 \left(r^3-4 r+8\right) \sigma +\right.
\\\label{Eq:delta-a-delta2}
&&\left.+(r (r (r (r (r+2)+4)-8)+16)-32) \sigma ^2+16 (r-2)\right)-16 r^4 \sigma=0,
\\&&\nonumber
\\\nonumber
&&\mathbf{r_{\pounds}:}\quad  a^2 r^6 \sigma ^2+2 a^2 r^5 \sigma ^2+4 a^2 \left(a^2-2\right) r^3 \sigma ^2-2 a^2 r (\sigma -1) \left(a^2 \sigma -4\right) \left[\sigma  \left(a^2 (\sigma +1)-4\right)+4\right]+
\\\nonumber
&&r^4 \sigma  \left[a^4 \sigma  (3-2 \sigma )+4 a^2 (\sigma +2)-16\right]+a^2 r^2 \left(\sigma  \left(a^4 (\sigma -3) (\sigma -1) \sigma +\right.\right.\\\label{Eq:delta-a-delta3}
&&\left.\left.4 a^2 (4-3 \sigma )+16 (\sigma -2)\right)+16\right)+a^4 (\sigma -1)^2 \left(a^4 \sigma ^2+8 \left(a^2-2\right) \sigma +16\right)=0,
\eea
respectively, whereas
the limiting spin $a_{\delta}$ is a solution of the following equation
\bea&&\label{Eq:delta-a-delta}
\mathbf{a_{\delta}:}\quad a^{10} (\sigma -1)^2 \sigma ^6-2 a^8 (\sigma -1) \sigma ^4 (3 \sigma  (5 \sigma -28)+68)+a^6 \sigma ^2 [\sigma  (\sigma  [\sigma  (201 \sigma +1160)-2784]+1408)+16],
\\\nonumber
&&+8 a^4 \sigma  ([\sigma  (\sigma  [(1045-411 \sigma ) \sigma -1228]+588)+16]+16 a^2 (\sigma -1) (\sigma  [\sigma  (97 \sigma +447)-312]-16)-6912 (\sigma -1)^2 \sigma=0.
\eea
We show the behavior of these quantities and the constraints for the existence of orbits with frequency of the inner horizon in the
{outer region of the extended plane $a>a_+$ in Fig.\il(\ref{Fig:Pnoveplot})}.
 A more detailed study of the frequency ratio of \textbf{MBs} as horizon frequencies is  given in  Sec.\il(\ref{Sec:allea-5Ste-cont}).

\begin{figure}
  \includegraphics[width=5cm]{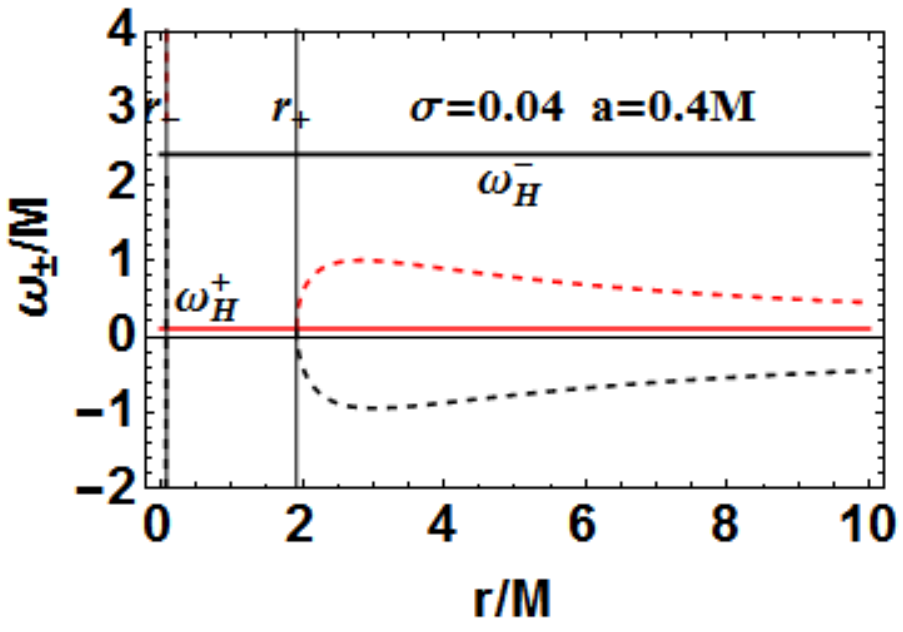}
\includegraphics[width=5cm]{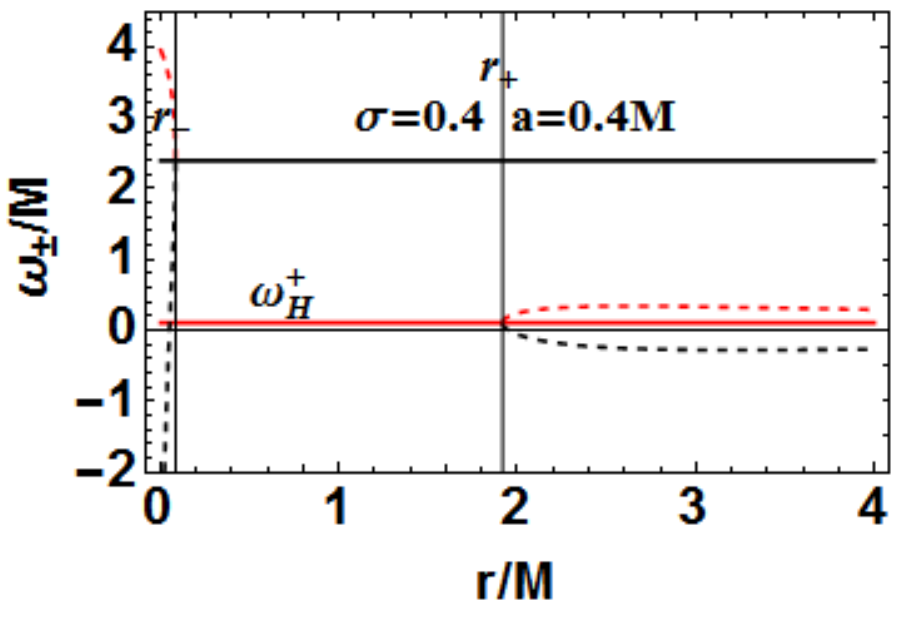}
\includegraphics[width=5cm]{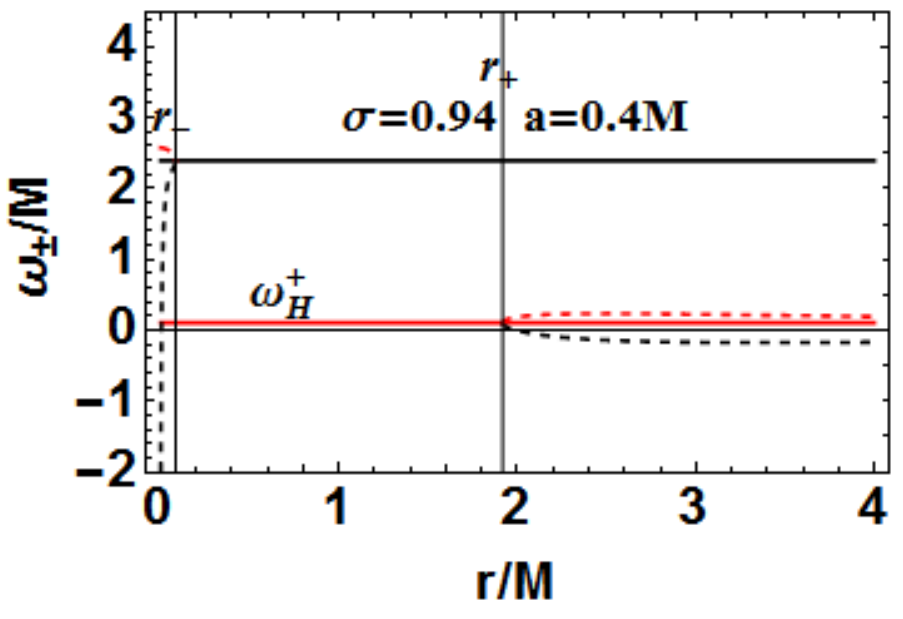}
\includegraphics[width=5cm]{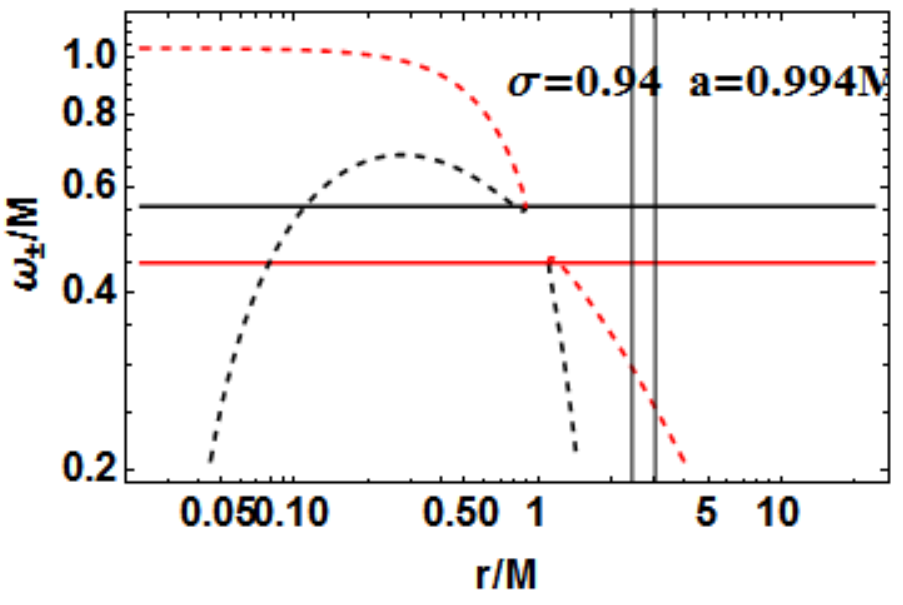}
\includegraphics[width=5cm]{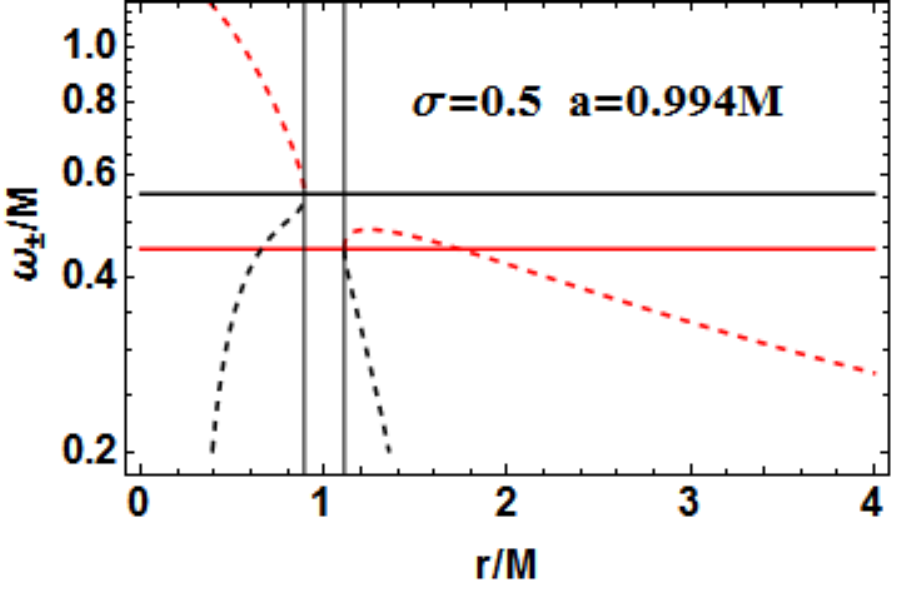}
\includegraphics[width=5cm]{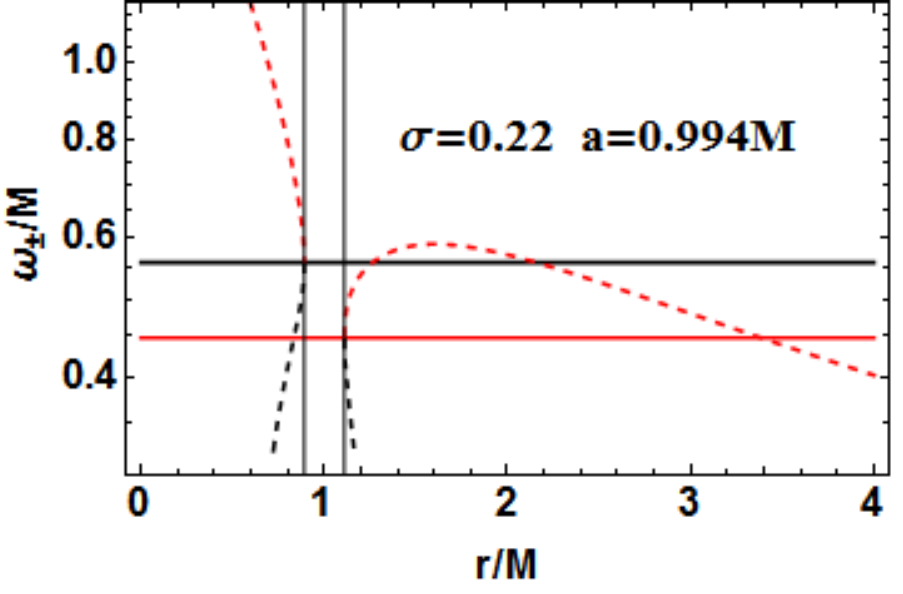}
\includegraphics[width=5cm]{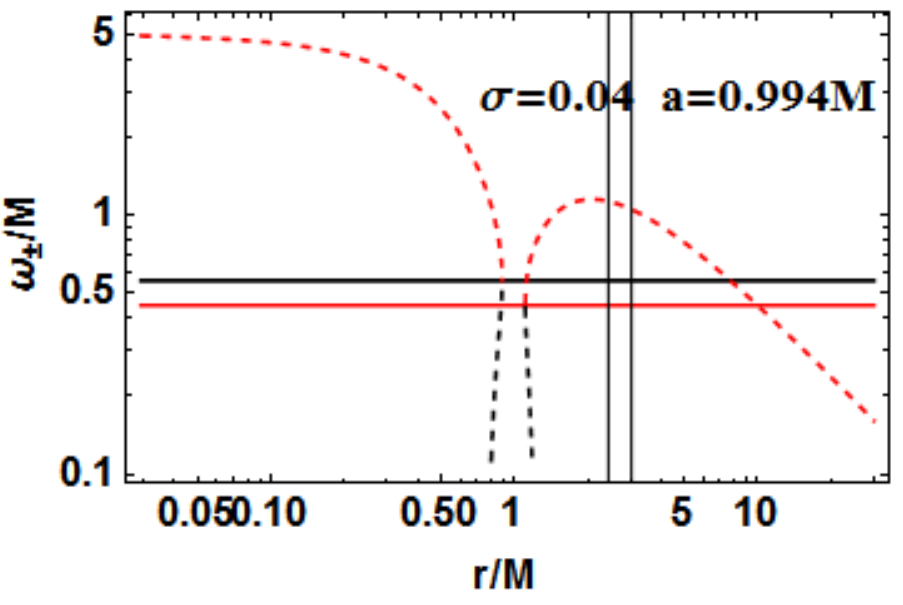}
  \caption{Frequencies $\omega_{\pm}$ for different spins $a/M$ and planes $\sigma$, as functions of $r/M$. The horizontal lines are horizon frequencies $\omega_H^{\pm}$, vertical lines are the inner and outer horizons $r_{\pm}$. There are solutions $\omega=\omega_H^{\pm}$ particularly for $r>r_+$, according to Eqs.\il(\ref{Eq:unirmm},\ref{Eq:unirmm1}).}\label{Fig:Werepers1}
\end{figure}

\begin{figure}
  \includegraphics[width=5cm,angle=90]{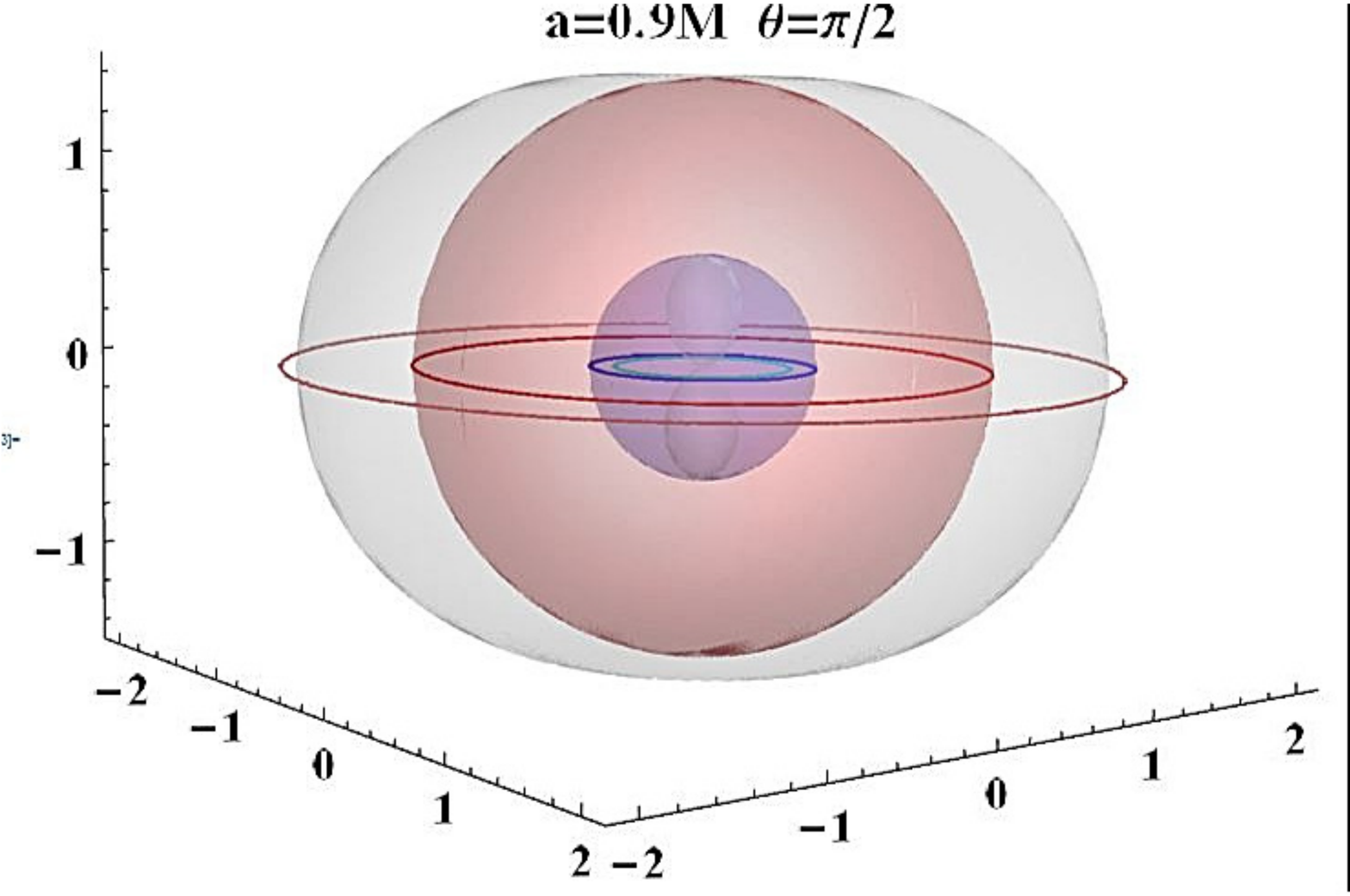}
\includegraphics[width=5cm]{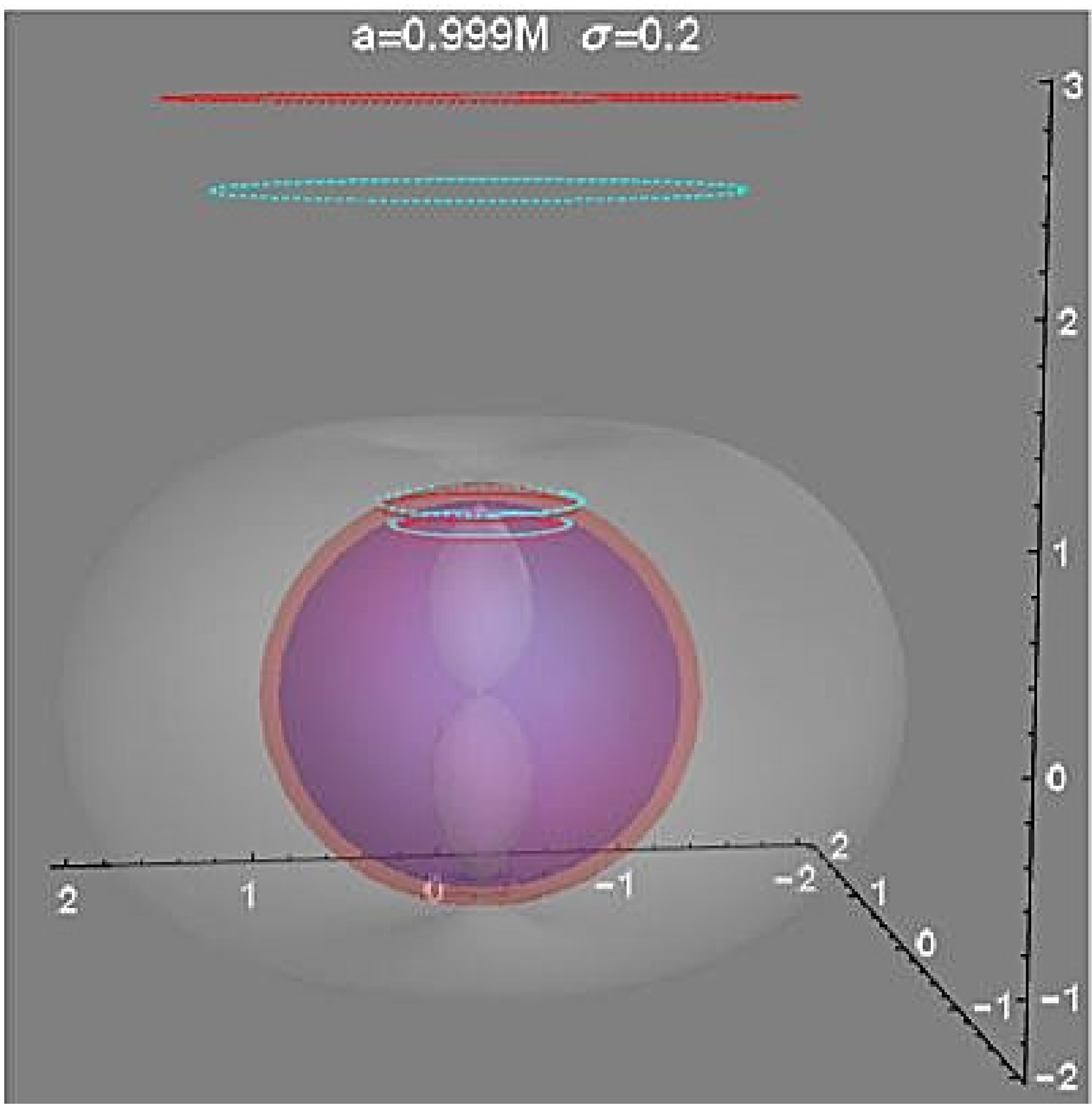}
\includegraphics[width=6cm]{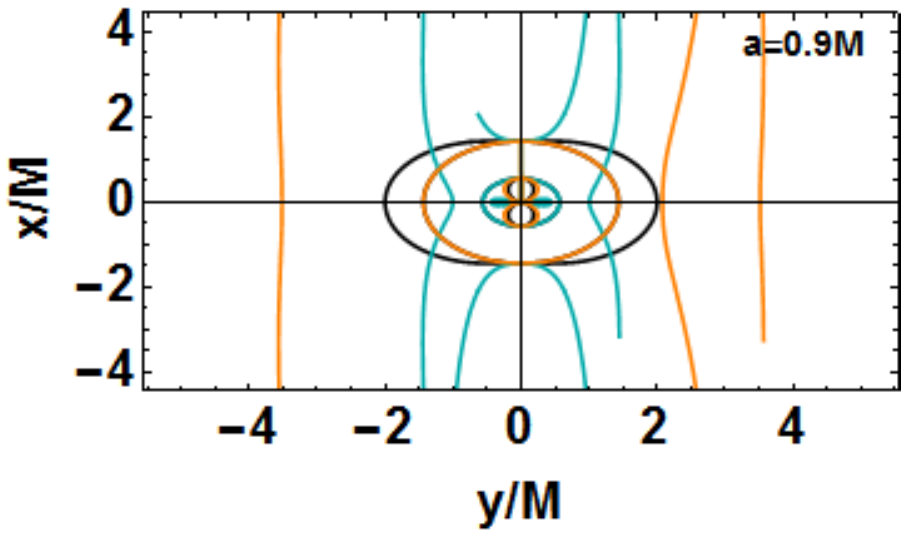}
\includegraphics[width=6cm]{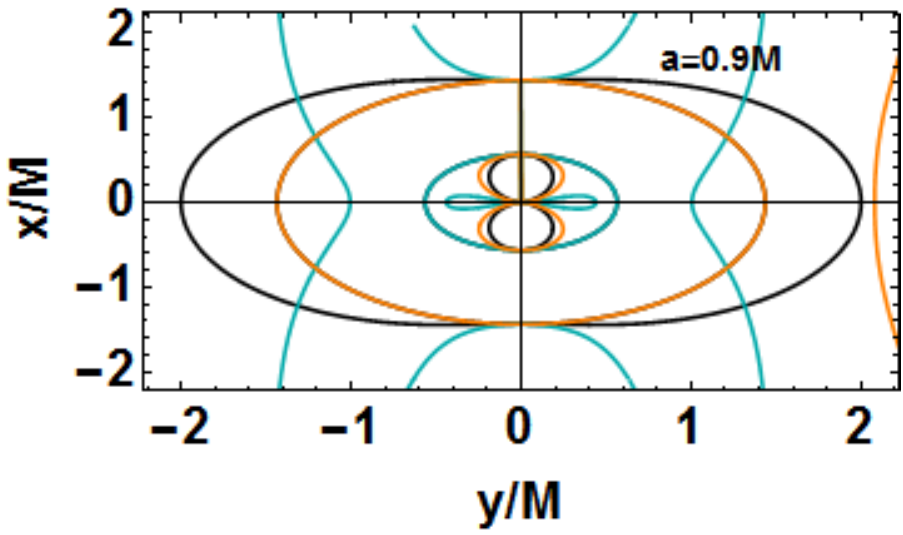}
\includegraphics[width=5cm]{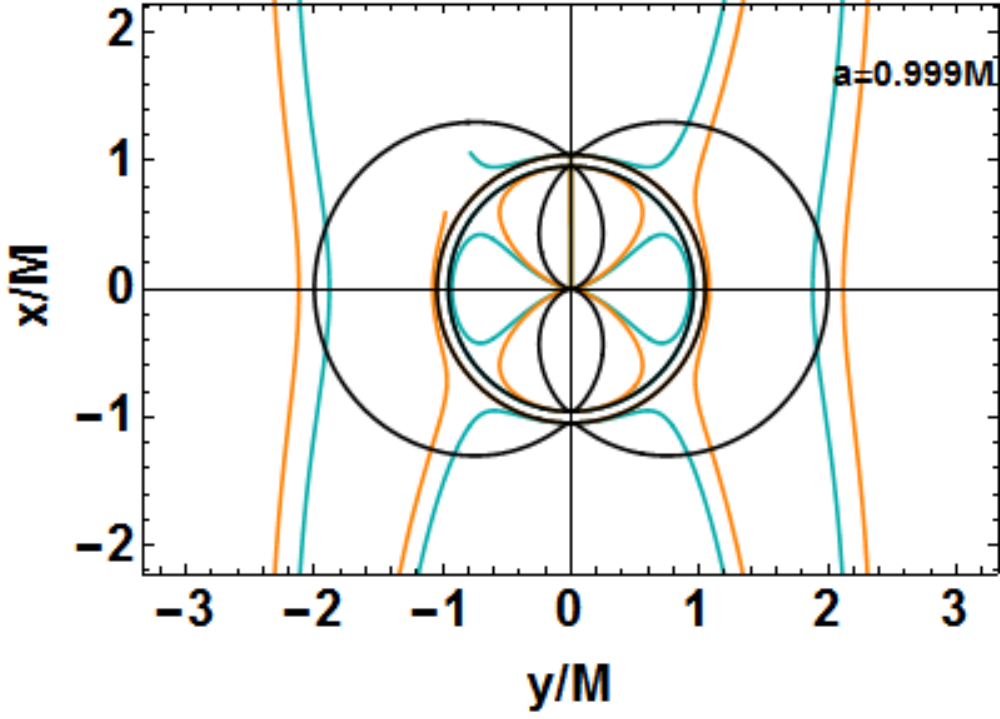}
  \caption{Horizon frequency solutions of the \textbf{MBs}  in the exterior region are horizontal lines of the extended plane in Figs\il(\ref{Fig:Ly-b-rty}). The plots show,  for fixed plane $\sigma\equiv(\sin\theta)^2$, the inner and outer horizons, the inner and outer ergosurfaces, the circle in the outer horizon, (red) on the inner horizon (blue) and  orbital solutions with frequency of the inner (blue) and outer (red) horizon. In the  frame we use,   the  \textbf{BH} Kerr  ring  singularity  is at
	$r=0$. In the lower panels, the light surface solutions
$\{r_s^1,r_s^2,r_s^4\}$ of  Eq.\il(\ref{Eq:rs1234})  are plotted  for $\omega=\omega^+_H$ (orange) in $\omega=\omega^-_H$ (darker cyan). Black curves are the inner and outer ergosurfaces and the inner and outer Killing horizons.
These orbits contain {replicas} of the horizons as defined  in Eq.\il(\ref{Eq:repliches}). }\label{Fig:JirkPlottGerm}
\end{figure}

As we shall see in detail also in
Sec.\il(\ref{Sec:allea-5Ste-cont}), bundles with  {origin spin in the weak black holes  (\textbf{WBH}) region of the extended plane}, i.e., with $\la_0=a_0\sqrt{\sigma}\leq4/5\leq0.8M$, for sufficiently large  values $\sigma\lessapprox1$ and particularly on the equatorial plane, are  \emph{entirely confined} in the \textbf{BH} region (the inner region of the extended plane as in Fig.\il\ref{Fig:Ly-b-rty}).
{It follows that their characteristic frequencies  (with tangent spin $r_g\in]0,2M/5[$ and characteristic frequencies  $\omega_b\geq1$)) can never be "replicated"   in those planes in  the outer region of the extended plane. However,  constraints can be found  in terms of the variable  $\la_0$. Frequencies $\omega_b\geq0.5$ (range of $\omega_H^-$) for   orbits in the outer region are shown in Figs\il(\ref{Fig:Werepers1},\ref{FIG:funzplo},\ref{FIG:disciotto1},\ref{FIG:rccolonog},
\ref{FIG:Aslongas},\ref{Fig:JirkPlottGerm}).}

 This is a consequence of the fact that if  the $(a_g, r_g)$ relations can be expressed  as functions of  $\la=a\sqrt{\sigma}$, this does not hold for  the quantity $\mathcal{L}_{\mathcal{N}}$; consequently, it follows that the bundles can extract ``information" on the inner horizons frequencies  near the rotation axis and, therefore, in this sense, the  inner region is not entirely confined.
Equally, \textbf{MBs} with weak naked singularities  (\textbf{WNS}) origins, i. e.,  with  $\la_0=a_0\sqrt{\sigma}\in ]M,2M]$, which complete the inner horizon with  tangency radius  $r_g\in]2M/5,M[$, spin $a_g\in]0.8M,M]$ and characteristic frequency $\omega_b\in[0.5,1[$),
for small $\sigma$ can be found  in the outer region of the extended plane.  We shall see this in details in Sec.\il(\ref{Sec:allea-5Ste-cont}).

For the equatorial plane, $\sigma=1$, this condition was considered in detail in  \cite{remnants}, where  the concept of inner horizon confinement was introduced. We  resume this issue  as follows (see Fig.\il\ref{Fig:Werepers1}). It was shown that there are  two radii
$r_\pm^\pm$ such that
$r_-^-<r_-<r_+<r_+^+$ almost everywhere (exceptions are at $a_g=0$ and  $a_g=M$) and
$\omega_*(r_\pm^\pm)=\omega_H^\pm$, respectively.

The problem of finding solutions of $\mathcal{L}_{\mathcal{N}}=0$ for $\omega_H^{\pm}$ is clearly related to the characterization of   the  \textbf{MBs} $\Gamma_{a_g,\omega_b}$,  constituted by the \textbf{MBs} $\Gamma_{a_g}$   with equal tangent spin $a_g$ and
already analyzed in Eqs.\il(\ref{Eq:enagliny}), \emph{and} equal frequency $\Gamma_{\omega_b}$
(and equal tangent radius $r_g$) also analyzed in Eqs.\il(\ref{Eq:trav7see}). Therefore, the  particular relations given
in Eqs\il(\ref{Eq:weth-mome}) are valid.

 There  are  then  two classes of \textbf{MBs} to be considered.
Firstly, we focus on   equal $\sigma$ solutions, discussed in
Eq.\il(\ref{Eq:weth-mome}). Solutions $r: \omega_{\pm}(r)=\omega_H^{\pm}(a_g)$ belong to the same bundle  and, therefore,
depend on the bundle curves bending (curvature) on the extended plane. In fact, as follows from
Figs\il\ref{FIG:funzplo},\ref{FIG:disciotto1},\ref{FIG:rccolonog} and \ref{FIG:Aslongas}, the situation is  simple in the case $\sigma=1$,
 where there are two bundle branches (closed on the horizons) {corresponding to} the two solutions $r_{\pm}^{\pm}$, respectively.
Alternatively, the analysis can be performed by considering the frequency solutions  (\ref{Eq:bab-lov-what}) directly
(see Fig.\il(\ref{Fig:Werepers1}),
where the horizon-frequency solutions   are given by horizontal lines {(corresponding  to $\omega=\omega_H^{\pm}$)}  on the extended plane.

Solutions for equal planes and equal tangent frequencies
 belong to the same metric bundle;
therefore, they must share the same origin $a_0$.
{(The relevance of the equatorial case $\sigma=1$ for the Kerr \textbf{MBs} analysis  lies also in the fact that  for the spherically symmetric, static and  electrically charged Reissner Nortstr\"om solution,   the \textbf{MBs} curvature properties in the extended plane are  similar to the off-equatorial case of the Kerr geometry, showing the interplay between symmetries and geometry
in defining the \textbf{MBs} properties-- \cite{remnants}.)}
Then, for a general  $\sigma$,  we consider  the two-faced problem for the orbits of frequencies $\omega_H^{\pm}$, considering the \textbf{MBs}
$\Gamma_{a_g}^-\equiv\{g_{\omega}(a_g,r_g^-)\}_{\sigma}$ and $\Gamma_{a_g}^+\equiv\{g_{\omega}(a_g,r_g^+)\}_{\sigma}$,
 where $r_{g}^{\pm}$  are, respectively, the tangent points of the two bundle families on the outer and inner horizons of  the \textbf{BH} with spin $a_g$.
To these \textbf{MBs} correspond, respectively, the horizon frequencies
$\omega_b^->\omega_b^+$ of the   \textbf{BH} spacetime with spin $a=a_g$.
In the second case, we consider  in the classes $\Gamma_{a_g}^\pm$  the  pairs
$(g_{\omega}(a_g,r_g^-,\sigma),g_{\omega}(a_g,r_g^+,\sigma))
\in\Gamma_{a_g,\sigma}^-\times\Gamma_{a_g,\sigma}^+$
i.e.   with the same $a_g$ on the same $\sigma$
(recall that the horizon frequencies
$\omega_b^{\pm}$  are  independent of $\sigma$, a property  linked to the \textbf{BHs} horizon rigidity).

\begin{figure}
    \includegraphics[width=5cm]{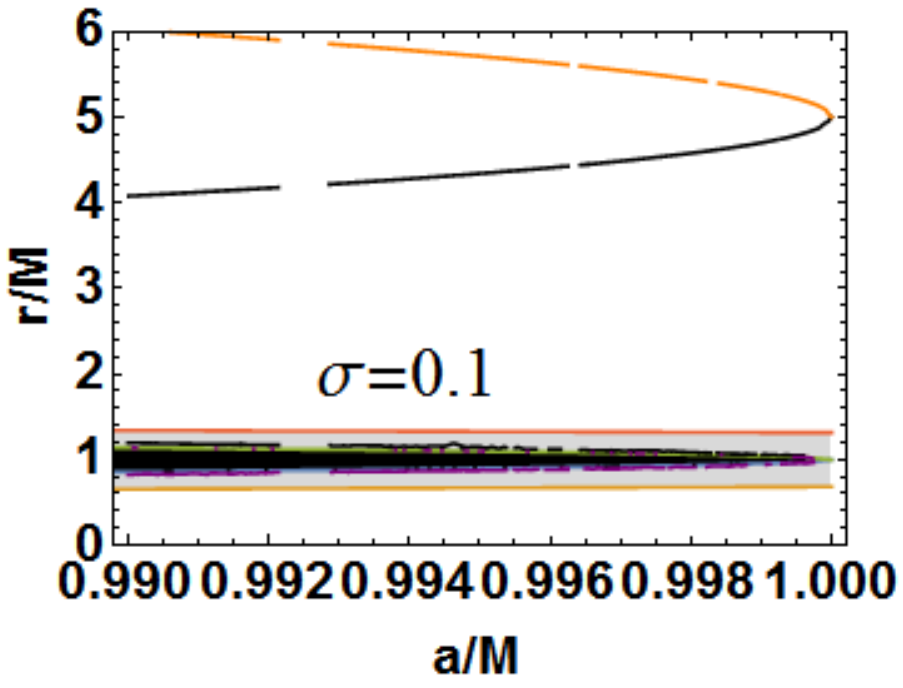}
  \includegraphics[width=5cm]{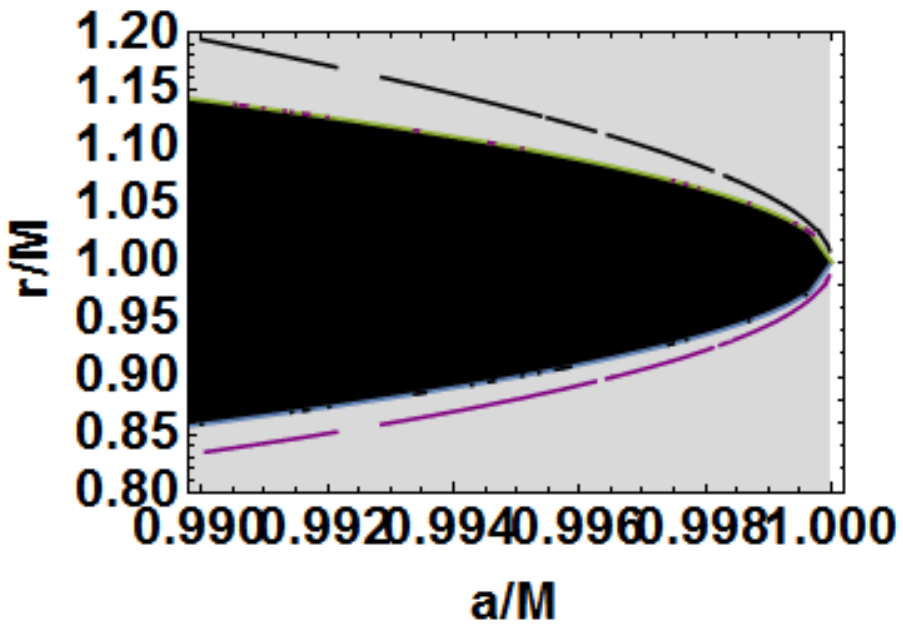}
   \includegraphics[width=5cm]{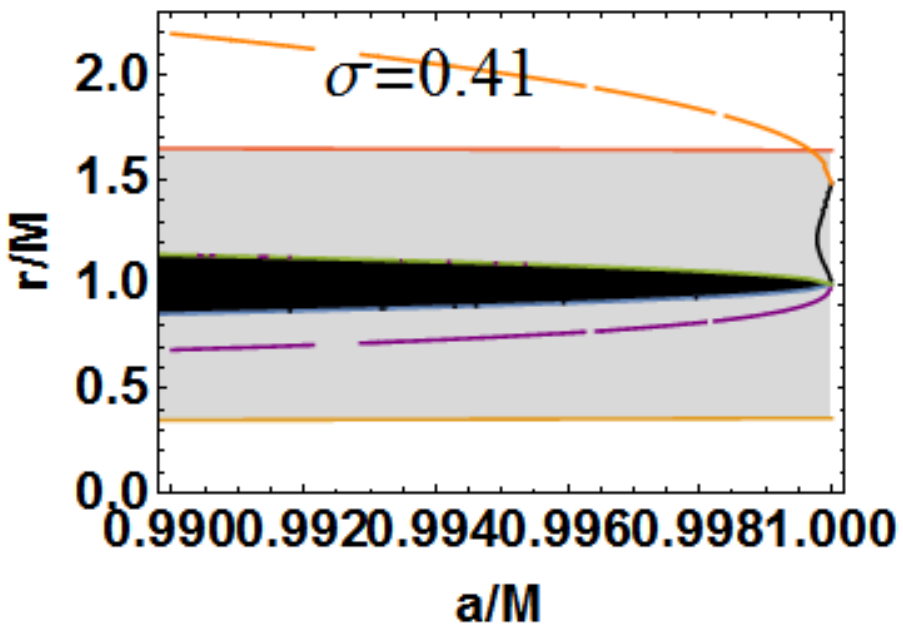}
  \includegraphics[width=5cm]{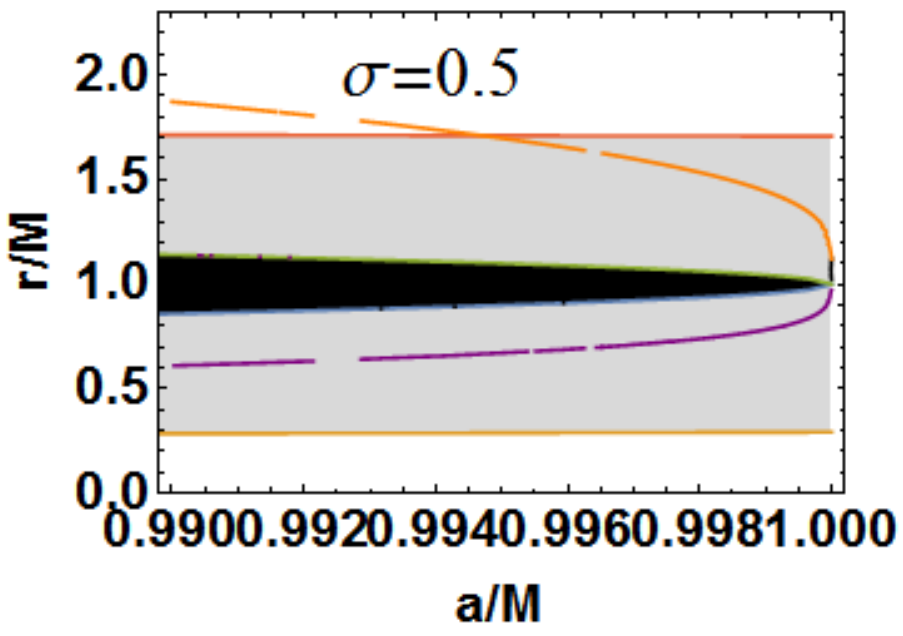}
  \includegraphics[width=5cm]{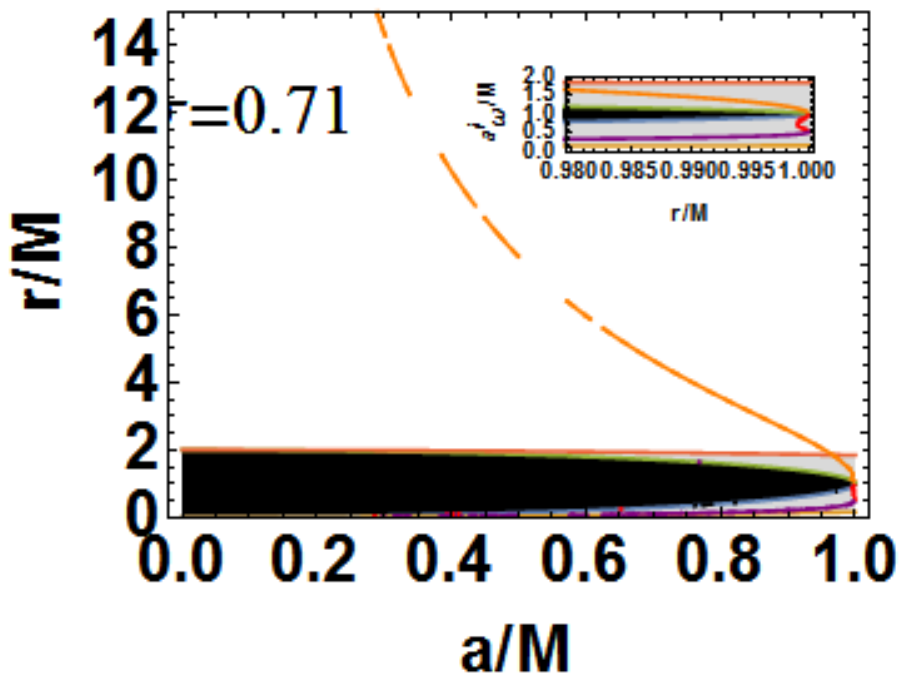}
      \includegraphics[width=5cm]{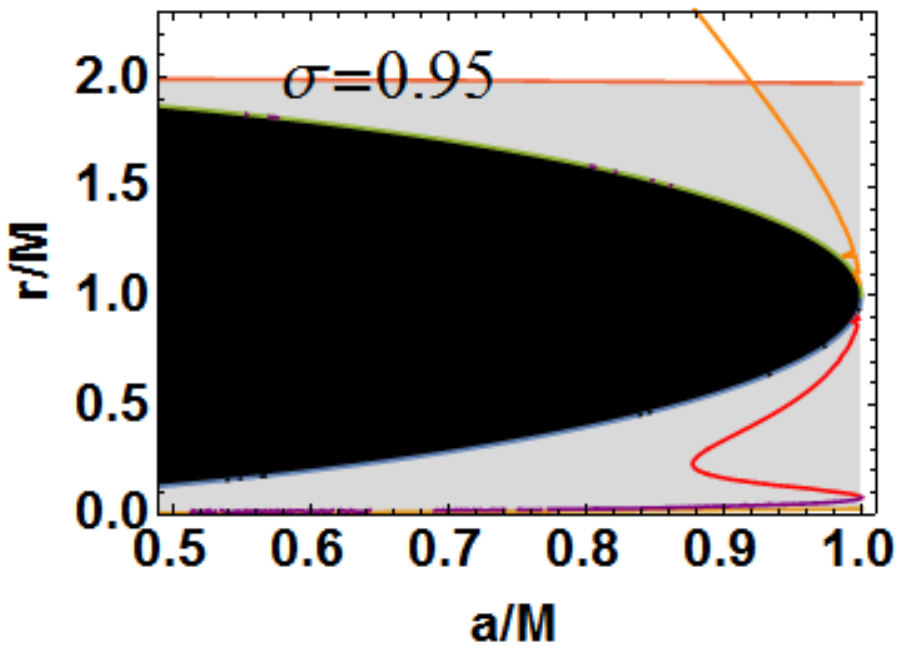}
  \includegraphics[width=5cm]{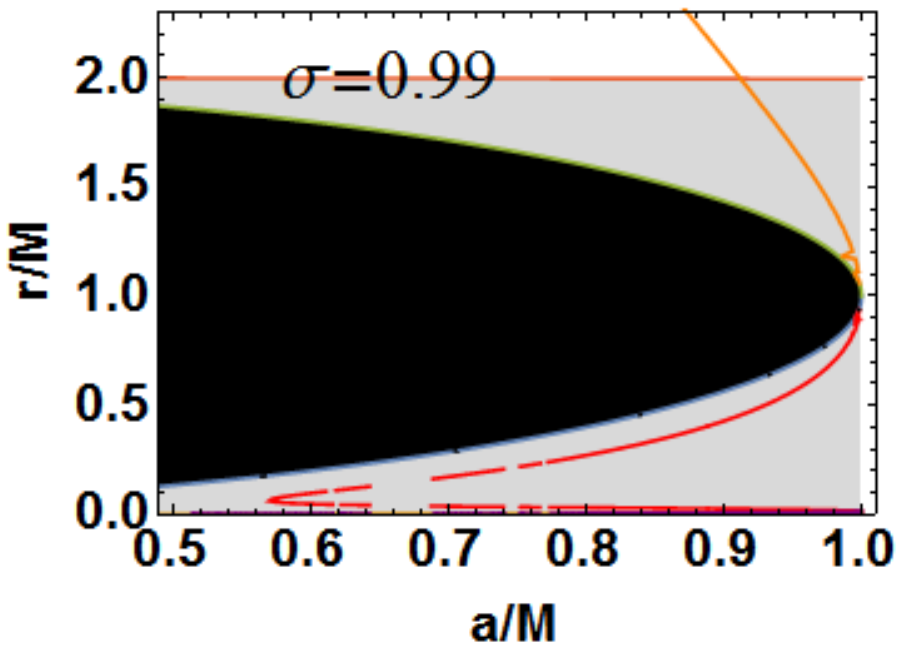}
   \includegraphics[width=5cm]{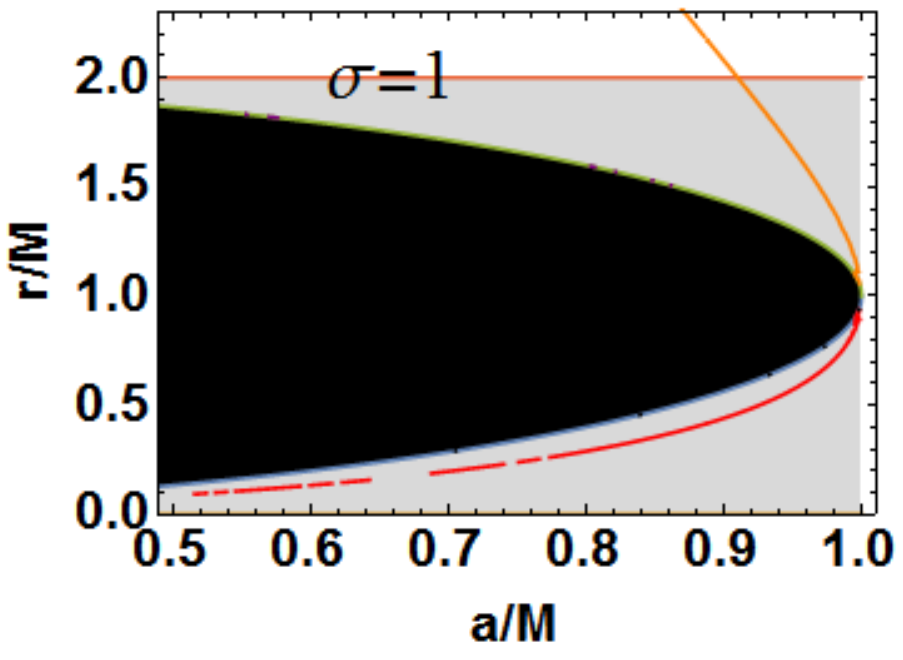}
   \caption{Photon orbital frequencies and horizon frequencies. In the plane $(a/M, r/M)$, for different planes $\sigma$,
	the solutions $\omega_+=\omega_+(r_-)$  red curve,
$\omega_-=\omega_-(r_-)$  black curve,
$\omega_-=\omega_-(r_+)$ orange curve,
$\omega_+=\omega_+ (r_+)$ purple curve. The black region is a \textbf{BH} on the extended plane $r\in[r_-,r_+]$, the gray region is the ergoregion.}\label{FIG:ARESE1}
\end{figure}

We now focus on the range  $a\in[-M,M]$ of \textbf{BH} spacetimes.
In fact,  for  different $a$, solutions are provided as  \textbf{MBs} intersections which, for each plane $\sigma$ at each point
$r$ (except for the origin), involve  only two bundles.  This case is considered in Fig.\il(\ref{FIG:ARESE1}), where we see  solutions
for $r\neq r_{\pm}$ for  small values of $\sigma$ and  for  $r_-^o>r_+$ with frequencies $\omega_{H}^-$, that is,
 equal to the frequency of the inner  horizon which, therefore, is not confined  to $r<r_-$.

\subsubsection{Results}

We study the problem of finding solutions with a frequency that coincides with that of the inner  horizon $\omega_H^-$. There are two  possibilities:

\textbf{(1)} There are no solutions of
$\omega_+=\omega_+(r_-)$ for $r>r_-$
in $a\in[0,M]$.

\begin{figure}
  \includegraphics[width=5.cm]{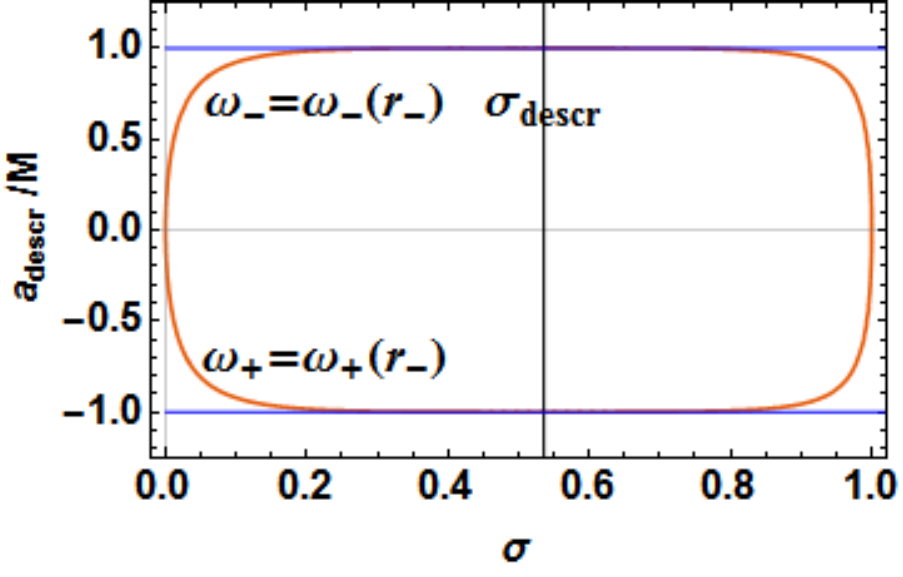}
    \includegraphics[width=5.6cm]{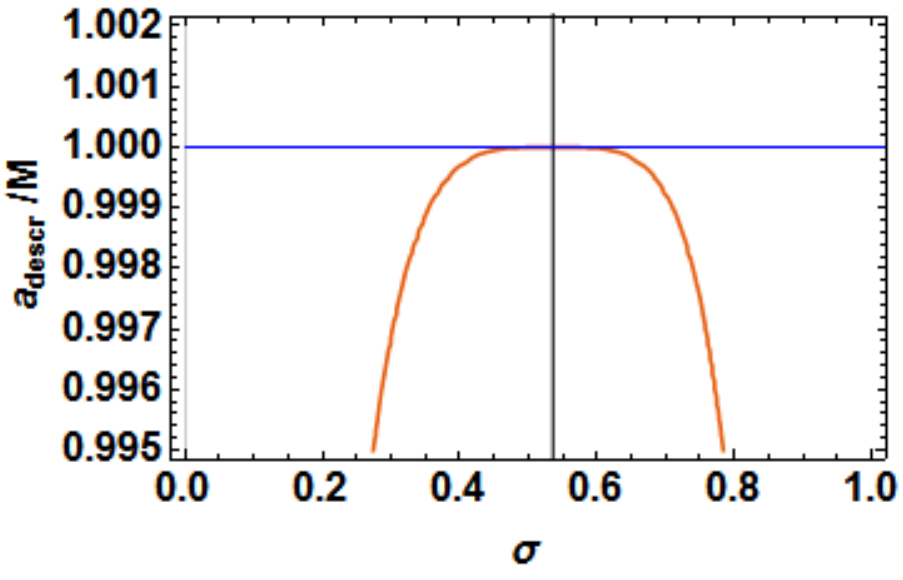}
  \includegraphics[width=6cm]{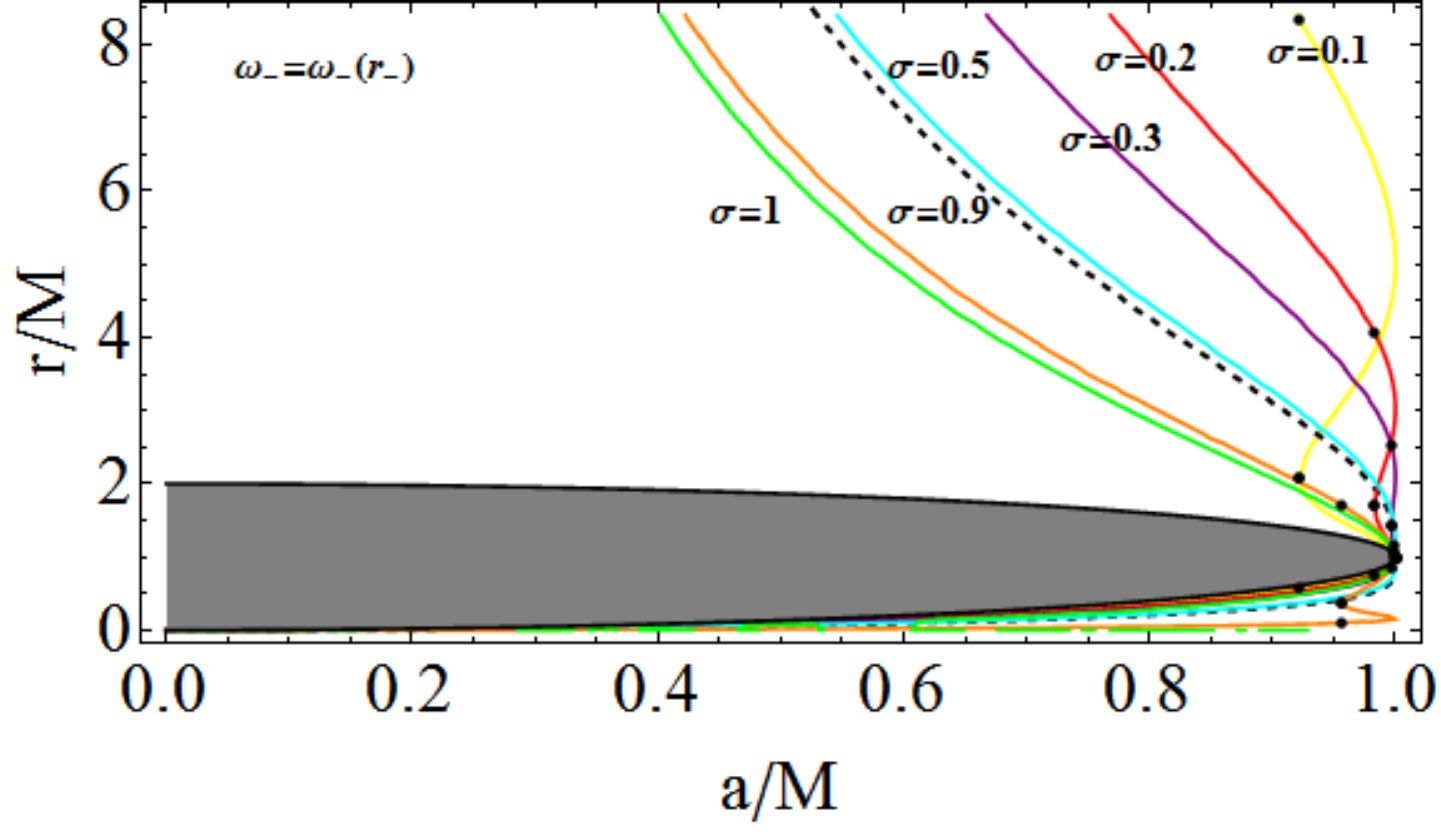}
  \includegraphics[width=5.6cm]{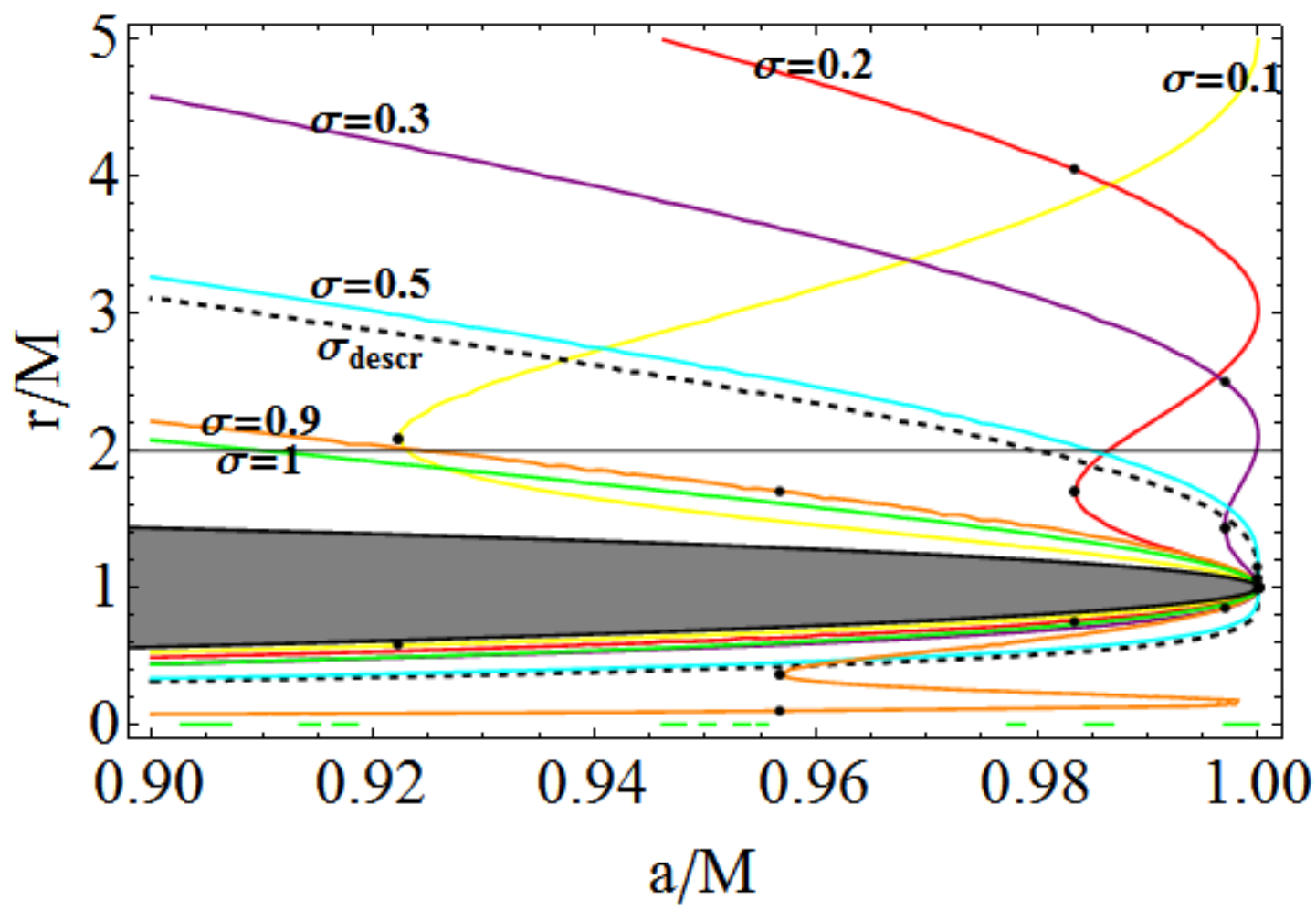}
  \includegraphics[width=5.6cm]{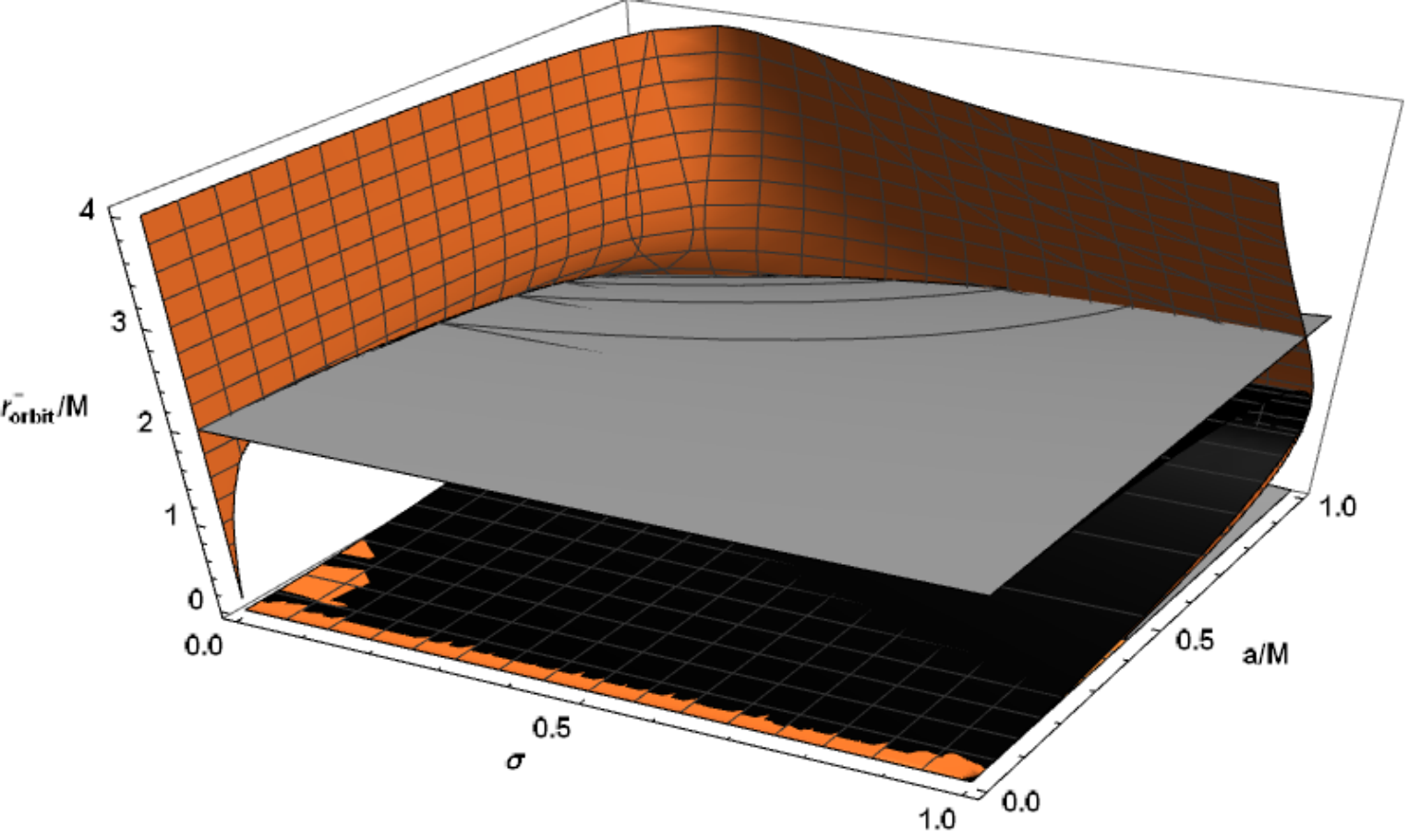}
  \includegraphics[width=6cm]{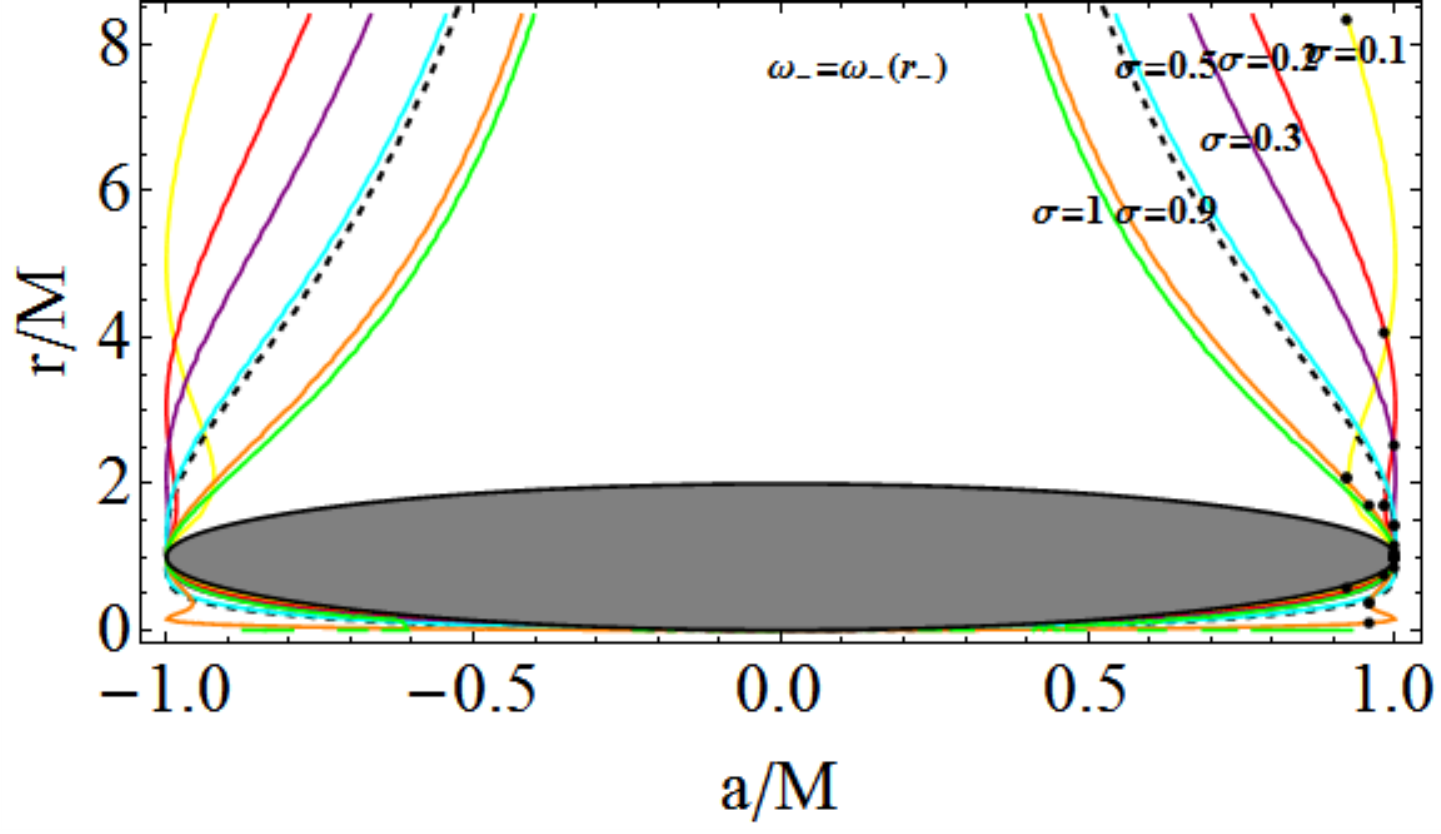}
 \caption{Photon orbital frequencies as horizons frequencies. Solutions  of
$\omega_-=\omega_-(r_-)$
for $a\in [0,M]$,
and of
$\omega_+=\omega_+(r_-)$
for $a\in[-M,0]$
 for $r>r_-$--Eqs\il(\ref{Eq:unirmm}) and  (\ref{Eq:5sta-r-dibatt}).
Left panel: $a_{descr}/M$ as function of $\sigma\equiv\sin^2\theta$.
Center panel:  solutions $\omega_-=\omega_-(r_-)$ on the plane $(r/M,a/M)$ for different planes $\sigma\in[0,1]$;
a zoom for $a\approx M$ is in the right panel. The black region is a \textbf{BH} on the extended plane ($r\in[r_-,r_+]$). Below panel: solutions $\omega_-=\omega_-(r_-)$ as functions of $r/M$ for $\sigma\in[0,1]$ and $a\in[0,M]$. The gray region represents the ergosphere on the extended plane.}\label{FIG:ARESE1Pola1}
\end{figure}

\textbf{(2)} There are instead solutions  of
$\omega_-=\omega_-(r_-)$. In this case, we express the solutions  in terms of the plane $\sigma$:
\bea\label{Eq:5sta-r-dibatt}
&&
\sigma_{\gamma}^\pm\equiv \frac{\mathcal{B}}{2}\pm\mathcal{LL}-\frac{\sqrt{\mathcal{B}^2+\mathcal{F}-\mathcal{Y}}}{2},
\quad\mbox{where}\quad
\mathcal{B}\equiv\frac{a^4+2 a^2 r^2-4 a^2+r^4-8 r+8}{a^2 \left(a^2+r^2-2 r\right)}, \quad\mbox{and}\quad \mathcal{F}\equiv-\frac{8 \left(a^2-2\right)}{a^4},
\\&&\nonumber
\mathcal{LL}\equiv\frac{1}{2} \sqrt{2 \mathcal{B}^2-\frac{8 \mathcal{B}^3-\mathcal{M}-\mathcal{R}}{4 \sqrt{\mathcal{B}^2+\mathcal{F}-\mathcal{Y}}}-\mathcal{F}-\mathcal{Y}};
\quad
\mathcal{M}\equiv\frac{64 \left(a^6+2 a^4 r^2-6 a^4+a^2 r^4-4 a^2 r^2+8 a^2 r-2 r^4\right)}{a^6 \left(a^2+r^2-2 r\right)};
\\\nonumber
&&\mathcal{R}\equiv8 \left(a^4+2 a^2 r^2-4 a^2+r^4-8 r+8\right)\cdot
\\\nonumber
&&\frac{\left(a^6+3 a^4 r^2+2 a^4 r-16 a^4+3 a^2 r^4+4 a^2 r^3-12 a^2 r^2-16 a^2 r+48 a^2+r^6+2 r^5+4 r^4-8 r^3+16 r^2-32 r\right)}{a^6 \left(a^2+r^2-2 r\right)^2};
\\\nonumber
&&\mathcal{Y}\equiv\frac{a^6+a^4 (r-2) (3 r+8)+a^2 [(r-2) r (r+2) (3 r+4)+48]+r [r (r [r (r (r+2)+4)-8]+16)-32]}{a^6+a^4 (r-2) r}.
\eea
The behavior of these quantities is illustrated  in Figs\il\ref{FIG:ARESE1Pola1} and \ref{FIG:ARESE1}.

\subsubsection{The discriminant spin $a_{descr}(\sigma)$}

For the problem under consideration an important quantity is the  {discriminant spin}
$a_{descr}(\sigma)\in[0,M]$,  which is a solution of the following equation
\bea&&
a^{10} (\sigma -1)^2 \sigma ^6-2 a^8 (\sigma -1) \sigma ^4 [3 \sigma  (5 \sigma -28)+68]+a^6 \sigma ^2 [\sigma  (\sigma  [\sigma  (201 \sigma +1160)-2784]+1408)+16]+
\\&&\nonumber+8 a^4 \sigma  [\sigma  (\sigma  [(1045-411 \sigma ) \sigma -1228]+588)+16]+16 a^2 (\sigma -1) [\sigma  (\sigma  [97 \sigma +447]-312)-16]-6912 (\sigma -1)^2 \sigma =0.
\eea
The radii $r$ corresponding to  frequencies equal to that of the horizon are solutions of the equation
\bea&&\label{Eq:par.notre}
a^2 r^6 \sigma ^2+2 a^2 r^5 \sigma ^2+4 a^2 \left(a^2-2\right) r^3 \sigma ^2-2 a^2 r (\sigma -1) \left(a^2 \sigma -4\right) \left[\sigma  \left(a^2 (\sigma +1)-4\right)+4\right]+
\\\nonumber
&&+r^4 \sigma  \left[a^4 \sigma  (3-2 \sigma )+4 a^2 (\sigma +2)-16\right]+a^2 r^2 \left(\sigma  \left(a^4 (\sigma -3) (\sigma -1) \sigma +4 a^2 (4-3 \sigma )+\right.\right.
\\\nonumber
&&+\left.\left.16 (\sigma -2)\right)+16\right)+a^4 (\sigma -1)^2 \left(a^4 \sigma ^2+8 \left(a^2-2\right) \sigma +16\right)=0.
\eea
More precisely, there are solutions of the problem  $\omega_-=\omega_-(r_-)$ as follows:
\bea&&\label{Eq:unirmm}
\textbf{Solutions $\omega_-(r)=\omega_-(r_-)$ for $r>r_-$:}
\\\nonumber
&&
\mbox{For}\quad\sigma\in]0,\sigma_{descr}[ \quad\mbox{where}\quad \sigma_{descr}\approx0.535898:
\\&&\nonumber
\mbox{and  \textbf{(a)}} \quad a=M \quad \mbox{with} \quad  r=\hat{r}_-^-\quad\mbox{or \textbf{(b)}}\quad\mbox{for} \quad a\in[a_{descr},M[\quad r=\check{r}_-^-\quad\mbox{and \textbf{(c)}} \quad a\in]a_{descr},M[ \quad \mbox{for} \quad r=\hat{r}_-^-
\\
&&\nonumber
\\
&&\label{Eq:unirmm1}
\textbf{Solutions $\omega_-(r)=\omega_-(r_-)$ for $r<r_-$:}
\\\nonumber
&&
\mbox{For}\quad
\sigma\in[\sigma_{descr},M[\quad\mbox{and}\quad a=a_{descr}\quad\mbox{with}\quad r=r_{\star}\quad\mbox{or}\quad a\in]a_{descr},M[\quad\mbox{with}\quad  r\in(r_{\star}, r_{\circ})
\quad
\mbox{or}\quad (a=M, r=r_{\diamond})
\\
&&\nonumber \mbox{For}\quad
\sigma =1\quad\mbox{and}\quad a\in]0,M[ \quad\mbox{with }\quad r=r_-^-.
\eea
This implies  that the occurrence of orbits $r>r_-$ with  photon orbital frequency  equal to that of the \emph{inner} horizon is expected for  small $\sigma$ ($\sigma\in[0,\sigma_{descr}]$) and larger  spins  ($a\in[a_{descr},M]$)--Figs\il\ref{FIG:ARESE1Pola1},
\ref{FIG:ARESE1} and \ref{Fig:Werepers1}.
It should be noted, as evident also from Fig.\il\ref{Fig:Werepers1}, that
$a_{descr}\approx M$ for $\sigma\lnapprox M$ and $\sigma\gnapprox 0$. Finally, the radii $(r_-^-,r_{\diamond},r_{\star}, r_{\circ},\hat{r}_-^-,\check{r}_-^-)$ of
Eqs.\il(\ref{Eq:unirmm} and (\ref{Eq:unirmm1}), are solutions of Eq.\il(\ref{Eq:par.notre}),
  where $r_-^-$ is provided in  \cite{remnants}.

We  now consider  negative spins for retrograde orbits, i.e., $a\in[-M,0]$.
There are solutions of
$\omega_+=\omega_+(r_-)$  for $r>r_-$ similarly to  the conditions (\ref{Eq:unirmm}) and (\ref{Eq:unirmm1})
in $a\in[-M,0]$--Fig.\il(\ref{Fig:Werepers1}).  We also address this analysis in more details in Sec.\il(\ref{Sec:contro-orbits-omega}).

There are no solutions of  $\omega_{-}=\omega_{H}^{+}$ for  $r>r_-$. Moreover, for
 $\omega_{+}=\omega_{H}^{+}$, we obtain
\bea
&&\label{Eq:unirmm2}
\textbf{Solutions $\omega_+(r)=\omega_+(r_+)$ for $r\geq r_-$:}
\\\nonumber
&& \mbox{For}\quad
\sigma\in]0,\sigma_{descr}[\quad\mbox{and \textbf{(a)}}\quad a\in]0,a_{descr}[\quad r\in(r_+,{\hat{\hat{r}}})\quad\mbox{or \textbf{(b)}}\quad a\in[a_{descr},M[\quad r\in(r_+,r,{\tilde{r}})\quad\mbox{or \textbf{(c)}}\quad  (a=M, r={\breve{\breve{r}}})
\\\nonumber
 &&\mbox{For}\quad\sigma=\sigma_{descr} \quad\mbox{and}\quad a\in]0,M[\quad  r\in(r_+,r_+^+).
  \\
&&\nonumber\mbox{For}\quad\sigma\in[\sigma_{descr},M[  \quad\mbox{and \textbf{(a)}}\quad  a\in]0,a_{descr}[\quad r\in(r_+,{\hat{\hat{r}}})\quad \mbox{or \textbf{(b)}}\quad a\in[a_{descr},M[\quad r\in(r_+,{\tilde{r}}).
\\
&&\nonumber \mbox{For}\quad\sigma =1\quad \mbox{and}\quad a\in]0,M[\quad r\in (r_+, {\tilde{r}}).
\\&&\nonumber\\
&&\label{Eq:unirmm3}
\textbf{Solutions $\omega_+(r)=\omega_+(r_+)$ for $r<r_-$:}
\\\nonumber
&&\mbox{For}\quad \sigma\in]0,\sigma_{descr}]\quad \mbox{and}\quad a\in]0,M[\quad r=\tilde{\tilde{r}}.
\\
&&\nonumber\mbox{For}\quad \sigma\in[\sigma_{descr},M[\quad \mbox{and}\quad   a\in]0,M]\quad r=\tilde{\tilde{r}}.
\eea
Note that the radii $(\hat{\hat{r}},\tilde{r},\breve{\breve{r}},\tilde{\tilde{r}})$ are solutions of Eq.\il(\ref{Eq:par.notre}), where
$r_+^+$ is given in \cite{remnants}.
For  $a\in[-M,0]$, there exist solutions  for $\omega_{-}=\omega_{H}^{+}$, similarly to  the former case
(in fact, the equations have some remarkable symmetries).
The second frequency at  point  $r$ is
\bea\label{Eq:re-set}
\omega_+(r)=\omega_H^\pm(r)+\Delta\omega(r),\quad \omega_-(r)=\omega_H^\pm(r)-\Delta\omega(r),
\quad
\Delta \omega(r)\equiv\frac{4 \sqrt{\sigma  \Delta \Sigma^2}}{\sigma  \left[a^2 \sigma  \Delta-\left(a^2+r^2\right)^2\right]}.
\eea
%
In Sec.\il(\ref{Sec:replicas}), we complete this analysis by introducing the definition of \emph{horizon replicas}. We investigate the vertical lines of the extended plane crossing the horizon curve, in other words,
 the metric bundles with characteristic frequencies $\omega_b(a)\in\{\omega_H^+(a_p),\omega_H^-(a_p)\}$ with orbits  located
exactly in $r_\pm(a_p)>r_+(a)$, that is, on  the horizon with frequency $\omega_b(a)=\omega_H^\pm(a_p)$.

\subsubsection{On retrograde solutions with frequencies equal to the horizon frequencies.}

We also investigate photon orbits,  solutions of $\mathcal{L}\cdot\mathcal{L}=0$,  with frequencies  equal and opposite to the
 horizon frequencies.
{Because of the symmetries we can  restrict ourselves to the case
$\omega>0$ and $a\in]-M,0[$. We recall that the condition $\mathcal{L}_{\mathcal{N}}=0$  is symmetric under the  inversion $a\rightarrow -a$ and
$\omega\rightarrow -\omega$, i.e.
$\mathcal{L}_{\mathcal{N}}(a,-\omega)=\mathcal{L}_{\mathcal{N}}(-a,\omega)$ and thus
$\mathcal{L}_{\mathcal{N}}(-a,-\omega)=\mathcal{L}_{\mathcal{N}}(a,\omega)$.}
Our analysis is not {focused on} the retrograde case (negative frequencies); however, this case is relevant because  the corresponding  frequencies
are equal in magnitude  to the horizon frequencies. A systematic analysis of this case  is considered in
Sec.\il(\ref{Sec:contro-orbits-omega}).

\section{Concluding Remarks}\label{Sec:conclu}
In this work, we presented a complete characterization  of the \textbf{MBs} of the Kerr geometries.
Metric bundles are curves of the extended plane, which represent collections  of   \textbf{BH} geometries or \textbf{BH} and \textbf{NS} geometries.  All the geometries of a bundle   have the same limiting photon (orbital) frequency $\omega=\omega_b$, the characteristic frequency of the metric bundle, which coincides with  the horizon frequency $\omega_H^{\pm}$  corresponding to the   tangent point $a_g$  of the metric bundle with the horizon curve $a_{\pm}$ on the extended  plane.  All the  \textbf{MBs} are tangent to the horizon curve on the extended  plane and, viceversa, the horizon curve emerges as  the envelope surface of all the  \textbf{MBs}. In this sense, the horizons are constructed by  all the \textbf{MBs} on the extended  plane throughout the tangency property.
In the sense of the  frequencies coincidence, the horizon curve on the extended  plane contains the information   of all the Kerr geometries.  The \textbf{MBs} frequencies (and the  limiting frequencies at any point of any spacetime of the bundle)    are all and only the horizon frequencies defined on the extended  plane.

We summarize below the main steps of this analysis, briefly commenting  on the  main results regarding the bundle structures. We conclude this final section of our work with  comments on some aspects  relating \textbf{MBs} to  various  astrophysical contexts involving \textbf{BHs} in interaction with its   environments of matter and fields and, more generally, in processes of  formation and evolution of singularities. This work is also  part of the more general study of the  significance and interpretation of naked singularity geometries.

\medskip
\textbf{Analysis and results}
\medskip

\begin{description}
\item[\textbf{MBs} in the extended plane]
In Sec.\il(\ref{Sec:bundle-description}), we analyzed special sets of metric bundles    $\Gamma_x$,  which are characterized by the constant parameter $x$, such  as the bundle origin $a_0$ or the bundle frequency $\omega_b$, and  provide a division of the extended plane into parts. We used this particular split to  construct  the Kerr extended plane in the representation of
Fig.\il(\ref{Fig:Ly-b-rty}), through transformations  presented and discussed in Eq.\il(\ref{Eq:A0-A0-Atanget}) and
Fig.\il(\ref{Fig:meniangu}), where we pointed out the particular role of the  parameter
$\la \equiv a\sqrt{\sigma}$--Fig.\il(\ref{FIG:raisemK}).
\item[Horizons confinement]
We analyzed in detail  the conditions for the inner horizon confinement, as seen   from the exterior region of the extended plane. The inner horizon confinement is a condition involving, in general,  the entire extended plane; therefore, we also addressed  this problem  in the case of naked singularities and,  in particular, in the bottleneck region of the extended plane, which is characterized by the horizon remnants region in each \textbf{NS} geometry, as seen in
Fig.(\ref{Fig:Ly-b-rty}).
It turned out that the conditions are similar to the case of the equatorial plane ($\sigma=1$), if we  consider the  quantities parameterized by $\la\equiv a\sqrt{\sigma}$--Sec.\il(\ref{Sec:allea-5Ste-cont}) and Fig.\il(\ref{FIG:raisemK}).
We studied in detail black hole spacetimes.
The confinement of the inner horizon was discussed in various   parts of this work, for example, in
Fig.\il(\ref{Fig:Pnoveplot}) by considering the dependence on the  plane $\sigma$  and different parametrizations  for the \textbf{MBs}. We proved, however, that the confinement    can be overcome by the horizons replicas {(leading to    "frequencies extraction" through bundles in the outer region)}.
There are  solutions  with frequencies equal to the inner (and outer) horizon, which are regulated by  constrains on
$\la=a\sqrt{\sigma}$.
\item[Horizons replica] Our study
led to the  introduction of   the  horizon replicas  by considering the bundle orbits $r$ in  a specific geometry,   other  than that of \textbf{BH} horizons,  but characterized by  photons  with orbital frequency  equal to the (inner or outer) horizon frequency  of the \textbf{BH} spacetime.
In fact, this constitutes an aspect of information extraction of the \textbf{BH} properties into the exterior region of the extended plane. {Examples of  replicas are given in Figs\il(\ref{Fig:JirkPlottGerm}).}
The existence of  horizon replicas depends on the angle $(\sigma\in[0,1])$ with respect to  the rotational axis, while the frame-dragging of the ergoregion plays a limited role in this aspect of the analysis.
The frame-dragging is related to the bending (curvature) of the curves in the positive section of the extended plane. This aspect was first  evidenced   in \cite{remnants}, where it was proved that the  frame dragging affects   the bending of the \textbf{MBs} curves in the extended plane.
 This result was obtained by the comparing the same situation in  the static   Schwarzschild and Reissner-Nordstr\"om geometries and by analyzing  the balance  between the effects of the electric charge
(the dimensionless parameter $Q/M$ related to the  sphericity and staticity  of the  geometry in the sense described in
\cite{remnants}) and the rotational charge (the spin $a/M$) of the Kerr-Newman geometries.
We could say that frame-dragging "closes"  the \textbf{MBs} curves on the equatorial plane, whereas they are open for  large spin   and close to the rotational axis (small $\sigma\in[0,1]$)--see Figs\il\ref{FIG:funzplo}, \ref{FIG:disciotto1},
\ref{FIG:rccolonog}, and \ref{FIG:Aslongas}. {We also proved that  replicas    reveal to be  tools for the exploration of the  region closed to  $(\theta\approx0, \sigma \approx0)$.}
\item[Extracting information]
 From the observational view-point, we established that the rotational axis of a Kerr \textbf{BH}  may have important implications for the knowledge of spacetimes structures closed to the singularity and horizons.
The other significant  aspect of the horizons confinement  concerns the intriguing possibility of extracting  information from counter-rotating orbits (negative frequencies with respect to the positive frequencies in the positive section of the extended plane), which we analyzed considering a supplement of the extended plane as seen in Fig.\il(\ref{Fig:mess-sicur}).
\item[Significance of the extended plane]
The introduction  of an extended plane for the  representation of Kerr solutions has proved to be very significant since it
leads to the establishment of a \textbf{BHs}--\textbf{NSs} connection and allows us to highlight   important properties  of the Kerr geometries {and to formulate an alternative definition  of the horizons for the Kerr geometries}.
\item[\textbf{NSs}, bottleneck and extended plane]
We  pointed out that \textbf{MBs} and horizon remnants of the  bottleneck region  ($\la\in [M,2M]$)  on the extended  plane  appear to be related to the concept of
spacetime pre-horizon regimes,
indicating the  existence of detectable  mechanical effects allowing circular orbit  observers to
recognize the close presence of an event horizon.
Pre-horizon regimes  were introduced in  \cite{de-Felice1-frirdtforstati}
and worked out  for the Kerr geometry in \cite{de-FeliceKerr,de-Felice-anceKerr}.
A rather  intriguing implication of this property is that an object like a gyroscope could observe
 a connected phenomenon and  interpret  it as
 a \emph{memory} of the static (Schwarzschild-like) case
in the Kerr metric--\cite{de-Felice-first-Kerr}. This interpretation  could be perfectly fitted in an extended plane representation of the Kerr metric family, providing the interpretation of a (black hole) spacetime as a line of the plane as illustrated in Fig.\il(\ref{Fig:Ly-b-rty}).
The concept of remnants  evokes  a sort of spacetime ``plasticity'' (or memory),
which naturally  can be read through    the concept of the  extended plane. Actually,
we  consider the  entire
family of  solutions    as a  \emph{unique geometric object}, and we propose a metric representation where we consider a spacetime as a single part of this plane.
The relevance  of this new representation, together with the possible conceptual significance, lies in its usefulness.
In this new framework, we found several properties  of   the spacetime geometries: considering the Kerr family  as a single object,  the  geometric {quantities} (for example,
the horizons) defined for a single solution acquire a completely different significance, when considered for the entire family. The naked singularity solutions, for example,  play a specific role in the construction of the horizon curve on the extended  plane,  in the sense of the tangency property.
Moreover, these structures could play an important role
for the description of  the  black hole formation and for testing the possible existence of naked  singularities.
 Horizon remnants could be of  relevance during the gravitational
collapse \cite{de-Felice3,de-Felice-mass,de-Felice4-overspinning,Chakraborty:2016mhx}.
 Similar  interpretations have been presented in
\cite{de-Felice1-frirdtforstati,de-FeliceKerr,de-Felice-anceKerr,de-Felice-first-Kerr}, by using the concept of pre-horizons,
 and in \cite{Tanatarov:2016mcs}, by analyzing the so-called whale  diagrams.
  We note that the physical  evolution of the Kerr spacetime,  during  unchanged symmetry  process stages, have to  occur along parts  of the extended plane.
 This also leads to  an analysis of the causal structure  as determined by stationary observers of  the Kerr geometries.
 {(For an analysis of gravitational instabilities and unstable axisymmetric modes in Kerr spacetimes, we refer to
\cite{Dotti:2008yr}.)}
\\
\item[\textbf{MBs} frequencies and horizons frequencies]
Finally, a further relevant aspect, very closely connected  to the spacetime thermodynamical properties, concerns the horizon frequencies,  which we proved are all and only all the bundle characteristic frequencies.
For the horizon frequency  $\omega_x\in \omega_H^{\pm}$ there is the set    $\Gamma_{\omega_x}$ of Eq.\il(\ref{Eq:trav7see}) of the \textbf{MBs}, which was  considered in several parts of this analysis and, in  particular, in  Sec.\il(\ref{Sec:allea-5Ste-cont}).
 The extended plane is essentially equivalent to a function relating  the frequency (bundle and horizon frequencies) to  the
(bundle origin) spin. Then, an important part of this analysis fixes also  the \textbf{MBs} characteristic frequencies in relation to the   photon orbital frequency and the  horizons frequencies--see Eq.\il(\ref{Eq:enagliny}) and
Fig.\il(\ref{FIG:rccolonog}).  Notably,
the \textbf{MBs} associated to the extreme Kerr \textbf{BH} corresponds to a regular curve, tangent to the horizon with
bundle origin $a_0=2M$ when $\sigma=1$; otherwise, it corresponds to the class  $\Gamma_{a_g=M}$ of  bundles  with characteristic frequency  $\omega_b=1/2$ and tangent radius $r_g=a_g=M$.
Thus, according to Eq.\il(\ref{Eq:summ-colo-quiri}),  the origin $a_0$ of each \textbf{MB} of $\Gamma_{a_g=M}$ is  $a_0=2/\sqrt{\sigma}$ ($\la=2M$); then, the minimum value of the bundle origin $a_0$ in $\Gamma_{a_g=M}$, which is used
to construct the limiting case of the  extreme Kerr \textbf{BH} horizon,  is  the \textbf{NS}
with $a_0=2M$ occurring for $\sigma=1$--Fig.\il(\ref{FIG:raisemK}).
\\
\item[\textbf{BHs} thermodynamics and \textbf{NSs}]Conditions relating the  tangent point $a_g$ to  the origin spin $a_0$ are given in Eq.\il(\ref{Eq:summ-colo-quiri}).
	   In this sense, \textbf{NSs} are ``necessary'' for the  construction of horizons.
	Properties of horizons and bundles and different frequency relations are given  in Sec.\il(\ref{Sec:partic-freq-ratios}),
	considering some notable frequency ratios related to the bundles of the  classes  $\Gamma_x$.
	On the other hand, we had pointed out  in \cite{remnants} the horizon  frequency  relations through the metric bundles for
	$\sigma=1$, that is, the bundles of  $\Gamma_{\sigma}$ for $\sigma=1$. (The horizons and, therefore, the tangency frequencies are obviously independent of $\sigma$ since  horizons are represented by spheres  on the extended  plane).
The possible thermodynamic implications of the results discussed here, particularly,
 in relation to the possibility of formulating the  \textbf{BH} thermodynamic laws   in terms of metric bundles are the  focus of a planned future work, where  we  also exploit the several symmetries found on the extended  plane in the \textbf{BH} representation of  Fig.\il(\ref{Fig:Ly-b-rty}).
\end{description}
\medskip
\textbf{Final remarks}
\medskip

\begin{description}
\item[-]
 Significant  for the transformations  from one solution to another, \textbf{MBs} represent  a global frame  for the analysis of  \textbf{BHs}. Extended  planes  and metric bundles allow us to  connect different points of  one geometry,  but also different Kerr geometries,   providing  a new  and global frame for the  interpretation of these metrics and, in particular, of \textbf{NSs} solutions.
In this respect, \textbf{MBs}  in the extended plane enlighten   some properties  of the horizons  connecting  different \textbf{BHs} and \textbf{NSs}  geometries.
In this context, it has to be seen the relevance of the horizon replicas, or the opposite effect of causal balls, which are defined after the horizon confinement.
Through the notion of replication we highlight those properties of a \textbf{BH} horizon which can be replicated  in other points of the same or different spacetimes; otherwise,  there is a confinement. We proved this,   for example,  for a portion of the Kerr inner horizon curve.
Replicas (an therefore the confinement) are studied with
the analysis of self-intersections of the bundles curves in the extended plane, in the same geometry (horizon confinement)  or
intersection of bundles curves  in different geometries.
(Note that, from \textbf{MBs} definition,
 there are  horizons replicas in different  geometries.)
Viceversa, a local  causal ball, a concept related to the {horizon confinement},  as a region of  the extended plane $\mathcal{P}-r$ where \textbf{MBs} are entirely confined,  implies   that there are no horizons replicas  in any other region of the extended plane.
\item[-]
 \textbf{MBs} utility  lies in enlightening   spacetime properties    emerging in the extended plane,  related to the local causal structure and \textbf{BHs}  thermodynamics.
The fact that the bottlenecks and remnants, the   pre-horizons regime and  whale diagrams have a clear interpretation in terms of metric bundles adds a further significance to the use of \textbf{MSs} in the analysis of \textbf{NSs}.
These structures essentially explicate some properties of the Killing horizons in  axially symmetric spacetimes and event horizons in the spherically symmetric case.  The horizons in this last case  are  limiting surfaces of the \textbf{MBs} in the extended plane.  We note that the  Kerr  \textbf{MBs} and the extended plane parametrization were conveniently settled  using the dimensionless spin of the Kerr geometry $a/M$  or $\la=a \sqrt{\sigma}/M$. In the more general case  and, for example, in  spherically symmetric spacetimes, this choice is not immediate. We see some aspects of this in \cite{LQG-bundles}.
\item[-]
Investigating  metric bundle,s  we explore in
an alternative way some aspects of the geometries defining  the bundle as measured by an observer at infinity.
As such, \textbf{MBs}  are of importance for the analysis of \textbf{NSs} solutions,
horizons and \textbf{BH} thermodynamics.
One way to see the connection between  \textbf{MBs}  and   \textbf{BHs} astrophysics (including  \textbf{BHs} thermodynamics) passes through the appreciation that these  conformal invariant structures  are, in fact, expressed   in terms of the light surfaces, particularly those light-surfaces from time-like  stationary observers.
 The orbital light-like frequency $\omega(r)$, the bundle characteristic frequency,  can be measured  by an observer at a point $r$ of the extended plane $\mathcal{P}-r$, where the \textbf{MBs} are defined as curves having equal limiting light-like orbital frequency, which is also an asymptotic limiting value for time-like stationary observers. These stationary observers and their limiting frequencies  are related to the analysis of many aspects of \textbf{BHs} physics, such as \textbf{BH} images,   and of several  processes of energy extraction such as the  jet emission and  collimation, Blandford-Znajek process, accretion disks or of the  Grad-Shafranov equation for the force free magnetosphere around \textbf{BHs},  where limiting frequencies  (here the bundle frequencies) are used as  limiting conditions.
\end{description}

\appendix
\section{Characteristics of the  metric Killing bundles}
{In these appendix sections, we have included a detailed and extensive analysis  of metric bundles. }

\subsubsection{\textbf{Metric bundles $\Gamma_{x}$}}\label{Sec:gamma-x-y}
In this section we study in details the \textbf{MBs} $\Gamma_{\mathbf{x}}$ for different quantities $\textbf{x}$
as listed in Table\il(\ref{Table:pol-cy-multi}), for each  $\Gamma_{\mathbf{x}}$  we consider also some notable sets of \textbf{MBs} $\Gamma_{\mathbf{x;y}}$  for different quantities $\textbf{y}$ as defined in Sec.\il(\ref{Sec:definitiosn}). We conclude this subsection introducing the concept of \emph{corresponding bundles}
and with the investigation of the  relation between origin spin and tangent spin
\begin{description}
\item[
\textbf{Metric bundles $\Gamma_{a_0}$  with equal spin origin $a_0$}
]
Metric bundles with equal spin origin $a_0$ have, according to  Eq.\il(\ref{Eq:bab-lov-what1}), in general, different bundle frequencies  $\omega_b$ and different tangent points at the horizon $r_g$; consequently, they also have different tangent spins $a_g$  for any polar plane $\theta$ (or $\sigma\in[0,1]$).
This case is studied in Figs\il(\ref{FIG:disciotto1}).
\begin{figure}
  \includegraphics[width=3cm]{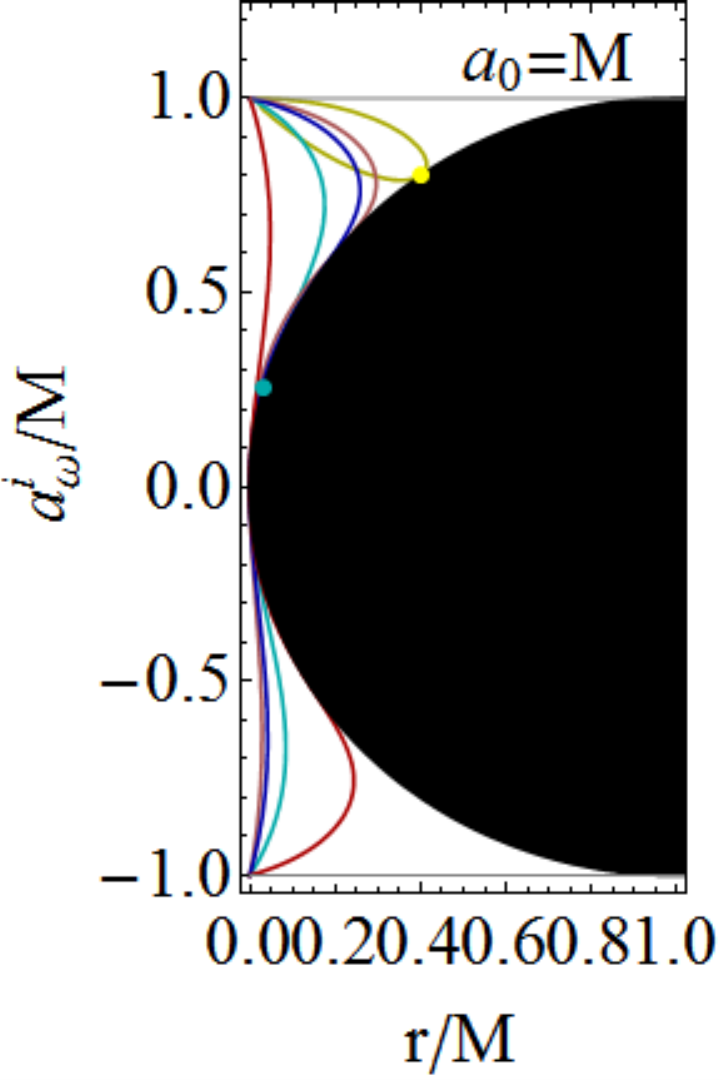}
  \includegraphics[width=3cm]{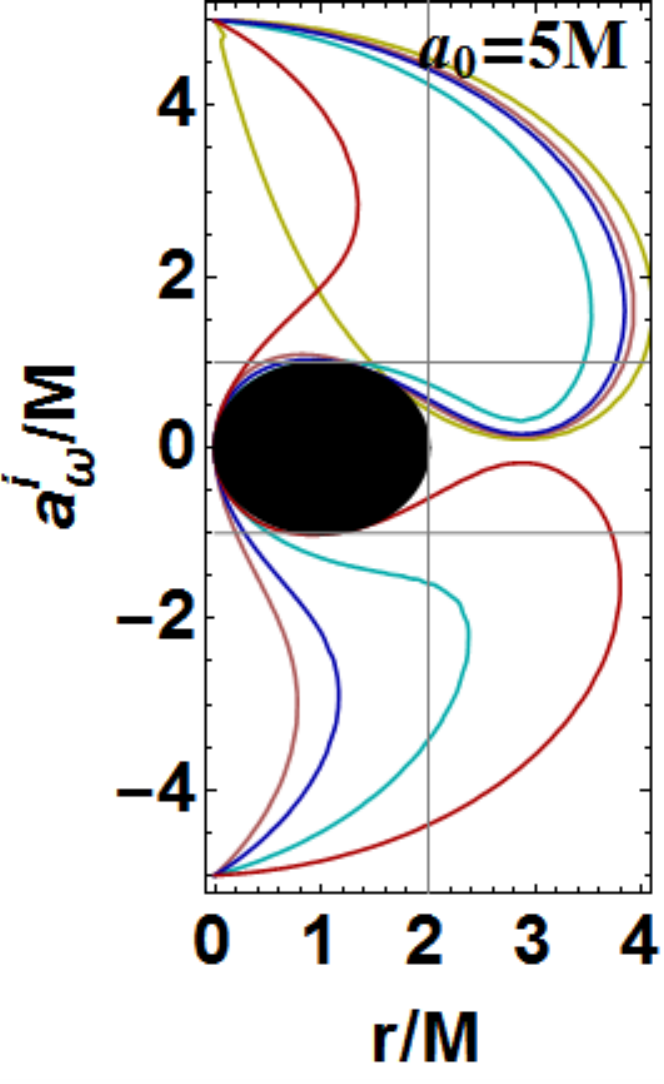}
      \includegraphics[width=3cm]{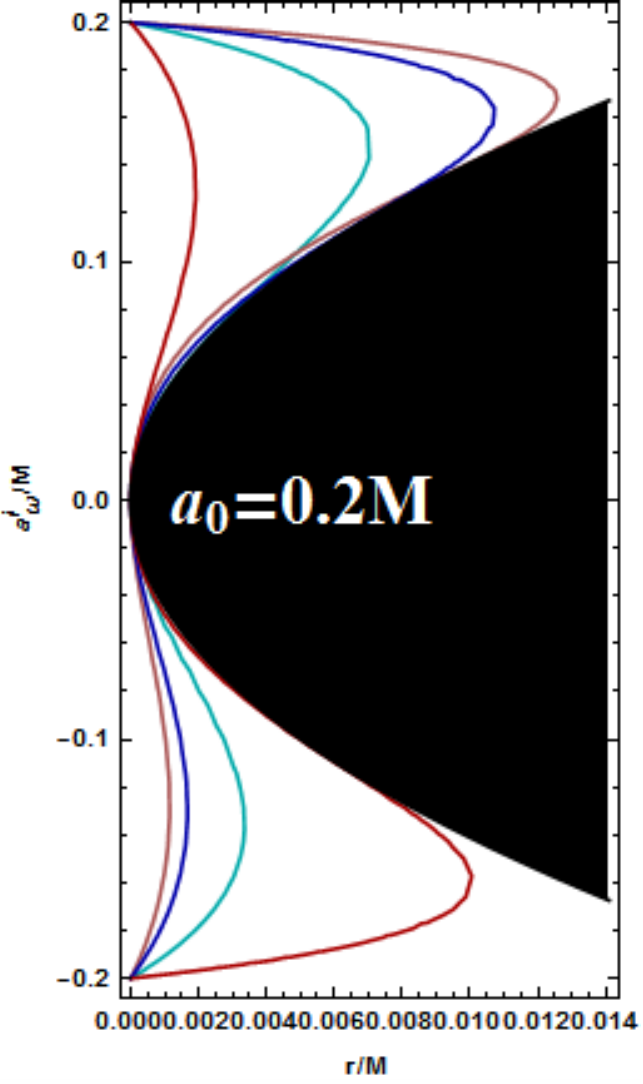}
  \includegraphics[width=3cm]{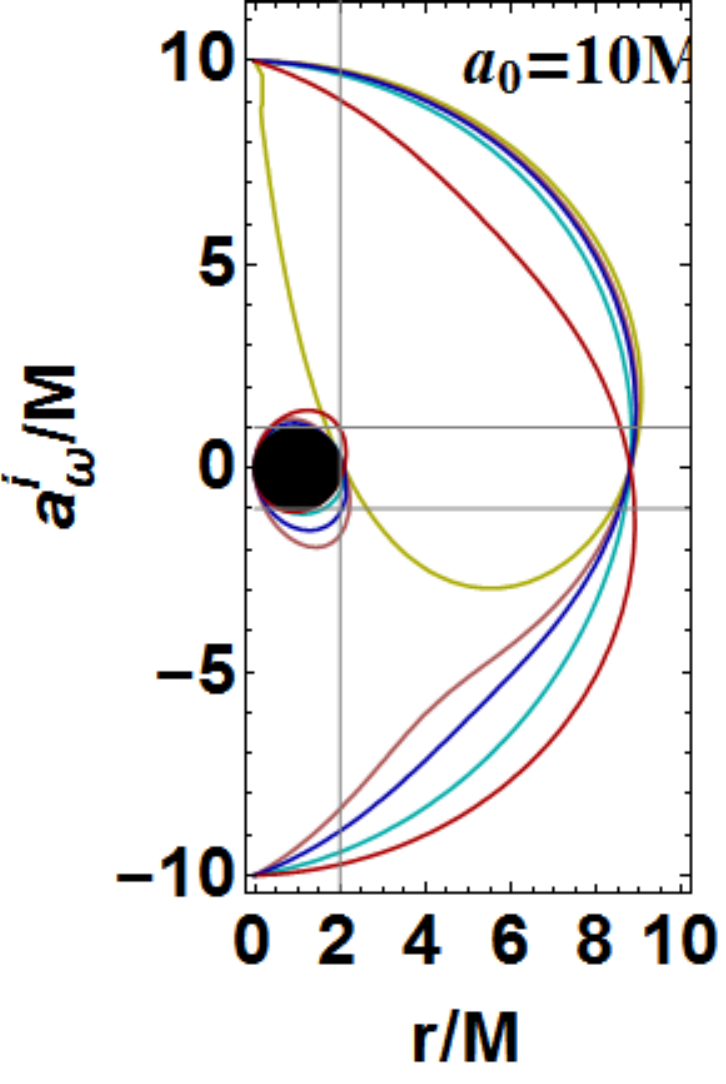}
   \caption{Metric bundles $\Gamma_{a_0}$ with equal origins $a_0$, different planes $\sigma$ and different frequencies
	$\omega_b={1}/a_0\sin (\theta)$. The black region is    the \textbf{BH}  $a<a_{\pm}$ on the extended plane, gray region is the ergoregion.
	There  $\theta=\pi/2$ (yellow curve),  $\theta=\pi/12$ (cyan curve), $\theta=\pi/4$ (pink curve), $\theta= 2.2 \pi$ (blue curve), $\theta=3.82\pi$ (red curve). These results follow from the analysis of Eqs.\il(\ref{Eq:inoutrefere}).}\label{FIG:disciotto1}
\end{figure}
However, according to the relation $a_0(\omega,\sigma)$, the equal-origin bundles frequencies  are
$\omega(a_0)=\frac{1}{a_0\sqrt{\sigma}}$ with a minimum   (in magnitude)  $\omega(a_0)=1/a_0$ on the equatorial plane. In this case
the bundle is closed, as shown in \cite{remnants}. Thus,  $|\omega(a_0)|\in[1/a_0,+\infty[$.

{The dependence of the  tangent point $r_g$    on  the bundle origin $a_0$ changes with   the polar angle $\theta$.
 Explicitly, the tangent points $r_g$ for bundles having   the same origin  spin $a^M_0$ is  $r_g(a^M_0)$, for a first fixed bundle
with the same origin frequency  is
$r_g(\omega^M_0)$ and, moreover, $a_g(a^M_0)$ is the tangent spin  corresponding  to the origin of the first bundle:}
\bea\label{Eq:inoutrefere}
&&r_g(\omega^M_0)=\frac{2}{4(\omega_0^M)^2 \sigma^{-1}+1},\quad r_g(a^M_0)=\frac{2}{\frac{4}{\sigma (a_0^M)^2}+1},
\quad a_g(a_0)=\frac{4 a_0^M \sqrt{\sigma }}{(a_0^M)^2 \sigma +4}
\\\nonumber
&&\mbox{where}\quad \omega_b(a_0^M)=\frac{1}{a_0^M\sqrt{\sigma}},\quad \mbox{and}\quad a_0=\frac{2 \sqrt{r_g}}{\sqrt{\sigma (2 -r_g)}}\ .
\eea
Note that the definition $r_g(\omega^M_0)$ holds for the maximum origin, while
$\omega_b(a_0^M)$ is obviously the frequency of the bundle with origin  $a_0^M$; consequently,
  it holds at each  point of the bundle particularly at the tangency point   and its origin located at  $a_0$.

The tangent spin has a maximum for $\sigma$ and  for $a_0$ at $\sigma={4}/{a_0^2}$ where $a_0=M$.
However, according to Eq.\il(\ref{Eq:tae-fraieuy}),  the tangent point to the horizon is  bounded in the range
$r_{g}(\theta)\in[0,r_{g}^*]\in[0,2M]$, where $r_{g}^*=\left.r_{g}^*\right|_{\theta=\pi/2}={2}/{{4}/({a_0^2})+1}$.
The bundle frequency  and its tangent point are fixed points  {under variations of  the   plane in $\sigma\in[0,1]$}.

We  now consider  some special cases  of metric bundles with  equal tangent spin $a_g$ or radius $r_g$ or plane
$\sigma$. Some of these cases will be detailed below in dedicated paragraphs, in which we  consider two  bundles with
$(a_0, \sigma, a_g, \omega_b)$ and  $(a_0, \sigma_p,a_g^p,\omega_b^p)$, respectively.
\bea&&\label{Eq:cli.chage}
\mbox{\textbf{Metric bundles with equal $a_0$ and $a_g$}}
\quad (a_g=a_g^p): \mbox{In this case}
\\\nonumber
&&
\bullet\quad a_0\in]0, 2M[,\quad\mbox{for}\quad\sigma_p\in[0, 1]\quad\mbox{and}\quad \sigma =\sigma_p\\\nonumber
&&
\bullet\quad a_0>2M\quad\mbox{\textbf{(i)}}\quad\sigma_p\in\left[0,\frac{16}{a_0^4}\right[,\quad\mbox{and}\quad \sigma =\sigma_p\quad\mbox{\textbf{(ii)}}\quad
\sigma_p\in\left[\frac{16}{a_0^4},\frac{4}{a_0^2}\right[,\quad\mbox{and}\quad  \sigma\in(\sigma_p,\sigma^p_{\ell});
\\\nonumber
&&\mbox{\textbf{(iii)}}\quad \sigma_p=\frac{4}{a_0^2}\quad\mbox{and}\quad  \sigma =\sigma_p,\quad\mbox{\textbf{(iv)}}\quad\sigma_p\in\left]\frac{4}{a_0^2}, 1\right]\quad\mbox{and}\quad  \sigma\in(\sigma_p,\sigma^p_{\ell})
\\\label{Eq:freq-ratio}
&&
\mbox{where}\quad \sigma^p_{\ell}\equiv\frac{16}{a_0^4 \sigma_p},\quad\mbox{frequency ratio}\quad
s=\frac{\omega_b}{\omega_b^p}\in\{1,\frac{\sqrt{\sigma_p}}{\sqrt{\sigma }}=\frac{4}{a_0^2 \sigma }=\frac{a_0^2 \sigma_p}{4}\}
\\\nonumber
&&
\mbox{\textbf{Metric bundles with equal $a_0$ and $\sigma$: }} \mbox{In this case}
\quad a_g=a_g^p\quad\mbox{and}
\quad \omega_b=\omega_b^p
\\\nonumber
&&
\mbox{\textbf{Metric bundles with equal $a_0$ and $r_g$ $(\omega_b)$: }}
\mbox{In this case}
\quad \sigma=\sigma_p \quad\mbox{and }
\quad a_g=a_g^p
\\\nonumber
&&
\mbox{\textbf{Metric bundles with equal $a_0$ and $a_g$ and $r_g$ $(\omega_b)$: }}
\mbox{In this case}
\quad \sigma=\sigma_p.
\eea
Metric bundles with equal $a_0$ \emph{and} $\sigma$, \emph{or} equal  origin \emph{and}   tangent radius  $r_g$ (or frequency)
are equal.
In the case of equal origin and tangent spin  $a_g$, except for the trivial case of coincidence of the bundle  $\sigma=\sigma_p$, it is noteworthy that  only two distinct bundles can  exist for \textbf{NS} origins with spin $a_0>2M$  and large plane
$\sigma_p>{4}/{a_0^2}$, with a fixed ratio  between the $\sigma$ planes and the frequencies. In Sec.\il(\ref{Sec:allea-5Ste-cont}), we will particularly focus on the relations between the frequencies, which are the horizon frequencies  $\omega_H^{\pm}$  (see also Figs\il\ref{Fig:stypolitic}).
 Note that the  ratio in Eq.\il(\ref{Eq:freq-ratio})  depends on the spin because it is parameterized by the bundle origin  $a_0$, which is related to the bundle frequency   through  $\sigma$.
\\
\item[
\textbf{Metric bundles $\Gamma_{\sigma}$  on the same plane $\theta$}]
We focus on metric bundles on an equal plane $\sigma$. The  case $\sigma=1$ has been analyzed in  detail in  \cite{remnants}.
We will also take into account the results given in  Eq.\il(\ref{Eq:inoutrefere}).
However, we present here some of the  relations (\ref{Eq:inoutrefere}) as follows:
\bea\label{Eq:rgrbomegab}
&&\textbf{Tangent radius}\quad r_g(\omega_b)=\frac{2}{4 \omega_b^2+1},\quad r_g(a_0)=\frac{2}{\frac{4}{a_0^2 \sigma }+1},
 \\\label{Eq:mple-pan}
&&\textbf{Tangent spin}\quad
a_g(\omega_b)=4 \sqrt{\frac{\omega_b^2}{\left(4 \omega_b^2+1\right)^2}},\quad a_g(a_0)=\frac{4 a_0\sqrt{\sigma }}{a_0^2 \sigma +4},
 \\&&\label{Eq:in-our-hea}
\textbf{Bundle origin}\quad  a_0(\omega_b)=\frac{1}{\sqrt{\sigma } \omega_b},\quad a_0(r_g)=\frac{2 \sqrt{r_g}}{\sqrt{2 \sigma -r_g \sigma }},
\\\nonumber
&& \mbox{where}\quad r_g\in[0,2M],\quad a_0\in[0,+\infty[\cup]-\infty,0],\quad
a_g\in[0,M]\quad \omega_b=\omega_H^{\pm}\in]0,+\infty[
\eea

See the Figs\il(\ref{FIG:raisemK},\ref{FIG:toa11},\ref{FIG:toa8},\ref{FIG:funzplo}), where $(r_g(\omega_b)$ and $r_g(a_0))$ is the tangent radius as function of the bundle frequency $\omega_b$ and of the bundle origin spin $a_0$, respectively.
The radius $r_g(a_0)$ is maximum (for $\sigma$ and $a_0$) at $a_0=\sigma=0$, where $r_g=2M$.
$(a_g(\omega_b),a_g(a_0))$  is the tangent spin as function of the bundle frequency $\omega_b$ and  bundle origin spin $a_0$, respectively.
$(a_0(\omega_b),a_0(r_g))$ is the bundle origin spin as function of the bundle frequency $\omega_b$ and  bundle tangent radius  $r_g$, respectively.
The function $a_g(\omega_b)$ has an extremum at $\omega_b=1/2$, where $a_g(\omega_b)=1$.
\begin{figure}
  \includegraphics[width=3cm]{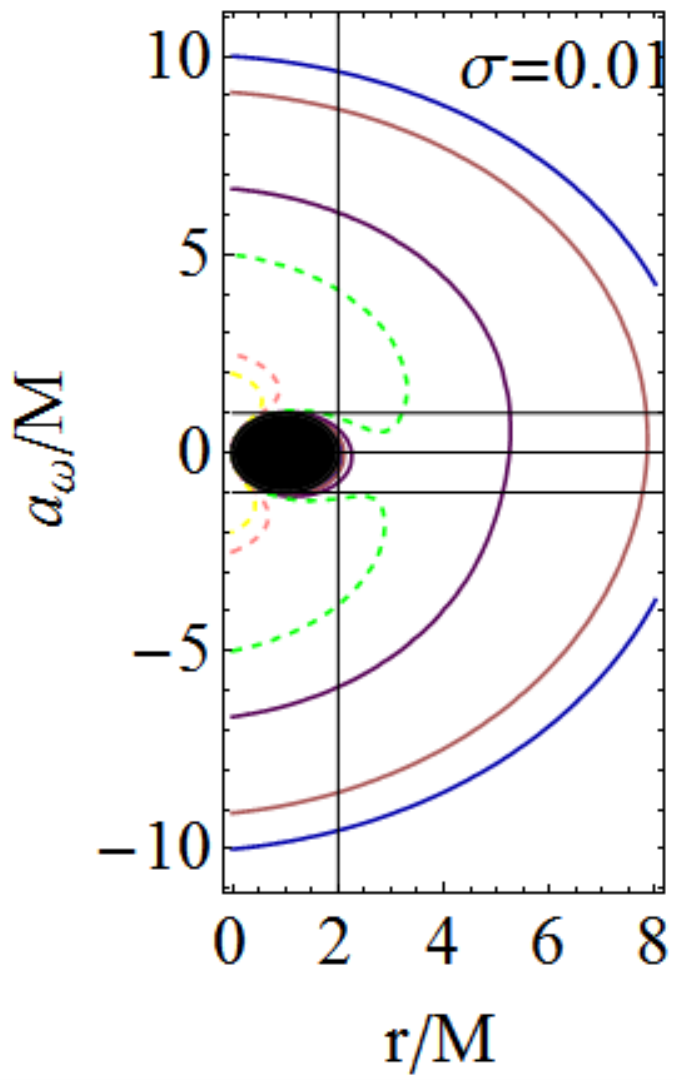}
  \includegraphics[width=3cm]{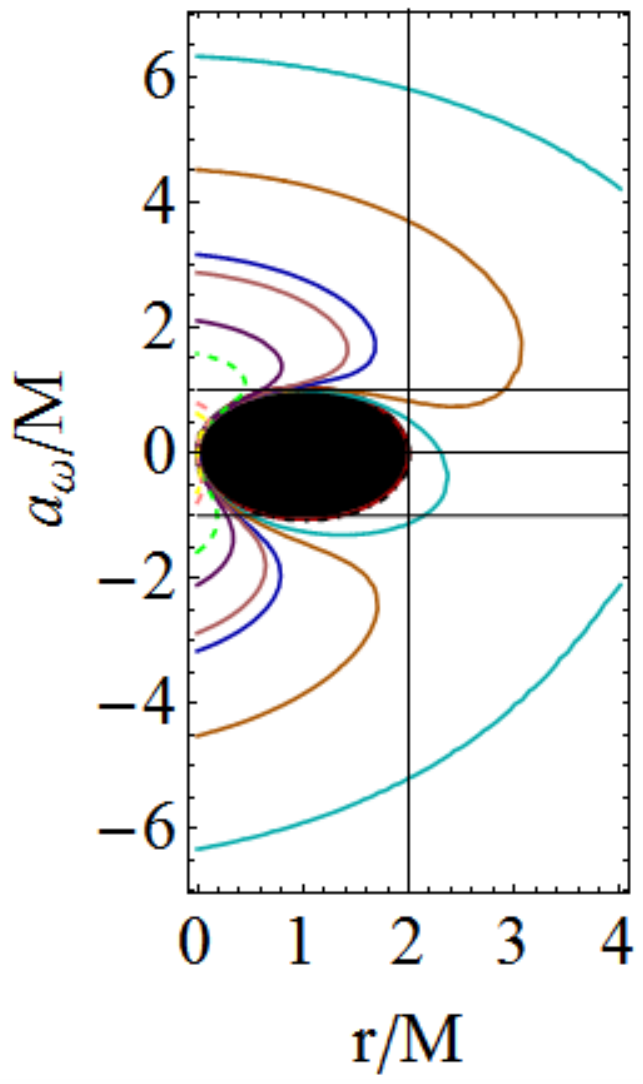}
      \includegraphics[width=3cm]{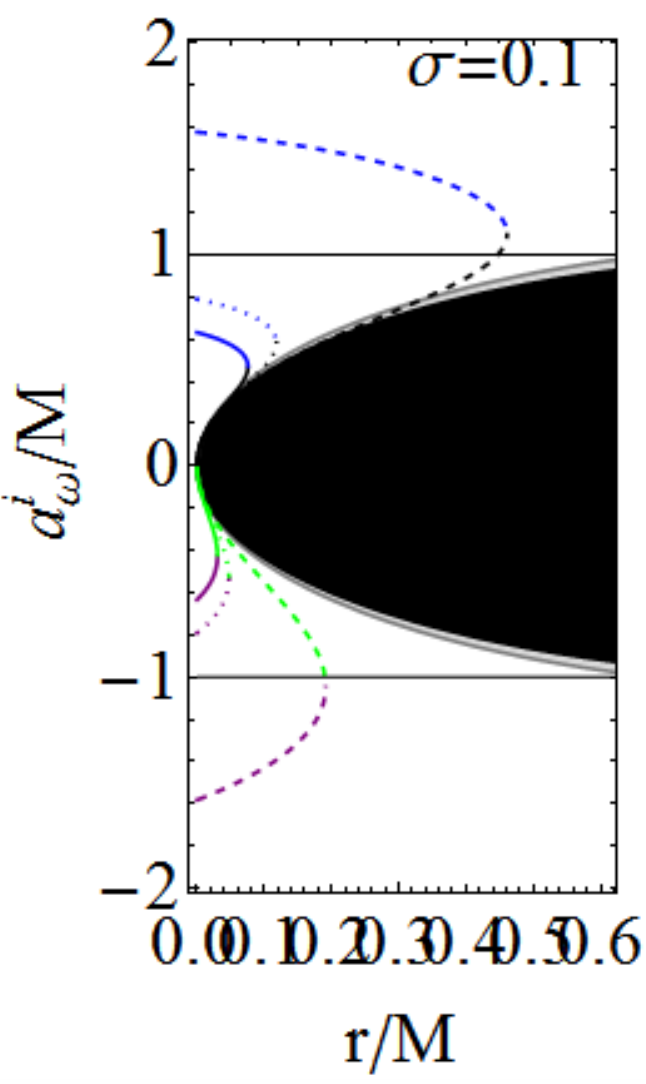}
  \includegraphics[width=3cm]{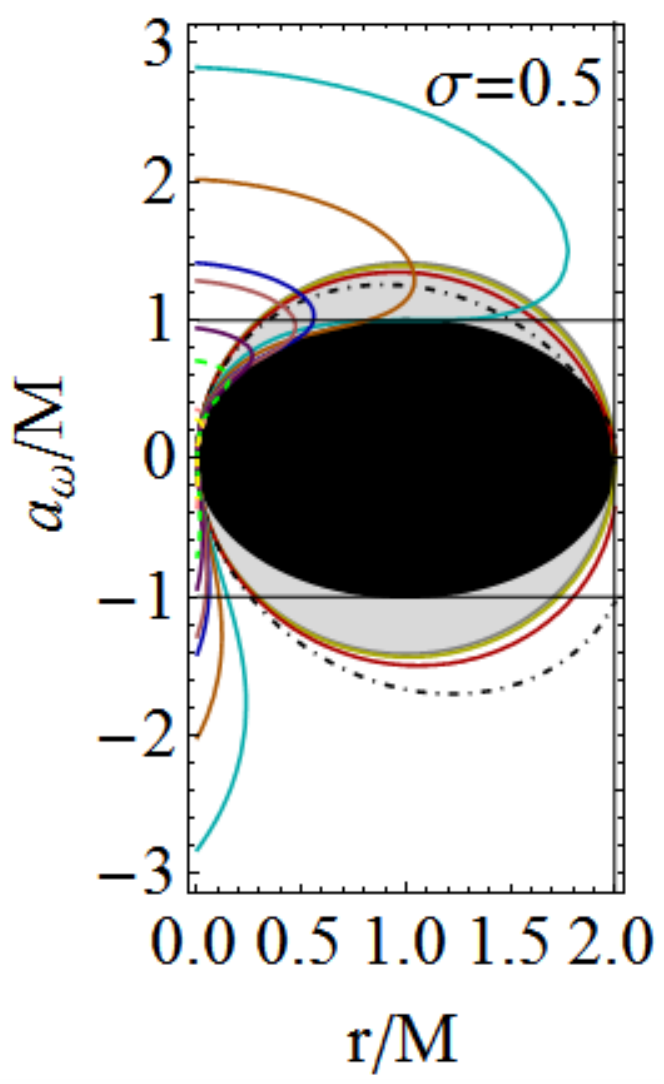}
  \includegraphics[width=3cm]{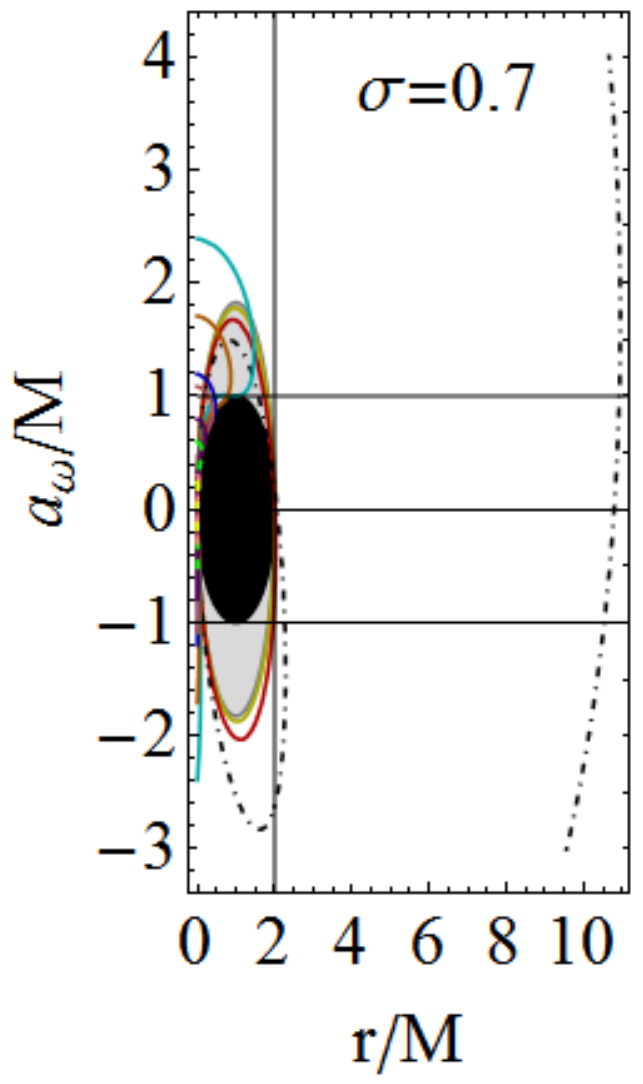}
  \includegraphics[width=3cm]{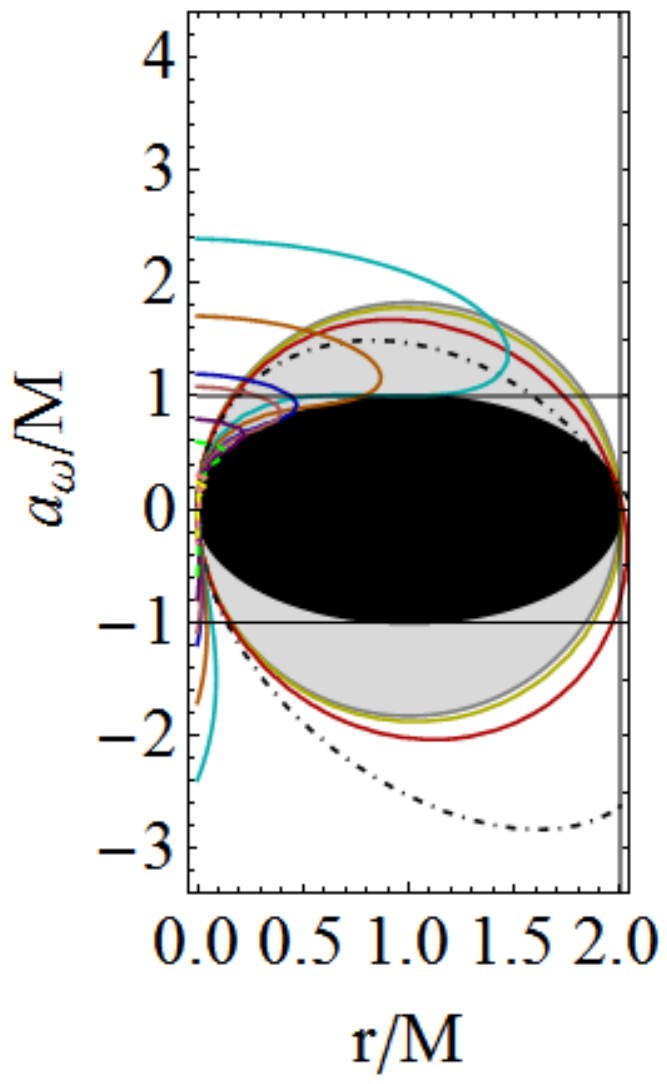}
  \includegraphics[width=3cm]{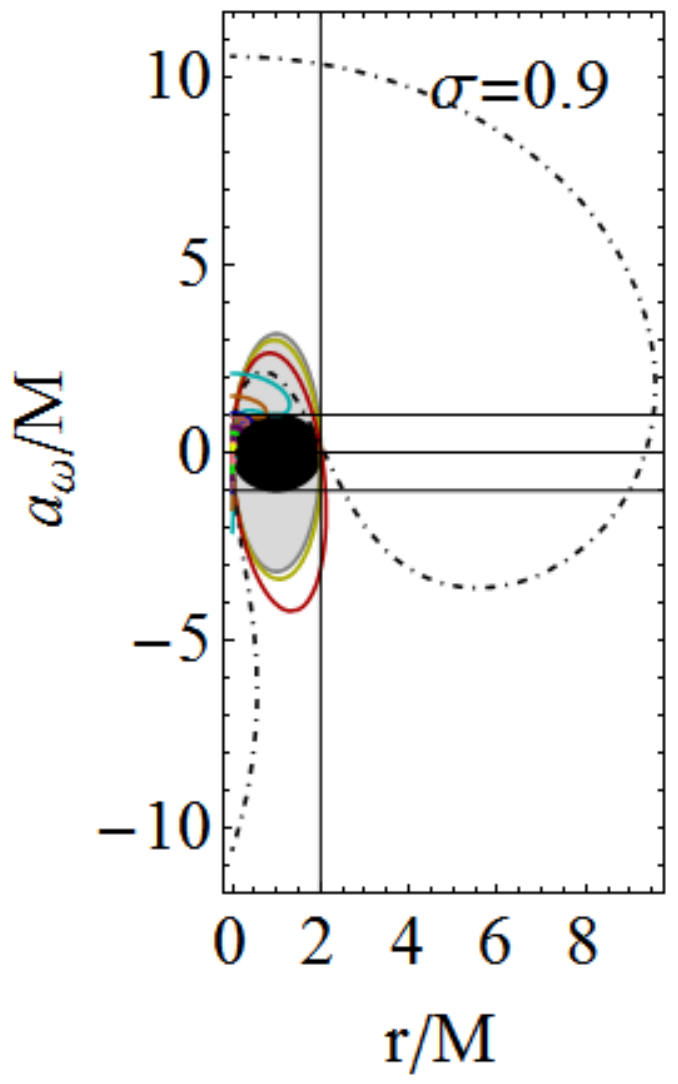}
  \includegraphics[width=3cm]{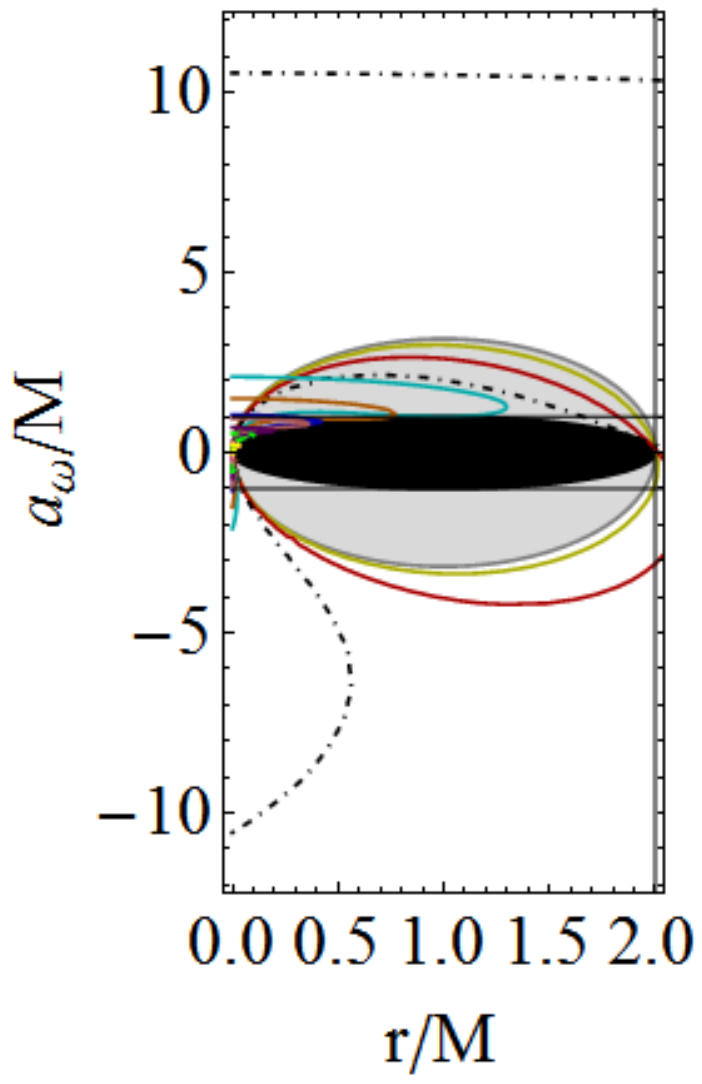}
  \includegraphics[width=3cm]{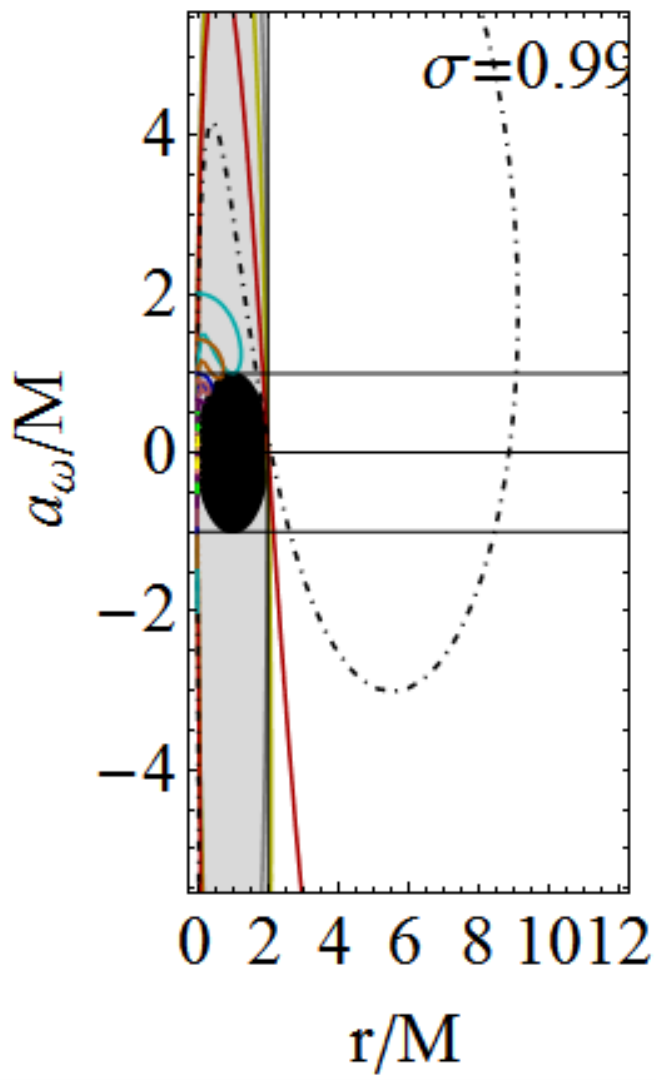}
  \includegraphics[width=3cm]{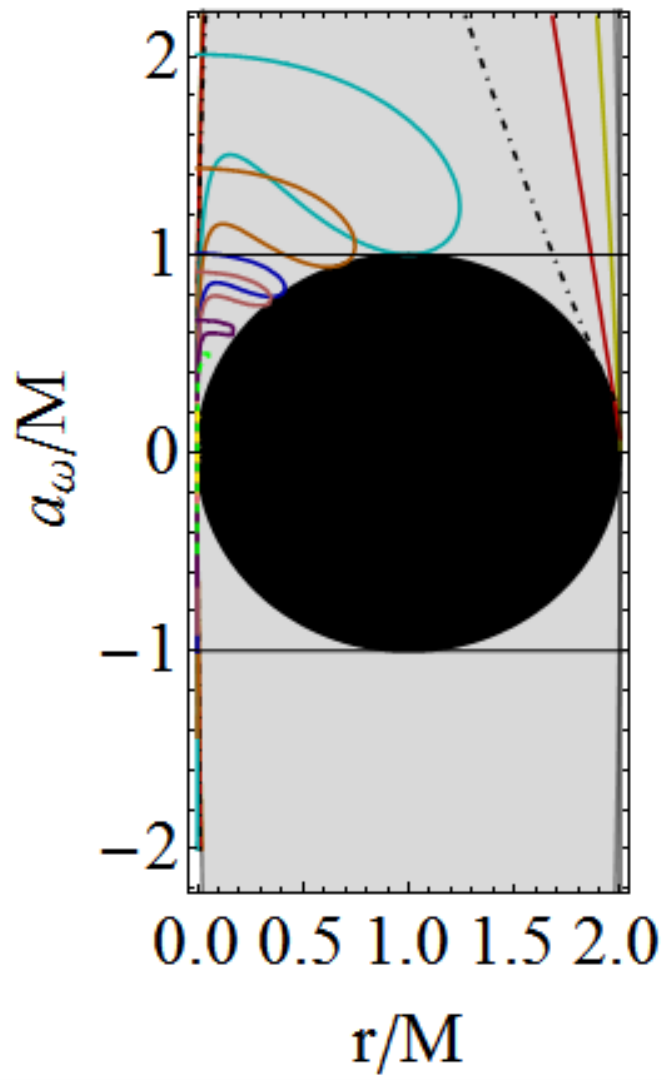}
  \includegraphics[width=2cm]{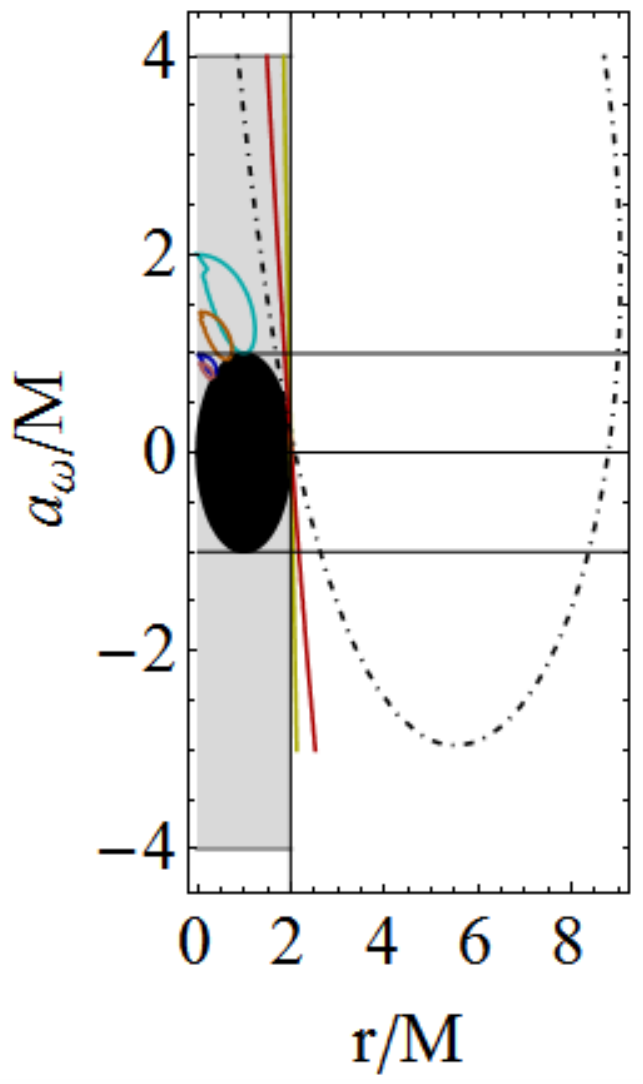}
   \caption{Metric bundles $\Gamma_{\sigma}$ for the same plane $\sigma$ and different bundle frequencies
	$\omega_b\in\{0.01$ (yellow), 1/27 (Red), 0.1 (dotted-dashed), 0.5 (cyan), 0.7 (orange),
  1 (blue), 1.1 (pink), 1.5 (purple), 2 (green), 4 (dashed-pink),
 5 (yellow-dashed)\}. The black region represents  a \textbf{BH}  ($a<a_{\pm}$) on the extended plane. It follows the analysis of Eq.\il(\ref{Eq:rgrbomegab}).}\label{FIG:funzplo}
\end{figure}
\item[
\textbf{Metric bundles $\Gamma_{\omega_b}$  with equal bundle frequencies $\omega_b$}]

According to  Eq.\il(\ref{Eq:rgrbomegab}), metric bundles characterized by the same  frequency $\omega_b$
 have the same point of tangency with horizon $r_g$  and tangent spin
$a_g$,  but different origin spins $a_0(\omega_b,\theta)$  for different planes $\theta$.
The relation between the origins of the equal-frequency bundles $\omega_b$ and, therefore, also the same point of tangency $(a_g,r_p)$,
  is
\bea\label{Eq:trav7see}
\textbf{bundles  with equal} \quad \mathbf{\omega_b \quad(r_g,a_g)}\quad
\frac{a_0^p}{a_0}=\sqrt{\frac{\sigma }{\sigma_p}}, \quad\mbox{in particular}\quad
{a_0^p}={a_0}\quad\mbox{iff}\quad \sqrt{{\sigma }}=\sqrt{{\sigma_p}}
\eea
This case is studied in Figs\il(\ref{FIG:rccolonog}).
\begin{figure}
  \includegraphics[width=3cm]{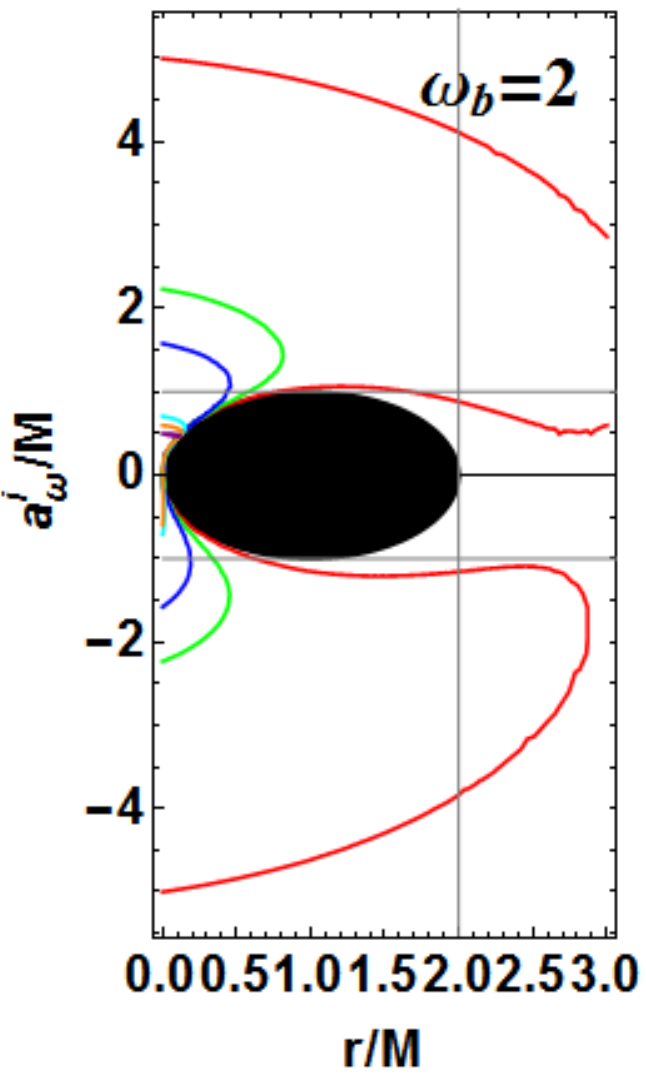}
  \includegraphics[width=3cm]{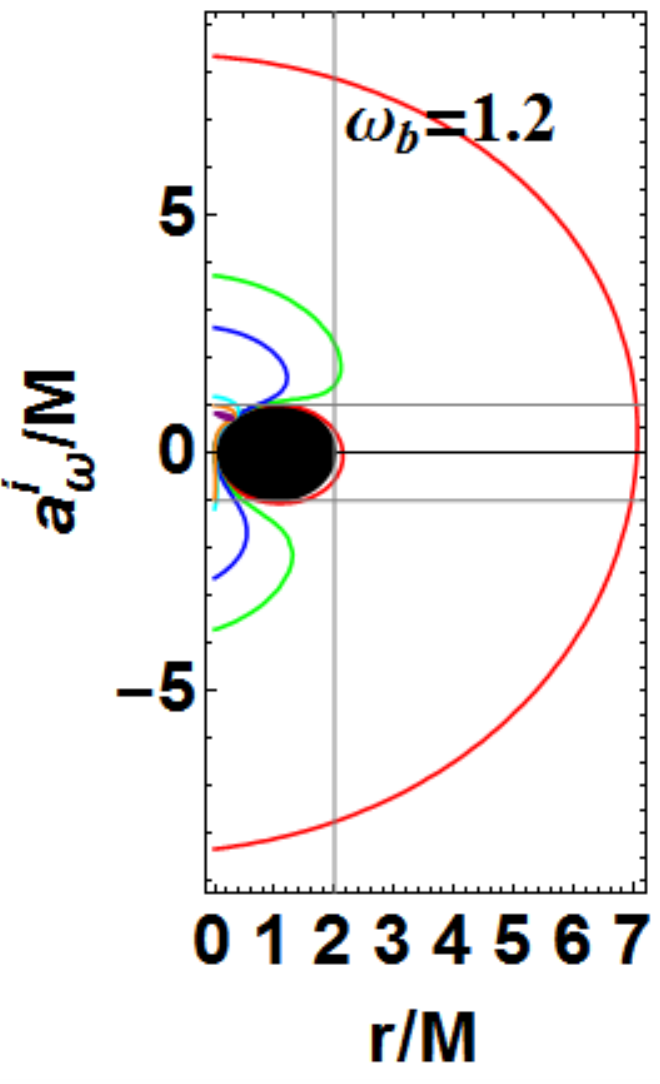}
      \includegraphics[width=3cm]{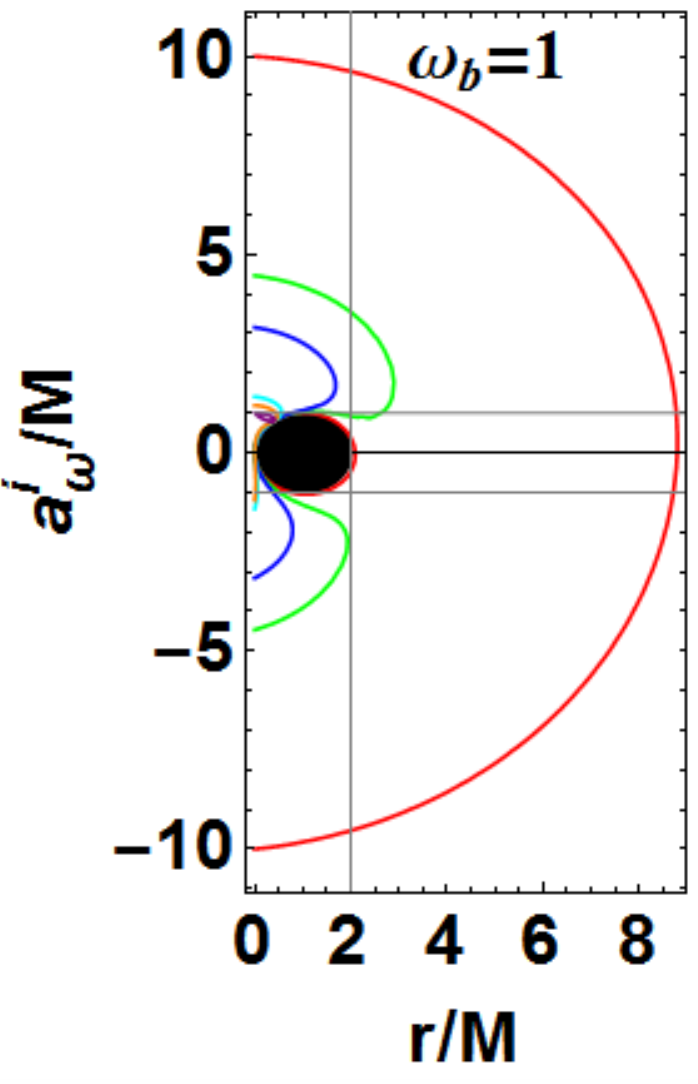}
  \includegraphics[width=3cm]{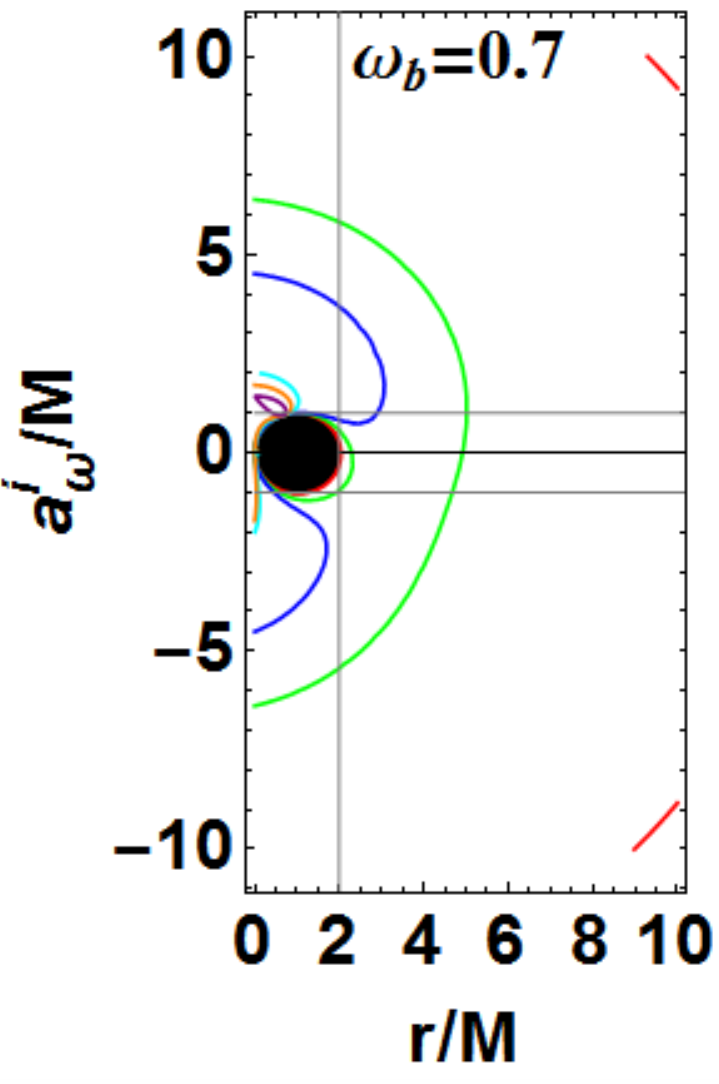}
 \includegraphics[width=3cm]{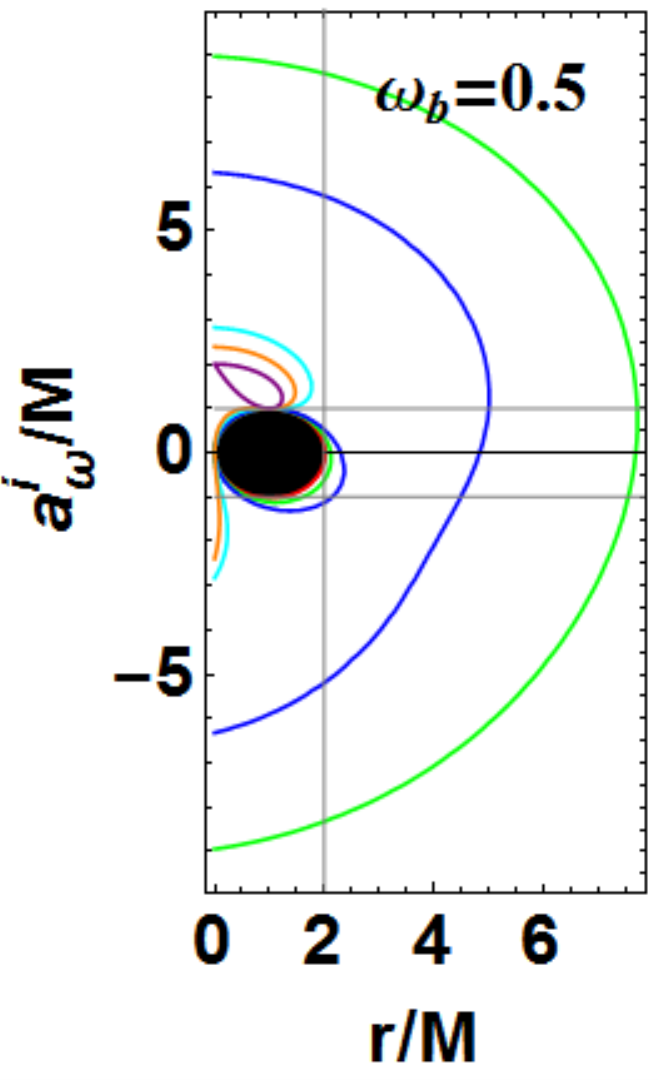}
    \includegraphics[width=3cm]{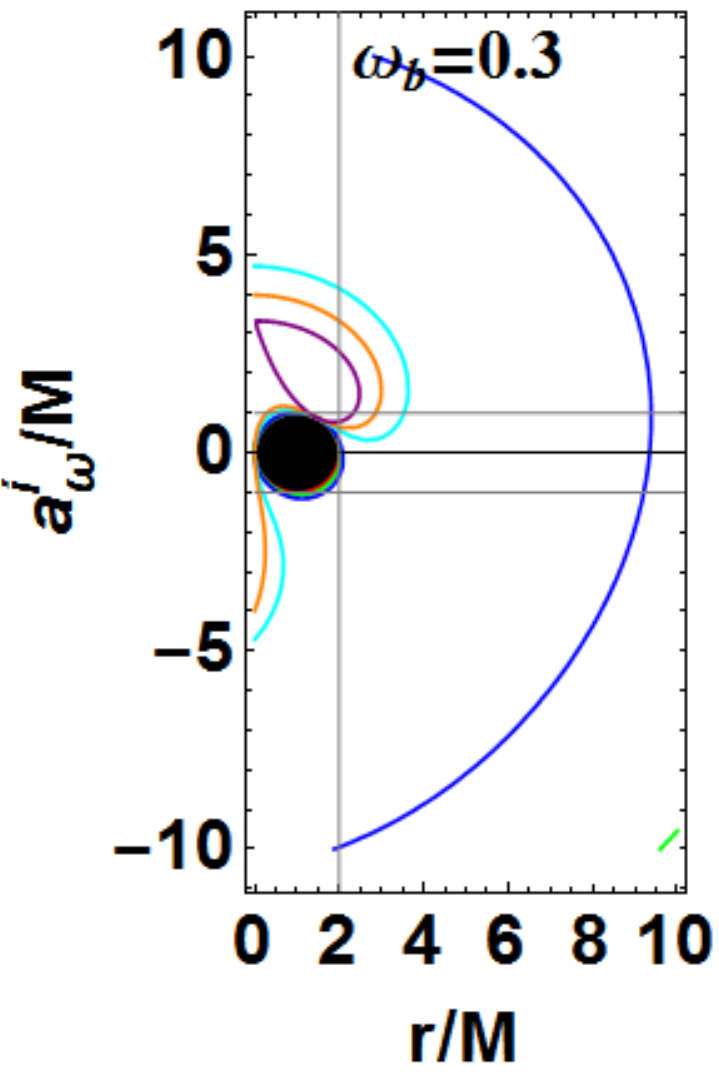}
      \includegraphics[width=3cm]{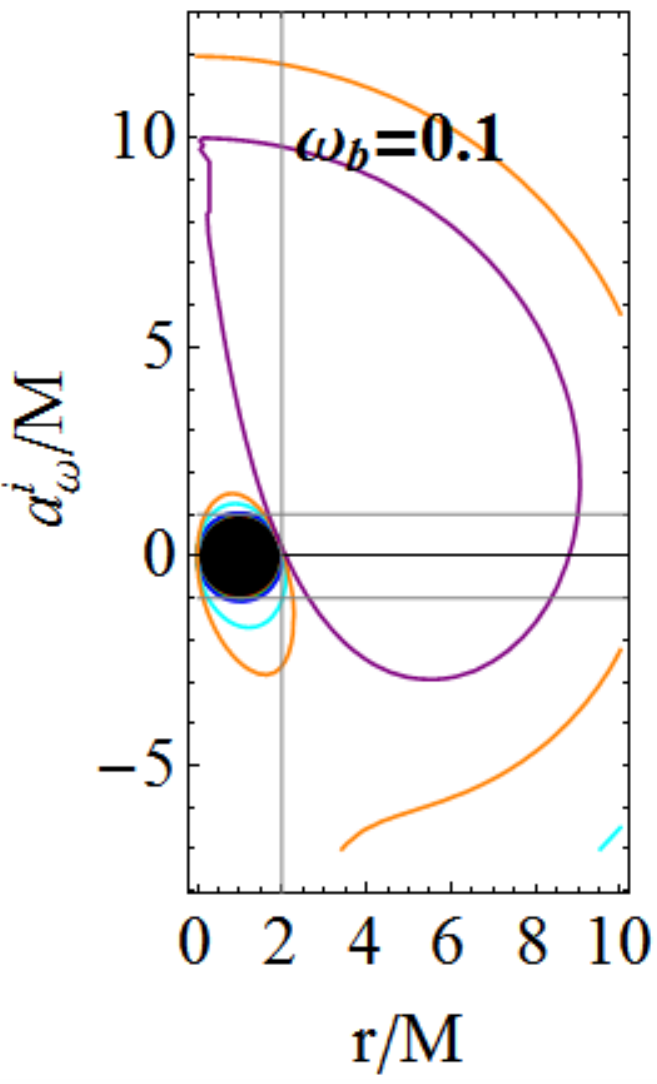}
        \includegraphics[width=3cm]{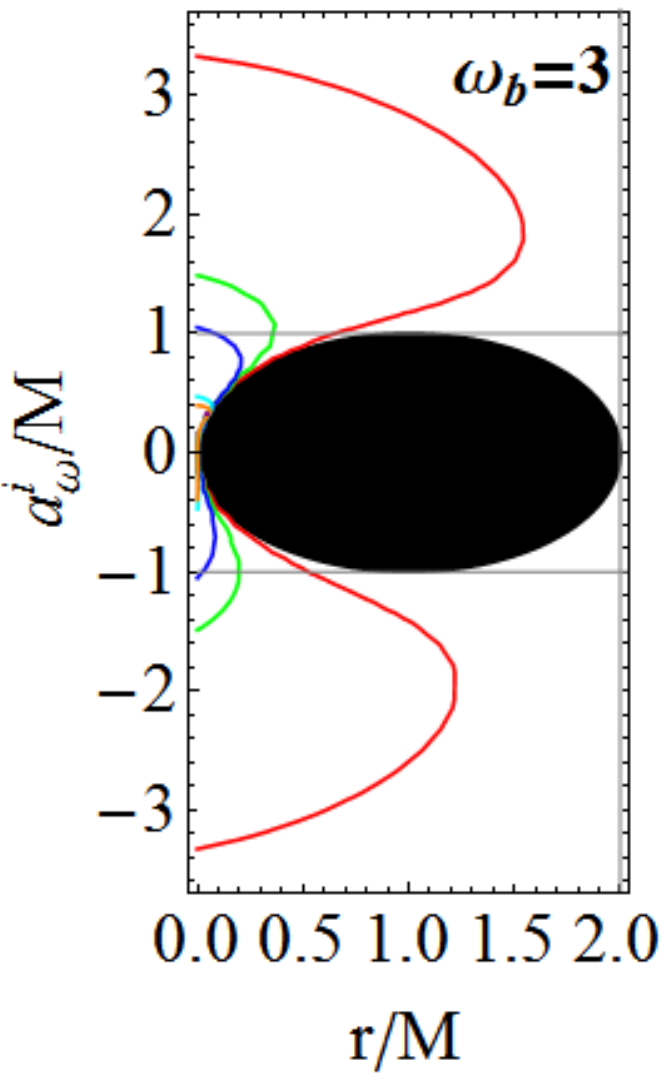}
   \caption{Metric bundles $\Gamma_{\omega_b}$ with the same frequency $\omega_b$ and, therefore, also the same spin tangent
	$a_g$ and  tangent radius $ r_g$. The bundle origins $a_0$ depend on the plane $\sigma$: the {blue}  curve is for $\sigma=0.1$;  \text{cyan} curve is for $\sigma=0.5$, {green} is for  $\sigma=0.05$;  {orange} curve $\sigma=0.7$, \text{purple} curve is $\sigma=1$,  \text{red} curve is $\sigma=0.01$. The black region corresponds to a \textbf{BH}  ($a<a_{\pm}$) on the extended plane. It follows from the analysis of Eq.\il(\ref{Eq:trav7see})}\label{FIG:rccolonog}
\end{figure}
This confirms that for each  plane $\sigma$ there is one and only one metric bundle associated with
the  horizon frequency   $\omega_b$  and the pair $(r_g,a_g)$  (recall that the tangent radius $r_g$ and, therefore, the tangent spin
$a_g$ are determined by the bundle frequency only, while the bundle origin $a_0$ depends on the plane $\sigma$).
In addition, on different planes, there may be different bundles all tangent to the point of the horizon with a radius  $r_g$ and
frequency $\omega_b$,  but different  $a_0$  for different $\sigma$, as follows from Eq.\il(\ref{Eq:trav7see}).
This relation implies the following fact: if the origin  $a_0$  corresponds to a  \textbf{BH}, (i.e. $a_0\in[0,M]$),
it is always tangent to the inner horizon (this will be proved explicitly later especially in Sec.\ref{Sec:allea-5Ste-cont}). Then, a bundle with the same frequency as the inner horizon and therefore the same point of tangency can be generated by an origin  $a_0^p$ in \textbf{NS} related to the frequency by $a_0^p=a_0\sqrt{\sigma/\sigma_p}$ if   $\sqrt{\sigma_p}\leq \la_0\equiv a_0\sqrt{\sigma}$.
If  $\sqrt{\sigma_p}=\la_0$, then $a_0^p$ is always tangent to  the point of the horizon that corresponds to the extreme \textbf{BH}.
Particularly, for the equatorial plane where $\sigma=1$ and
$a_0^p=a_0/\sqrt{\sigma_p}$, we conclude that for a fixed point of tangency the \emph{minimum} origin spin   occurs  \emph{always}  on the equatorial plane; every other bundle has a higher origin.
Then, the bundle with origin in $a_0=M$ on the equatorial plane has other bundles with equal tangent point necessarily located in the
\textbf{NS} region with  $a_0^p=1/\sqrt{\sigma_p}$.
\item[
\textbf{Metric bundles $\Gamma_{a_g}$  with equal bundle tangent spin $a_g$:  Construction of horizons}
]

We focus on the case of metric bundles with the same tangent spin  $a_g$. This  condition allows us to construct the horizons in the extended plane {corresponding  to a \textbf{BH}} with with spin $a_g$.  We recall that if two metric bundles have the same spin tangent $a_g$, then
irrespectively of the plane $\sigma$ or origin $a_0$, they can have either
\textbf{(1)} the same frequency  $\omega_b$, i.e., they belong to the family  $\Gamma_{\omega_b}$ studied  in Eq.\il(\ref{Eq:trav7see}),
 or \textbf{(2)} different  $\omega_b^1$; in any case, i holds that
$(\omega_b(a_g),\omega_b^1(a_g))=\omega_H^{\pm}(a_g)$, that is, they must both have one of the  horizon
frequencies with spin \textbf{BH} spin $a_g$.

To obtain the relation between these bundles, we use  Eq.\il(\ref{Eq:rgrbomegab}). We divide the problem by considering
\textbf{(1)} firstly,  a fixed spin $a_g$ with  equal  $\sigma$,  and then \textbf{(2)} a fixed $a_g$ with arbitrary $\sigma$.
This case is studied in  Figs\il(\ref{FIG:Aslongas}).
For fixed spin tangent  $a_g$, the two tangent radii  are obviously  $r_g=r_{\pm}(a_g)=1\pm \sqrt{1-a_g^2}$, which coincide
only for extreme \textbf{BH} case  $a_g=M$. As this a notable case, we will analyze it separately.
The metric bundles with the same point of contact  $p_g=(a_g,r_g)$ have the same frequency $\omega_b$,
but pertain to different planes  $\sigma$ and origins $a_0$; the \emph{larger} origin spin (for fixed frequency) corresponds to
the equatorial plane ($\sigma=1$).
 Thus we obtain a family of metric bundles  with the same  $(a_g,\omega_b,r_g)$, but different  $a_0(\sigma)$ for different $\sigma$;
each $\sigma$ corresponds to one and only one bundle with  $(a_g,\omega_b,r_g)$.

For fixed $\sigma$, we now consider metric bundles with equal  $(a_g,\omega_b)$ which have equal $r_g$ related by the horizon curve, i.e.,
$r_g^\pm=1\pm \sqrt{1-a_g^2}$. The bundle frequencies  $(\omega^1_b,\omega_b)$ and the bundle origins  $(a^1_0,a_0)$ are related as follows
 \bea&&\label{Eq:enagliny}
 \mbox{\textbf{Metric bundles with equal $(a_g,\sigma):$}}\quad
 \omega^1_b=\frac{1}{4\omega_b},\quad a_0=\frac{1}{\omega_b\sqrt{\sigma}},\quad
 a^1_0=\frac{4\omega_b}{\sqrt{\sigma}}=4\omega_b^2 a_0=\frac{4}{a_0\sigma }
 \\\nonumber
&&\qquad\qquad\mbox{from which }\quad
 a^1_0 a_0=\frac{4}{\sigma },\quad \frac{a^1_0}{a_0}=4\omega_b^2
 \\&&\label{Eq:weth-mome}
\mbox{\textbf{Metric bundles with equal $(a_g,\omega_b,r_g):$}}\quad a_0^c=\frac{1}{\omega_b\sqrt{\sigma}}=\frac{a_0 \sqrt{\sigma }}{\sqrt{\sigma_c}}\quad \mbox{from which}\quad  \frac{a_0^c}{a_0}=
\sqrt{\frac{\sigma }{\sigma_c}}.
 \eea
 The behavior of these quantities is illustrated in Fig.\il(\ref{Fig:Cmapobaseplot}).
\begin{figure}
  \includegraphics[width=7cm]{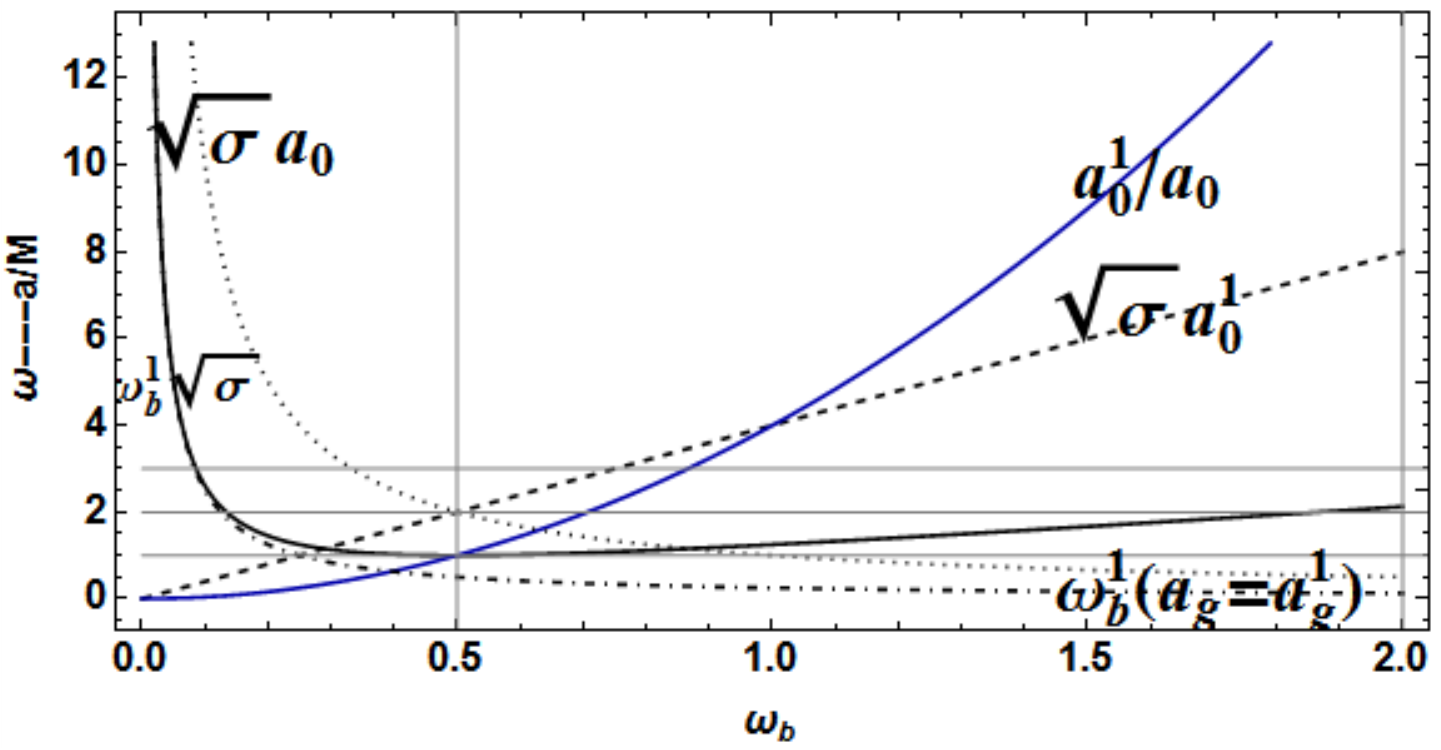}
  \caption{Th black curve is $\omega_b^1(\omega)$ from Eq.\il(\ref{Eq:enaglinya}), the dashed curve is $a_0(\omega_b)\sqrt{\sigma}$ for equal $(a_g,\sigma)$ of Eq.\il(\ref{Eq:enagliny}), the blue curve is $a_0^1/a_0$  for equal $(a_g,\sigma)$
	from Eq.\il(\ref{Eq:enagliny}),  the dotted curve is for $a_0\sqrt{\sigma}$ for equal $(a_g,\sigma)$ of Eq.\il(\ref{Eq:enagliny}), the dotted-dashed curve is $\omega_b^1(\omega_b)$ of  Eq.\il(\ref{Eq:enaglinyb}) for
$a_{g}(\omega_b^1)=a_{g}^1(\omega_b)$. }\label{Fig:Cmapobaseplot}
\end{figure}
It is clear that for the class  (\ref{Eq:enagliny}),  composed by
metric bundles with equal $(a_g,\sigma)$, the two bundles have  tangent points on the horizon $(r_g,r_g^1)=(r_-(a_g),r_+(a_g))$,
generating in this way the horizon of the \textbf{BH} spacetime with spin $a_g$. We will see also more details below.
\\
\item[
 \textbf{Corresponding  bundles}]
As studied in  \cite{remnants}, there are corresponding bundles with $a_0^1=a_g$: the tangent point of a bundle is the origin  of a second bundle  with $a_0^1=a_g$,  which must  have a
 \textbf{BH} origin. Then, we can identify an entire class of corresponding  bundles
  $\Gamma_{a_g}$  with the same tangent spin, studied in Eq.\il(\ref{Eq:enagliny}), and the class $\Gamma_{a_0^1}$  of metric bundles  with the same origin, studied in  Eq.\il(\ref{Eq:inoutrefere}).
 The frequencies of these bundles are related by
 \bea\label{Eq:enaglinya}
\mbox{\textbf{Frequency relations for corresponding bundles $a_0^1=a_g$:}}\quad \omega_b^p=\frac{4\omega_b^2+1}{4 \sqrt{\sigma{\omega_b^2}}}
 \eea
See Fig.\il(\ref{Fig:Cmapobaseplot}). There is an extremum  at  $\omega_b^p={1}/{\sqrt{\sigma}}$ for $\omega_b=1/2$.
In general, the following relations hold:
\bea&&\label{Eq:enaglinyb}
\textbf{Corresponding metric  bundles $a_0^1=a_g$: }
\\\nonumber
&&
\mbox{It holds}\quad
a_{g}(\omega_b^1)=a_{g}^1(\omega_b), \quad \mbox{for}\quad \omega_b^1=\pm \frac{1}{4 (\omega_b)}.
\\\nonumber&&\mbox{It holds}\quad
a_{g}(a_0^1)=a_{g}(a_0), \quad \mbox{and}\quad\sigma=\sigma_1\quad  \mbox{for} \quad a_0=\frac{4}{a_0^1 \sigma }
\quad\mbox{ and }
 \\
 &&\nonumber\sigma\neq\sigma_1\quad\mbox{ for  }\quad a_0=\frac{4}{a_0^1 \sqrt{\sigma \sigma_1}}\quad\mbox{and}\quad a_0=\frac{a_0^1 \sqrt{\sigma_1}}{\sqrt{\sigma }}.
\\\nonumber
&&\mbox{It holds }\quad
a_{g}(a_0,\sigma)=a_{g}^1(a_0,\sigma_1)\quad \mbox{for}\quad\sigma_1= \frac{16}{a_0^4 \sigma } .
\eea
See Fig.\il(\ref{Fig:Cmapobaseplot}). The tangent radii for these bundles are
\bea\label{Eq:enaglinyc}
&&
\textbf{Tangent radii}\quad
r_g(\omega_b)=r_\flat\quad \mbox{for}\quad a=\sharp \frac{4 \omega_b}{4 \omega_b^2+1},\quad\mbox{where}\quad \flat=\pm,\quad \sharp=\pm,
\\\label{Eq:bimb-per}
&&
r_g(\omega_b)=r_g(\omega^p_b)\quad \mbox{for}\quad\omega^p_b=\pm\omega_b;\quad r_g(a_0)=r_\flat\quad \mbox{for}\quad a=\sharp \frac{4a_0 \sqrt{\sigma }}{a_0^2 \sigma +4},
\\\nonumber
&&
r_g(a_0^1,\sigma^1)=r_g(a_0,\sigma)\quad \mbox{for}\quad a_0^1=\pm\frac{a_0\sqrt{\sigma }}{\sqrt{\sigma_1}}
\eea
(the use of  $\flat$ and $\sharp$ for the sign convention  means that there is no correspondence between the two options).
The following special points determine some particular values of the bundles:
\bea\label{Eq:enaglinyd}
&&\mbox{}\quad a_g=M\quad \mbox{for}\quad\omega_b=\pm\frac{1}{2},\quad a_g=M\quad \mbox{for}\quad a_0^2 = \frac{4}{\sigma},\quad\mbox{and}\quad a_g(a_0=M)=\frac{4 \sqrt{\sigma }}{\sigma +4},\quad a_g(0)=0.
\\\nonumber
&&
\mbox{}\quad
r_g=M\quad \mbox{for}\quad\omega_b=\pm\frac{1}{2} \quad\mbox{ and}\quad
r_g(a_0)=M\quad a_0=\frac{2}{\sqrt{\sigma }}.
\\\nonumber
&&
\quad \mbox{For}\quad a_0=M \quad \mbox{it holds}\quad  \omega \to \frac{1}{\sqrt{\sigma }}
\quad \lim_{a_0\rightarrow+\infty}a_g= \lim_{a_0\rightarrow+0}a_g=\lim_{a_0\rightarrow+\infty}\omega_b=0,\quad
\quad \lim_{a_0\rightarrow+0}\omega_b=\infty
\\&&\nonumber
\lim_{a_0\rightarrow+0}r_g=0,\quad\lim_{a_0\rightarrow+\infty}r_g=2M,\quad\mbox{and}\quad r_g=2M \quad\mbox{for}\quad\omega_b=0,
\eea
see Fig.\il(\ref{FIG:Aslongas}). These limiting frequencies will be found also  in the frequency relations of
Sec.\il(\ref{Sec:allea-5Ste-cont}).
\begin{figure}
  \includegraphics[width=3cm]{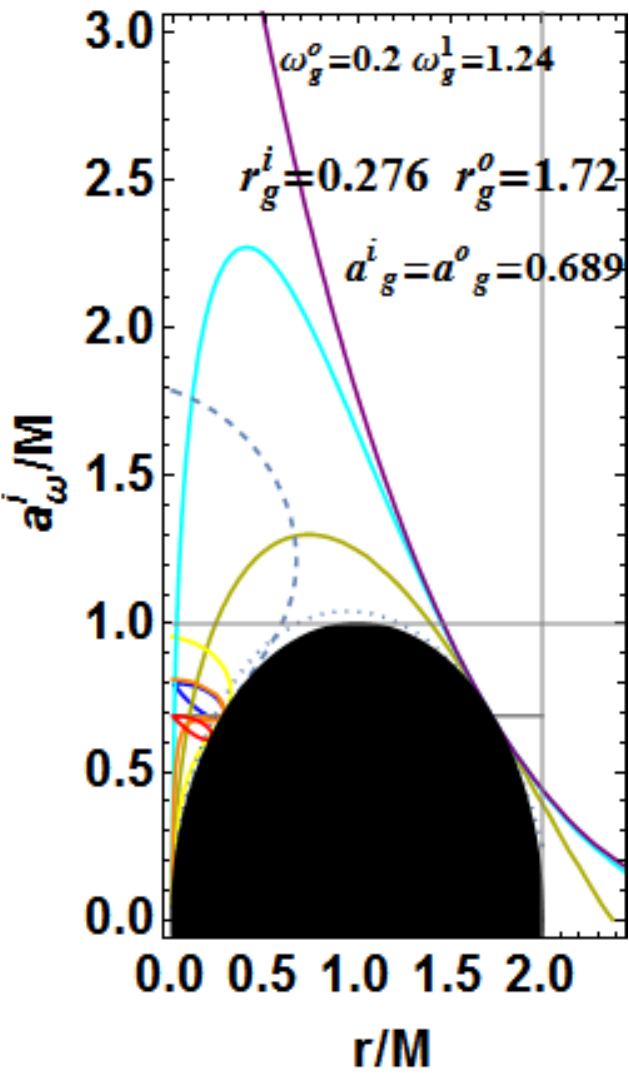}
  \includegraphics[width=3cm]{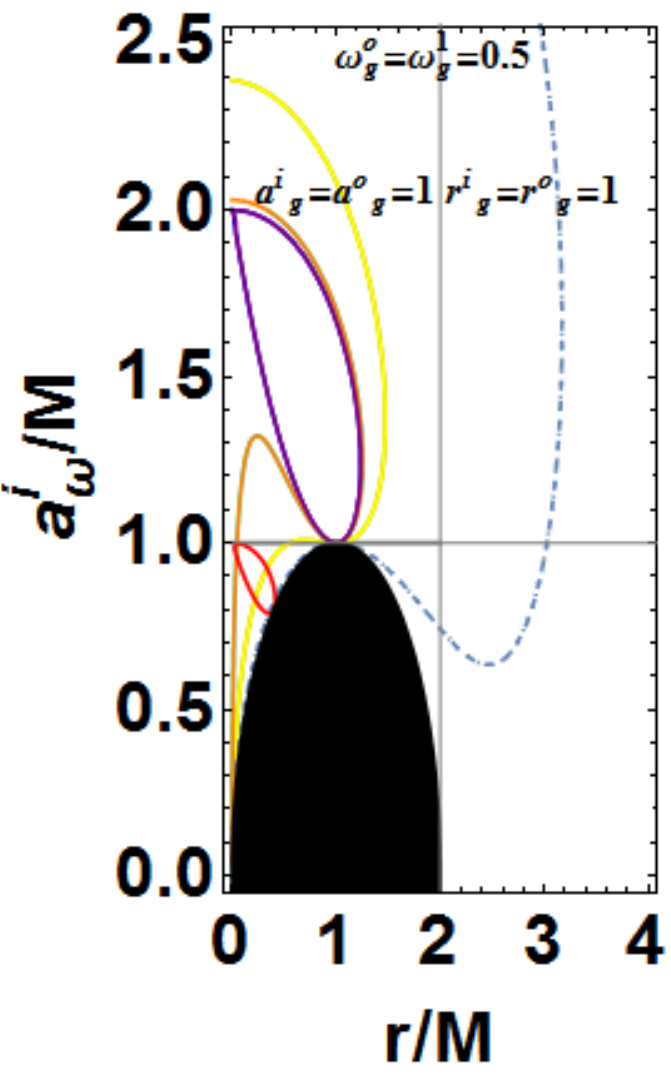}
      \includegraphics[width=3cm]{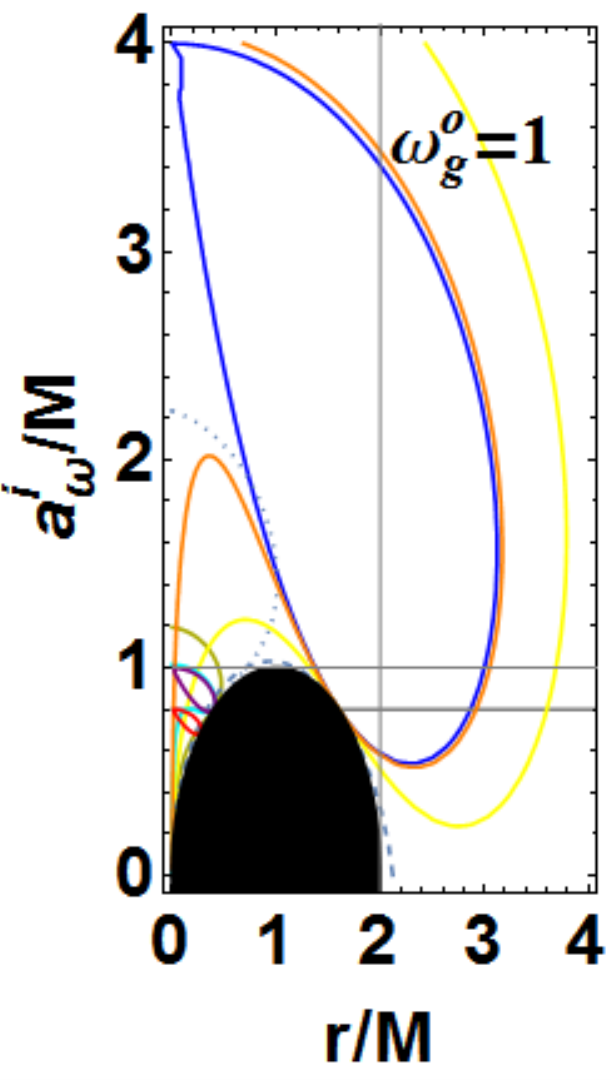}
  \includegraphics[width=3cm]{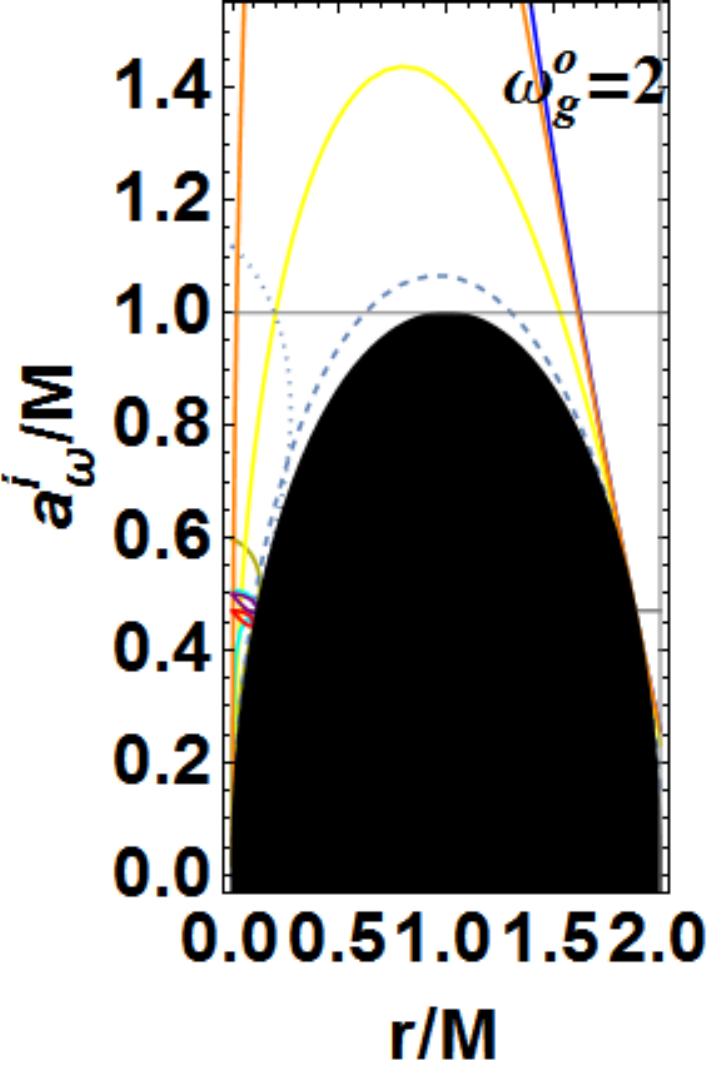}
   \caption{Horizon construction:  metric bundles $\Gamma_{a_g}$ with equal tangent  spin  $a_g$ and tangent frequencies
	$\omega_g=\omega_b$ and, consequently, equal tangent radius $r_g$. Construction of horizons
   $r_+=r_g^o$ and $r_-=r_g^i$ for a spacetime with $a=a_g$ through the corresponding bundle $a_g^1=a_g^2$   is also shown as the metric bundles with origin  $a_g^i=a_0$. The black region is a \textbf{BH} on the extended plane. Purple  and blue curves  are for $\sigma=1$;  orange  and cyan curves are for $\sigma=0.97$,
darker-yellow  and yellow curves are for $\sigma=0.7$,
 dotted and dashed curves are for $\sigma=0.2$, and the red curve is for $\sigma=1$.
It follows from the discussion of Eqs.\il(\ref{Eq:enagliny},\ref{Eq:enaglinya},\ref{Eq:enaglinyb},\ref{Eq:enaglinyc},\ref{Eq:enaglinyd}).}
\label{FIG:Aslongas}
\end{figure}
\\
\item[
\textbf{On the relation between origin spin and tangent spin}]

Consider the  origin frequency $\omega_0$, associated with a spacetime with   origin spin coincident with the  tangent  spin  ($a_0=a_g$) of a  bundle with  frequency $\omega^p_b$, plane $\sigma_p$ and origin $a_0^p$,  there is then
\bea\label{Eq:own-poi-tn-spea}
&&
\mbox{for}\quad\sigma=\sigma_p\quad
\omega_0^{\pm}(a_g)=\frac{1}{a^p_0 \sigma }+\frac{a^p_0}{4},\quad \omega_0^{\pm}(a_g)=\frac{4 (\omega^p_b)^2+1}{4 \sqrt{\sigma } \omega^p_b},\quad\mbox{where}\quad a_g=a_g(a_0^p)
\\\nonumber
&&
\sigma\neq\sigma_p\quad
\omega_0^{\pm}(a_g)=\frac{a_0^p}{4} \sqrt{\frac{\sigma_p}{\sigma}}+\frac{1}{a_0^p \sqrt{\sigma  \sigma_p}},\quad \mbox{where the tangent spin of the bundle with origin in $a_g$ is}
\\\nonumber
 &&a_g^p=\frac{4 a_0^p \sqrt{\sigma \sigma_p} \left[(a_0^p)^2 \sigma_p+4\right]}{(a_0^p)^2 \sigma_p \left[(a_0^p)^2 \sigma_p+4 \sigma +8\right]+16}.\quad
\eea
In this way, we also obtained a relation between the origins $(a_0,a_0^p)$ and the characteristic frequencies of the two corresponding  \textbf{MBs}.
The  relation between the frequencies  $(\omega_b,\omega_b^p$) is clearly independent of the plane  $\sigma_p$, because it is
included in the form of $\omega_b^p$ in terms of the origin spin.
The tangent spin $a_g^p$ has a maximum in terms of the first \textbf{MB} origin as $a_0$ as $\sigma_p={4}/{(a_0^p)^2}$--see Fig.\il(\ref{Fig:celtycplot}).
\begin{figure}
  \includegraphics[width=5cm]{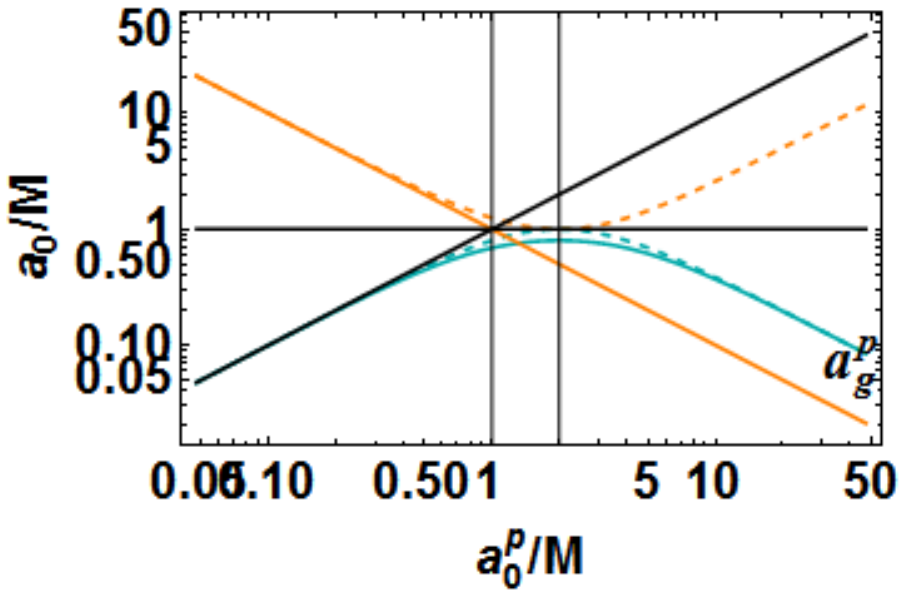}
 \includegraphics[width=5cm]{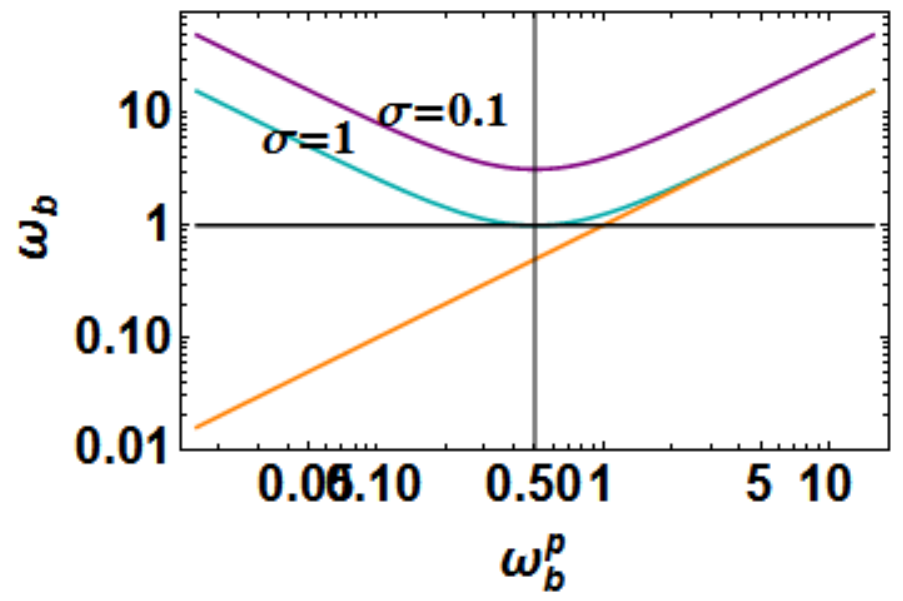}
  \caption{Analysis of Eq.\il(\ref{Eq:own-poi-tn-spea}). Left panel: for $\sigma=1$, $a_g^p$ of the corresponding bundle  (cyan), origin of the corresponding  \textbf{MBs} (cyan, dashed),
frequency $(1/a_0^p)$ (orange),
frequency of the corresponding \textbf{MBs} (orange dashed),
origin $a_0^p$ (black) as function of $a_0^p$. The right panel illustrates the frequencies relations $\omega_b^p$ as functions of the bundle frequencies of  Eq.\il(\ref{Eq:own-poi-tn-spea}) for different planes $\sigma$. The orange line is the $\omega_b=\omega_b^p$. }\label{Fig:celtycplot}
\end{figure}
\end{description}
\section{Horizon frequencies as bundle frequencies}
\label{Sec:allea-5Ste-cont}
The fact that the bundle frequency   $\omega_b$ coincides with the horizon frequencies $\omega_H^{\pm}(a_g)$  with spin  $a_g$ and, consequently, each photon orbital frequency, $\omega_+(a)$ or $\omega_-(a)$, at any point  $(r,a,\sigma)$ and any azimuthal angle
$\varphi=$constant {in any geometry $a$ of the bundle $a_{\omega}^{\pm}$} is a horizon frequency in the extended plane,  allows us   to analyze the bundle properties and, therefore,
 the Kerr spacetime causal structure  in terms of horizon frequencies\footnote{{More precisely, we can characterize the spacetime {causal structure} by using the fact that the condition $\mathcal{L}\cdot\mathcal{L}=0$  determines lightlike, causal,  boundary   delimiting stationary observer orbits.}}.

This analysis also allows us to better understand the \textbf{MBs} meaning   and specifically the relations between  the origin
$a_0$ and the bundle tangent spin $a_g$ as given in Eqs.\il(\ref{Eq:rom-a-witho-felix}) and
(\ref{Eq:summ-colo-quiri}).
The relation $a_g(a_0)$ given in Eq.\il(\ref{Eq:rom-a-witho-felix}),  as function of
$\la \equiv a \sqrt{\sigma}$,  is similar to that of the case $\sigma=1$ as follows from Fig.\il(\ref{FIG:raisemK}).
Consequently, all the results concerning $(a_g,r_g,\omega_b)$  obtained for the equatorial case   are valid also for
$\sigma\in]0,1[$, in terms of  $\la \equiv a \sqrt{\sigma}$; see  Eq.\il(\ref{Eq:summ-colo-quiri}) and Fig.\il(\ref{FIG:raisemK}).

\medskip

 In \cite{remnants}, we  demonstrated  that for $\sigma=1$:

$\bullet$ \textbf{MBs}  with \textbf{BH}  origins $a_0\in[0,M]$  are \emph{all} tangent to the inner horizon and, therefore, have
tangent spin $a_g\in[0,0.8M]$ and tangent radius $r_g\in[0,2/5M]$ with frequencies $\omega_b=\omega_H^-\geq1$--see
Fig.\il(\ref{Fig:adisciottoc2}). These \textbf{MBs} are \emph{all confined} in  the region of the extended plane upper-bounded  by the inner horizon (because $a_0\geq a\geq a_g$ for any spin $a$ of the bundle)--(see the $\mathbf{BB}$ model of \cite{remnants}).

 $\bullet$ \textbf{MBs} with \textbf{WNS} origins  $a_0\in[M,2M]$ can be used to construct only a portion of the inner horizon
 with $a_g\in[0.8M,M]$, $r_g\in[2/5M,M]$, and $\omega_b\in[0.5,1[$. These bundles are made of \textbf{NSs} and \textbf{BHs}.

$\bullet$ Metric bundles with strong  naked singularities origins (\textbf{SNSs}),  $a_0> 2M$, can be used to construct only the outer horizon on the extended  plane.

\medskip

These results have some remarkable implications:

\textbf{(1)} On the equatorial plane and, more generally, for sufficiently large values of $\sigma\in[0,1]$, \textbf{MBs} with \textbf{BH} origins are all \emph{confined} in the inner region of the extended plane (\emph{bundles confinement})--see
Fig.\il(\ref{Fig:Ly-b-rty}). Consequently:

\textbf{(2)} All the  frequencies   $\omega_b\geq1$ belonging to these bundles  cannot be found in the \textbf{NS} regime nor in the outer region  ($r>r_+$) on the extended  plane for large values of $\sigma$ (\emph{frequencies confinement}).
Results \textbf{(1)} and  \textbf{(2)},   therefore, hold  if we consider the variable   $\la_0=a_0\sqrt{\sigma}$.

\textbf{(3)} The  bottleneck region of Fig.\il(\ref{Fig:Ly-b-rty}), including \textbf{NS} metric bundle origins
$\la_0\equiv a_0\sqrt{\sigma}\in[M,2M]$, with bundle characteristic frequencies  $\omega_b\in[0.5,1[$ remain confined to
the \textbf{BH} inner region  or to the \textbf{NS} region, i.e.,  they are outside the region $r>r_+$ and $a\in[0,M]$, for large values of  $\sigma$.

\begin{figure}
 \includegraphics[width=5cm]{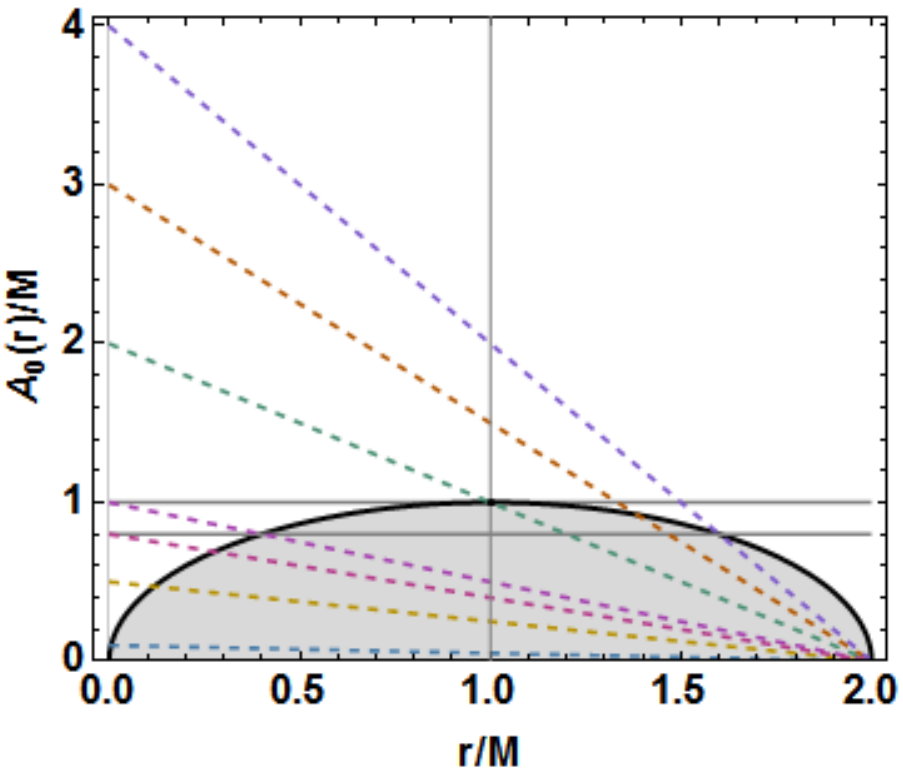}
  \includegraphics[width=5cm]{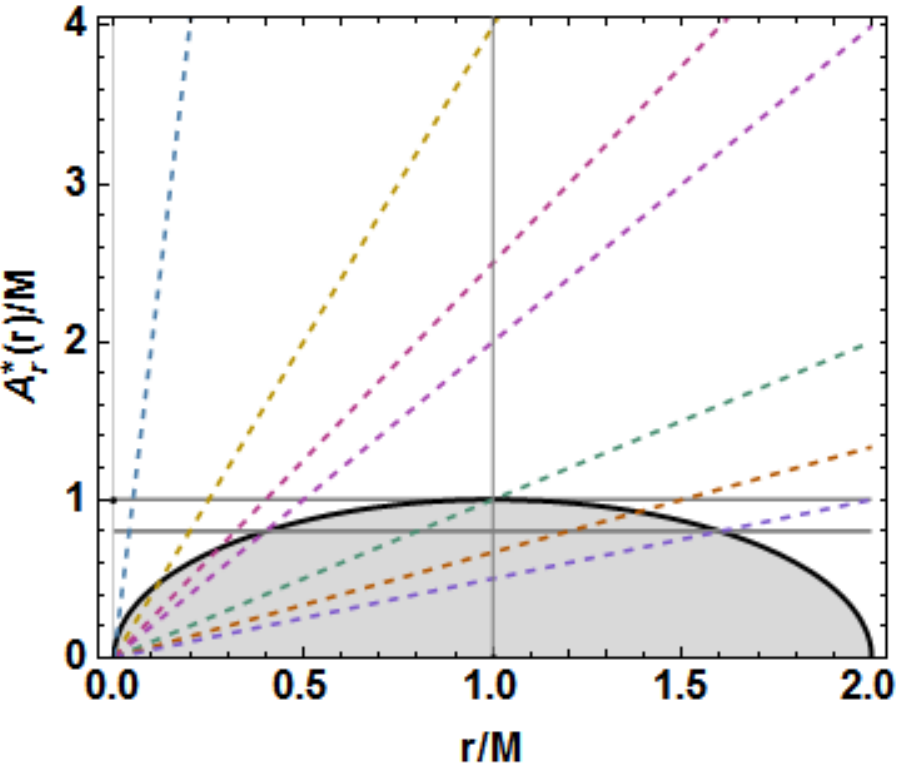}
 \includegraphics[width=5cm]{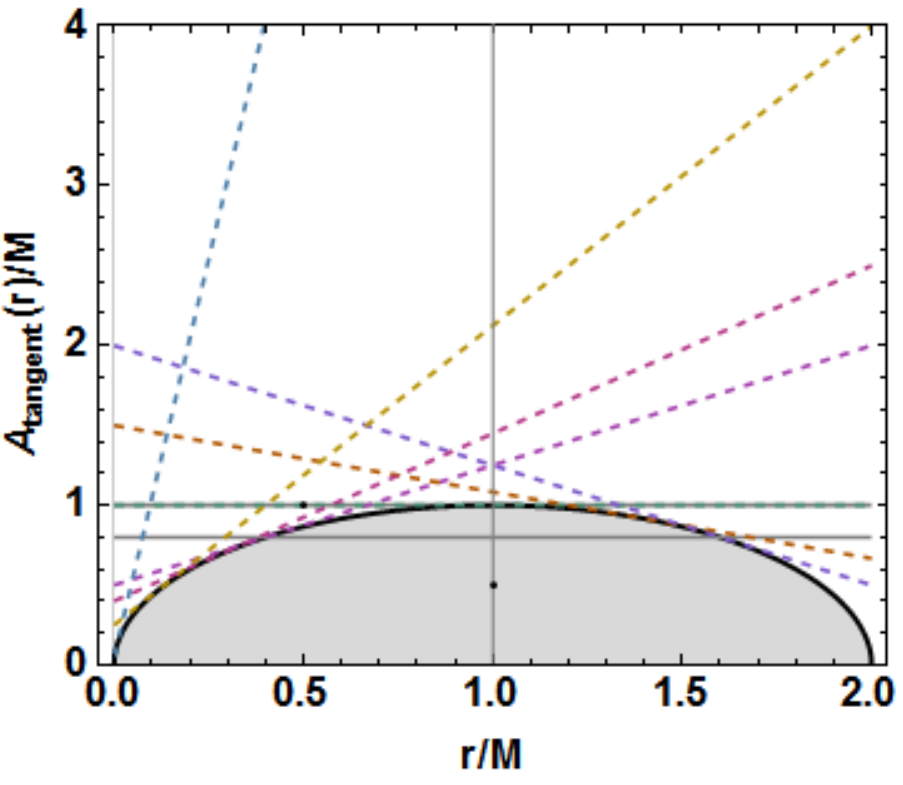}
  \caption{
	Gray region is the black hole on the extended  plane. Here  $\la \equiv a\sqrt{\sigma}$. The relations between  tangent spin  and origin spin of the bundles are shown--see Eq.\il(\ref{Eq:A0-A0-Atanget}).  Third panel defines the interior region and the bottleneck region of
	Fig.\il(\ref{Fig:Ly-b-rty}).}\label{Fig:meniangu}
\end{figure}

\textbf{(4)} All \textbf{MBs} with an origin spin of a {\bf NS} have characteristic frequencies of the outer horizon on the extended  plane,  \emph{apart} from the {\emph{bottleneck}} region ($\la_0\in[M,2M]$), where \textbf{MBs} have the frequencies of the inner horizon. We can see this relation, in terms of  $\la$ between tangent spin and tangent radius also in
Fig.\il(\ref{Fig:meniangu}). As it is  clear from the analysis of
Sec.\il(\ref{Sec:allea-5Ste-cont}), {there are bundles (and therefore orbits) with characteristic  frequencies equal to  the outer horizon  frequencies,  which are located in the inner
region of the extended plane}. Vice versa, outer horizon frequencies and are all and only those of \textbf{NS} origins
$\la_0=a_0\sqrt{\sigma}\geq2M$ for \emph{any} plane $\sigma$ (thus, $a_0\geq2M/\sqrt{\sigma}\geq2M$).

The results \textbf{(4)} imply that the outer horizon can be constructed only by strong naked singularity origins with
$\la_0\geq2M$ and,  therefore,  with  $a_0\geq2M/\sqrt{\sigma}\geq2M$.

More precisely, concerning  the inner horizon confinement,
 as  the relations  $a_g(a_0,\sigma)$, $r_g(a_0,\sigma)$ and $\omega_b(a_0,\sigma)$ can be read as
$a_g(\la_0)$, $r_g(\la_0)$ and $\omega_b(\la_0)$, then all the relations between these quantities  valid for the equatorial plane $\sigma=1$, hold also for any $\sigma\neq0$
(from the symmetry properties it follows  also the scheme presented in Fig.\il\ref{Fig:Ly-b-rty}). On the contrary, the
quantity  $\mathcal{L}_{\mathcal{N}}$ cannot be written as a function of $\la$ only, the explicit expression
for the \textbf{MBs} given  in Eq.\il(\ref{Eq:Gro-e-nNoth}), or for frequencies given in Eq.\il(\ref{Eq:bab-lov-what}). Therefore,
they do not depend only on the variable  $\la$, but depend explicitly on the plane $\sigma\in[0,1]$.
This has important consequences for the confinement of  \textbf{MBs} tangent to  the inner horizon and  of the inner horizon frequencies.
For each point  $r_g\in a_{-}$, tangent point belonging to the  inner horizon on the extended  plane, it is possible to find a plane $\sigma<\sigma_{descr}$ such that solutions of the condition   $\mathcal{L}\cdot\mathcal{L}=0$ exist for  $a\in[0,M]$ and
$r>M$, with $\omega_b\geq0.5$ and, particularly, $\omega_b\geq1$ (frequency range of the inner horizon $a_g<M, r<M$ and $a_g<0.8M$, $r<2.5 M$). This condition, moreover, had already been discussed differently in  Eq.\il(\ref{Eq:unirmm}). This result implies that there are no orbits confined in the inner horizon and that close to the rotational axis $\sigma\ll1$,  for any spacetime $a$ it is possible to find an orbit in the outer region,  $r>r_+$, with frequency equal to that of the internal horizon for $a$ close to the rotation axis,
see also Figs\il\ref{Fig:Werepers1}, \ref{FIG:funzplo}, \ref{FIG:disciotto1}, \ref{FIG:rccolonog}, and \ref{FIG:Aslongas}.

\begin{figure}
 \includegraphics[width=3.4cm]{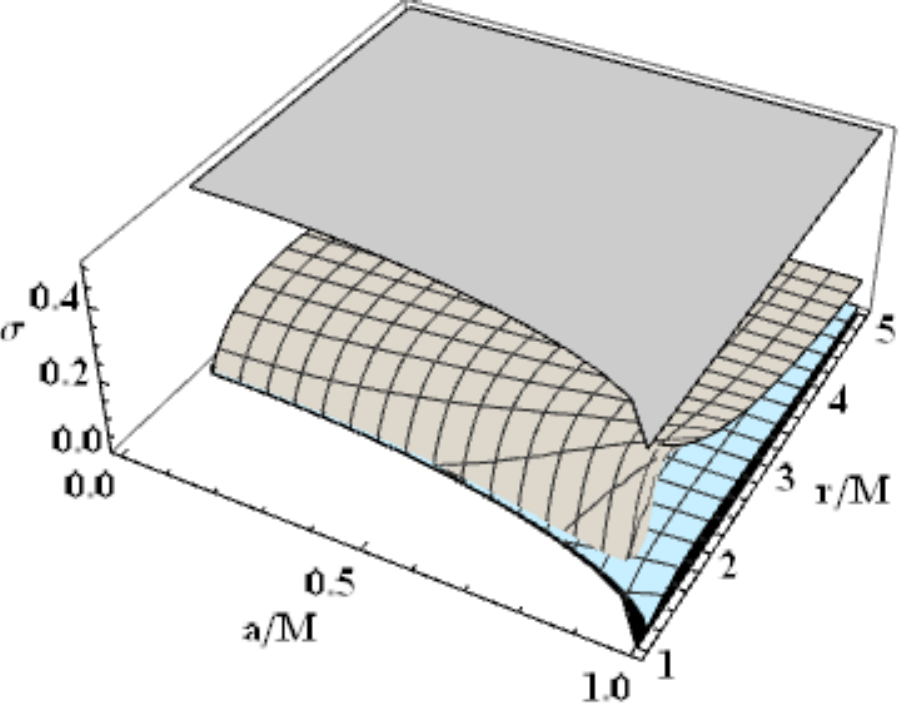}
  \includegraphics[width=4cm]{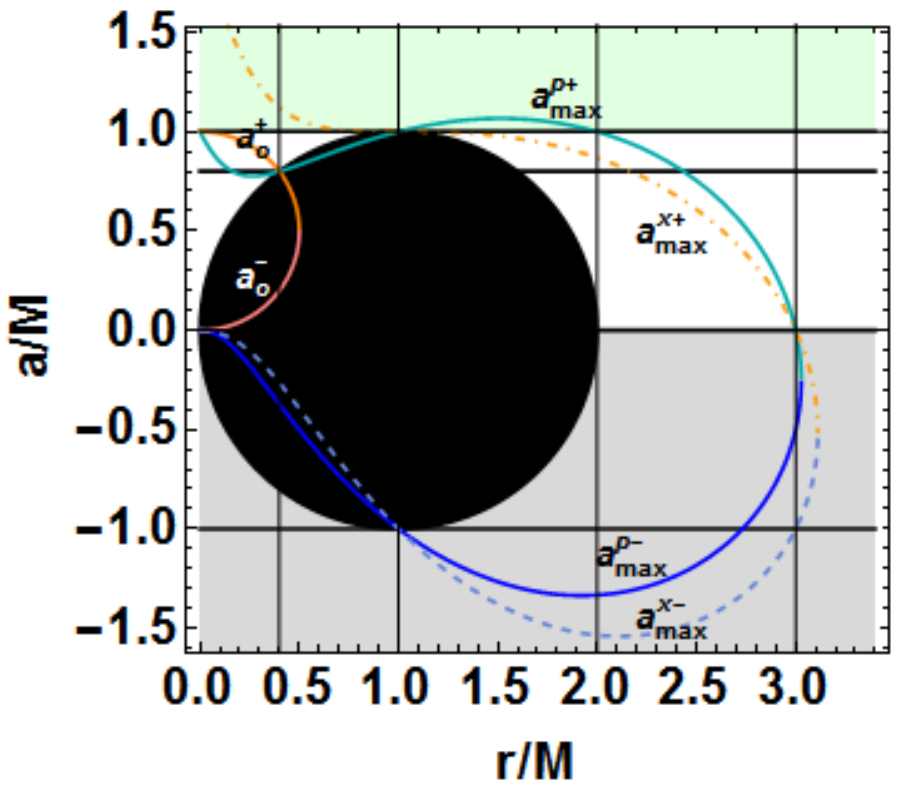}
 \includegraphics[width=3.3cm]{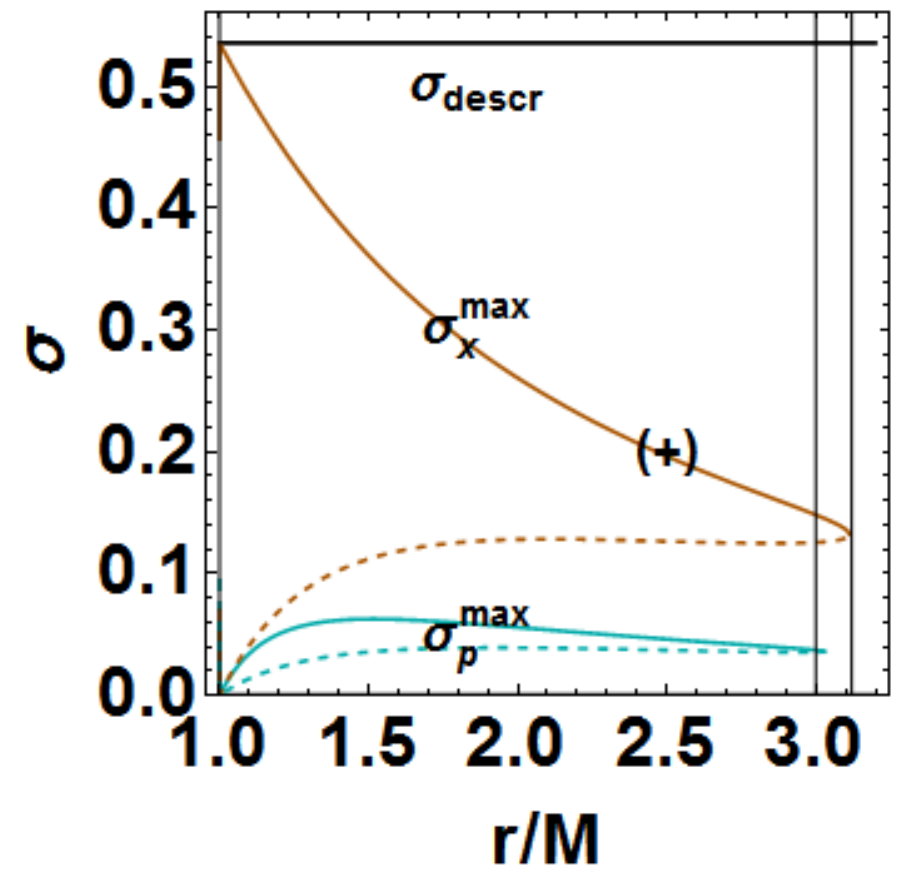}
 \includegraphics[width=3.4cm]{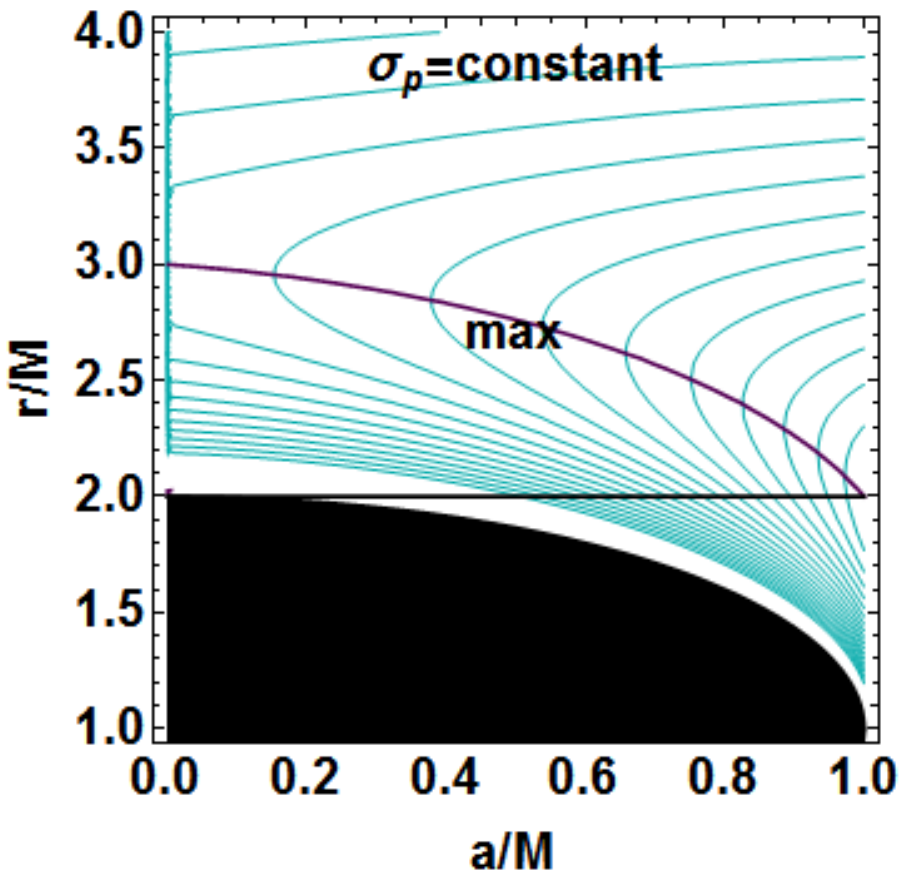}
 \includegraphics[width=3.4cm]{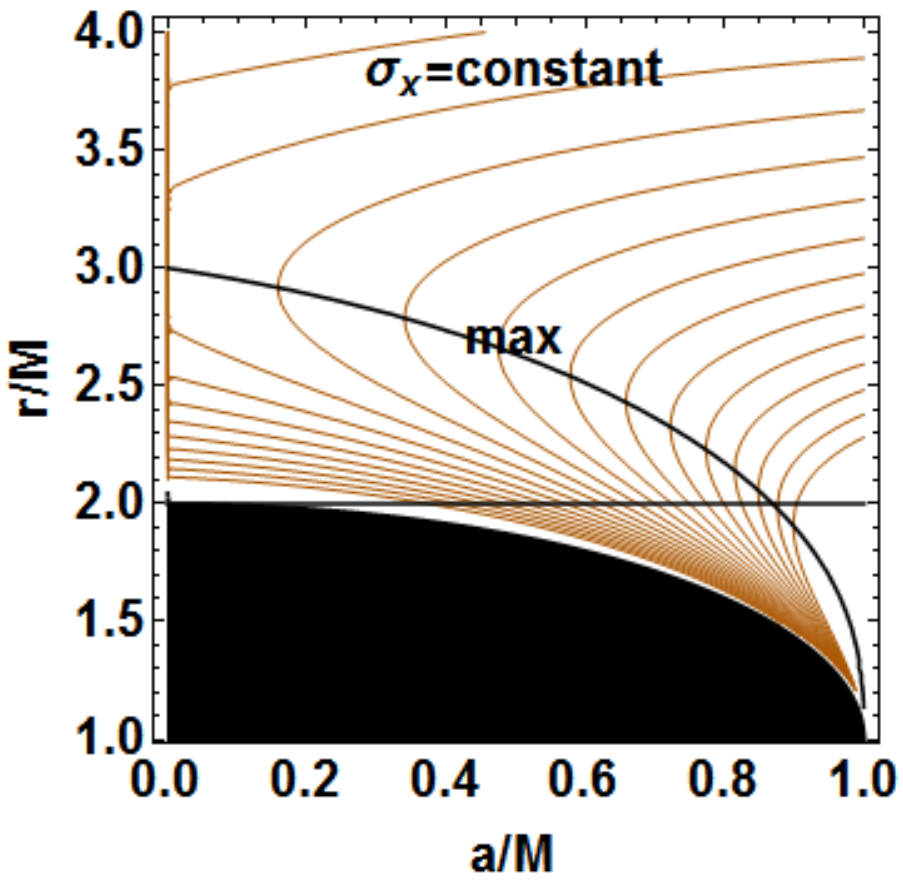}
  \caption{Analysis of Eqs.\il(\ref{Eq:rtdl225}) and (\ref{Eq:rtdl2251}) on the confinement of the metric bundles with a frequency equal to that of the internal horizon $\omega_H^-$ on the extended  plane. This topic is also addressed in
	Eq.\il(\ref{Eq:unirmm}). It is shown that it is always possible to find bundle frequencies $\omega>1$ (equal to the  inner horizon frequencies) in the exterior region as in  Fig.\il(\ref{Fig:Ly-b-rty}) for  large spin $a$   and small $\sigma$.  The frequency $\omega_b\geq1$ refers to  $a_g\in[0,0.8]$, and $\omega_b\in[0.5,1[$ is related to  $a_g\in]0.8,1]$.
	Left panel: 3D plot of  planes $\sigma_{p}<\sigma_x$ as functions of $a\in[0,M]$ and $r>M$. Planes $\sigma\in\{0,1\}$ are also shown. Second panel: black region is the \textbf{BH} on the extended  plane, the gray region contains negative values of frequencies (corresponding  to retrograde motion or $a<0$). The  maximum  values of $\sigma_{p}<\sigma_x$ are also shown. Third panel: maximum values of $\sigma_{p}<\sigma_x$  for positive and negative $a$ are shown as functions of $r/M$.
	The value $\sigma_{descr}$ as in Eq.\il(\ref{Eq:unirmm}) and Fig.\il(\ref{FIG:ARESE1}) is also shown.
	Fourth and fifth panels: planes $\sigma_p=$constant and $\sigma_x=$constant for $(a/M, r>r_+)$}\label{Fig:FrencPlot}
\end{figure}

More in detail, we have that
\bea&&\label{Eq:rtdl225}
(\textbf{SBH})\quad\mathcal{L}\cdot\mathcal{L}=0,\quad\mbox{for}\quad\mbox{\textbf{(1)}}\quad a\in]0,M],\quad r\geq M,\quad \sigma\in]0,1],\quad\mbox{\textbf{(2)}}\quad \omega >1\quad\mbox{for}
\\\nonumber
&&\omega=\omega_-\quad \sigma\in ]0,\sigma_p[\quad\mbox{and}\quad\bullet\quad r\in]M,2M],\quad a\in]a_{\pm},M]\quad\mbox{or}\quad\bullet r>2M,\quad a\in]0,M].
\\\nonumber
&&
\\\nonumber
&&
(\textbf{WBH})\quad\mathcal{L}\cdot\mathcal{L}=0,\quad\mbox{for}\quad\mbox{\textbf{(1)}}\quad a\in]0,M],\quad r\geq M,\quad \sigma\in]0,1],\quad\mbox{\textbf{(2)}}\quad \omega\in[1/2,1]\quad\mbox{for}
\\\nonumber
&&\omega=1/2\quad\mbox{for}\quad a=M\quad r=M,\quad
\\\nonumber
&&\omega=\omega_-\quad \sigma\in [\sigma_p,\sigma_x]\quad\mbox{and}\quad\bullet\quad r\in]M,2M],\quad a\in]a_{\pm},M]\quad\mbox{or}\quad\bullet r>2M,\quad a\in]0,M].
\\\nonumber
&&\mbox{where}\quad\sigma_p\equiv \frac{1}{2} \left[\frac{a^4+a^2 \left(2 r^2+1\right)-4 a r+r^4}{a^4+a^2 (r-2) r}-\sqrt{\frac{\left[(a-1) a+r^2\right]^2 \left[a^2 (a+1)^2+2 a (a+1) r^2-8 a r+r^4\right]}{a^4 \Delta^2}}\right]\\\nonumber
&&\sigma_x\equiv\frac{1}{2} \left[\frac{a^4+2 a^2 \left(r^2+2\right)-8 a r+r^4}{a^4+a^2 (r-2) r}-\sqrt{\frac{\left[(a-2) a+r^2\right]^2 \left[a^2 (a+2)^2+2 a (a+2) r^2-16 a r+r^4\right]}{a^4 \Delta^2}}\right].
\eea
The frequency $\omega_b\geq1$ refers to  $a_g\in[0,0.8M]$, $\omega_b\in[0.5,1[$ refers to tangent spins $a_g\in]0.8M,M]$.
This analysis is in agreement with Sec.\il(\ref{Sec:pri-photon-fre}) and particularly Eq.\il(\ref{Eq:unirmm}), as can be seen  also from  Fig.\il(\ref{FIG:ARESE1}), where it is clear the role of $\sigma_{descr}\geq\sigma_x\geq\sigma_p>0$.
There are maximum points for $(\sigma_p,\sigma_x)$ as functions of $r/M$:
\bea&&\label{Eq:rtdl2251}
\partial_r\sigma_p=0,\quad\mbox{for}\quad a_{\max}^{p \pm}\equiv
\frac{1}{r+1}-\frac{1}{2}\pm\frac{\sqrt{r [r (13-4 (r-2) r)-2]+1}}{2 \sqrt{(r+1)^2}},\quad
a_o^{\pm}\equiv\frac{1}{2} \left(1\pm\sqrt{1-4 r^2}\right)
\\\nonumber
&& \partial_r\sigma_x=0,\quad\mbox{for}\quad a=a_o^{\pm},\quad\mbox{and}\quad a_{\max}^{x \pm}\equiv
\frac{2}{r+1}-1\pm\frac{\sqrt{1-r (r [(r-2) r-4]+2)}}{\sqrt{(r+1)^2}},
\eea
--see Fig.\il(\ref{Fig:FrencPlot}).

\medskip

Before moving on to the analysis of frequencies, we conclude this part by considering the tangent lines to the horizon curve  on the extended  plane, which also provide  the  transformations used in the scheme of  Fig.\il(\ref{Fig:Ly-b-rty}).
Note that as they relate $a_g$ and $r_g$ to $a_0$, then these relations can be parameterized in terms of the variable $\la \equiv a \sqrt{\sigma}$.
Therefore,
\bea\nonumber
&&
\la_0(r)\equiv \la_0-\frac{\la_0}{2} r,\quad \la_{r}^*(r)\equiv \frac{2}{\la_0}r,\quad \la_{tangent}(r)\equiv -\frac{\left[\sqrt{\frac{\la_0^2}{\left(\la_0^2+4\right)^2}} \left(\la_0^4-16\right)\right]}{4 \la_0^2} r+\frac{\left(\la_0^4-16\right) \sqrt{\frac{\la_0^2}{\left(\la_0^2+4\right)^2}}+8 \la_0}{2 \left(\la_0^2+4\right)},
\\
&&\nonumber \mbox{and  }\quad \la_0(r)=\la_{r}^*(r)\quad\mbox{for}\quad r= r_g;\quad \la_0(r)=\la_{tangent}(r)
\quad\mbox{for}\quad r=r_g\quad\mbox{or}\quad \la_0=-\frac{2}{\sqrt{3}},
\\\label{Eq:A0-A0-Atanget}
&&\mbox{and}\quad  \la_{tangent}(r)=\la_{r}^*(r) \quad\mbox{for}\quad r=r_g\quad\mbox{or}\quad  \la_0=-2 \sqrt{3},
\eea
see Fig.\il(\ref{Fig:meniangu}).
\medskip

\subsection{On the bundle frequency relations}

In the following we consider the frequency relations among metric bundles.  We start by  considering the following quantities
\bea&&\label{Eq:cora-W}
s\equiv\frac{\omega_H^+}{\omega_H^-}=\frac{r_-}{r_+}=\frac{1}{4(\omega_H^-)^2}=4(\omega_H^+)^2\quad \mbox{see also Eq.\il(\ref{Eq:enagliny}).}\quad
\mbox{It holds }\quad {\omega_H^+}{\omega_H^-}=\frac{1}{4}\quad r_+ r_-=a^2\quad\mbox{and}
 \\
&&\partial_a^{(2)}\ln s=0\quad\mbox{for}\quad a= a_{\partial}\equiv\frac{1}{\sqrt{2}}\quad\mbox{where}\quad s_{\partial}=3-2 \sqrt{2}\approx0.171573,\quad (\ln s_{\partial}\approx-1.76275),
\eea
We see that $s=1/4$ (and then $s=4$) is a remarkable frequency ratio (note that for a  fixed tangent  spin $a_g$,
the inequality  $s\leq 1$ is valid).
This quantity is  showed in  Fig.\il(\ref{Fig:AlbertsDOME}). The spin $a_{\partial}$  provides  a remarkable frequency ratio for the physics of black holes;  this is a saddle point  of the function $\ln s$ for  $a$. Clearly, $s$  is minimal in the limiting spherically symmetrical case where $a=0$ and $s=0$.
We also note that, from the analysis of the  previous sections, it is evident that, for  small values of $\sigma$
($\sigma<\sigma_{descr}<0.5)$ and, {in general, for  large spins $a/M$, the  inner horizon frequencies can exist on orbits of the outer region--see particularly Fig.\il(\ref{Fig:Pnoveplot})}.

\begin{figure}
\includegraphics[width=4cm]{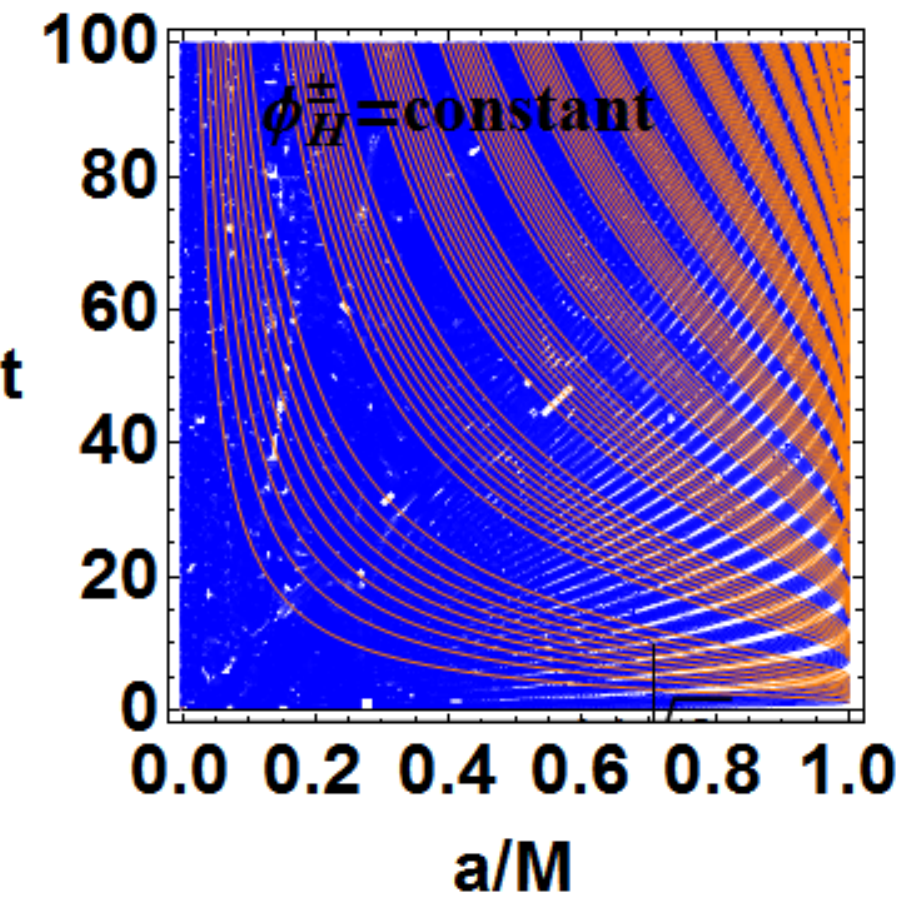}
\includegraphics[width=5cm]{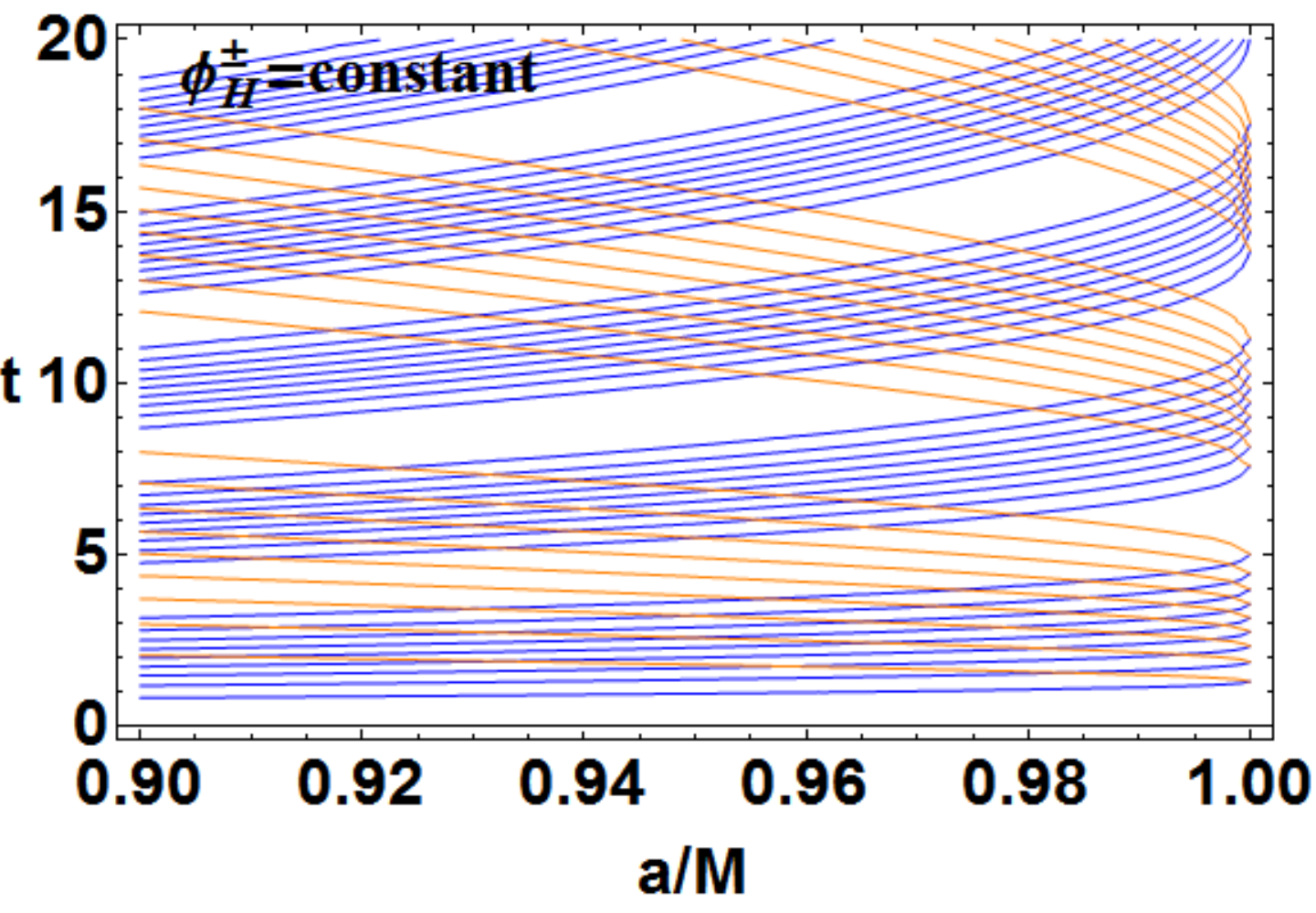}
\caption{Function $\phi_H(\omega_H^{\pm} t)=$constant in the plane $(t,a/M)$. $\omega_H^{\pm} $are the frequencies of the outer and inner horizons, respectively. Based on the analysis of Sec.\il(\ref{Sec:allea-5Ste-cont}). See also Figs\il(\ref{Fig:exppLOTINVAS1}).}\label{Fig:dainseriD}
\end{figure}

Note  that if  there would be  no outer orbits (orbits in the exterior region $r>r_+$ of the extended plane) with frequency equal to the inner horizon frequency $\omega_H^-$, it would imply that there are not metric bundles tangent to the inner horizons in the exterior region;  therefore, the \textbf{MBs} tangent to the inner horizon would be all confined in the interior region.
This is also  the problem addressed  in  Sec.\il(\ref{Sec:pri-photon-fre}). 
To understand the relations between different   bundle frequencies   as horizon frequencies for  \textbf{MBs} of the classes
$(\Gamma_{a_g},\Gamma_{r_g},\Gamma_{a_0})$ at equal $a_g$, $r_g$, or $a_0$ respectively, it is convenient to study the frequency ratio
$s=(\omega_b/\omega_b^1)$. The parameter  $s$  is also understood, in general, as the bundle frequency ratio, in which case the dependence from $a$ in $(\omega_b,\omega_b^1)$ will obviously be different. This allows us to consider both  $s$ and  $1/s$ as ratios.
It can also be considered as a wave signal $\phi_{H}^{\pm}(t)$  with frequencies $\omega_b=\omega_H^{\pm}$, respectively, that gives an immediate impression of the photon frequency difference in those orbits and  geometries included in  the metric bundles --Figs\il\ref{Fig:dainseriD} and \ref{Fig:exppLOTINVAS1}.

It is clear that the ratio $s$ is a  function of the tangent spin $a_g\in a_{\pm}$. \textbf{MBs} tangent   to the inner horizon on the extended  plane  will always have a greater characteristic  frequency  $\omega_b=\omega_H^-(a_g)$ than \emph{any} \textbf{MB} tangent to the outer horizon--see also Figs\il(\ref{FIG:raisemK}) and \il(\ref{FIG:toa8}).
However,  we can express this ratio in terms of the bundle origins to enlighten  the relation between different \textbf{MBs} frequencies   related by   Eq.\il(\ref{Eq:rom-a-witho-felix}) and
Eq.\il(\ref{Eq:summ-colo-quiri}). In Sec.\il(\ref{Sec:partic-freq-ratios}), we analyze particular cases.

We now  go back to the frequency analysis,  introducing the following ``resonance'' solutions:
\bea\label{Eq:a-beta-i-ii}
&&a_{\beta}^{i}:\quad\omega_H^\pm(a_{\beta}^{i})=s\omega_H^-(a)  \quad\mbox{where}\quad
a_{\beta}^i\equiv\frac{2 a s \left[r_-(a) s^2+r_+(a)\right]}{a^2 \left(s^2-1\right)^2+4 s^2},
\\
&&
a_{\beta}^{ii}:\quad\omega_H^\pm(a_{\beta}^{ii})=s\omega_H^+(a)  \quad\mbox{where}\quad a_{\beta}^{ii}\equiv\frac{2 a s \left[r_+(a) s^2+r_-(a)\right]}{a^2 \left(s^2-1\right)^2+4 s^2},
\\
&&\mbox{then}\quad a_{\beta}^{i}=a_{\beta}^{ii}\quad\mbox{for}\quad \lambda=\pm 1\quad\mbox{or}\quad
\lambda=0\quad\mbox{where}\quad \lambda=\{a/M,s\}.
\eea
Note that the equations with solutions $a_{\beta}=(a_{\beta}^{i},a_{\beta}^{ii})$ allow us to analyze and characterize separately   both the solutions for  $(s, s^{-1})$ since $a_{\beta}^i(s)=a_{\beta}^{ii}(1/s)$--see Fig.\il(\ref{Fig:AlbertsDOME}).

The spins $a_{\beta}=(a_{\beta}^i,a_{\beta}^{ii})$ have a maximum with  respect to the  spin
$a=a_g\in[0,M]$ for  $a_{\mathcal{s}}(s)$, which is a   function of $s$ as:
\bea&&\label{Eq:short-a-s.sta}
 a_{\mathcal{s}}(s)\equiv\pm\frac{2 s}{\left(s^2+1\right)}:\quad \partial_a a_{\beta}=0,\quad \mbox{where}\quad a_{\beta}=(a_{\beta}^i,a_{\beta}^{ii}),\quad\mbox{and }\\\nonumber
&&
a_{\beta}^{ii}(a_{\mathcal{s}})=\frac{s^4+\left[\sqrt{\left(s^2-1\right)^2}+2\right] s^2-\sqrt{\left(s^2-1\right)^2}+1}{2 \left(s^4+1\right)},\quad
a_{\beta}^{i}(a_{\mathcal{s}})=\frac{s^4-\left[\sqrt{\left(s^2-1\right)^2}-2\right] s^2+\sqrt{\left(s^2-1\right)^2}+1}{2 \left(s^4+1\right)},
\eea
--see Fig.\il(\ref{Fig:AlbertsDOME}).
The following remarkable feature clarifies some significant aspects of the metric bundles:
\bea\label{Eq:bre-Siti-dea-lno}
\mbox{there is}\quad\textbf{(1)}\quad a_{\mathcal{s}}(s)=a_g(a_0,\sigma) \quad\mbox{for}\quad s\rightarrow \frac{2}{a_0 \sqrt{\sigma }},\quad\mbox{and}\quad\textbf{(2)}\quad
a_{\mathcal{s}}(s)=a_g(\omega_b)\quad\mbox{for}\quad s\rightarrow 2 \omega_b
\eea
where $a_g(a_0,\sigma) $ and $a_g(\omega_b)$ are defined in  Eq.\il(\ref{Eq:rom-a-witho-felix})--see Fig.\il(\ref{Fig:AlbertsDOME}).
These relations ensure that the maximum point spin $a_{\mathcal{s}}(s)$ of Eq.\il(\ref{Eq:short-a-s.sta}) is the tangent
spin $a_g$ (that is the horizon  $a_{\pm}$ on the extended  plane)  for  fixed frequency ratios $s$ that are twice those of the bundle or, equivalently,
which depend on the bundle spin  origin (in all the analysis of Eq.\il(\ref{Eq:bre-Siti-dea-lno}) we are considering equal spin $a$).
\begin{figure}
\includegraphics[width=5cm]{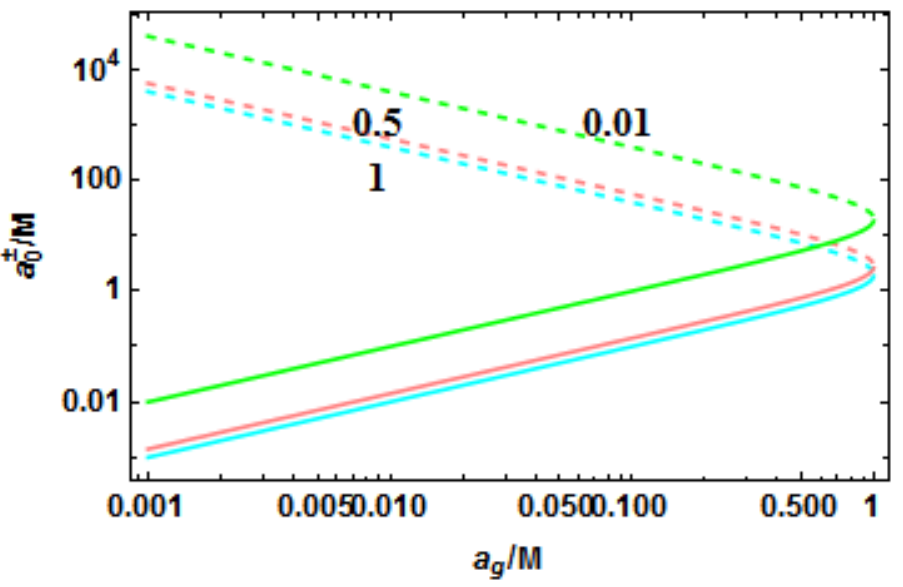}
\includegraphics[width=5cm]{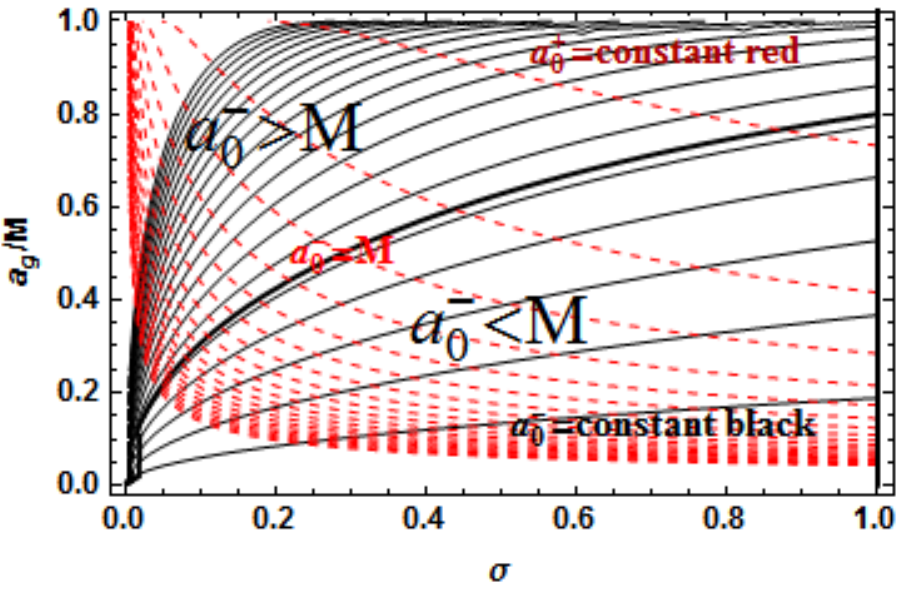}
  \includegraphics[width=5cm]{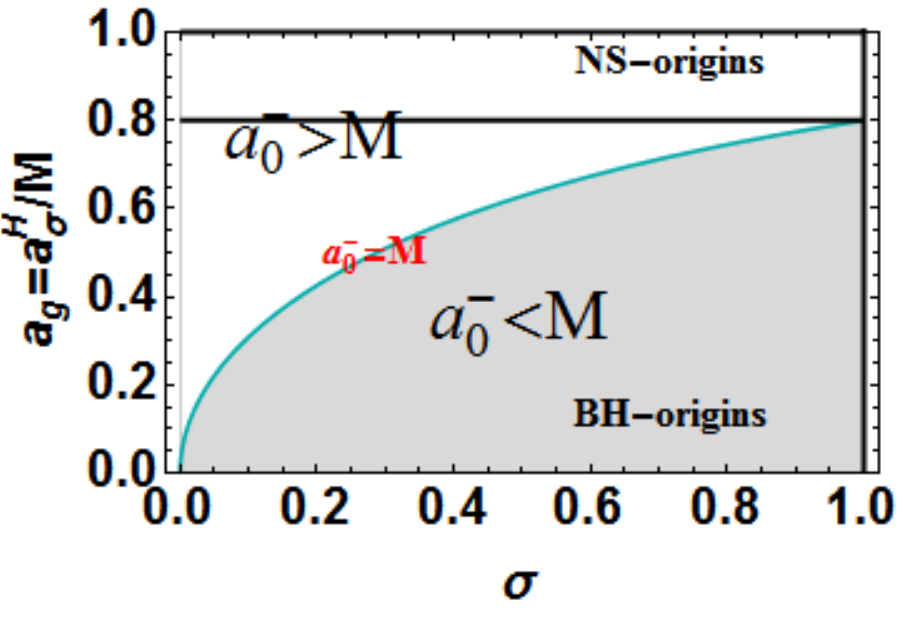}
\includegraphics[width=5cm]{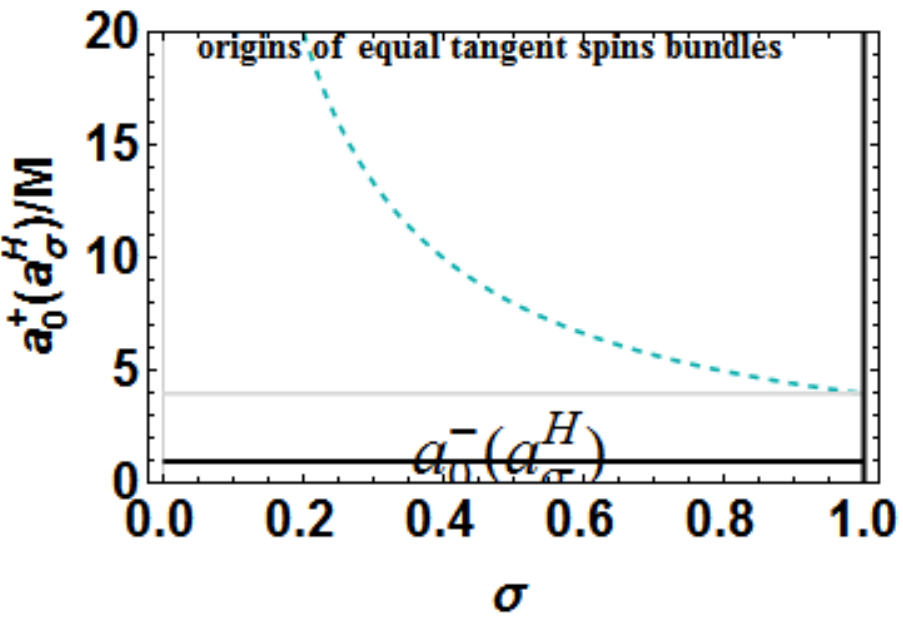}
\includegraphics[width=5cm]{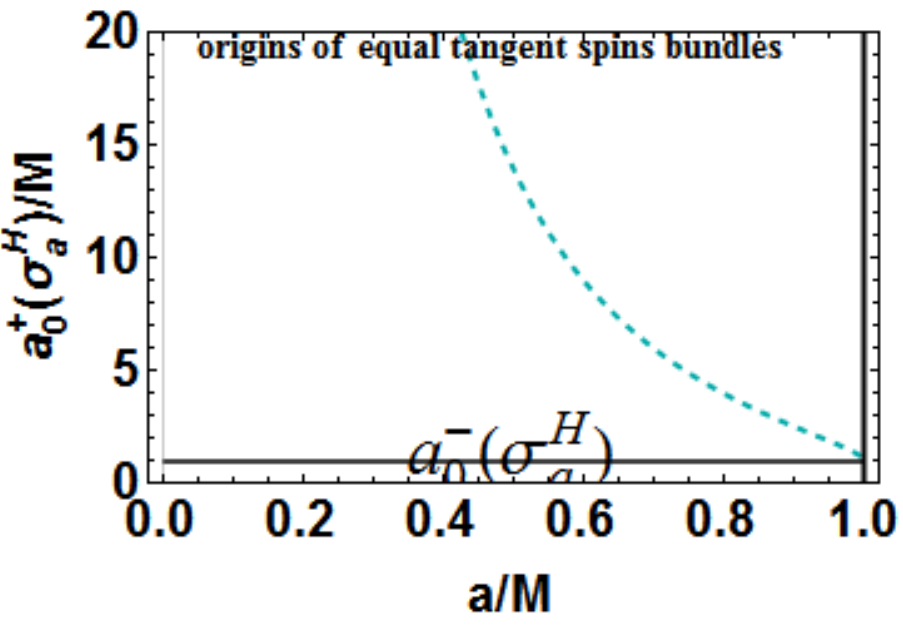}
  \caption{Study of origins $a_0^{\pm}$ in terms of the outer and inner horizon frequency. See Sec.\il(\ref{Sec:allea-5Ste-cont}) and Eqs.\il(\ref{Eq:condgi-ut-ura})  and \il(\ref{Eq:gov-vote}).
	Upper panels left: plots of the bundle origin $a_{0}^{\pm}$ as functions of the tangent spins $a_0$  on the outer and inner horizon, respectively, for different planes  $\sigma$. The logarithm plot shows the logarithm identity as discussed in the text.
	Here $a_0^+>a_0^-$ and $a_0^+=a_0^-$ for tangent spin in $a_g=M$.
Bundle tangent spin $a_g$ as  horizon spins $a_{\pm}$. }\label{Fig:thelas}
\end{figure}
Clearly, $a_{\mathcal{s}}(s)$   is  invariant under the transformation $s\rightarrow1/s$, i.e.,
$a_{\mathcal{s}}(s)=a_{\mathcal{s}}(s^{-1})$.   The maximum  $a_{\beta}(a_{\mathcal{s}})$ obviously  presents the discriminating value  $s=\pm1$. This is also clear from Fig.\il(\ref{Fig:AlbertsDOME}).
We are particularly interested in the integer values of $s$, considering then the symmetries under the transformation
$s\rightarrow1/s$.

Similarly, both  $a_{\beta}=(a_{\beta}^i,a_{\beta}^{ii})$ have a maximum for the value:
\bea\label{Eq:s-beta-toni}
s_{\beta}^\pm\equiv \sqrt{\frac{2 r_{\pm}}{a^2}-1}=\sqrt{\frac{4\omega_H^{\mp}}{a}-1}.
\eea
The function   $s_{\beta}^\pm=s$ coincides, in fact, with the solution of $a=a_{\mathcal{s}}$ of Eq.\il(\ref{Eq:short-a-s.sta}); this is  also clear from the three-dimensional  plots of  Fig.\il(\ref{Fig:AlbertsDOME}).
Moreover, the following  bi-logarithmic identity holds
$\ln s_{\beta}^+(\ln a)=-\ln s_{\beta}^-(\ln a)$ (i.e., under the transformation $a\rightarrow\ln s$ and $a\rightarrow\ln a$). This identity will be found in many of the quantities related to the horizon frequencies  we  analyze in this section. This is also   clear from  the  bi-logarithmic plots.

\subsection{Origin and tangent spins  relations: metric bundles confinement}

With regard to the relation between the origin-spin  $a_0$ and the tangent-spin $a_g$, we analyze the conditions on the \textbf{MBs} origin for the metric bundle to be tangent to the inner horizon $r_g=r_-$ (equivalently, $\omega_b=\omega_H^-$) \emph{or} the outer horizon $r_g=r_+$  (equivalently, $\omega_b=\omega_H^+$) on the extended  plane. {Particularly, we discuss the  conditions  for the \textbf{MBs} tangent
to the inner horizon curve to remain confined  in
the inner region  of the extended plane,  bounded by the inner horizon,  where  $a\in [0,M]$ and $r\in [0,M]$.}

The situation for $\sigma=1$, equatorial plane,  was  analyzed in \cite{remnants}. For a general plane the situation is analogue to the equatorial plane, if we consider the variable $\la=a\sqrt{\sigma}$ as shown in  Fig.\il(\ref{FIG:raisemK}).
We know that the \textbf{MBs} (and therefore the  bundle frequencies)   with tangent $a_g\in[0,0.8]$, $r_g\in[0,2/5]$ and $\omega_b\geq1$,  origin $\la_0=a_0\sqrt{\sigma}\in[0,M]$ remain confined  for large $\sigma$s or, equivalently,
$a_0\in[0,1/\sqrt{\sigma}]\supseteq[0,M]$;  the  {bottleneck} region with
  $\la_0=a_0\sqrt{\sigma}\in]M,2M]$ or, equivalently,
$a_0\in[1/\sqrt{\sigma}],2/\sqrt{\sigma}]\supseteq]M,2M]$, where $a_g\in]0.8M,M]$, $r_g\in]2M/5,M]$ and $\omega_b\in[0.5,1[$. Then, depending on the value of $\sigma$, they are confined to the exterior region of the extended plane
($a\in[0,M]\cup r>r_+$), i.e.,  there are no  photon orbital  frequency equal to the inner horizon for $r>r_+$.
We  point out, however, that the analysis of Sec.\il(\ref{Sec:pri-photon-fre}) was based on the condition  $a<M$ because it is related to  the tangent spin $a_g=a<M$.

\begin{figure}
  \includegraphics[width=7cm]{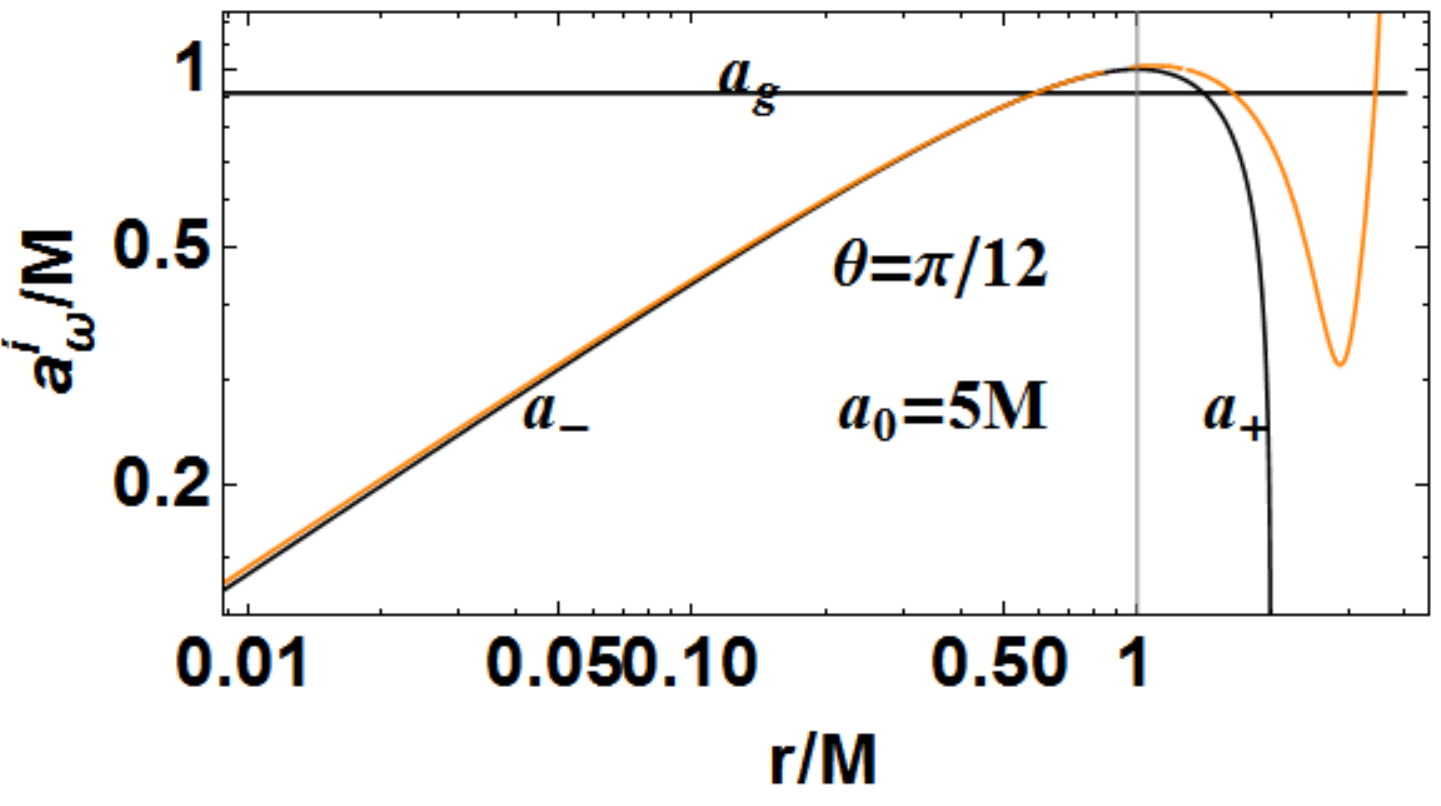}
\includegraphics[width=6cm]{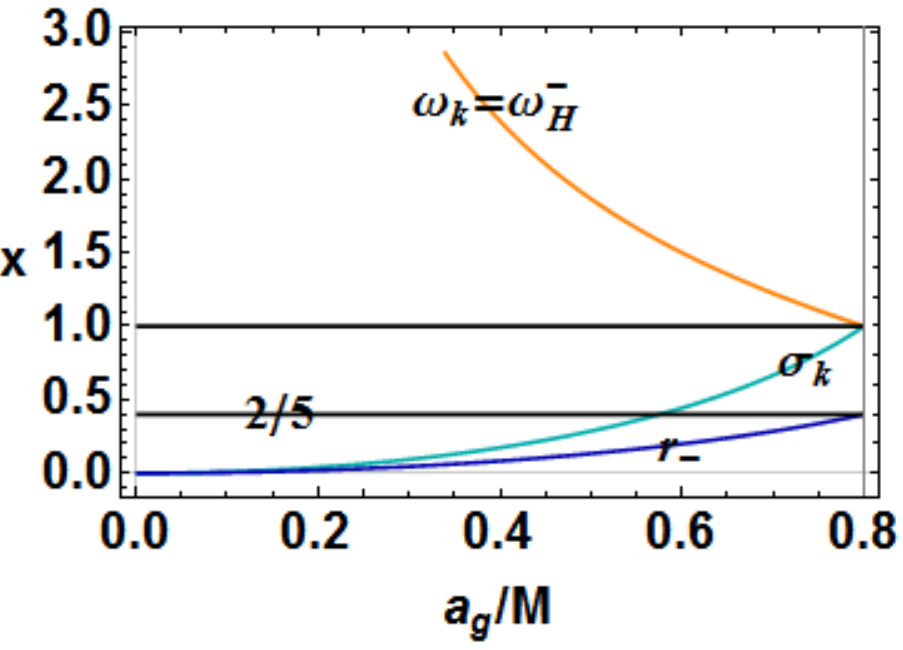}
  \caption{Left: metric bundle for the origin spin $a_0=5M$ and plane $\theta=\pi/12$. The black line is the tangent spin of the bundle (orange) to the horizon curve $a_{\pm}$ (black) on the extended  plane. It shows also the existence of minimum of the bundle curve with spin lower than the tangent spin. The full extension of this bundle is depicted in
	Fig.\il(\ref{FIG:disciotto1}). Right panel: limiting plane $\sigma_k$, frequency $\omega_k=\omega_{H}^-$ (inner horizon frequency on the extended  plane), the inner horizon $r_-$ as function of the tangent spin $a_g$.  Based on the analysis
	of Eqs.\il(\ref{Eq:exitin-Ue}) and (\ref{Eq:cu-r-ren-t-proce}). }\label{Fig:adisciottoc2}
\end{figure}

In general, however,  \textbf{MBs} spins $a$ satisfy the relation $a_g\leq a_0>0$; therefore, on the equatorial plane, there is $a_g\leq a\leq a_0>0$, that is, the  origin spin is a supremum of the \textbf{MBs} spins. We should note that some  with \textbf{MBs} origin $a_0>0$ have, for some plane $\sigma$ solutions $(r=0,a=0)$ and different maximum points, which are evident from Figs\il\ref{FIG:funzplo},\ref{FIG:disciotto1},\ref{FIG:rccolonog}, and \ref{FIG:Aslongas}. We do not consider here  these situations focusing specifically on the relation between the bundle origin spin $a_0$ and the tangent spin $a_g$, where the relationship $a_g<a_0$ holds.
This problem concerns the following three aspects of the metric bundle properties:
\textbf{(1)} the confinement of bundles in the region of the inner horizons,
\textbf{(2)} the presence of mixed \textbf{NS} and \textbf{BH} bundles,  depending on the origin (or tangency spin) and
\textbf{(3)} the number of orbits for a given spacetime with the same frequency of the inner horizons, that is,
 the intersection of the horizontal lines of the extended plane $a=$constant
(this problem concerns the curvature of the bundle solutions on the extended  plane).
We see an example of bundles with extremum in Fig.\il(\ref{Fig:adisciottoc2}).

The following property, however, holds
\bea\label{Eq:exitin-Ue}
&&
\mbox{\textbf{Confinement I}: There are \emph{no} solutions of the equation $\mathcal{L}\cdot\mathcal{L}=0$ in the following case:}
\\\nonumber
&&
\sigma\in[0,1],\quad r\geq 0,\quad a>M, \quad\mathbf{(1)}\quad 0<\frac{1}{\sqrt{\sigma } \omega }\leq 1,
\quad\mathbf{(2)}\quad  \omega >\frac{1}{2}
\\\nonumber
&&
\mbox{\textbf{Confinement II}: There are \emph{no} solutions of the equation $\mathcal{L}\cdot\mathcal{L}=0$ in the following case:}
\\\nonumber
&&
\sigma\in[0,1],\quad r\geq M,\quad 0<a<M, \quad\mathbf{(1b)}\quad 0<\frac{1}{\sqrt{\sigma } \omega }\leq 2,
\quad\mathbf{(2b)}\quad  \omega >\frac{1}{2}
\eea
The condition  \textbf{(1)} on $\sqrt{\sigma } \omega$ means the  \textbf{BH} origin  $a_0\in]0,M]$;
 the condition  \textbf{(2)} on the frequency requires that  $\omega=\omega_H^-$, that is,
the inner  horizon frequency on the extended  plane  ($\omega=1/2$ for $a_g=M$).
Finally, the condition $a>M$  is the assumption that  \textbf{MBs} with origin in \textbf{BH}  $a_0\in \mathbf{BH}$ are \emph{not} confined in the \textbf{BH} region, which means that $a>M$. This case does not  occur under those conditions.
This demonstrates the \emph{confinement} of the \textbf{MBs} occurs in the interior region,
if  the origin  is $a_0\in]0,M]$.
These confined  \textbf{MBs}  with origin in the \textbf{BH} region have tangent spin $a_g/M\in[0,4/5]$.  Specifically,
\bea&&\label{Eq:cu-r-ren-t-proce}
a_g\in]0,{4M}/{5}[\quad\mbox{and}\quad \sigma\in[\sigma_{k},1]\quad\mbox{where}\quad \sigma_{k}\equiv \frac{8}{a_g^2}-8 \sqrt{\frac{1}{a_g^2}\left(\frac{1}{a_g^2}-1\right)}-4\quad\mbox{and}
 \\
&&\omega =\omega_k\equiv\frac{a_g \sqrt{\frac{1}{a_g^2}-1}+1}{2 a_g},\quad r_g\in]0,2M/5]\quad(\mbox{for}\quad a_g=\frac{4M}{5},\quad  \sigma =1,\quad \omega =1)
\eea
--Fig.\il(\ref{Fig:adisciottoc2}).

Therefore, this has the  important consequence that if the origin is   $a_0\in \mathbf{BH}$, then the metric bundle will be confined in the  \textbf{BH} region for \emph{any} plane $\sigma\in]0,1]$.
The outer horizon  on the extended  plane is generated \emph{only} by \textbf{NS} bundle origins with  $a_0>2M$.
Viceversa, according to the value of the plane $\sigma$, the inner horizon can be generated either by \textbf{NS} or only by \textbf{BH} origins (as it was also for the $\sigma=1$ case),
\textbf{BH} origin \textbf{MBs} generate \emph{only} the inner  horizon on the extended  plane.
This holds also according to  $a_0>a_g(a_0,\sigma)$, which is always verified.
The  \textbf{MBs} with \textbf{NS} origins,  tangent to the inner horizon can also be at $a_0>2M$:
$\la_0>2M$  generate only the   outer horizon, viceversa $\la_0<2M$  generate only the inner horizon,
as  $\la=a\sqrt{\sigma}$; this implies    $a_0>2/\sqrt{\sigma}\geq 2$ (equivalently,
$\sqrt{\sigma}>2/a_0$ with $a_0\geq2$) with   $\sigma\in[0,1]$, for the construction of only the outer horizon.
For the inner horizon,  there is
$a_0<2/\sqrt{\sigma}\geq 2$ (equivalently,
$\sqrt{\sigma}<2/a_0$) thus there is no upper limit for $a_0$; it can be larger, depending on  how close the {orbit} $r$ is
to the rotation axis ($\sigma\approx 0$), whereas the bottom limit  is  $2$, i.e., for $a_0<M$; in particular,  the origin is  always in the  \textbf{BH} region. 
Therefore, portions of the internal horizon are generated by different \textbf{BH} origins in agreement also with the analysis of \cite{remnants}.
We shall investigate these limits more precisely  below.

\begin{figure}
  \includegraphics[width=5cm]{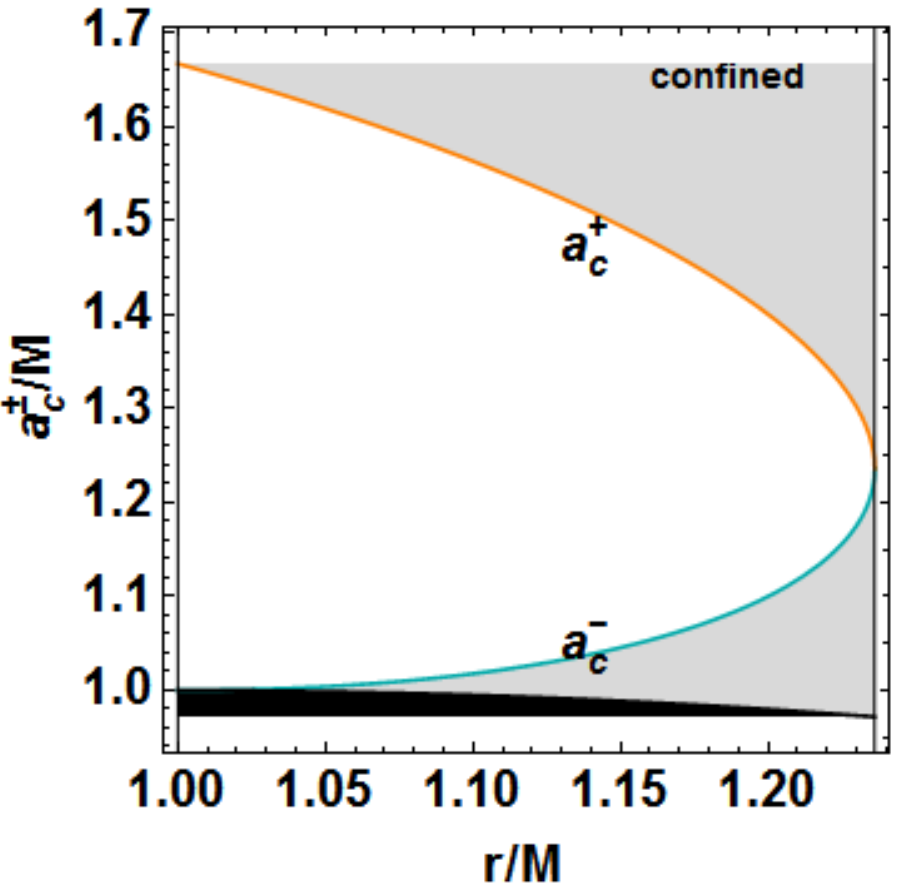}
\includegraphics[width=5cm]{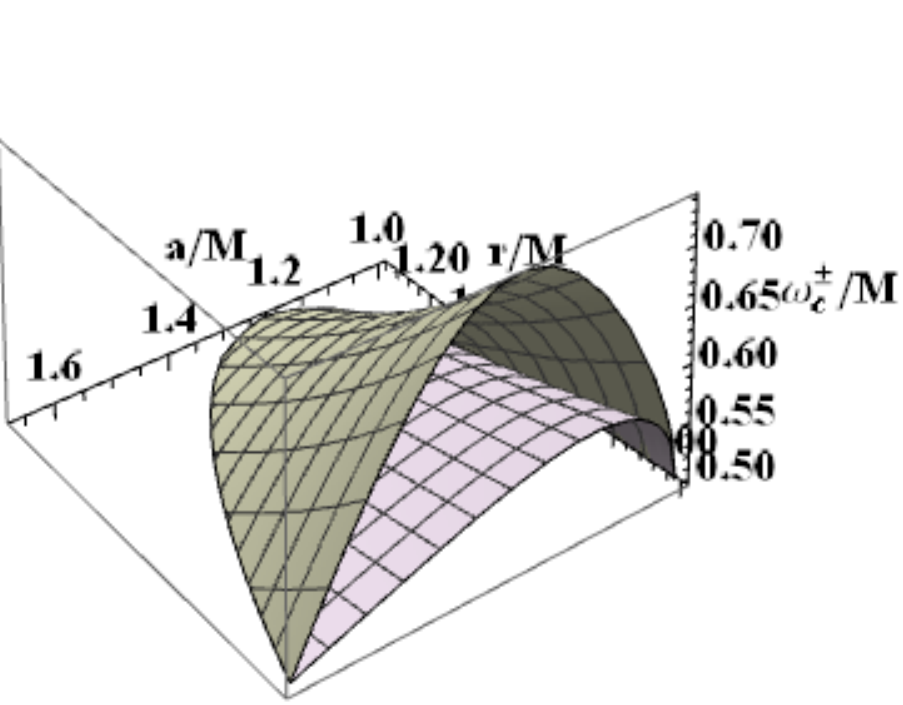}
 \includegraphics[width=5cm]{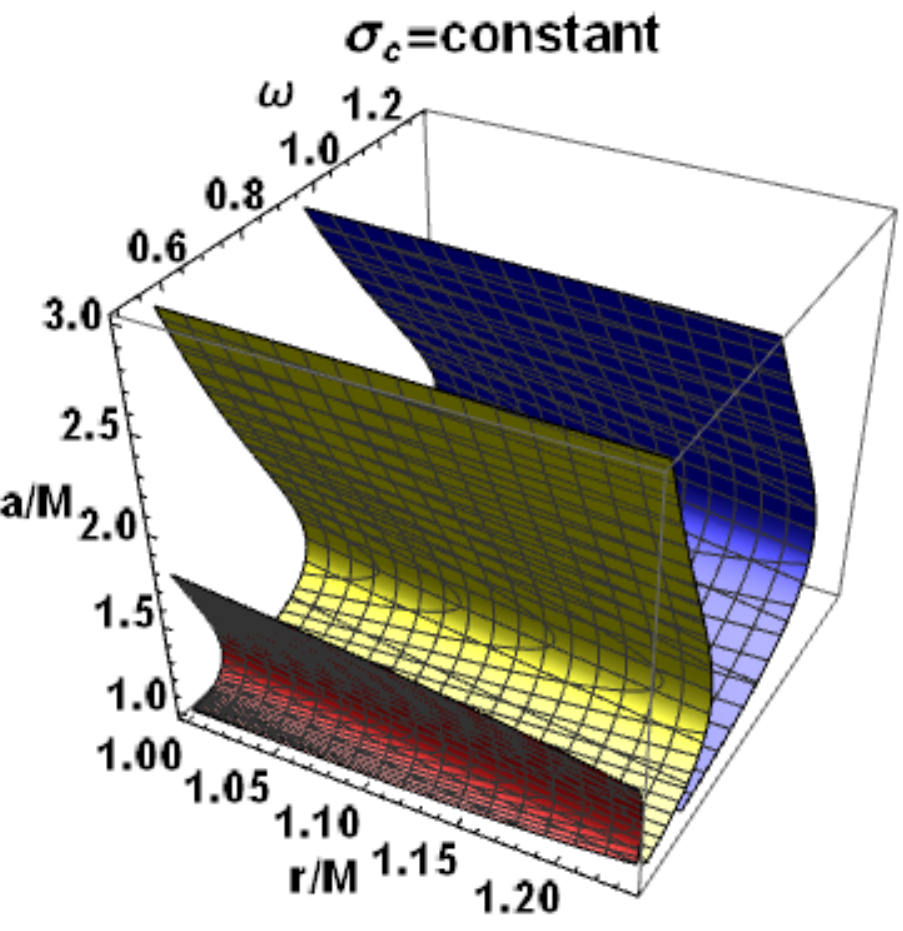}
  \caption{Confinement of the  inner horizon frequencies based on the analysis  of Eq.\il(\ref{Eq:bisc-tt-etelleb}).
	Left: limiting spins $a_c^{\pm}$ as functions of $r/M$.
	Center: limiting frequencies $\omega_c^{\pm}$ as functions of $r/M$ and $a/M$.
	Right: plane solution $\sigma_c=$constant for $a/M$ and  $r/M$ and frequency  $\omega$ (blue is for $\sigma=0.1$, yellow is for $\sigma=0.3$, red for $\sigma=0.9$).}\label{Fig:Plorzo3Db9}
\end{figure}

\begin{figure}
  \includegraphics[width=5cm]{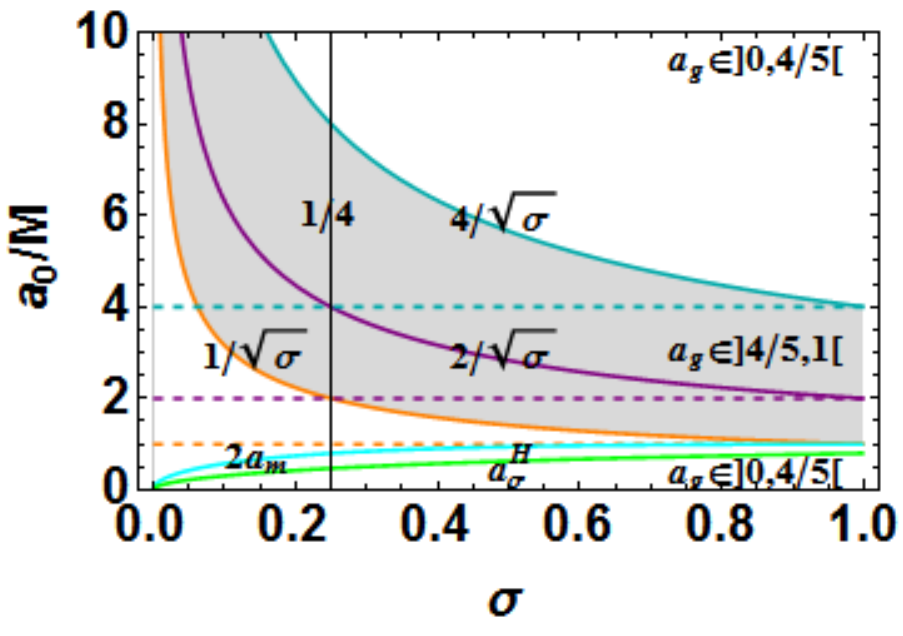}
  \includegraphics[width=5cm]{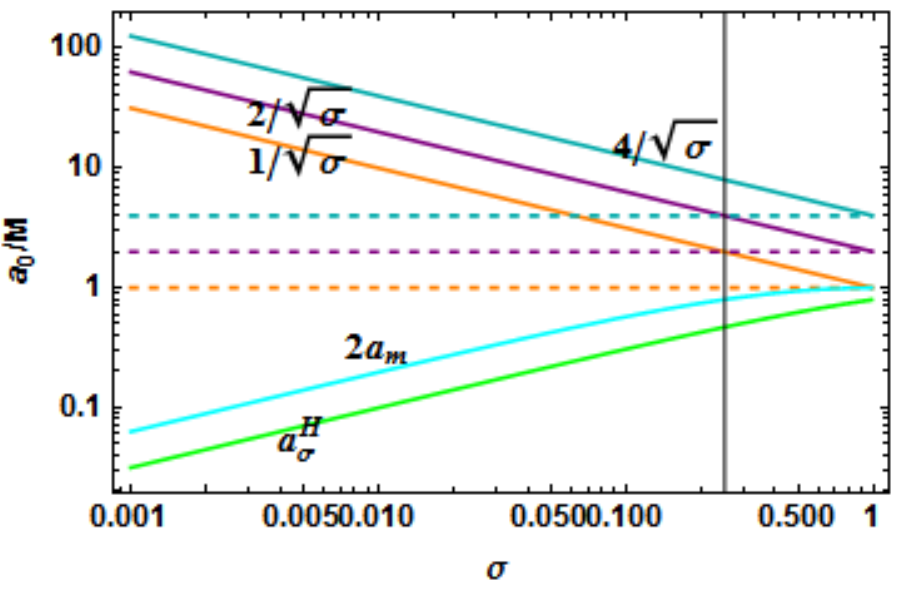}
 \includegraphics[width=5cm]{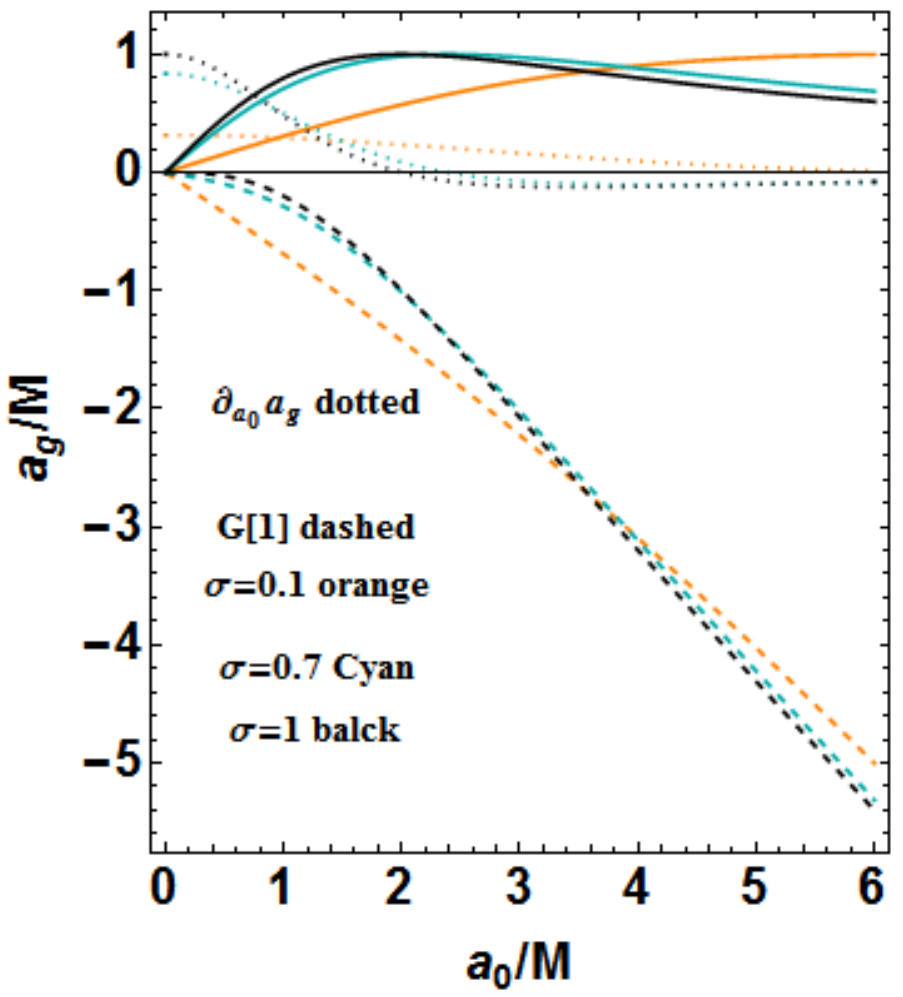}
\includegraphics[width=5cm]{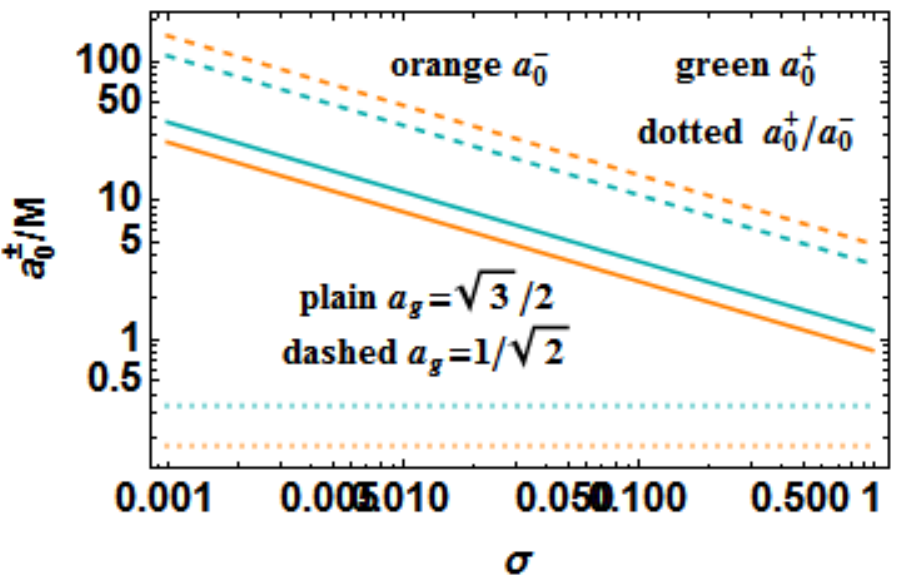}
 \includegraphics[width=5cm]{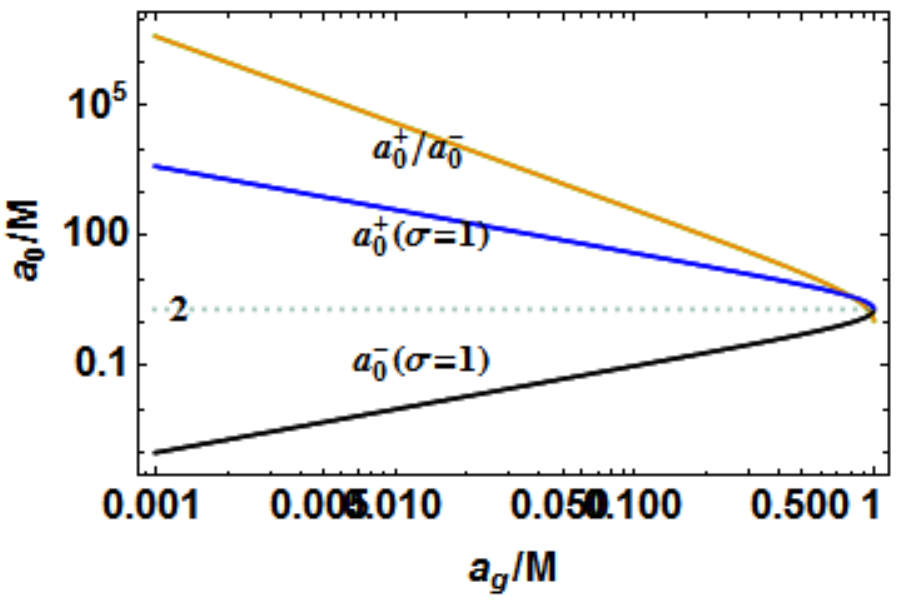}
  \caption{Uperr line. Left panel: relation $a_0(a_g)$ and regions \textbf{BH-NS} origin $a_0$ to \textbf{BH} tangent spins $a_g$ as function of the plane $\sigma\in[0,1]$. The limiting values $a_0\in \{1,2,4\}$ are shown explicitly.
	In the gray region $a_0$ corresponds to $a_g\in[0.8M,M]$.   It refers to the analysis of  Eqs.\il(\ref{Eq:fre-tevo-te},
\ref{Eq:cab-i-net}) and   (\ref{Eq:condgi-ut-ura}). Center panel: logarithmic plot shows the logarithm identity discussed in the text. Right panel: tangent spin  $a_g(a_0)$ and bundle origin spin for different planes $\sigma$, dotted line is the  $a_g$ gradient. Dashed  curve is $G[1]\equiv a_g(a_0)-a_0$ for different $a_0$. Crossing $a_g$  curves  represent bundles with equal tangent spin $a_g$  and origin $a_0$, consequently, with different planes $\sigma$ and eventually different frequencies
$\omega_H^+(a_g)$ or $\omega_H^-(a_g)$ --see Eq.\il(\ref{Eq:cli.chage}).
The origins are related to the frequencies and planes by $a_0=1/(\omega_b\sqrt{\sigma})$,
it follows for two bundles planes  $\sigma_p= \frac{16}{a_0^4 \sigma }$ or $\sigma=\sigma^p$ with frequencies related by $\omega_b^p/\omega_b={\sqrt{\sigma }}/{\sqrt{\sigma_p}}$. This is also related to bundle frequency ratios
$s=\omega_b^p/\omega_b={\sqrt{\sigma }}/{\sqrt{\sigma_p}}$. Based upon the analysis of Eq.\il(\ref{Eq:leav-29-chans}).}\label{Fig:stypolitic}
\end{figure}

There are  solutions  of the condition $\mathcal{L}\cdot\mathcal{L}=0$ in the following cases:
\bea&&\nonumber
\sigma \in]0,1],\quad a>M,\quad 1<\frac{1}{\sqrt{\sigma } \omega }\leq 2,\quad\omega>\frac{1}{2},\quad r>M\quad\mbox{more precisely}
,\quad
\frac{r}{M}\in ]1,\sqrt{5}-1[\quad a\in]a_c^-,a_c^+[\quad \omega \in[\omega_c^-,\omega_c^+[,
\\
&&\label{Eq:bisc-tt-etelleb}\mbox{where}\quad
a_c^{\mp}\equiv\frac{4}{r+2}\mp\sqrt{\frac{-r^4-2 r^3+4 r^2}{(r+2)^2}},\quad\omega_c^-\equiv\frac{2 a}{a^2 (r+2)+r^3}+\sqrt{\frac{r^2 \Delta}{\left[a^2 (r+2)+r^3\right]^2}},\\\nonumber
&& \omega_c^+\equiv\frac{1}{2} \left[\frac{4 a r}{a^4+2 a^2 \left(r^2-2\right)+r \left(r^3-4 r+8\right)}+\sqrt{\frac{a^2 \Delta \left(a^2+r^2-4\right)^2}{\left[a^4+2 a^2 \left(r^2-2\right)+r \left(r^3-4 r+8\right)\right]^2}}\right],
\\\nonumber
&&
\sigma_c\equiv\frac{a^4 \omega ^2+2 a^2 r^2 \omega ^2+a^2-a^2 \omega ^2\Delta \sqrt{\frac{\left[a-\omega  \left(a^2+r^2\right)\right]^2 \left[\omega ^2 \left(a^2+r^2\right)^2+2 a \omega  \left[a^2+(r-4) r\right]+a^2\right]}{a^4 \omega ^4 \Delta^2}}-4 a r \omega +r^4 \omega ^2}{2 a^2 \omega ^2 \Delta},
\eea
--Fig.\il(\ref{Fig:Plorzo3Db9}).
Firstly, we made use of the recursive relation considered in  \cite{remnants} and  Eqs.\il(\ref{Eq:enaglinya})
 and (\ref{Eq:own-poi-tn-spea}) to study the corresponding \textbf{MBs} with $a_0^p=a_g$, i.e., the origin of the  second  bundle is the tangent spin of the first bundle. This analysis also shows another aspect of the confinement of metric bundles. From the condition  $a_g<a_0$,  it follows that the spin  $a^{(1)}_g(a_g)$  corresponds to a \textbf{BH}.
Consider the origins $a_0^\pm$ of the \textbf{MBs} tangent to the inner and outer horizons,  respectively, and we express them in terms of the tangency frequency. Then,
\bea&&\label{Eq:condgi-ut-ura}
a_0^\pm\equiv\frac{1}{\omega_H^{\pm}\sqrt{\sigma}},\quad\mbox{where}\quad a_0^+\geq a_0^-.\\
&&\nonumber\mbox{\textbf{Outer horizon tangency:} }
 \\
&&\nonumber a_0^+\geq 2M \quad \mbox{for any}\quad  a_g\in ]0,M]\quad \mbox{and}\quad    \sigma\in ]0,1], \quad (a_0^+=2M)\quad for \quad a_g=M
\\\label{Eq:truff-pola}
&&\mbox{\textbf{Inner horizon tangency:} }
 \\\nonumber
&&\bullet\quad a_0^-=1 \quad \mbox{for}\quad  a_g=a_{\sigma}^H,
\quad\bullet\quad a_0^->M \quad \mbox{for}\quad  a_g\in]a_{\sigma}^H,M],
\quad\bullet\quad   a_0^-\in ]0,M[ \quad \mbox{for}\quad  a_g\in]0,a_{\sigma}^H[.
\\\nonumber
&& \mbox{Equivalently}
 \\\nonumber
&&
\bullet\quad a_0^-=M \quad \mbox{for}\quad  a_g/M\in]0,4/5]\quad\mbox{and}\quad \sigma=\sigma_a^H,
\\ &&\nonumber
\bullet\quad a_0^->M \quad \mbox{for}\quad  a_g/M\in]0,4/5[\quad\mbox{and}\quad \sigma\in]0,\sigma_a^H[,
\quad  a_g/M\in]4/5,1]\quad\mbox{and}\quad\sigma\in]0,1]
\\\nonumber
&&
\bullet\quad  a_0^-\in ]0,M[ \quad \mbox{for}\quad   a_g/M\in]0,4/5]\quad\mbox{and}\quad \sigma\in]\sigma_a^H,1],
\\\nonumber
&&\mbox{where}\quad a_{\sigma}^H\equiv4 \sqrt{\frac{\sigma }{(\sigma +4)^2}}<\frac{4}{5}=0.8,\quad \sigma_a^{H}\equiv\frac{4 \left(2 r_--a^2\right)}{a^2},
\eea
(we consider mainly $a\geq0$ and $\sigma\geq0$)--Figs\il(\ref{Fig:thelas})-(\ref{Fig:stypolitic}).
The spin $a_0^+(a_g^H)$  is the origin of the bundle tangent  at $a_{\sigma}^H$ to the outer  horizon  on the extended  plane of a \textbf{BH} spacetime with spin $a_g^{H}$; the origin of the bundle relative to the construction of the inner horizon,
$r_-(a_{\sigma}^H)$ is obviously, in agreement with  Eq.\il(\ref{Eq:truff-pola}), $a_0^-(a_{\sigma}^H)=M$, where
\bea\label{Eq:gov-vote}
a_0^+(a_\sigma^H)\equiv\frac{(\sigma +4) \left[\sqrt{\frac{(\sigma -4)^2}{(\sigma +4)^2}}+1\right]}{2 \sigma },\quad a_0^-(a_\sigma^H)=M;\quad
a_0^+ (\sigma _a^H)\equiv\frac{a}{r_- \sqrt{\frac{2r_--a^2}{a^2}}},\quad  a_0^-(\sigma _a^H)=M
\eea
--Fig.\il(\ref{Fig:thelas}).
We note that these relations  depend on the plane $\sigma$  {(dependence  on the origins $\la_0$)}.
Note also that
$a_0^+(a_{\sigma}^H)=4M$ for $\sigma=1$ where $a_{\sigma}^H/M=4/5$.
Clearly,  $a_0^{\pm}(a_g)$ is the inverse function of $a_g(a_0,\sigma)$ represented in Fig.\il(\ref{FIG:raisemK}) and
studied in different parts of this article.
We have already seen that this function can be set for variable $\la \equiv a \sqrt{\sigma}$. Now,
 the explicit form of $a_0(a_g)$  under the restriction
$a_0>0$ is
\bea\label{Eq:leav-29-chans}
&&
a_0^\mp(a_g)\equiv \sqrt{\frac{4({2}-a_g^2)}{\sigma a_g^2 }\mp8 \sqrt{\frac{1-a_g^2}{a_g^4 \sigma ^2}}},
\quad \mbox{(equivalently} \quad a_0^\mp=\frac{2 r_\mp(a_g)}{a_g \sqrt{\sigma }}),\quad  \mbox{where} 
\\\nonumber
&& \sigma =0\quad a_g=0,\quad a_0>0; \quad a_g=1\quad a_0=\frac{2}{\sqrt{\sigma }},\quad\frac{a_0^-(a_g)}{a_0^+(a_g)}=\sqrt{-\frac{4 \left[r_+(a_g)+1\right]}{a_g^2}+\frac{8 r_+(a_g)}{a_g^4}+1}, \quad\mbox{and }
\\\nonumber
&&
 \bullet\quad\partial_{a_g}^{(2)}a_0^+(a_g)=0\quad\mbox{for}\quad
a_g=\frac{\sqrt{3}}{2},\quad\mbox{where}\quad \left(\omega_H^-=\frac{\sqrt{3}}{2}, \omega_H^+=\frac{1}{2 \sqrt{3}}\right),\quad \mathbf{s=\frac{\omega_H^+}{\omega_H^-}=\frac{1}{3}}=
\mathbf{\frac{a_0^{-}}{a_0^{+}}}\quad\mbox{and}
\\\nonumber
&& \qquad \qquad a_0^{\mp}=\frac{2 \sqrt{\frac{5}{\sigma }\mp4 \sqrt{\frac{1}{\sigma ^2}}}}{\sqrt{3}},\quad
\mbox{for}\quad \sigma=1\quad \left(a_0^{-}=\frac{2}{\sqrt{3}}, a_0^{+}=2 \sqrt{3}\right),\quad\mbox{and }
\\\nonumber
&&\bullet\quad\partial_{a_g}^{(2)}a_0^-(a_g)=0\quad\mbox{for}\quad
a_g=\frac{1}{\sqrt{2}},\quad\mbox{where}\quad \omega_H^\mp=\frac{1}{2}\pm\frac{1}{\sqrt{2}},\quad \mathbf{s=\frac{\omega_H^+}{\omega_H^-}=3-2 \sqrt{2}}=\mathbf{\frac{a_0^{-}}{a_0^{+}}}\quad\mbox{and}
\\
&&\qquad \qquad a_0^\mp=2 \sqrt{\frac{3}{\sigma }\mp2 \sqrt{2} \sqrt{\frac{1}{\sigma ^2}}},
\quad
\mbox{for}\quad \sigma=1\quad a_0^{\mp}=2 \sqrt{3\mp2 \sqrt{2}}
\eea
--Fig.\il(\ref{Fig:stypolitic}). The origin spin
$a_0={2}/{\sqrt{\sigma }}$  is a limiting  value that is often found in this analysis as this is  related to the extreme Kerr geometry.
Other remarkable spins are  $a_g={\sqrt{3}}/{2}$ and $a_g={1}/{\sqrt{2}}$. Notably these are independent from the plane
$\sigma$; furthermore,  we recall that these spins correspond to particular bundle frequencies and particular  frequency ratios.
It is important to note that the ratio  ${a_0^-(a_g)}/{a_0^+(a_g)}$, that is, the bundle spin origins related to the construction of the inner  and outer  horizon of the same geometry
with spin $a_g$ (the tangent spin of the two bundles) is \emph{independent} of the plane $\sigma$.

These relations allow us to establish for a  given  tangent spin if  the origin is a  \textbf{BH} or a \textbf{NS} and, vice versa,  given a bundle origin we can establish the tangent spin $a_g$.
Obviously,  there is always  $a_g\in \mathbf{BH}\equiv[0,1]$.
However, as was the case in the equatorial plane studied in  \cite{remnants},
there are  particular classes of \textbf{BHs} and \textbf{NSs}.

We see that $a_{0}^{+}$,  i.e., the origin of the \textbf{MBs}f tangent to the outer horizon is necessarily a
\textbf{NS} in agreement also with Eq.\il(\ref{Eq:summ-colo-quiri}).
Therefore, for each plane $\sigma$  the outer horizon is built by \textbf{MBs} with \textbf{NS} origins and particularly for
$a_0>2M$.
In the case of the inner horizon, as it was for $\sigma=1$, for  $a_g\in]0.8M,M]$ there is a  \textbf{NS}  origin
$a_0\in]M,2M[$ for any plane.  Therefore, the  portion    $a_g\in]0.8M,M]$  of the inner horizon is always constructed by  bundles with \textbf{WNS}   spin origin $a_0\in]M,2M[$ for any plane. For $a_g<0.8M$ there may be   \textbf{BH} or \textbf{NS}
origins,  depending  on the plane (if the angles are considerably small).

In other words, if  $a_0$ is a \textbf{BH}, then the tangent point is certainly a \textbf{BH} with $a_g\in [0,0.8M]$.
An  extreme spacetime $a_g=M$ corresponds to an origin $a_0=2 /\sqrt{{\sigma }}$; particularly,
if $a_g\in ]0,0.8M[$, then $a_0\in ]0,0.8M[$ and, viceversa, for the range $]0,0.8M[$ the bounding range for the spin is
$a\in\{a_0,a_g\}$, which confines the \textbf{MBs} in the region bounded by the inner horizon--see also\cite{remnants}.

Then,
\bea&&\nonumber
\bullet\quad a^-_0>2M\quad \text{for}\quad a_g\in]2a_m,M];\quad\bullet\quad a^-_0\in]M,2M[\quad \text{for}\quad a_g\in]a_{\sigma }^H,2 a_m];\quad\bullet\quad a^-_0\in]0,1[\quad \mbox{for}\quad a_g>a_{\sigma }^H
\\&&\label{Eq:fre-tevo-te}
\mbox{where}\quad\frac{a_m}{M}\equiv \sqrt{\frac{\sigma }{(\sigma +1)^2}}.\quad\mbox{It holds }\quad
 \partial_{x}a_g=0 \quad \text{for}\quad a=\frac{2}{\sqrt{\sigma }}
\quad \text{where}\quad x\in\{a_0,\sigma\},
\eea
(we note that $a_g\neq \la \equiv a\sqrt{\sigma}$, and $a_g=\la$  only for $\sigma=0$ and $a_g\neq a_0$)--Fig.\il(\ref{Fig:stypolitic}).

\begin{figure}
\includegraphics[width=5cm]{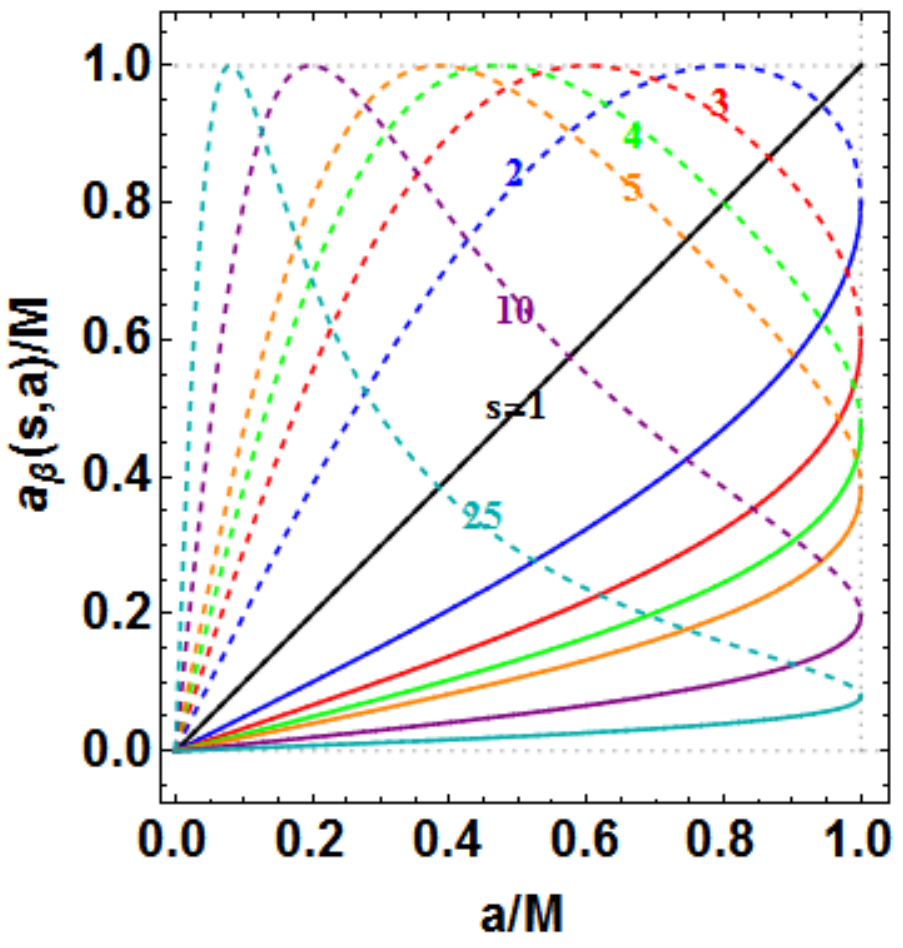}
\includegraphics[width=5cm]{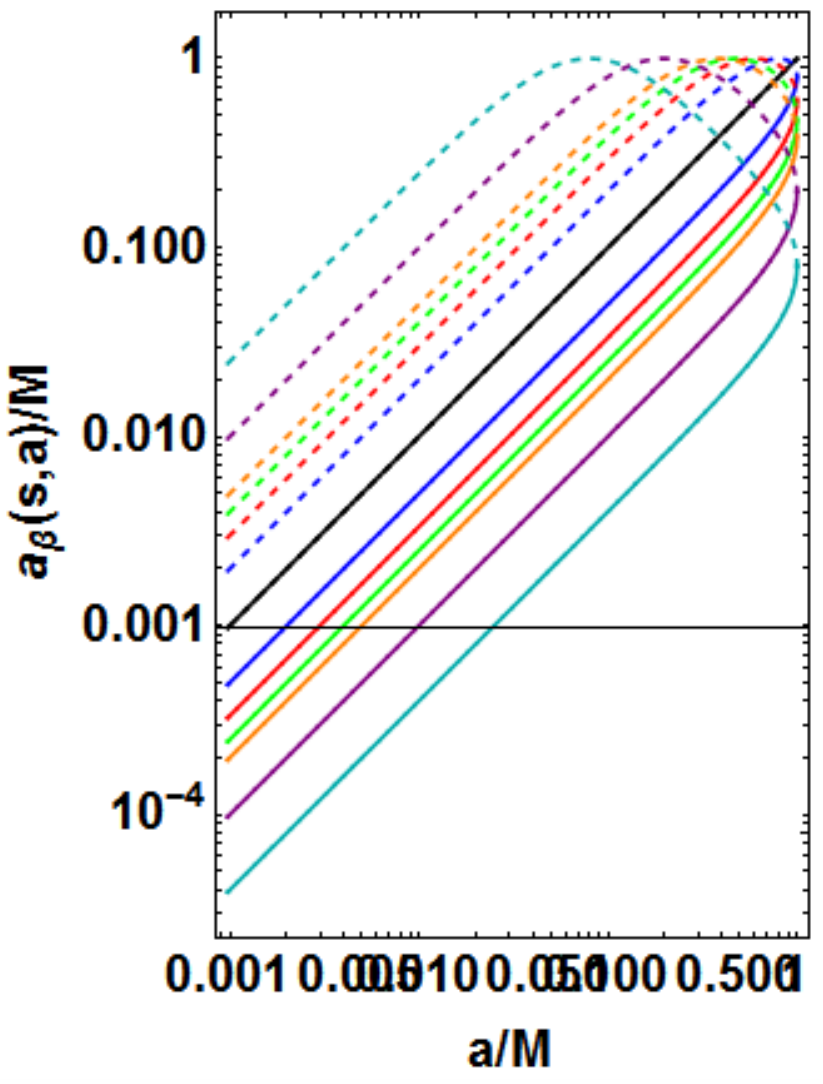}
\includegraphics[width=5cm]{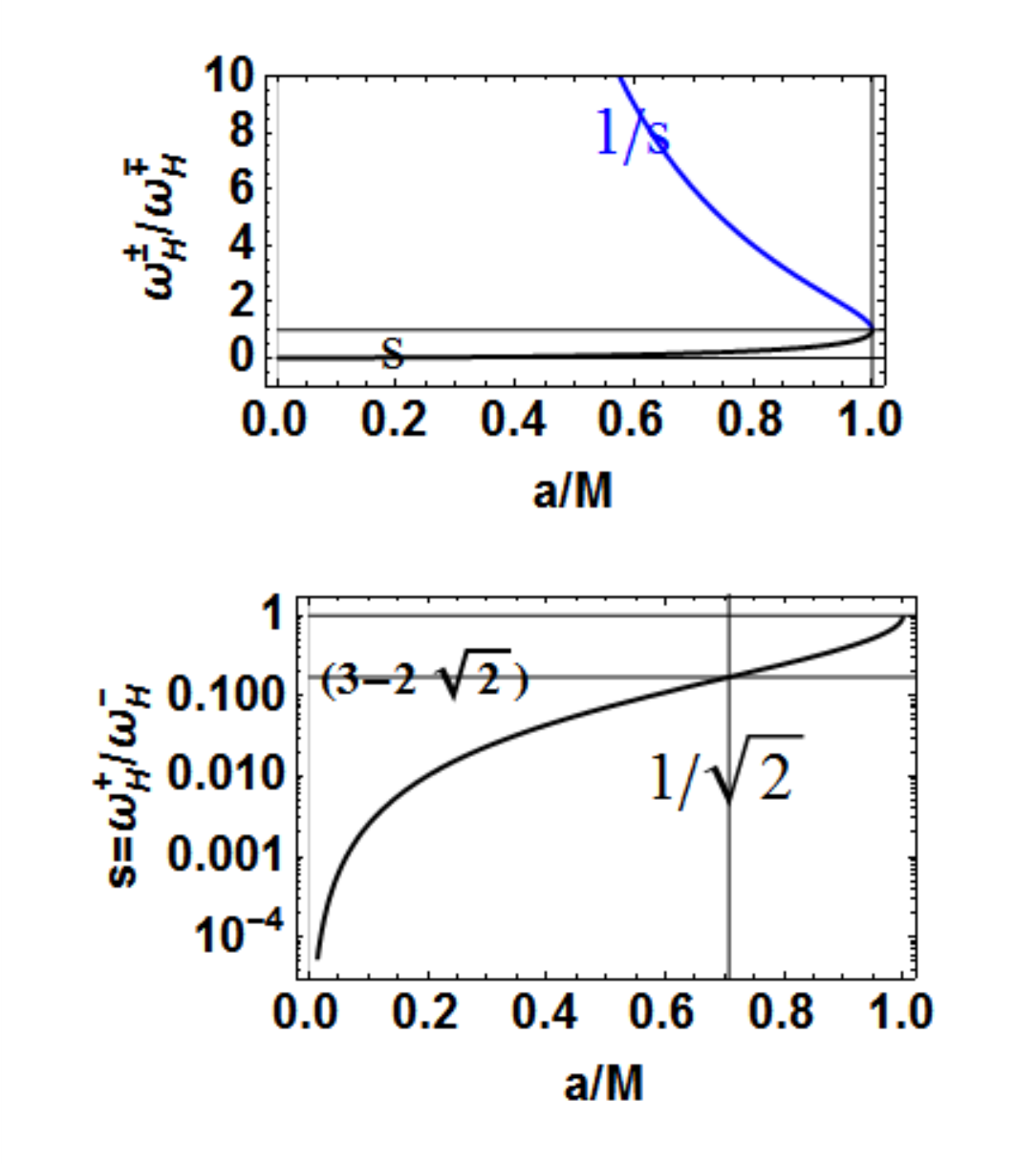}
\includegraphics[width=5cm]{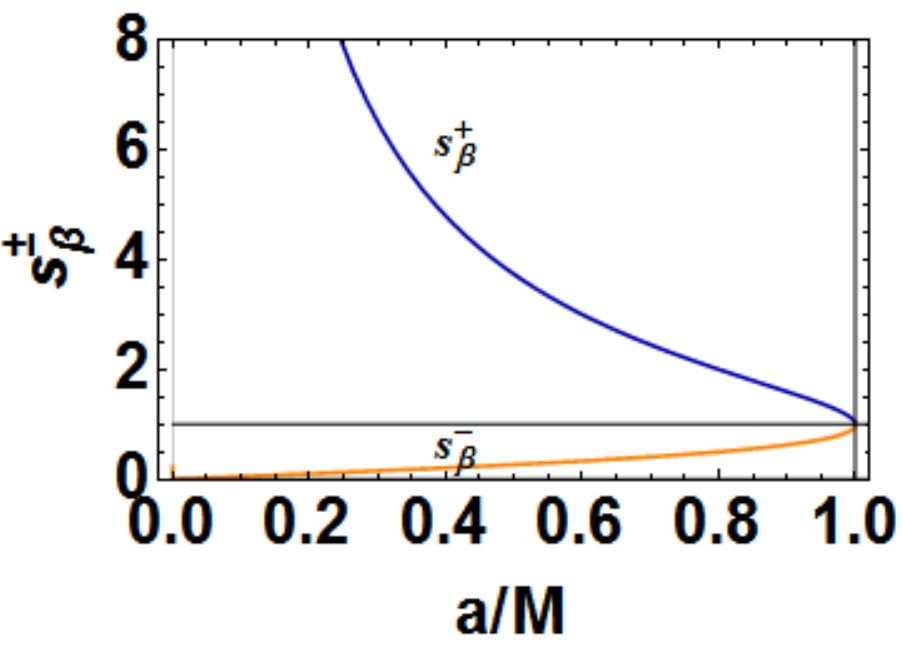}
\includegraphics[width=5cm]{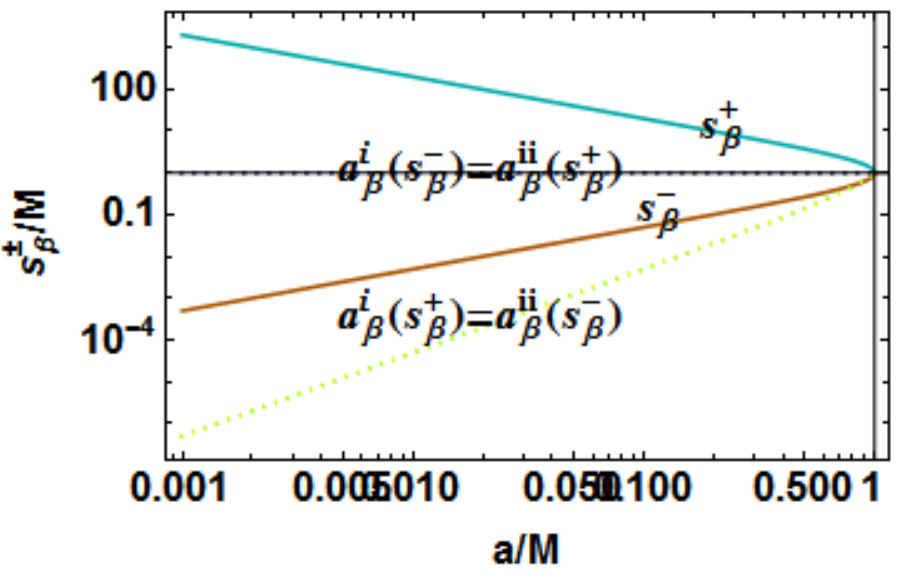}
 \includegraphics[width=5cm]{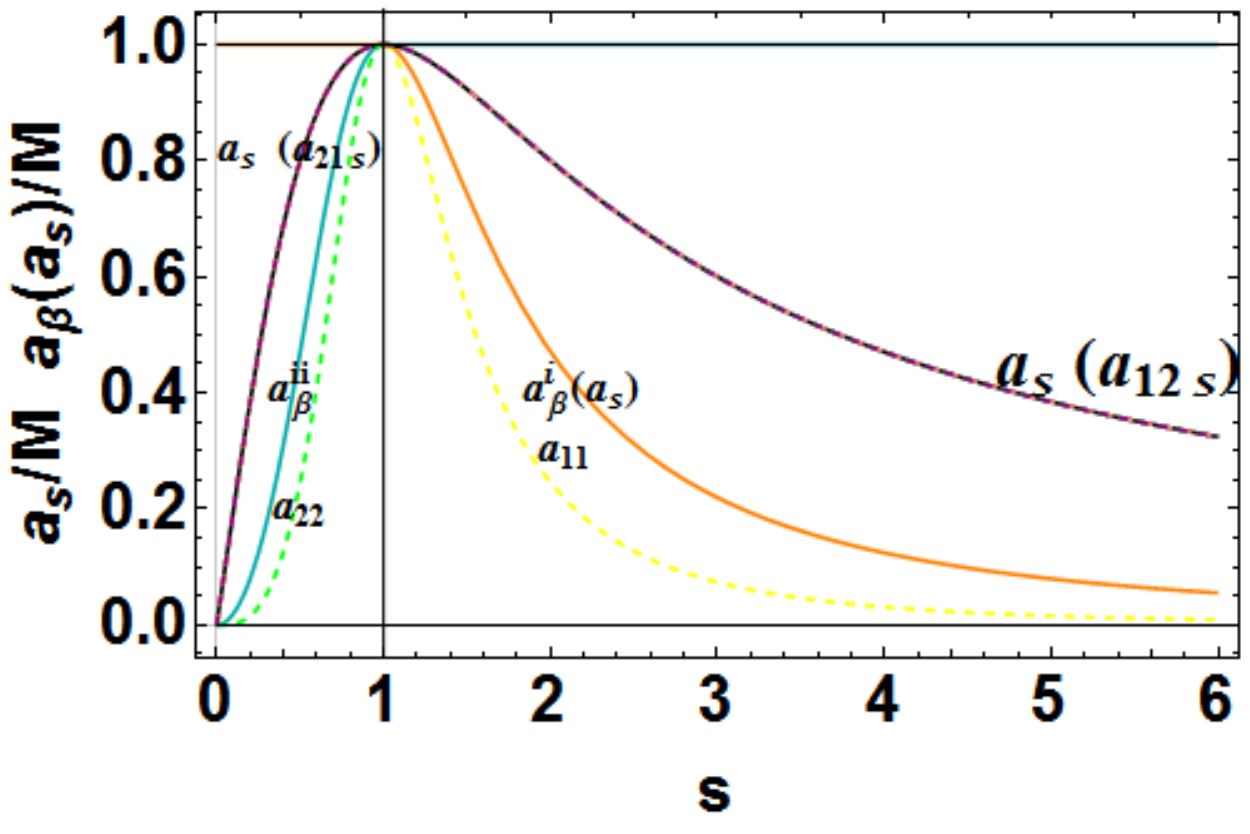}
\includegraphics[width=5cm]{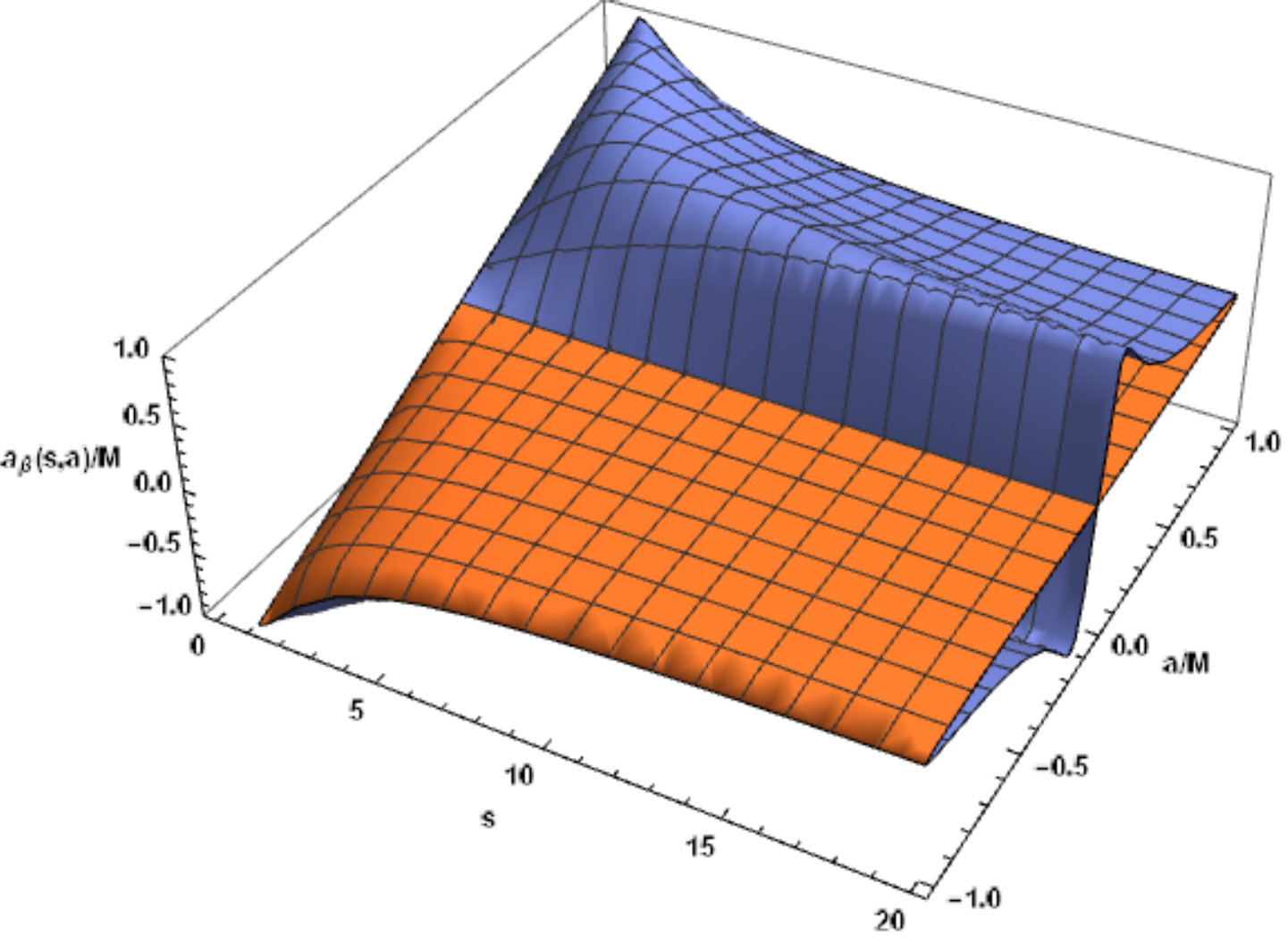}
\includegraphics[width=5cm]{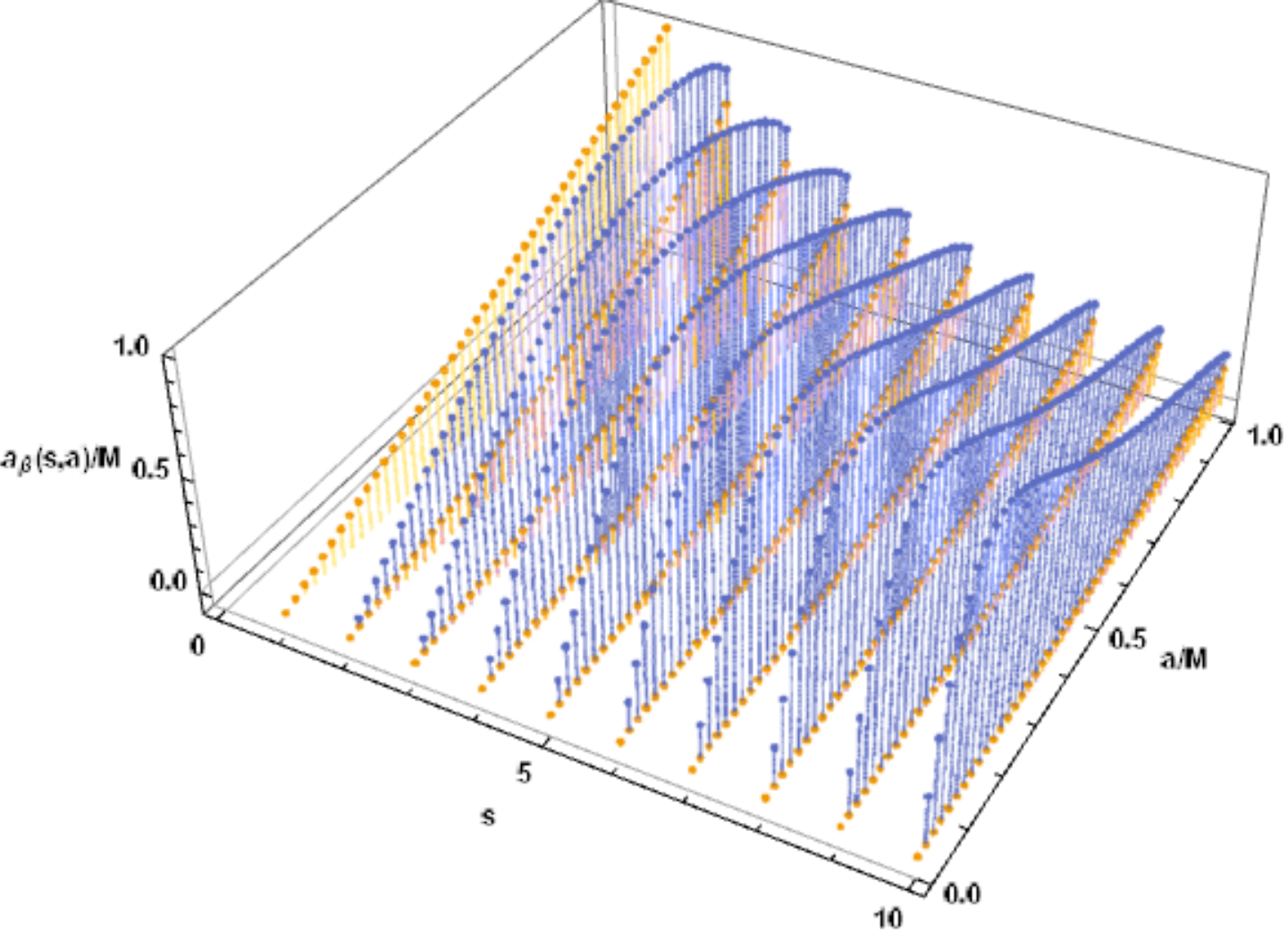}
  \caption{Upper line. Left panel: spins $a_{\beta}=\{a_{\beta}^{i},a_{\beta}^{ii}\}$  of Eq.\il(\ref{Eq:a-beta-i-ii}) for different resonance solutions $s\equiv\omega_H^+/\omega_H^-$  as  functions of the tangent spin  $a$ of the second frequency bundle $\omega_H^{\pm}$, in accordance with  Eq.\il(\ref{Eq:a-beta-i-ii}). The identity for $s=1$ and  the symmetry for
	$s\rightarrow1/s$ are shown. Note the behavior as $s$ increases.  Plots  provide also the  classes of metric bundles with  same resonant frequencies. The situation for a fixed \textbf{BH} geometry with spin   $a_g$  is described by vertical $a_g=$constant line of the graph (which is clearly independent of the plane $\sigma$). Central panel: logarithmic plot shows the logarithmic identities discussed  in the text (generally in the form $\ln \Qa_1(\ln q_1)=-\ln \Qa_2(\ln q_2)$ for any quantities $\{\Qa_1, \Qa_2, q_1, q_2\}$).  Right panel: $s(a_g)$ as function of the bundle tangent spin--Eq.\il(\ref{Eq:cora-W}).
	Note the relation  between $(s,s^{-1})$ and the role of the saddle point $a_{\partial}=M/\sqrt{2}$. Here  $s=1$ for $a_g=M$. Center panel line. Left: ratios $s_{\beta}^{\pm}$ of Eq.\il(\ref{Eq:s-beta-toni}) as functions of the tangent spin $a_g/M$ and maximum of $a_{\beta}=(a_{\beta}^i,a_{\beta}^{ii})$  with respect to $s$. The role of $s=1$ for $a=M$ and the symmetries of
	$(s_{\beta}^{+},1/s_{\beta}^{-})$ are explicitly shown.
	Center panel: di-logarithmic plot of $s_{\beta}^{\pm} $ and the functions $(a_{\beta}^i,a_{\beta}^{ii})$ on
	$s_{\beta}^{\pm}$. Right panel: maximum  spin $a_{\mathcal{s}}(s)$ of Eq.\il(\ref{Eq:bre-Siti-dea-lno})  as function of $s$. The spins maximum $a_{\beta}(a_{\mathcal{s}})$, Eq.\il(\ref{Eq:bre-Siti-dea-lno}),  and   $a_{11}=a_{\beta}^{i}(a_{\beta}^{i})$, $a_{22}=a_{\beta}^{ii}(a_{\beta}^{ii})$,  $a_{12}=a_{\beta}^{i}(a_{\beta}^{ii})$, $a_{21}=a_{\beta}^{ii}(a_{\beta}^{i})$ are also
	shown, where
$a_{12}\cup a_{21}=a_{\mathcal{s}}$.
Bottom panel lines: 3D plots of $a_{\beta}^{\pm}$ as functions of $a_g\in[-M,M]$ and $s$  (left panel)
for $s\in \mathds{N}$ (right panel).}\label{Fig:AlbertsDOME}
\end{figure}

\begin{figure}
  \includegraphics[width=4.35cm]{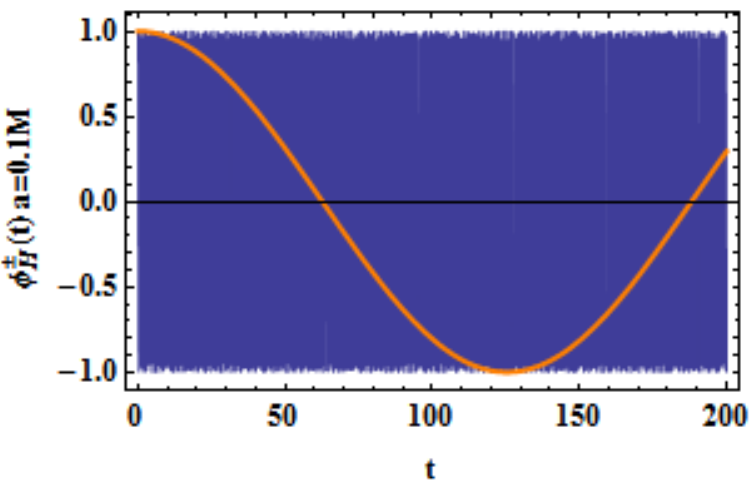}
\includegraphics[width=4.35cm]{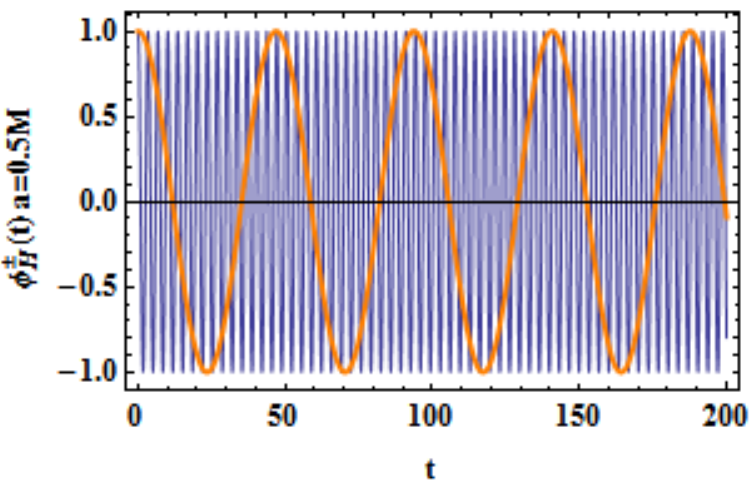}
\includegraphics[width=4.35cm]{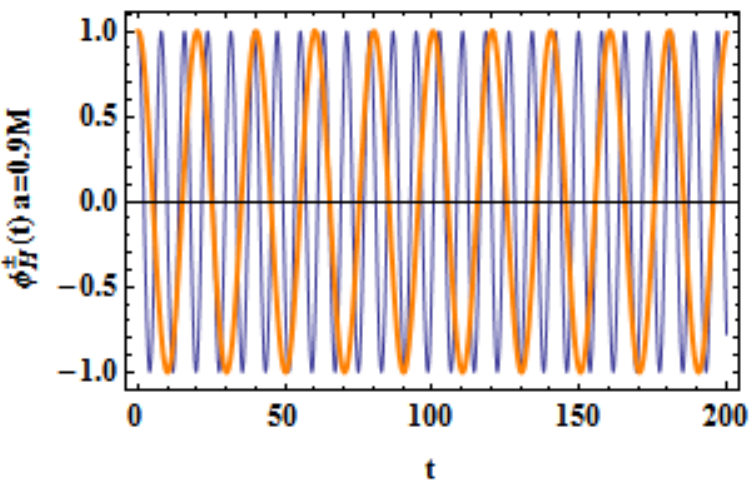}
\includegraphics[width=4.35cm]{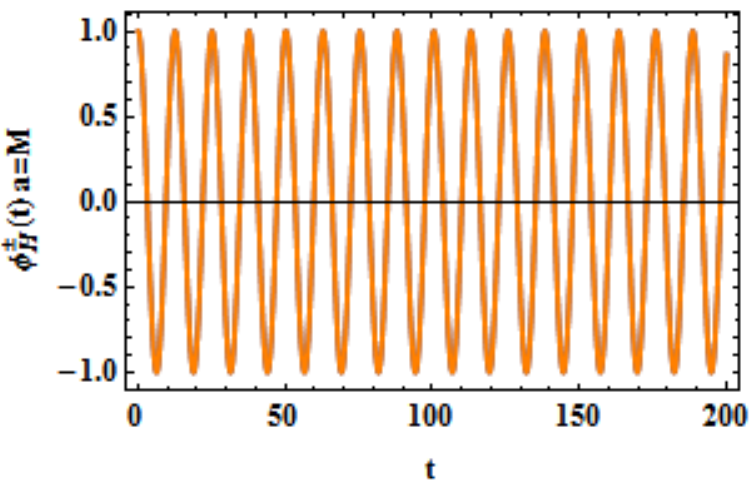}\\
 \includegraphics[width=4.35cm]{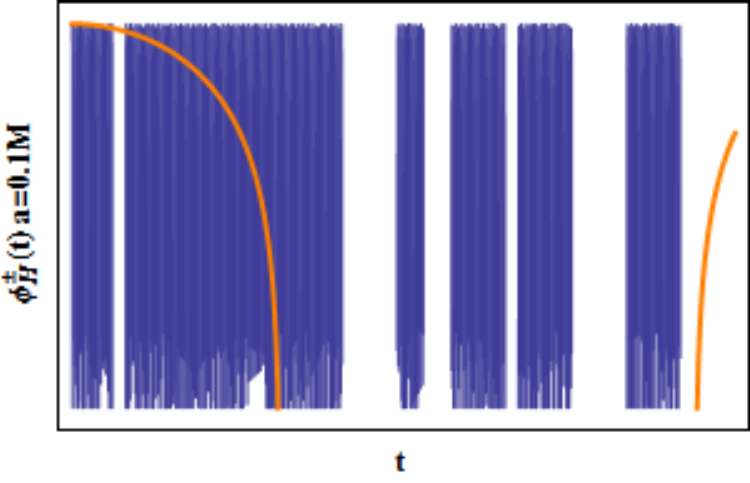}
\includegraphics[width=4.35cm]{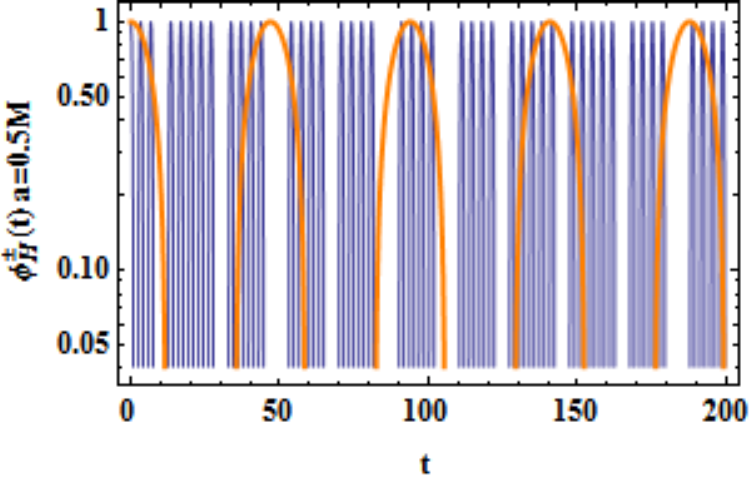}
\includegraphics[width=4.35cm]{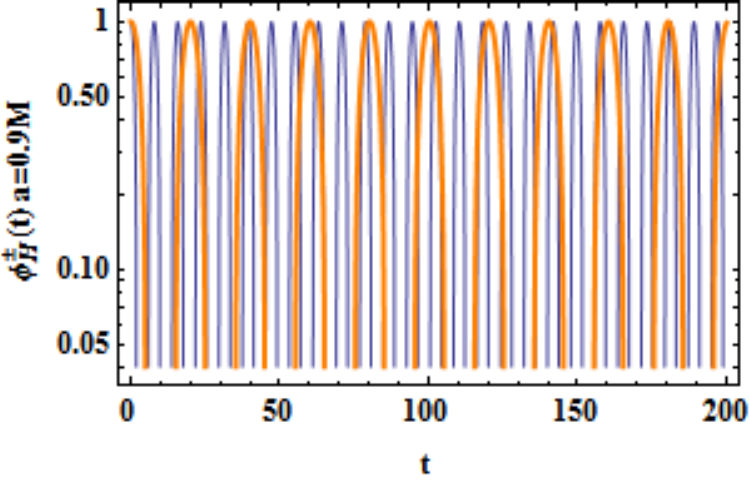}
\includegraphics[width=4.35cm]{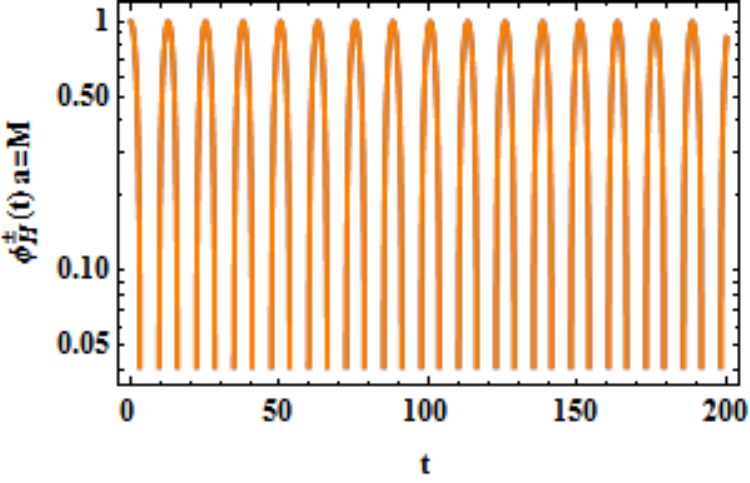}\\
 \includegraphics[width=4.35cm]{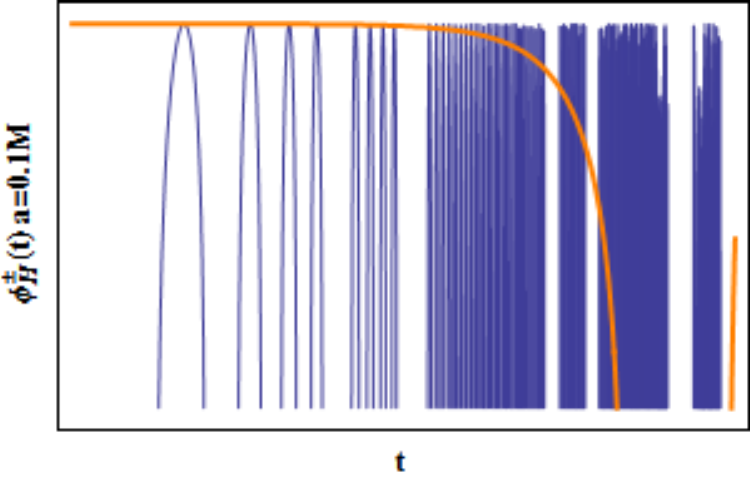}
\includegraphics[width=4.35cm]{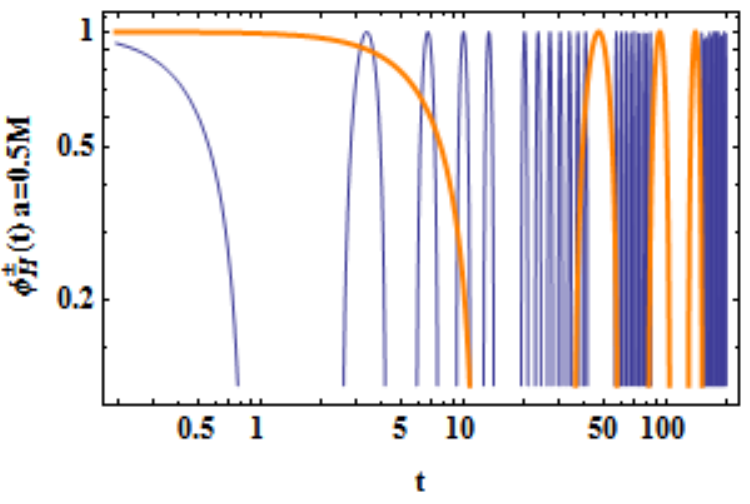}
\includegraphics[width=4.35cm]{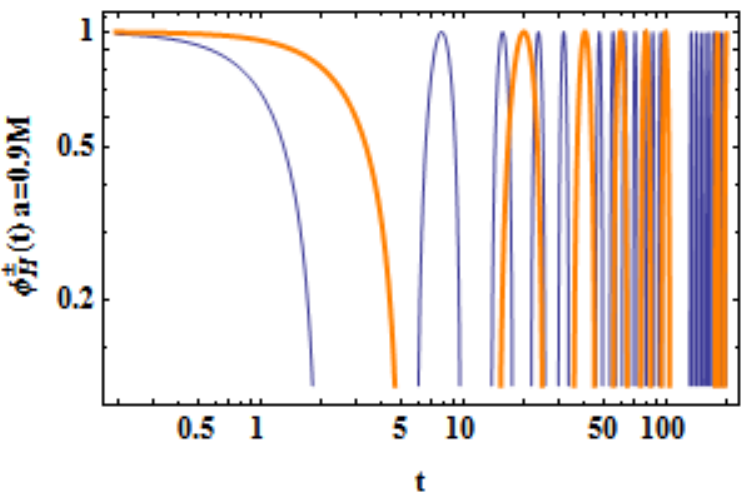}
\includegraphics[width=4.35cm]{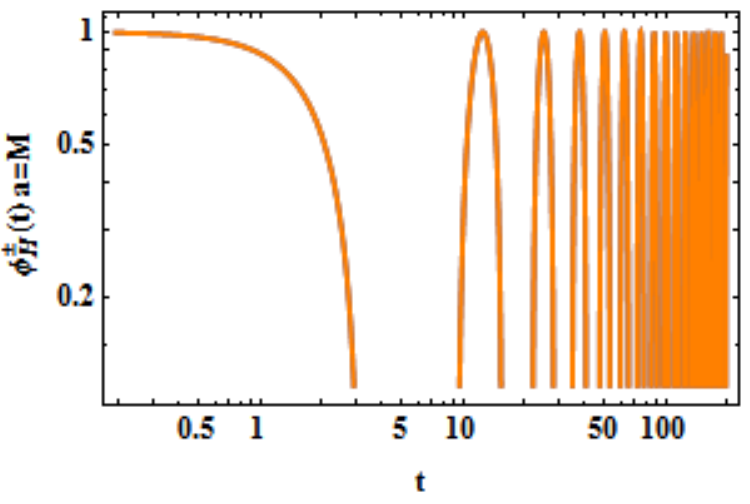}
  \caption{Cosinusoidal signal  $\phi_H^{\pm}(\omega_H^{\pm})$ associated with metric bundles  $\Gamma_{a_g}$ with equal tangent spin. $\omega_H^{+}<\omega_H^-$ are the frequencies of the outer and inner horizon, respectively,  where $\omega_H^{+}=\omega_H^-$  for $a=M$. Plots are for different tangent spins $a_g$  and logarithmic (central panels) and di-logarithms scales
	(below panels). Based on the analysis of Sec.\il(\ref{Sec:allea-5Ste-cont}). See also Figs\il(\ref{Fig:dainseriD}).}\label{Fig:exppLOTINVAS1}
\end{figure}

Then, $a_0^{(1)}=a_g\in]0.8M,M[$ (implying a \textbf{NS} bundle origin  $a_0$ according to  Eq.\il(\ref{Eq:truff-pola}),
 if $a_0\in]1/\sqrt{{\sigma }},4/\sqrt{{\sigma }}[$, and  for  $a_0=2/\sqrt{{\sigma }}$ we obtain that $a_g=M$.
Otherwise, we have that
 $a_0^{(1)}=a_g\in]0,0.8M[$, if  $a_0\in]0,1/\sqrt{\sigma}[$ and $a_0>4/\sqrt{\sigma}$, that is,
a \textbf{BH} and  \textbf{NS} origin--see  Fig.\il(\ref{Fig:stypolitic}).
We see below some essential features of the bundles concerning the  relation $a_g(a_0)$:
\bea\label{Eq:cab-i-net}
&&
\mbox{\textbf{BH origins}}\quad a_0\in]0,M[\quad \mbox{then}\quad\bullet \quad a_g\in ]0,0.8M[,\quad ({r_g}=r_-\in]0,0.5M[)
\\\nonumber
&&
\mbox{\textbf{NS origins}}\quad a_0>M
\\\nonumber
&&
\quad \bullet \quad a_g\in ]0.8M,M[\quad \mbox{for}\quad \mbox{\textbf{(i)}}\quad
a_0\in]M,2M[\quad  \sigma\in]1/{a_0^2},1],
\\\nonumber
&&\quad\quad\mbox{\textbf{(ii)}}\quad  a_0\in[2M,4M[\;  \sigma\in\left]{1}/{a_0^2}, 1\right];\quad \mbox{\textbf{(iii)}}\quad a_0\geq 4M\quad   \sigma\in \left]1/{a_0^2},16/{a_0^2}\right[
\\&&\nonumber
\mbox{{equivalently}}\quad {a_0}\in \left]1/{\sqrt{\sigma }},4/{\sqrt{\sigma }}\right[
\\\nonumber
&&
\\\nonumber
&&
\quad \bullet \quad a_g\in ]0,0.8M[\quad \mbox{for}\quad\mbox{\textbf{(i)}}\quad
a_0\in]M, 4M]\quad \sigma\in\left] 0,{1}/{a_0^2}\right[,\quad\mbox{\textbf{(ii)}}\quad a_0>4M,\quad\sigma\in\left] 0,{1}/{a_0^2}\right[\cup\sigma\in\left[ 16/{a_0^2},1\right],
\\&&\nonumber
\mbox{{equivalently}}\quad a_0\in\left]1,1/\sqrt{\sigma}\right[\cup a_0>4 /\sqrt{\sigma}
\\\nonumber
&&
\mbox{\textbf{Inner horizon tangency:}}\quad a_g\in]0,M]\quad r_g=r_-\in\left]{\sqrt{a_g^2 \sigma }}/{2}, 1\right[\quad a_0={2 r_g}/{a_g \sqrt{\sigma }}
\eea
--Figs\il(\ref{Fig:stypolitic},\ref{Fig:AlbertsDOME},\ref{Fig:exppLOTINVAS1}).

\subsubsection{Relations between the frequencies of particular bundles seen as horizon frequencies}
\label{Sec:partic-freq-ratios}
We now focus on particular pairs of bundles, introduced in Eq.\il(\ref{Eq:enaglinyb}), which are related through their frequencies.
Following the considerations of  Sec.\il(\ref{Sec:allea-5Ste-cont}), we know that each  photon orbital frequency at any  point
of any Kerr (\textbf{NS} or \textbf{BH}) geometry is a horizon frequency $(\omega_H^{\pm}(a_g), \omega_H^{\pm}(a^p_g))$,
where $(a_g,a_g^p)$ are the tangent spins of two  bundles crossing in the point $\bar{r}$. Therefore, the following relations hold
\bea&&
 \mbox{\textbf{Metric bundles with equal $(a_g,\sigma):$}}\quad
\omega_b^1=\frac{1}{4\omega_b}= \frac{a}{2 r_{\pm}} \quad \mbox{respectively for} \quad \omega_b=\omega_H^{\mp}--\mbox{see Eq.\il(\ref{Eq:enagliny})}
 \\
 &&
\mbox{\textbf{corresponding bundles $a_0^1=a_g$:}}\quad \omega_b^1=\frac{1}{4 \sqrt{\sigma } \sqrt{\frac{\omega_b^2}{\left(4 \omega_b^2+1\right)^2}}}=\frac{1}{a \sqrt{\sigma }}=\frac{1}{\la}\quad for \quad \omega_b=\omega_H^{\mp},\quad\mbox{see Eq.\il(\ref{Eq:enaglinya})}
\\\nonumber
 &&
\textbf{Bundles  with equal} \quad \mathbf{\omega_b \quad(r_g,a_g)}\quad a_0^p=a_0 \sqrt{\frac{\sigma }{\sigma_p}}\quad(\la_0^p=\la_0),\quad \omega_H^\pm=\frac{1\mp\sqrt{1-\frac{a_0^2 \sigma }{\sigma_p}}}{2 a_0 \sqrt{\frac{\sigma }{\sigma_p}}}\quad \mbox{see Eq.\il(\ref{Eq:trav7see})}
 \\
&&\textbf{Tangent spin} \quad  a_g=\frac{4 \la_0 }{\la_0^2  +4}\quad\mbox{see Eqs\il(\ref{Eq:rom-a-witho-felix},\ref{Eq:inoutrefere},\ref{Eq:mple-pan},\ref{Eq:bimb-per})}
\quad \quad \omega_H^\mp(a_g)=\frac{\la_0^2\pm\sqrt{\left(\la_0^2  -4\right)^2}+4}{8 \la_0}
 \\
&&\textbf{Bundle  origin} \quad a_0=\frac{1}{\sqrt{\sigma } \omega_b}.\quad \mbox{For} \quad \omega_b=\omega_H^-
\quad
a_0=\frac{2 a}{r_+\sqrt{\sigma }}. \quad \mbox{For}  \quad \omega_b=\omega_H^+\quad a_0=\frac{2 r_+}{a \sqrt{\sigma }}.
\eea

\section{Solutions of the condition $\mathcal{L_{\mathcal{N}}}\equiv\mathcal{L}\cdot\mathcal{L}=0$}
\label{Sec:explci-exstremAomega}

\subsection{Explicit expressions of metric bundles}
\label{Sec:explci-exstremAomega}
Metric bundles $a_{\omega}$ for $\sigma\neq1$ can be written as
\bea\nonumber
&&a_{\omega}^1\equiv\frac{\Upsilon_1-3 \sqrt{\Upsilon-\Gamma-\Upsilon_3}}{6},\quad a_{\omega}^2\equiv\frac{ \Upsilon_1+3 \sqrt{\Upsilon-\Gamma-\Upsilon_3}}{6},
\quad a_{\omega}^3\equiv\frac{-\Upsilon_1-3 \sqrt{\Upsilon+\Gamma-\Upsilon_3}}{6},\quad
a_{\omega}^4\equiv\frac{-\Upsilon_1+3 \sqrt{\Upsilon+\Gamma-\Upsilon_3}}{6};
\\\label{Eq:Gro-e-nNoth}
 &&
 \Upsilon\equiv\frac{4}{3} \left[\frac{r [2 \sigma -r (\sigma -2)]}{\sigma -1}+\frac{1}{\sigma  \omega ^2}\right],\quad
\Gamma\equiv\frac{8 \sqrt{3} r}{(\sigma -1) \omega  \sqrt{-2 r^2+\frac{2 (r+2) r}{\sigma -1}+4 r+3 \Upsilon_3+\frac{2}{\sigma  \omega ^2}}};
  \\\nonumber
 &&
\Upsilon_3\equiv\frac{4 r \left(r^3 \sigma  \omega ^2-r+2\right)}{\sqrt[3]{\Upsilon_4+\Upsilon_2}}-\frac{\left[r \sigma  \omega ^2 [2 \sigma -r (\sigma -2)]+\sigma -1\right]^2}{3 (\sigma -1) \sigma  \omega ^2 \sqrt[3]{\Upsilon_4+\Upsilon_2}}+\Upsilon_5
\\&&\nonumber
 \Upsilon_5\equiv-\frac{\sqrt[3]{\Upsilon_4+\Upsilon_2}}{3 (\sigma -1) \sigma  \omega ^2},\quad
\Upsilon_1\equiv\sqrt{\frac{-6 r^2 (\sigma -2)+12 r \sigma +9 \Upsilon_3 (\sigma -1)}{\sigma -1}+\frac{6}{\sigma  \omega ^2}};
 \\\nonumber
 &&
\Upsilon_4\equiv\frac{1}{2} \left(\left[72 r (\sigma -1) \sigma  \omega ^2 \left(r^3 \sigma  \omega ^2-r+2\right) \left[r \sigma  \omega ^2 [2 \sigma -r (\sigma -2)]+\sigma -1\right]-432 r^2 (\sigma -1) \sigma ^3 \omega ^4+\right.\right.
\\\nonumber
&&
\left.\left.2 \left[r \sigma  \omega ^2 [2 \sigma -r (\sigma -2)]+\sigma -1\right]^3\right)^2-4 \left(\left(r \sigma  \omega ^2 (2 \sigma -r (\sigma -2))+\sigma -1\right)^2-12 r (\sigma -1) \sigma  \omega ^2 \left(r^3 \sigma  \omega ^2-r+2\right)\right)^3\right)^{1/2},
 \\\nonumber
 &&
\Upsilon_2\equiv36 r (\sigma -1) \sigma  \omega ^2 \left[r^3 \sigma  \omega ^2-r+2\right] \left(r \sigma  \omega ^2 [2 \sigma -r (\sigma -2)]+\sigma -1\right)+
-216 r^2 (\sigma -1) \sigma ^3 \omega ^4+
\\\nonumber
&&\left[r \sigma  \omega ^2 [2 \sigma -r (\sigma -2)]+\sigma -1\right]^3.
   \eea
\subsection{Explicit expression for light surfaces}\label{Sec:wokers-super-cynd}
The expression (\ref{Eq:bab-lov-what}) for the frequencies of a stationary observer can be considered as an equation for the radii $r_{s}^{(i)}$ ($i\in\{1,...,4\}$) of the
light surfaces, i.e., solutions of the condition $\mathcal{L}\cdot\mathcal{L}=0$.
Solutions are then given as functions of the  frequency $\omega$ and the plane $\sigma$ as
\bea\label{Eq:rs1234}
&&
r_s^1\equiv\frac{\circleddash_{IV}-\sqrt{2 \circleddash-\circleddash_{III}-\circleddash_{II}}}{2},
\quad r_s^2\equiv\frac{\circleddash_{IV}+\sqrt{2 \circleddash-\circleddash_{III}-\circleddash_{II}}}{2},
\\
&&
r_s^3\equiv\frac{-\sqrt{2 \circleddash+\circleddash_{III}-\circleddash_{II}}-\circleddash_{IV}}{2},\quad r_s^4\equiv\frac{\sqrt{2 \circleddash+\circleddash_{III}-\circleddash_{II}}-\circleddash_{IV}}{2},\quad\mbox{where}
\\\nonumber
&&\circleddash_{IV}\equiv\sqrt{\circleddash+\circleddash_{II}};\quad \circleddash_{III}\equiv
\frac{4 (a \sigma  \omega -1)^2}{\sigma  \circleddash_{IV} \omega ^2};\quad \circleddash\equiv+\frac{2}{3} \left(a^2 (\sigma -2)+\frac{1}{\sigma  \omega ^2}\right),
\\\nonumber
 &&
 \circleddash_{II}\equiv\frac{a^2 \sigma  \omega ^2 \left[\sigma  \left(a^2 [(\sigma -16) \sigma +16] \omega ^2+14\right)-16\right]+1}{9 \circleddash_{V} \sigma ^2 \omega ^4}+\circleddash_{V};
 \\\nonumber
&&\circleddash_{V}\equiv\frac{\sqrt[3]{-\left[a^2 (\sigma -2) \sigma  \omega ^2+1\right]^3-36 a^2 (\sigma -1) \sigma  \omega ^2 \left(a^2 \sigma  \omega ^2-1\right) \left[a^2 (\sigma -2) \sigma  \omega ^2+1\right]+54 \sigma  \omega ^2 (a \sigma  \omega -1)^4+\frac{\circleddash_I}{2}}}{3 \sigma  \omega ^2}
\\\nonumber
&&
 \circleddash_I\equiv\left(4 \left[\left(a^2 (\sigma -2) \sigma  \omega ^2+1\right)^3+36 a^2 (\sigma -1) \sigma  \omega ^2 \left(a^2 \sigma  \omega ^2-1\right) \left[a^2 (\sigma -2) \sigma  \omega ^2+1\right]-54 \sigma  \omega ^2 (a \sigma  \omega -1)^4\right]^2+\right.\\
 &&\left.-4 \left(a^2 \sigma  \omega ^2 \left[\sigma  \left(a^2 [(\sigma -16) \sigma +16] \omega ^2+14\right)-16\right]+1\right)^3\right)^{1/2}.
\eea

\subsection{Planes  $\sigma$  in terms of  $(a,\omega,r)$}\label{Sec:boad-eny-isol}
Finally, in this section we report some  planes and limiting  values for the spins and frequencies
that are relevant for determining the existence of  \textbf{MBs}:
\bea&&\label{Eq:sunt-more}
\omega_{rad}=\sqrt{\frac{1}{r(r+2)}},\quad \omega_{\sqrt{2}} \equiv\frac{1}{2 \sqrt{2}},\quad a_{\sqrt{2}}\equiv\sqrt{2}\quad
a_{\lim}^{\sigma}\equiv\frac{1}{\sigma -1},\quad a_{\lambda}^{\sigma}\equiv4 \sqrt{-\frac{\sigma -1}{(\sigma -2)^4}}
\\\nonumber
&&
\sigma_{\omega}^{\pm}\equiv-\frac{4 a r \omega -a^2-r^4 \omega ^2-a^4 \omega ^2-2 a^2 r^2 \omega ^2\pm\left[a-\omega  \left(a^2+r^2\right)\right]\sqrt{\omega ^2 \left(a^2+r^2\right)^2+2 a \omega  \left[a^2+(r-4) r\right]+a^2}}{2 a^2 \omega ^2\Delta},
\\\label{Eq:sigma-omega-sol-pm}
&&
\mbox{where}\quad
\omega_{\pm}(a_{\epsilon})=\frac{\sqrt{1-\frac{(r-2) r}{\sigma -1}}\mp1}{2 \sqrt{\frac{(r-2) r}{\sigma -1}}}. \eea
In particular, $\sigma_{\omega}^{\pm}$ are solutions of  $\mathcal{L}\cdot\mathcal{L}=0$--
Fig.\il\ref{FIG:toa8}. Then, we obtain
\bea\label{Eq:wonder-wil}&&
\omega_{\pm}(a=0)=\mp \frac{(r-2) r}{\sqrt{(r-2) r^5 \sigma }},\quad\mbox{where}\quad\lim_{a\rightarrow\infty}
\omega_{\pm}=0,\quad\omega_{\pm}(r=0)=\mp\omega_0^{\pm}=\mp\frac{1}{a \sqrt{\sigma}},
\\\label{Eq:hid}
&&
\omega_{\pm}(r=2M,\sigma=1)=(0,\frac{a}{a^2+2}),\quad
\omega_{\pm}(r=M,\sigma=1)=\omega_{11}^{\pm}\equiv\frac{1}{2 a\pm\sqrt{a^2-1}},\quad
\frac{r_{\omega}}{M}\equiv\sqrt{\frac{\omega ^2+1}{\omega ^2}}-1
\eea
As a function of $\sigma$,  $\omega_{\pm}(a=0)$ has an extreme at $r=2M$, where it is $\omega_{\pm}(a=0)=0$,
and as a function of $r$ at $r=3M$,  where $\omega_{\pm}(a=0)={1}/{3 \sqrt{3} \sqrt{\sigma }}$. Moreover, the
frequency $\omega_{11}^{\pm}$ has an extreme for $a=\mp{2}/{\sqrt{3}}$, respectively, where
$\omega_{11}^{\pm}=\mp{2}/{\sqrt{3}}$. The frequency $\omega_{\pm}(r=2M,\sigma=1)$ has an extreme at $a=\pm\sqrt{2}$,
 where $\omega_{\pm}(r=2M,\sigma=1)={1}/{2 \sqrt{2}}$.
See Figs\il(\ref{FIG:toa11})--(\ref{FIG:toa8}).

\section{On  the metric bundles of the extended plane}
\label{Sec:gener}
\subsection{On  negative frequencies}\label{Sec:contro-orbits-omega}
\begin{figure}
  \includegraphics[width=5cm]{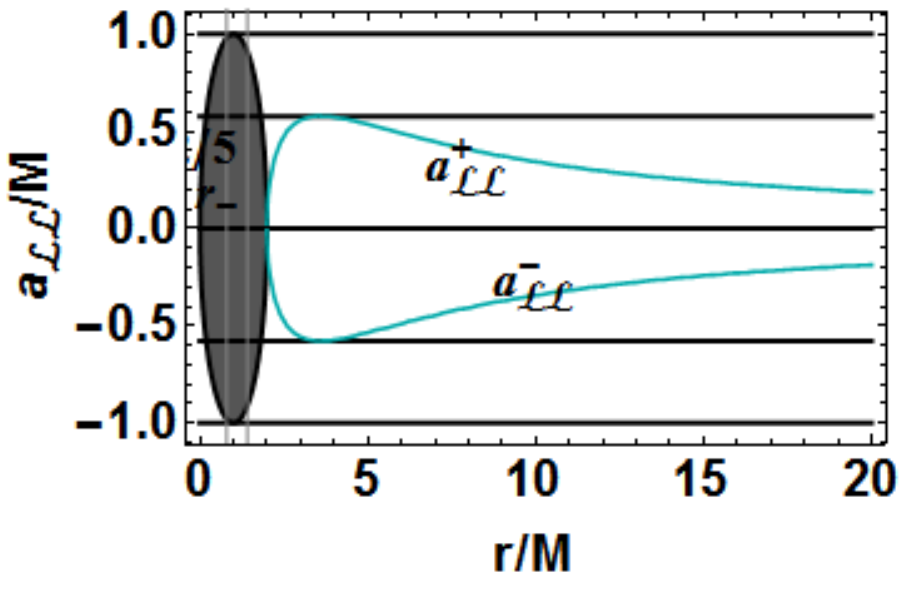}
\includegraphics[width=5cm]{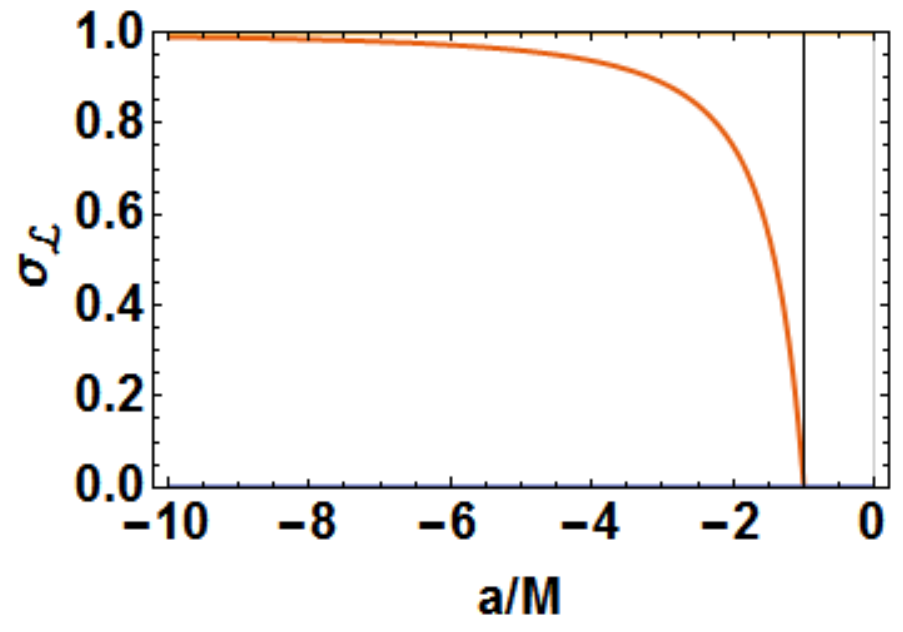}
\includegraphics[width=2.5cm]{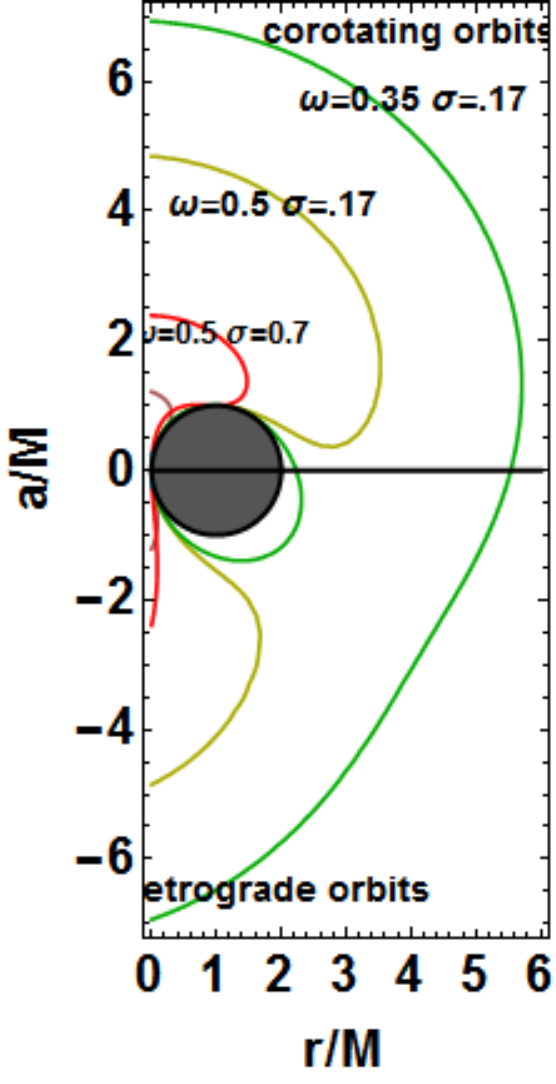}
\includegraphics[width=5cm]{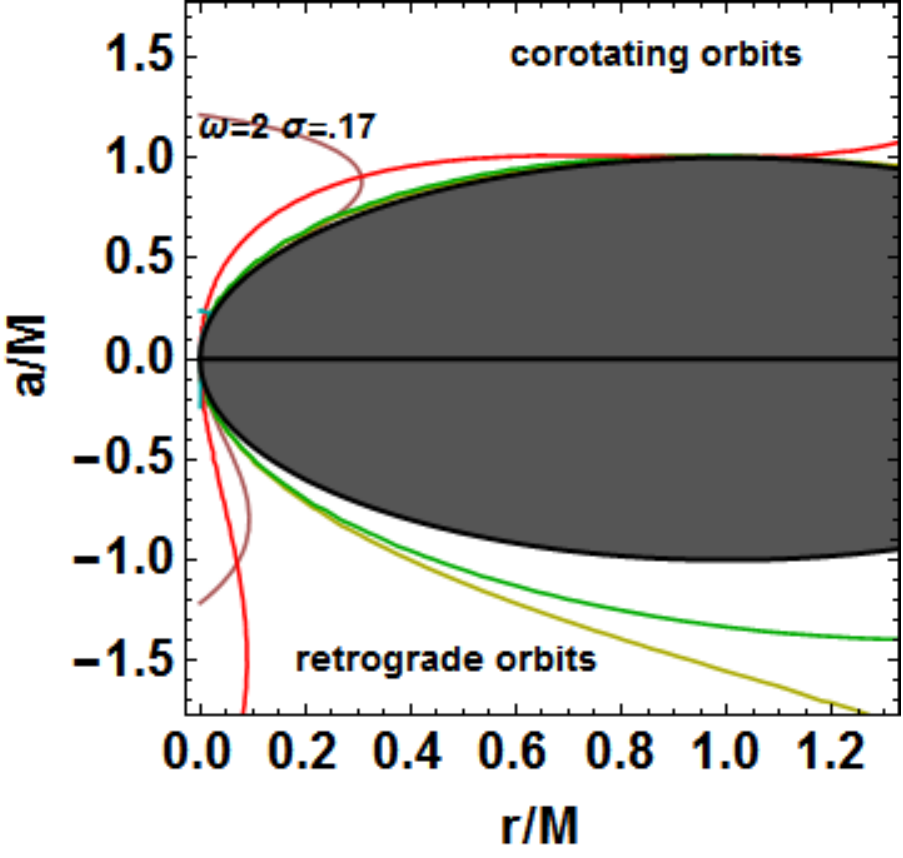}
  \caption{Study of negative frequencies based on the analysis of Sec.\il(\ref{Sec:contro-orbits-omega}). Left panel: plots of spin $a_{\mathcal{L}\mathcal{L}}$ of Eq.\il(\ref{Eq:all-mathc-windo}) as functions of $r/M$. The  black region is the black hole on the extended  plane (i.e. $a\in[-a_{\pm},a_{\pm}]$). Second panel: limiting spin $\sigma_{\mathcal{L}}$ of Eq.\il(\ref{Eq:ver-vit-toldto-barbe}) as function of $a<0$. Third panel: study of the metric bundles associated with retrograde photon motion represented by the solutions in the sector $a<0$ of the extended plane, assuming   $\omega>0$. It is clear the absence of tangency points with the horizon on the extended  plane and the coincidence with the bundles with equal magnitude  origin  $a_0^+=-a_0>0$ with equal frequency on the extended  plane sector $a_0>0$. }\label{Fig:mess-sicur}
\end{figure}

Solutions with frequencies  $\omega>0$ of  the condition  $\mathcal{L}\cdot\mathcal{L}=0$ with $a<0$
(counter-rotating, negative frequencies) exist for:
{\small{
\bea\nonumber
&&
\mathcal{L}_{\mathcal{N}}=0,\quad\mbox{with}\quad \omega=-\omega_H^+>0, \quad
  a<0, \quad \sigma\in]0,1], \quad r\geq0\quad\mbox{is $\sigma_{\mathcal{L}}$ for}
\\\nonumber
&&
  \bullet\quad r\in]0,2M[\quad
  \quad a\in[-1,-a_{\pm}[
\quad
  \bullet\quad  r>2M
  \quad a\in ]-1,a_\mathcal{LL}].
\\\nonumber
&&
\\\nonumber
&&
\mathcal{L}_{\mathcal{N}}=0,\quad\mbox{with}\quad \omega=-\omega_H^->0, \quad
  a<0, \quad \sigma\in]0,1], \quad r\geq0\quad\mbox{is $\sigma_{\mathcal{L}}$ for}:
\\\nonumber
&&
\bullet \quad r\in]0,2M[\quad a\in]-M,-a_{\pm}[\quad\bullet\quad r\geq2M\quad a\in]-M,0[
\\\label{Eq:ver-vit-toldto-barbe}
&&\mbox{where}\quad\sigma_{\mathcal{L}}\equiv\frac{a^2-1}{a^2},\quad\mbox{or equivalently}\quad a^{\pm}_{\mathcal{L}}\equiv \pm \frac{1}{\sqrt{1-\sigma}},\quad\mbox{where there is }
 \\&&\nonumber
\\\label{Eq:all-mathc-windo}
&&a_\mathcal{LL}\equiv\frac{1}{3} \left(6 \psi_\maltese \cos \left(\frac{1}{3} \cos ^{-1}\left[\left(r^{12}+6 r^{11}+24 r^{10}+80 r^9-216 r^8-2256 r^7-10304 r^6-27264 r^5-68736 r^4+\right.\right.\right.\right.\\\nonumber
&&\left.-127744 r^3-147456 r^2+24576 r+32768\right)\left.\left.\left.{(r+2)^{-6}\psi_\maltese^{-3}}\right]\right)-\frac{6 \left[r (r+2) \left(r^2+4\right)+32\right]}{(r+2)^2}\right)^{1/2}
\\\nonumber
&&
\mbox{where}\quad
\psi_\maltese\equiv\sqrt{\frac{r \left(r (r (r+4)+8) \left(r \left(r^3+4 r+16\right)+256\right)+512\right)+1024}{(r+2)^4}}.
\eea}}
In particular, there are no solutions with  inner  horizon frequency  for spins corresponding to \textbf{NSs}.
The limiting cases $2 M $ or $M$ can be studied separately ($\sigma_{\mathcal{L}}$ is linked to the definition of ergosurface).
Some relevant consequences of this analysis, evident also from  Fig.\il(\ref{Fig:mess-sicur}), are listed here:

\textbf{(1)}  Negative frequencies solutions are  intrinsically related to the positive frequency solutions, as they are the extension in the  extended plane of  the \textbf{MBs} with positive characteristic frequencies.

\textbf{(2)} Clearly, \textbf{MBs} with negative frequencies  are not tangent to the horizon curve in the negative frequencies region of the extended plane--Fig.\il(\ref{Fig:Ly-b-rty}).

\textbf{(3)} These correspond to  only one solution (with folding) for   $\omega_-$ frequency.

\textbf{(4)} More precisely, we have confined all the negative frequencies solutions  to the negative region of the  extended plane (assuming $a<0$ and $\omega>0$).

\textbf{(5)} The existence of negative frequencies solutions are constrained by several specific conditions depending also on  the ergosurfaces role.

\textbf{(6)} There are  \textbf{MBs} with frequency equal in magnitude to  the  horizon frequencies
(horizontal lines   $a=-\bar{a}<0$, such that there are  bundle portions    with frequency  $\omega_H^{\pm}(\bar{a})>0$
at the same plane as shown in the figures). Therefore, there are orbits different from the horizons with frequencies equal in magnitude  to the horizon's frequency --in the same geometry as confirmed by the analysis of Sec.\il(\ref{Sec:pri-photon-fre}).
Then,
\bea
&&\mathcal{L}\cdot\mathcal{L}=0\quad r>0\quad\sigma\in]0,1],\quad a<0\quad\omega>0
\\\nonumber
&&a<-M\quad  \omega =\omega_-\quad\bullet\quad r\in]0, 2M],\quad \sigma\in]0,\sigma_{\epsilon}^{\pm}[ \quad\bullet\quad
 r>2M\quad \sigma\in]0,M]\\\nonumber
&&a\in[-M,0[\quad\bullet\quad r\in][0,r_-[\quad \sigma\in]0,\sigma_{\epsilon}^{\pm}[ \quad\bullet\quad r\in ]r_+,2M]\quad \sigma\in]0,\sigma_{\epsilon}^{\pm}[\quad\bullet\quad r>2M\quad\sigma\in]0,1]
\\\nonumber
&&
\mbox{where}\quad
\sigma_{\epsilon}^{\pm}\equiv \frac{(r-2) r}{a^2}+1,\quad\mbox{equivalently}\quad
\\\nonumber
&&\mbox{for}\quad a<-M\quad \mbox{\textbf{Condition R} holds},
\mbox{for}\quad a\in[-M,0[\quad \mbox{\textbf{Condition S} holds}
\\\nonumber
&&
\\\nonumber
&&
\mbox{\textbf{Condition R}}:\quad \bullet\quad\sigma \in]0,\sigma_{\mathcal{L}}]\quad r>0\quad\bullet\quad \sigma\in]s,1]\quad r\in]0,r_{\epsilon}^-[\quad\cup\quad r>r_{\epsilon}^+
\\\nonumber
&&
\mbox{\textbf{Condition S}}:\quad \bullet\quad\sigma \in]0,1]\quad r\in]0,r_{\epsilon}^-[\quad\cup\quad r>r_{\epsilon}^+.
\eea
Conditions related to different boundary values and, particularly, for  $\sigma=\sigma_{\mathcal{L}}$ should be evaluated separately. In this analysis, we will no further investigate values (other notable spins are $a=-\sqrt{6 \sqrt{3}-9}M$ and
$a={\sqrt{2}}M/{3}$). The function $\sigma_{\epsilon}^{\pm}$  is linked to the existence of ergo-surfaces,
specifically,   it  is the  solution of $r_{\epsilon}^{\pm}=r$.
The function $\sigma_{\mathcal{L}}$ is a solution of the following equation:
\bea&&
16 \left[a^3+a (r-2) r\right]^2+a^6 \sigma ^4 \Delta^2-2 a^4 \sigma ^3 \Delta \left[a^4+2 a^2 \left(r^2-2\right)+r^4+8 r+8\right]+
\\\nonumber
&&8 \sigma  \Delta \left[a^6+2 a^4 \left(r^2-3\right)+a^2 r \left(r^3-4 r-8\right)-2 r^4\right]+a^2 \sigma ^2 \left[a^8+4 a^6 \left(r^2-4\right)+a^4 \left[6 r^4-32 r^2+48 r+48\right]+\right.
\\\nonumber
&&\left.4 a^2 r^2 \left[r \left(r^3-4 r+16\right)+24\right]+r^2 \left(r \left[r \left(r^4+16 r+32\right)-64\right]+64\right)\right]=0
\eea

\subsection{Consideration of areas}\label{Sec:areas}
\begin{figure}
  \includegraphics[width=5cm]{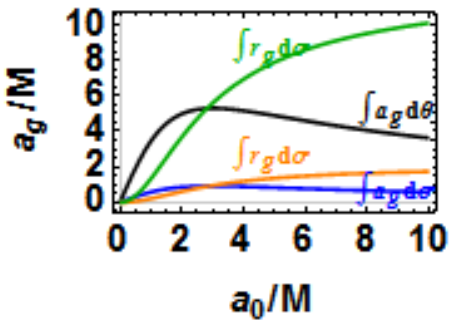}
  \caption{Areas of regions bounded by special curves. Based on the analysis of Eq.\il(\ref{Eq:det-traf-vua-sol-aatt}).}\label{Fig:nnsiplotsa}
\end{figure}

In this section, we focus on areas of extended plane  regions bounded  by remarkable curves;
these areas characterize  quantities invariants with respect to a  spin shift,
orbits translation or plane shift  $\sigma\in]0,1]$
(the integration region is on the extended  plane).
A part of this analysis has been done in Sec.\il(\ref{Sec:bundle-description}), considering  the linearized scheme of
Fig.\il(\ref{Fig:Ly-b-rty}),  where we showed the  linearized conditions on the horizon.
Therefore,
\bea&&\label{Eq:det-traf-vua-sol-aatt}
\textbf{Horizon curve }\quad\int_{0}^{2M}a_{\pm} dr=\pi/2;\quad\mbox{Origins}\quad \int_{0}^{2M}a_0(r) dr=2\pi, \quad\int a_{\epsilon}^{\pm} d\sigma dr=\pi,
\\\nonumber
&&\int a_g(a_0)d \theta=-\frac{16 \tan ^{-1}\left(\frac{a_0}{\sqrt{-a_0^2-4}}\right)}{\sqrt{-a_0^2-4}},\quad \int r_g(\omega) d\omega=\frac{1}{2}\int_{\omega_H^+} r_g(\omega) d\omega=\frac{\pi}{2},\quad\int \omega_H^+ da=\frac{1-\log (2)}{2},
\\
&&\int_{\omega_H^+} a_g(\omega_b) d\omega_b=\frac{\log (2)}{2},\quad
\int a_{g}(r_g) dr_g=\frac{2 \pi }{\sqrt{\sigma }}
\eea
see Fig.\il(\ref{Fig:nnsiplotsa}).

\subsection{Coincidence of radii: horizon replicas}\label{Sec:replicas}
The question we  investigate here is whether the special  orbits  investigated in  Sec.\il(\ref{Sec:pri-photon-fre}) and
Sec.\il(\ref{Sec:allea-5Ste-cont}) on the \textbf{MBs} with characteristic frequencies
$\omega_b(a)\in\{\omega_H^+(a_p),\omega_H^-(a_p)\}$, are located exactly in $r_\pm(a_p)>r_+(a)$, that is, on  the horizon with frequency $\omega_b(a)$. Such orbits are, therefore, called horizon replicas. It is clear that these are given by the vertical lines $r=r_\pm(a_p)$ on the extended  plane. Crossings with \textbf{MBs} of the  $\Gamma_{\omega_b}$  class are
clearly at  $(a=a_g=a_p,\,r=r_g=r_{\pm}(a_p))$. The other radii  are given by the intersections of the bundle branches with
the vertical line. Note that eventually there can be also solutions with negative frequencies, that is,
the crossing with the \textbf{MBs} branches  in the extension to negative frequencies of the extended plane. The case of negative characteristic bundle frequencies is considered in  Sec.\il(\ref{Sec:contro-orbits-omega}).

Consider, in a fixed  spacetime  of a bundle,  an orbit  $r^*_\pm(a)>r_+(a)$ with frequency  $\omega_H^{\pm}(a)$,
respectively. Clearly, there are orbits $r^*_\pm(a)>r_+(a)$ with frequency   $\omega_H^{\pm}(a_p)$ for a given spin
$a_p\neq a$, as all the limiting  frequencies $\omega_b=\omega_{\pm}$ are horizon frequencies  on the extended  plane.
Eventually, we could ask for  the conditions for the occurrence of $r^*_{\flat}(a)=r_{\natural}(a_p)$, where
$\{\flat,\natural\}=\pm$. For  a given spacetime with $a>0$,  there is an orbit, $r^*(a)$, with  limiting frequency
$\omega_H^*(a_p)$, that is, the frequency of  the inner  or outer  horizon,  respectively: $\omega_H^*(a_p)\in\{\omega_H^-(a_p),\omega_H^+(a_p)\}$.  The following cases answer this problem:

\textbf{(1)} If $a>M$, then the radius can be $r^*>0$ thus it could be in  $r^*_{-}(a)=r_-(a_p)\in]0,M]$ or also  $r^*_{+}(a)=r_+(a_p)\in[M,2M[$.

\begin{figure}
  \includegraphics[width=3.4cm]{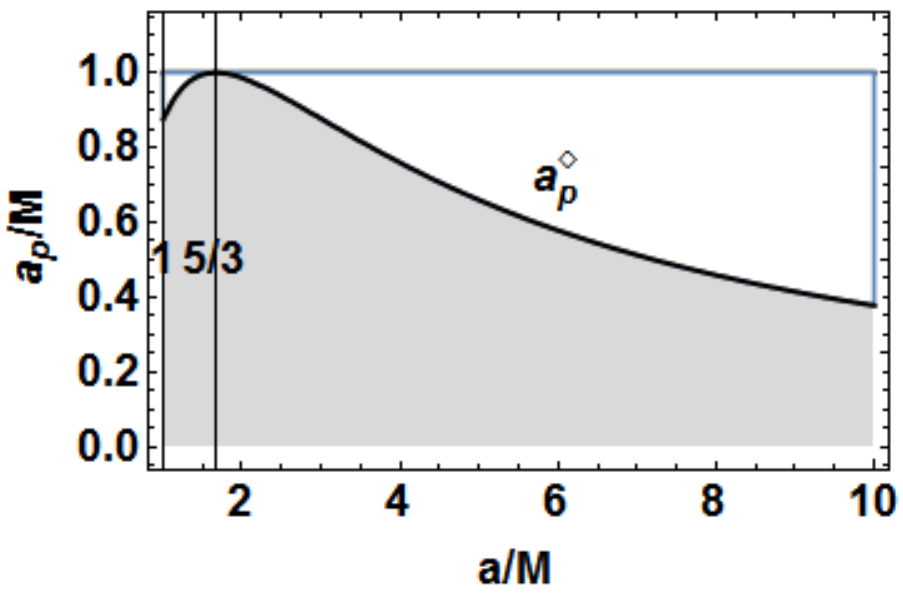}
 \includegraphics[width=3.4cm]{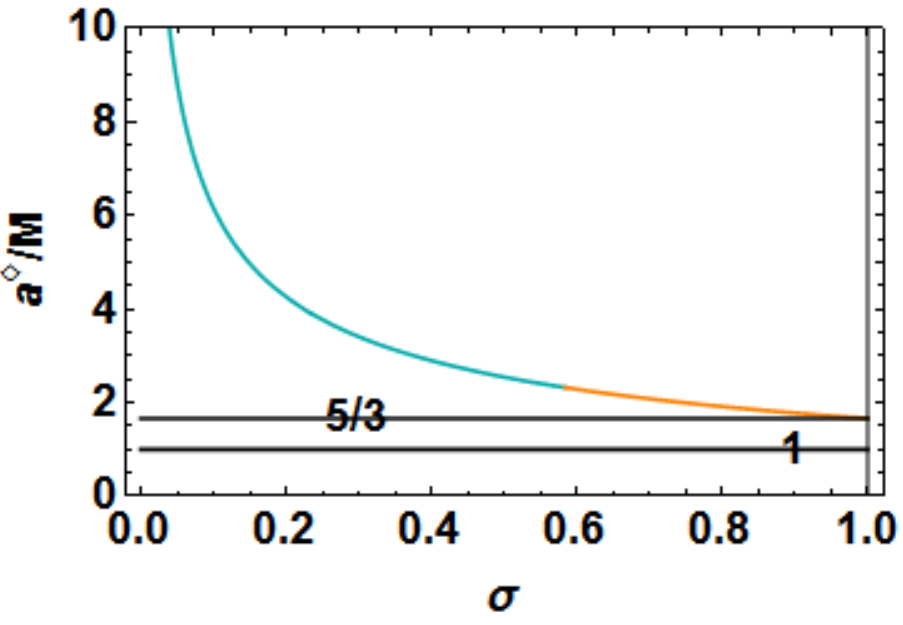}
 \includegraphics[width=3.4cm]{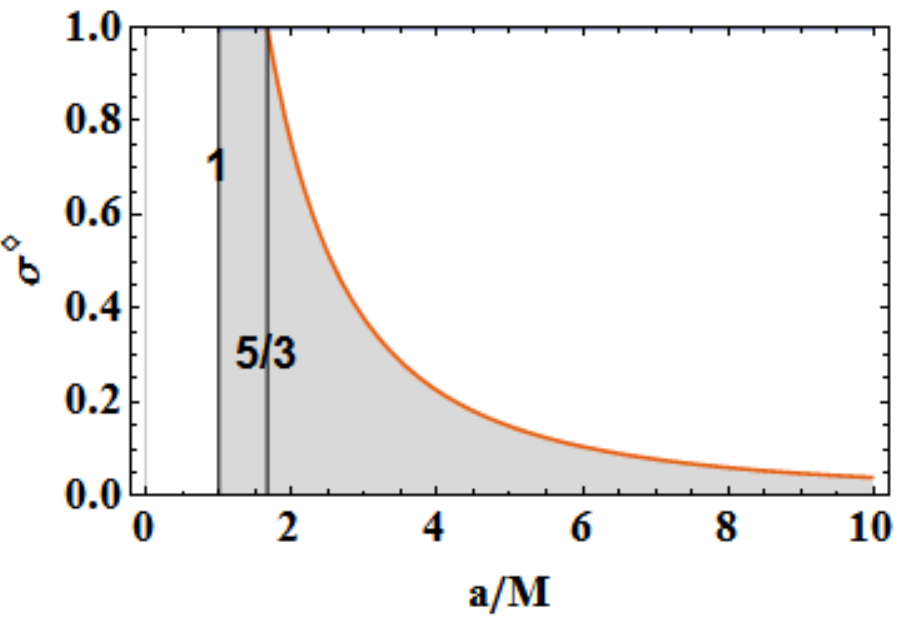}
 \includegraphics[width=3.cm]{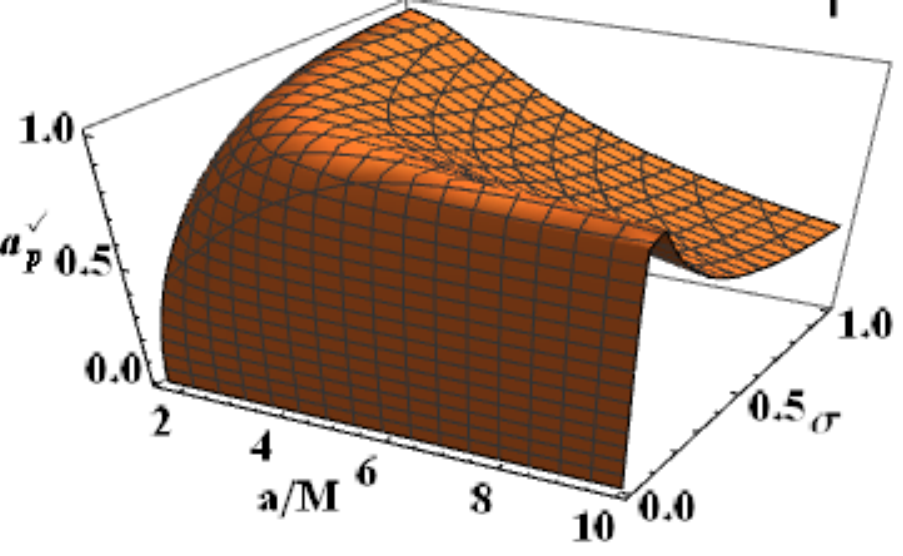}
\includegraphics[width=3.cm]{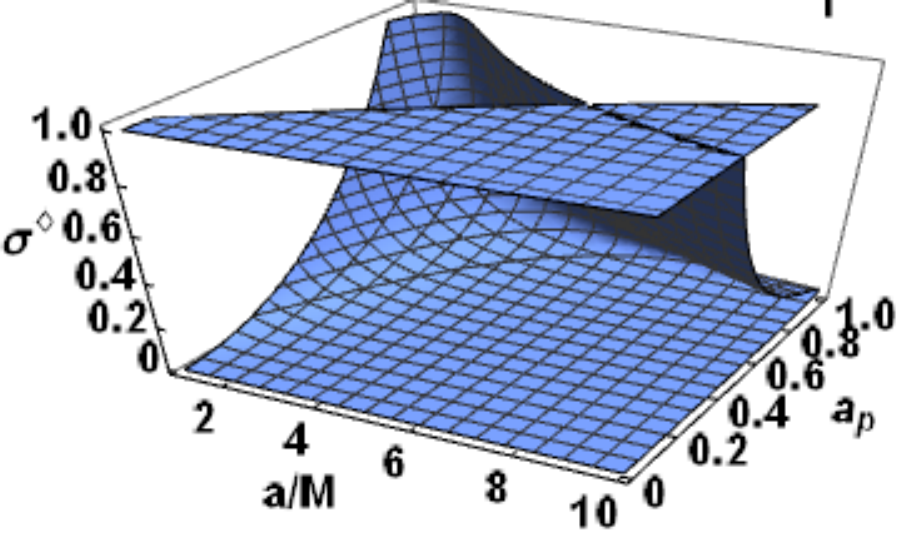}
  \caption{Analysis of Eq.\il(\ref{Eq:repliches}). Orbits $r(a)=r_\pm(a_p)$ with characteristic frequency  equal to that of the internal and external horizon  $\omega_H^{\pm}(a_p)$. They correspond to  solutions of  $\mathcal{L}\cdot\mathcal{L}=0$.
	Limiting spins $(a_p^{\diamond},a^{\diamond})$,  and plane $\sigma_p^{\diamond}$ are plotted. The solutions $(a^{\checkmark},\sigma^{\checkmark})$ are on 3D plots.}\label{Fig:Plotinsette4}
\end{figure}

\textbf{(2)} If, otherwise, $a\in[0,M]$, then the radius  must satisfy the relation $r^*_{+}(a)=r_+(a_p)>r_+(a)\in[M,2M[$ and, therefore, be only  the outer horizon frequency (which is also another aspect of the horizons confinement).
This condition implies  $a_p>a\in[0,M]$, with $\omega(r^*_{+}(a),a)=\omega_H^+(a_p)$.
More precisely, the results are as follows:
\bea&&\nonumber
\mathcal{L}\cdot\mathcal{L}=0\quad\mbox{for}\quad r=r_+(a_p),\quad \omega_b=\omega_H^+(a_p) \quad\mbox{for}\quad a_p=a_p^{\checkmark}
,\quad
a\geq 5M/3,\quad  (a\geq a^{\diamond})\quad a_p\in[a_p^{\diamond},M],\quad\sigma\in[\sigma^{\diamond},1[
\\\label{Eq:repliches}
&&\sigma^{\diamond}\equiv \frac{a \left[7-a (a+1) \left[\sqrt{\frac{(a-1)^3 [a (a (a+5)+11)-1]}{a^4 (a+1)^2}}-1\right]\right]-1}{2 a^2 (a+1)},
\\\nonumber
&& a^{\diamond}\equiv
\left\{\frac{1}{3} \left[2 \sqrt{\frac{12}{\sigma }+\frac{9}{\sigma -1}+1} \cos \left(\frac{1}{3} \sec ^{-1}\left[\frac{\sqrt{\frac{12}{\sigma }+\frac{9}{\sigma -1}+1} [12-\sigma  (\sigma +20)]}{\sigma  (\sigma +44)+36}\right]\right)-1\right],\right.
\\\nonumber
&&\qquad\qquad \left.\frac{1}{3} \left[-2 \sqrt{\frac{12}{\sigma }+\frac{9}{\sigma -1}+1} \sin \left(\frac{1}{3} \csc ^{-1}\left[\frac{\sqrt{\frac{12}{\sigma }+\frac{9}{\sigma -1}+1} [12-\sigma  (\sigma +20)]}{\sigma  (\sigma +44)+36}\right]\right)-1\right]\right\}
\\\nonumber
&&
a_p^{\diamond}: \quad \left(a^2-8\right) a_p^3+\left(8 a^2+32\right) a_p+2 a a_p^4-32 a+a_p^5=0.
\\\nonumber
&&
\mbox{There are \emph{no} solutions for $a\in[0,M]$ and $a_p\in[0,M]$.}
\eea
The spin  $a_p^{\checkmark}$ is a solution of the equation
\bea&&\nonumber
-16 a^4 (\sigma -1)^2 \sigma +32 a^3 a_p (\sigma -1) \sigma +a_p^6 \sigma  \left[a^2 \sigma  (2 \sigma -1)-8\right]+4 a a_p^5 (\sigma -1) \sigma  \left(a^2 \sigma +4\right)+
\\\nonumber
&&a_p^4 \left[\sigma  \left(a^4 \sigma  \left(\sigma ^2-1\right)+8 a^2 \sigma  (3-2 \sigma )+16\right)+16\right]+2 a a_p^3 \left[\sigma  \left(a^4 (\sigma -1)^2 \sigma -8 a^2 \left(\sigma ^2-1\right)-16 (\sigma +1)\right)+16\right]+
\\
&&a^2 a_p^2 (\sigma -1) \left[\sigma  \left(a^4 (\sigma -1) \sigma -8 a^2+16 (\sigma +3)\right)-16\right]+2 a a_p^7 \sigma ^2+a_p^8 \sigma ^2=0 .
\eea
See Fig.\il(\ref{Fig:Plotinsette4}).

\begin{figure}
  \includegraphics[width=3cm]{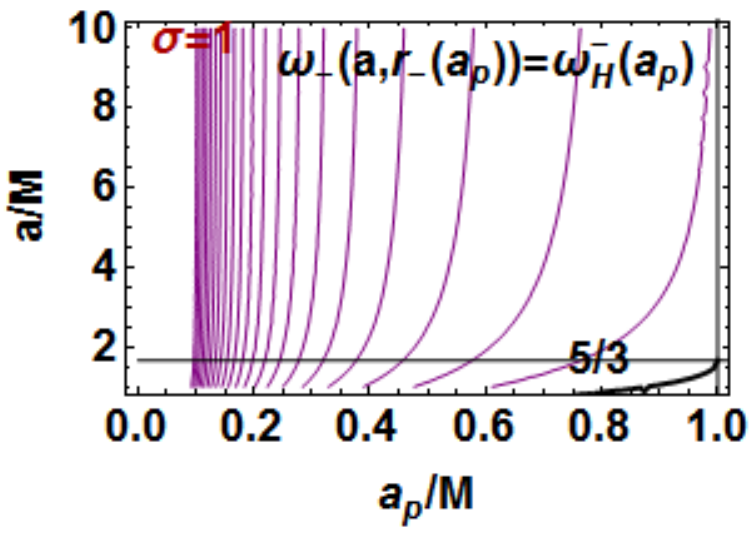}
\includegraphics[width=3cm]{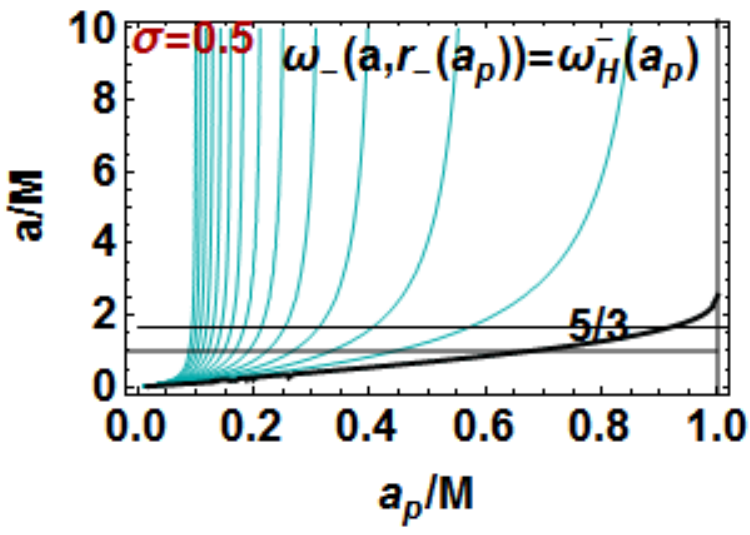}
\includegraphics[width=3cm]{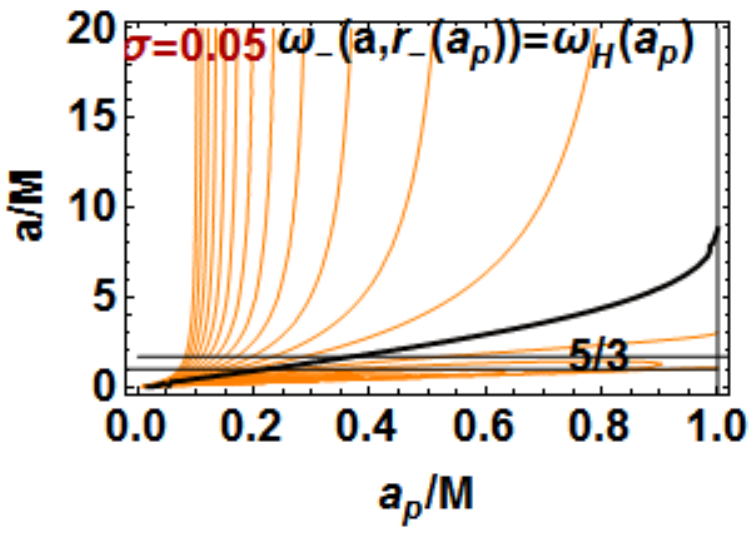}
\includegraphics[width=3cm]{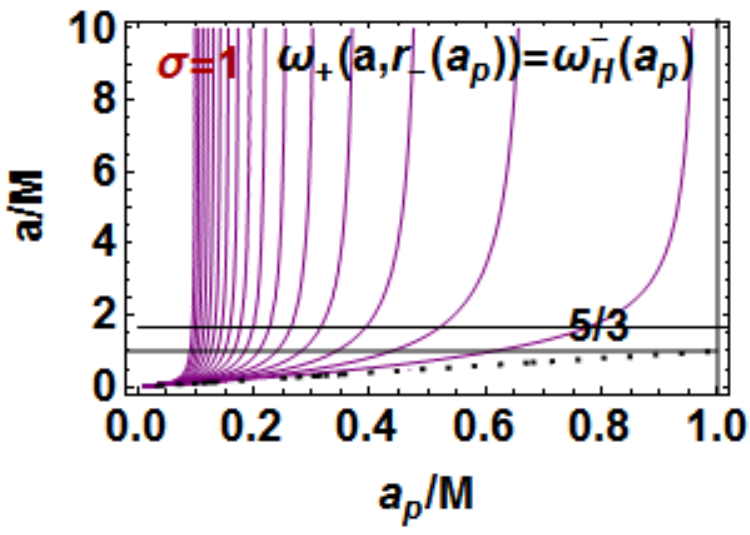}
\includegraphics[width=3cm]{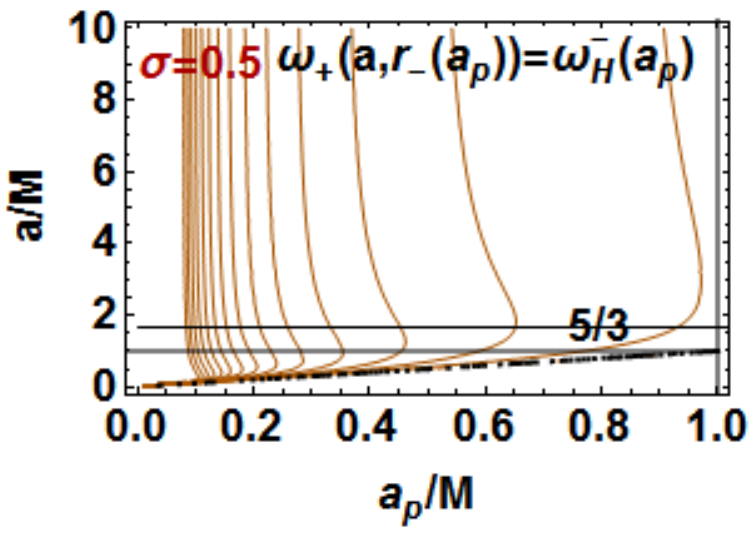}
\includegraphics[width=3cm]{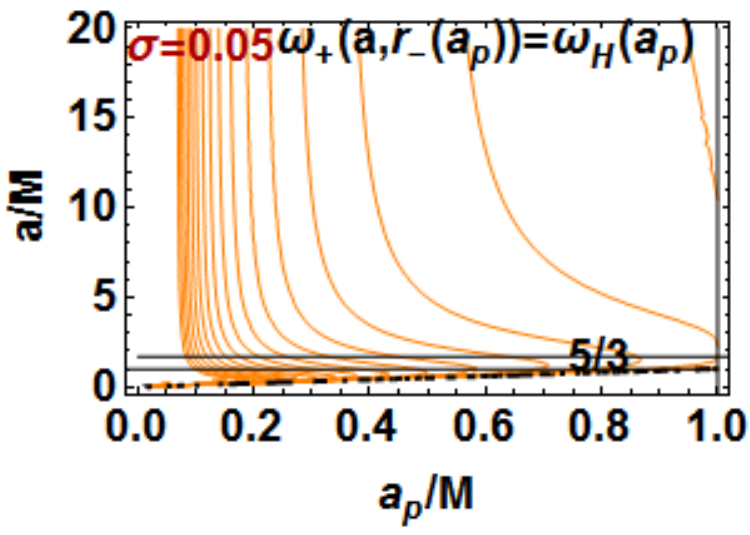}
\includegraphics[width=3cm]{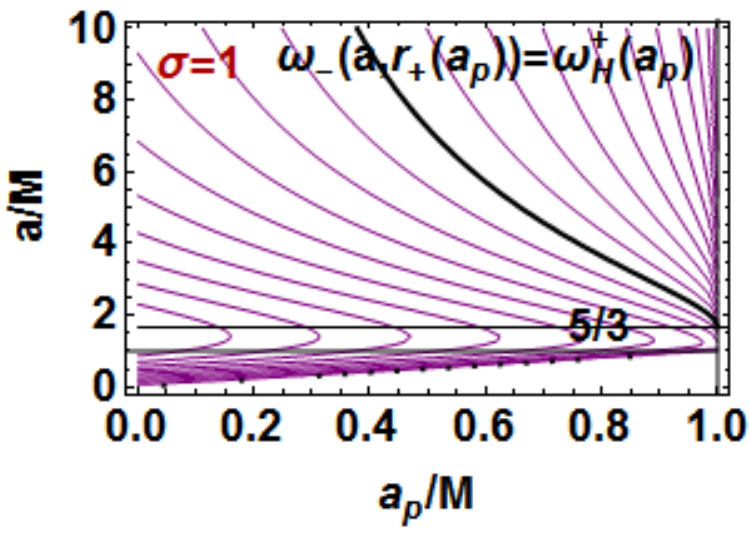}
\includegraphics[width=3cm]{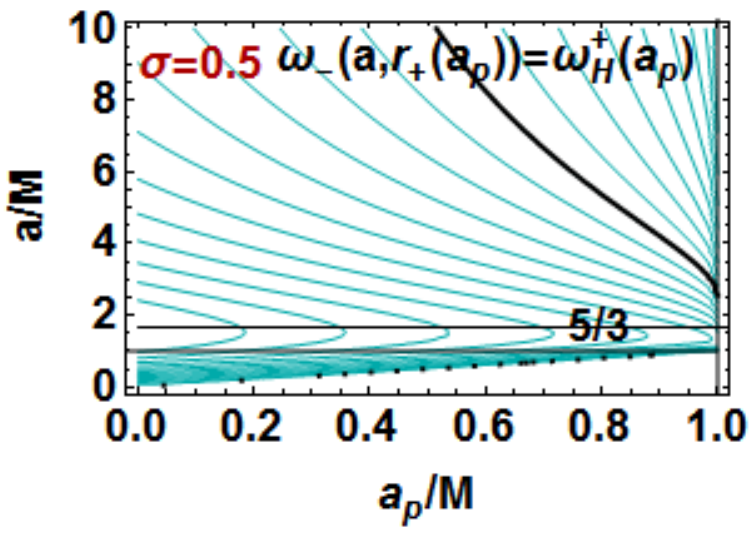}
\includegraphics[width=3cm]{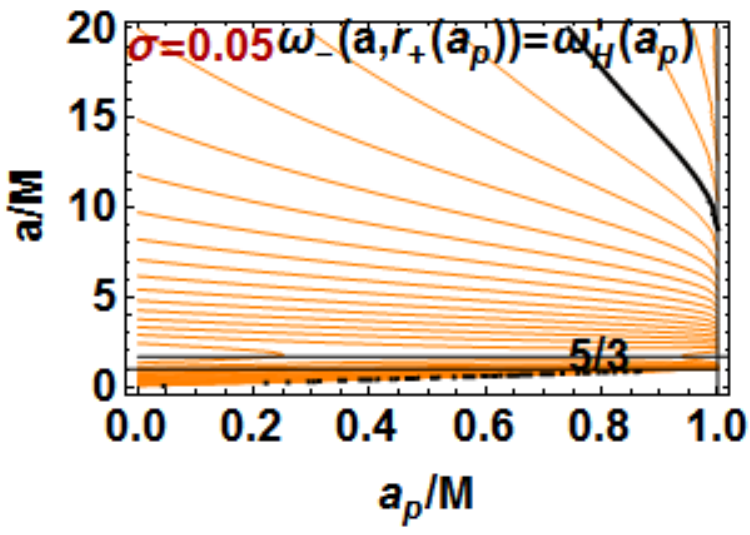}
\includegraphics[width=3cm]{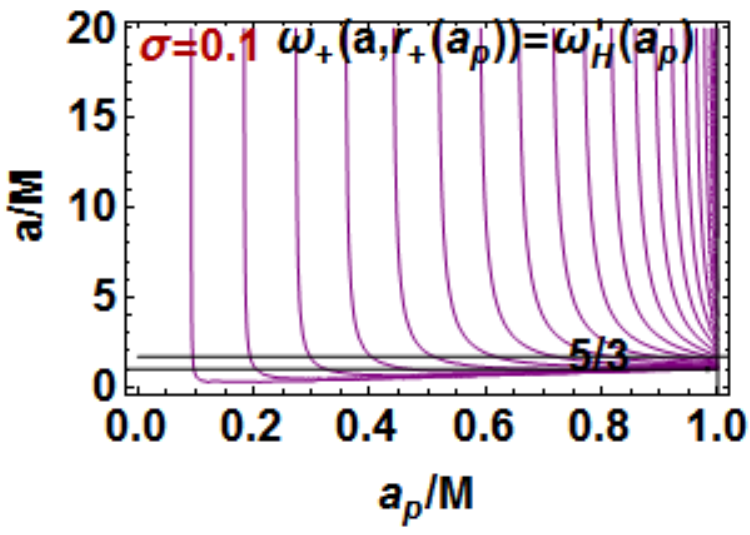}
\includegraphics[width=3cm]{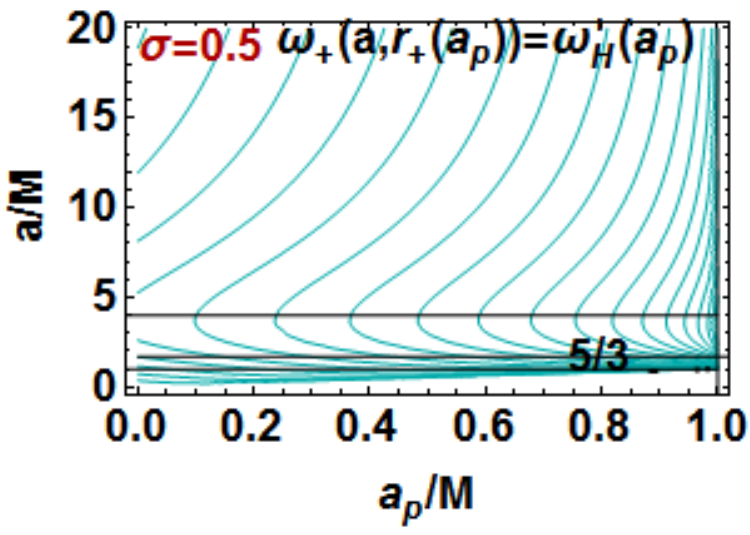}
\includegraphics[width=3cm]{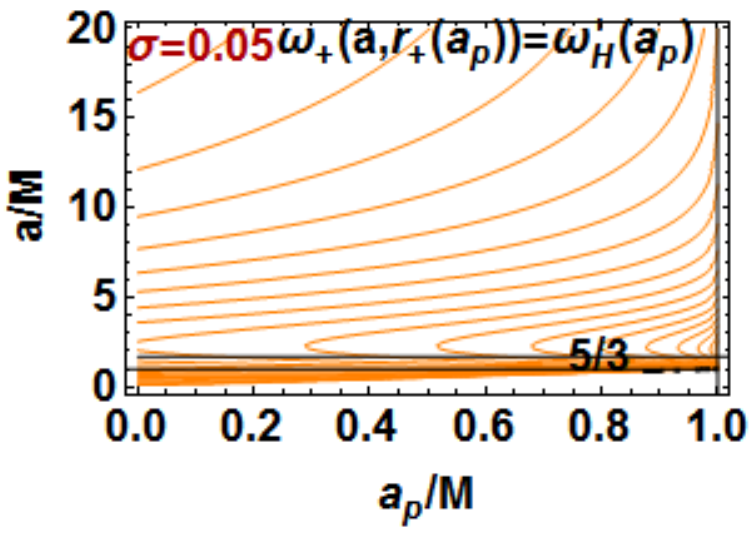}
  \caption{Differences  for the second frequencies  $\omega_\flat(a,r\pm(a_p))-\omega_H^{\pm}(a_p)$ for  $\flat=\pm1$,
	where  $\sigma\in\{0.05,0.5,1\}$ on the plane  $(a_p,a)$. The  limiting spins  $a/M=5/3$ and  $a=M$ are also shown.
	Solutions of $\omega_\flat(a,r\pm(a_p))=\omega_H^{\pm}(a_p)$
	--see Eqs\il(\ref{Eq:100-m-fiord-ancorar}).}\label{Fig:belviasetchiroma}
\end{figure}

We now consider the problem for the inner horizon:
\bea&&\nonumber
\mbox{}\quad \mathcal{L}\cdot\mathcal{L}=0\quad\mbox{for}\quad r=r_-(a_p),\quad \omega=\omega_H^-(a_p) \quad\mbox{for}\quad \sigma=\sigma^{\checkmark}\quad\mbox{and}
\\&&
\bullet\quad a\in]M,5M/3],\quad a_p\in]0,a_p^{\diamond}],\quad
\bullet\quad a>5M/3,\quad a_p\in]0,M].
\eea
The plane $\sigma^{\checkmark}$ is a solution of the equation
\bea&&\nonumber
\sigma^{\checkmark}:\quad a^4 a_p^2 \sigma ^4 (a+a_p)^2-2 a^2 \sigma ^3 (a+a_p) \left[a^3 a_p^2+\left(a^2+8\right) a_p^3-a a_p^4+8 a-a_p^5-8 a_p\right]+8 \sigma  \left[a^4 \left(a_p^2-2\right)+\right.
\\\nonumber
&&\left.2 a^3 a_p \left(a_p^2-2\right)-8 a^2 a_p^2-2 a a_p^3 \left(a_p^2+2\right)-a_p^4 \left(a_p^2-2\right)\right]+\sigma ^2 \left[a^6 a_p^2+2 a^5 a_p^3-a^4 \left(a_p^4+8 a_p^2-32\right)-4 a^3 \left(a_p^4-8\right) a_p+\right.
\\\label{Eq:100-m-fiord-ancorar}
&&\left.a^2 \left(-a_p^4+24 a_p^2+32\right) a_p^2+2 a \left(a_p^4+8 a_p^2-16\right) a_p^3+a_p^8\right]+16 a_p^2 (a+a_p)^2=0,
\eea
--see Fig.\il(\ref{Fig:Plotinsette4}). It is clear that for the causal structure (as determined by the condition
 $\mathcal{L}_{\mathcal{N}}$) the difference in frequencies  is relevant for the orbits considered here as horizon replicas.
This analysis has been also addressed in Eq.\il(\ref{Eq:re-set}), where a fixed geometric  with spin $a$, orbit  $r$ and plane
$\sigma$ was considered.
Here, we are interested in this aspect again considering the differences for
$\omega_\flat(a,r\pm(a_p))-\omega_H^{\pm}(a_p)$ and for $\flat=\pm1$--see Fig.\il(\ref{Fig:belviasetchiroma}).

\end{document}